\documentclass[twocolumn]{aastex701}
\usepackage[version=4]{mhchem} 
\usepackage{hyperref}

\newcommand{\KBD}{\citetalias{hainline2024a}}
\newcommand{\teff}{T$_{\mathrm{eff}}$}

\begin{document}
\defcitealias{hainline2024a}{H24} 

\title{JADES: An Abundance of Ultra-Distant T- and Y-Dwarfs in Deep Extragalactic Data}

\author[orcid=0000-0003-4565-8239,sname='North America']{Kevin N.\ Hainline}
\affiliation{Steward Observatory, University of Arizona, 933 N. Cherry Avenue, Tucson, AZ 85721, USA}
\email[show]{kevinhainline@arizona.edu}  

\author[orcid=0000-0003-4337-6211,sname='North America']{Jakob M.\ Helton}
\affiliation{The Department of Astronomy \& Astrophysics, The Pennsylvania State University, 525 Davey Lab, University Park, PA 16802}
\email[]{}  

\author[orcid=0000-0002-5500-4602]{Brittany E.\ Miles}
\affiliation{Steward Observatory, University of Arizona, 933 N. Cherry Avenue, Tucson, AZ 85721, USA}
\email[]{}  

\author[orcid=0000-0002-0834-6140]{Jarron Leisenring}
\affiliation{Steward Observatory, University of Arizona, 933 N. Cherry Avenue, Tucson, AZ 85721, USA}
\email[]{}  

\author[orcid=0000-0002-5251-2943]{Mark S. Marley}
\affiliation{Department of Planetary Sciences, Lunar \& Planetary Laboratory, University of Arizona, Tucson, AZ 85721, USA}
\email[]{} 

\author[orcid=0000-0003-1622-1302]{Sagnick Mukherjee}
\affiliation{Department of Astronomy and Astrophysics, University of California, Santa Cruz, 1156 High Street, Santa Cruz, CA 95064, USA}
\email[]{}  

\author[orcid=0000-0002-0413-3308]{Nicholas F. Wogan}
\affiliation{NASA Ames Research Center, Moffett Field, CA 94035}
\email[]{}  

\author[orcid=0000-0002-8651-9879]{Andrew J.\ Bunker}
\affiliation{Department of Physics, University of Oxford, Denys Wilkinson Building, Keble Road, Oxford OX1 3RH, UK}
\email[]{}  

\author[orcid=0000-0002-9280-7594]{Benjamin D.\ Johnson}
\affiliation{Center for Astrophysics $|$ Harvard \& Smithsonian, 60 Garden St., Cambridge MA 02138 USA}
\email[]{}  

\author[orcid=0000-0002-4985-3819]{Roberto Maiolino}
\affiliation{Kavli Institute for Cosmology, University of Cambridge, Madingley Road, Cambridge, CB3 0HA, UK.}
\affiliation{Cavendish Laboratory - Astrophysics Group, University of Cambridge, 19 JJ Thomson Avenue, Cambridge, CB3 0HE, UK.}
\affiliation{Department of Physics and Astronomy, University College London, Gower Street, London WC1E 6BT, UK}
\email[]{}  

\author[orcid=0000-0002-7893-6170]{Marcia Rieke}
\affiliation{Steward Observatory, University of Arizona, 933 N. Cherry Avenue, Tucson, AZ 85721, USA}
\email[]{}  

\author[orcid=0000-0002-5104-8245]{Pierluigi Rinaldi}
\affiliation{Space Telescope Science Institute, 3700 San Martin Drive, Baltimore, Maryland 21218, USA}
\email[]{}  

\author[orcid=0000-0002-4271-0364]{Brant Robertson}
\affiliation{Department of Astronomy and Astrophysics, University of California, Santa Cruz, 1156 High Street, Santa Cruz, CA 95064, USA}
\email[]{}  

\author[orcid=0000-0002-4622-6617]{Fengwu Sun}
\affiliation{Center for Astrophysics $|$ Harvard \& Smithsonian, 60 Garden St., Cambridge MA 02138 USA}
\email[]{}  

\author[orcid=0000-0002-8224-4505]{Sandro Tacchella}
\affiliation{Kavli Institute for Cosmology, University of Cambridge, Madingley Road, Cambridge, CB3 0HA, UK.}
\affiliation{Cavendish Laboratory - Astrophysics Group, University of Cambridge, 19 JJ Thomson Avenue, Cambridge, CB3 0HE, UK.}
\email[]{}  

\author[orcid=0000-0003-2919-7495]{Christina C. Williams}
\affiliation{NSF-DOE Vera C. Rubin Observatory/NSF NOIRLab, 950 N. Cherry Ave., Tucson, AZ 85719, USA}
\email[]{}  

\author[orcid=0000-0001-9262-9997]{Christopher N. A. Willmer}
\affiliation{Steward Observatory, University of Arizona, 933 N. Cherry Avenue, Tucson, AZ 85721, USA}
\email[]{}  

\accepted{for publication in the Astrophysical Journal, May 2026}

\begin{abstract}

Ultra-cool T- (T$_{\mathrm{eff}} \approx$ 500 - 1200 K) and Y-dwarfs (T$_{\mathrm{eff}}$ $\lessapprox 500$ K) have historically been found only a few hundred parsecs from the Sun. The sensitivity and wavelength coverage of the NIRCam instrument on board the James Webb Space Telescope offer a unique method for finding low-temperature brown dwarfs in deep extragalactic datasets out to multiple kiloparsecs. Here we report on the selection of a sample of 41 brown dwarf and brown dwarf candidates across the JWST Advanced Deep Extragalactic Survey (JADES) in the GOODS-S and GOODS-N regions. We introduce a new open-source Bayesian tool, Near-Infrared Fitting for T and Y-dwarfs (\texttt{NIFTY}), to derive effective temperatures, metallicities, and distances from JWST photometry. We find that 31 candidates have fits consistent with T-dwarf temperatures out to 5-6 kpc, and 10 candidates have fits consistent with Y-dwarf temperatures out to 1-2 kpc. The majority of the sources are best fit with sub-solar metallicity models. We report proper motions for ten brown dwarf candidates (three are newly presented), and calculate the number density of T- and Y-dwarfs as a function of temperature and distance above the Milky Way midplane. We further discuss how Y-dwarfs can serve as contaminants in the search for ultra-high-redshift galaxies. Together, these results demonstrate the power of deep JWST extragalactic imaging to probe the coldest substellar populations far beyond the solar neighborhood, providing new constraints on the Milky Way's structure and brown dwarf demographics.

\end{abstract}

\keywords{\uat{Brown dwarfs}{185} --- \uat{Halo stars}{699} --- \uat{Infrared astronomy}{786} --- \uat{James Webb Space Telescope}{2291}}

\section{Introduction}

The last several decades have seen the discovery of a large number of ultra-cool brown dwarfs, objects with effective temperatures lower than M dwarf stars (\teff{} $< 2500$ K). These elusive sources are key to understanding the low-mass end of the stellar initial mass function (M $< 0.07$ M$_{\odot}$), and as these objects lie at masses that range between stars and extrasolar planets, they are also of great interest for exploring both star and planet formation. Brown dwarfs have been broadly classified into three subclasses based on their observed spectral type: L-, T-, and Y-dwarfs,  which have effective temperatures of T$_{\mathrm{eff}} \approx 1200 - 2000$K, $\approx 500 - 1200$K, and $\lessapprox 500$K, respectively \citep{kirkpatrick1999, kirkpatrick2005A, burgasser2006, stephens2009, cushing2011, kirkpatrick2021, leggett2021}. At increasingly lower temperatures, the observed spectra of brown dwarfs change significantly based on the effects of molecular absorption, differing atmospheric chemistry, the disappearance of clouds and, eventually, the emergence of ices \citep{burrows1997, ackerman2001, marley2002, cooper2003, morley2012, leggett2013, morley2014, leggett2015, skemer2016}.

Brown dwarfs do not have enough mass to undergo sustainable nuclear fusion in their cores, and instead cool off over time as they radiate their gravitational potential energy \citep{kumar1962, kumar1963, hayashi1963}. As a result, they have low intrinsic luminosities, and based on their temperatures, emit much of their flux in the near-IR. To date, most known brown dwarfs have been discovered from large-area near-IR photometric surveys like the Two Micron All Sky Survey \citep[2MASS,][]{skrutskie2006}, The UKIRT Infrared Deep Sky Survey \citep[UKIDSS,][]{lawrence2007}, and the Wide-Field Infrared Survey Explorer \citep[WISE,][]{wright2010} All-Sky Survey \citep{cutri2012}, among others. The resulting brown dwarf samples span a wide range of effective temperatures, ages, masses, and metallicities, although the majority have distances within 100 pc, due to the relatively bright flux limits of these surveys. Currently, one of the coldest brown dwarfs known, WISE J085510.83-071442.5 (\teff{} $ \approx 250$ K), lies only 2 pc away \citep{luhman2014}. 

Searching for more distant brown dwarfs is of great interest given that there may be many more of these ultra-cool sources in the Milky Way thick disk or even the halo. There have been multiple low-metallicity subdwarfs discovered that are thought to be associated with the halo \citep{gizis1997, burgasser2003, loediu2010, pinfield2012, burningham2014}. \citet{schneider2020} reported the discovery of the 14 - 67 pc distant T dwarf WISE 1810055-1010023 (\teff{} $\sim 800$K) which was estimated to have both a low-metallicity atmosphere and a proper motion consistent with objects in the Milky Way halo. Additionally, \citet{meisner2023} announced the discovery of a new population of thick-disk halo T- and Y- dwarfs confirmed with ground-based Gemini near-IR photometry. The T subdwarf WISEA J153429.75-104303.3, more commonly known as ``The Accident'', has a metallicity consistent with [M/H] $ = -2.22 \pm 0.05$, T$_{\mathrm{eff}} = 502 \pm 6$ K \citep{faherty2025}, and a transverse velocity $>200$ km s$^{-1}$ \citep{kirkpatrick2021b}. Finding larger numbers of these sources is crucial for Milky Way kinematic studies, refining our current estimate of binary brown dwarf frequency, and exploring the evolutionary history of the Milky Way galaxy \citep{burgasser2007, joergens2008, fontanive2018, zhang2018, zhang2019, factor2023}. 

The unprecedented photometric sensitivity in the near-infrared offered by the James Webb Space Telescope (JWST) has allowed brown dwarfs to be found at \textit{kilo}parsec distances which probe the Milky Way thick disk and halo \citep{wilkins2014, holwerda2018, hainline2020}, following pioneering work from Hubble Space Telescope (HST) near-IR observations out to $\sim2\mu$m \citep{ryan2011, masters2012, aganze2022}. Brown dwarfs have often served as a contaminant in searches for high-redshift quasars \citep{warren2007} and Lyman-dropout galaxies \citep[][among others]{bunker2004, oesch2012a, wilkins2014, finkelstein2010, finkelstein2015}, but only with photometric detections out to 1 - 2$\mu$m. JWST's wavelength coverage allows for the estimation of the properties for these distant sources using observations at 1 - 12$\mu$m \citep{leggett2023, beiler2023, beiler2024}. 

While most early JWST studies of ultra-cool dwarfs have focused on previously discovered sources, a growing body of work has emerged which introduced \textit{new} samples of brown dwarf candidates found in JWST/NIRCam datasets, including GLASS, JADES, CEERS, UNCOVER, and COSMOS-Web \citep{nonino2023, hainline2024a, chen2025}. These photometrically-selected sources have best-fit temperatures consistent with T- and Y-dwarfs, and several exhibit high proper motions. A very small (currently $\sim 5$) subsample of these candidates have been spectroscopically observed, revealing their distances and low atmospheric metallicity \citep{langeroodi2023, burgasser2023, hainline2024b, tu2025, tu2025b, morrissey2026, li2026}. These studies demonstrate the ubiquity of ultra-cool brown dwarf candidates in line with Milky Way predictions, and, given the power of JWST pure parallel surveys with NIRCam observations \citep[e.g. PANORAMIC and SAPPHIRES, ][]{williams2025, sun2025}, it is likely that many hundreds of sources remain undiscovered across the wealth of archival extragalactic observations that exist. 

Understanding the selection and properties for these ultra-cool brown dwarfs is especially vital given current and planned large area surveys: The Dark Energy Survey \citep{DES2016}, SPHEREx \citep{bock2026}, LSST from the Vera Rubin Observatory \citep{ivezic2019}, Euclid \citep{euclid2025}, and the Nancy Grace Roman Telescope \citep{akeson2019} deep and wide area surveys, among others. Researchers using data from the Dark Energy Survey announced the discovery of \textit{11,745} L0 to T9 dwarfs selected photometrically \citep{carnerorosell2019}. \citet{domingueztagle2025} used EUCLID Data Release 1 observations to discover 11 late L and T dwarfs. LSST will help characterize substantial populations of cool L-type subdwarfs \citep{lsst2009}, while SphereX will find and characterize $> 40$ T and Y dwarfs \citep{dore2014} across the mission (see \citet{gagne2026} and \citet{tu2026} for early results). The Roman Space Telescope will allow for the discovery of brown dwarfs from data taken as part of the High Latitude Survey which can be used to explore the structure of the Milky Way \citep{holwerda2023}. Crucially however, many of these planned surveys will only attain observations at wavelengths short of $2\mu$m, and as the coldest Y dwarfs peak at $4-5\mu$m, observations with JWST/NIRCam and MIRI provide a critical complement to understand the full demographics of the ultra-cool dwarf population. 

This paper serves to expand the search for brown dwarfs across a significantly larger area from \citet[][hereafter \KBD{}]{hainline2024a}, who presented an initial 11 candidate brown dwarfs in GOODS-S,  three candidates in GOODS-N, and seven candidates in CEERS \citep{bagley2023, ceers_doi} selected photometrically from within the first JADES public data release area. These authors fit the photometry for these sources with brown dwarf atmospheric models and presented proper motions for seven sources across the sample which corresponded to transverse velocities from $\sim 30 - 120$ km/s. Subsequent JWST/NIRSpec spectroscopic observations of brown dwarfs from the \KBD{} sample have resulted in a number of confirmations, including the source JADES-GS-BD-9 presented in \citet{hainline2024b}. 

In this paper, we look for brown dwarfs across the full JADES Data Release 5 (DR5) footprint \citep{robertson2026, johnson2026, jades_doi} and present 25 sources in GOODS-S and 16 sources in GOODS-N using updated NIRCam color selection criteria designed to include the coldest brown dwarfs from the comprehensive JADES survey. The early public data release from which candidates were found in \KBD{} was only 125 square arcminutes, while the DR5 footprint is almost three times larger, 309.5 square arcminutes. In this paper, we combine the updated NIRCam data with longer-wavelength MIRI photometric data for eleven candidates to explore their SEDs and discuss how they separate from line emitting galaxies and ``little red dots'' \citep[LRDs,][]{matthee2023}. LRDs, thought to be powered by growing supermassive black holes, are observed to be very compact, and have similar colors to brown dwarf candidates, with blue 1 - 3$\mu$m slopes and very red 3 - 5$\mu$m slopes for those at $z > 5$. The sensitivity and wavelength coverage offered by JADES make this dataset ideal for finding ultra distant brown dwarfs. Unlike \KBD{}, we do not include the CEERS data in our present analysis, but instead focus on the JADES GOODS-S and GOODS-N areas. To fit the full sample, we also introduce \texttt{NIFTY}, a Bayesian brown dwarf photometric fitting package, which we use to derive properties and uncertainties for our sample of brown dwarfs and brown dwarf candidates from NIRCam and MIRI photometric fits. \texttt{NIFTY} is open-source, and we use three different model sets to explore the effective temperatures, distances, and metallicities of our sample. 

We have structured this search for low-temperature brown dwarfs as follows. We introduce the full JADES DR5 dataset and photometry in Section \ref{sec:observations}, and we discuss how we select brown dwarf candidates in Section \ref{sec:selecting_brown_dwarfs}. In Section \ref{sec:nifty}, we describe \texttt{NIFTY}, including the models we make use of and the details of our fitting procedure. We discuss the results from the fitting, properties of individual interesting sources, and present measured proper motions for a few additional candidates in Section \ref{sec:results}. We discuss source on-sky and number densities, and explore how the coldest Y-dwarfs might be mistaken for high-redshift ($z > 20$) galaxies in Section \ref{sec:discussion}, using the recently discovered source ``Capotauro'' \citep{gandolfi2025b} as an example. Finally, we conclude and point towards future avenues of research in Section \ref{sec:conclusions}. In Appendix \ref{sec:appendix_fluxes} we provide the fluxes for each source, and in Appendix \ref{sec:appendix_parameters}, we provide tables and SEDs for the output parameters for each object in our full sample.  Throughout the paper, we refer to the sources we select as ``brown dwarfs and brown dwarf candidates'' owing to the detection of proper motions and the spectroscopic confirmation for a number of these sources, including JADES-GS-BD-9 \citep{hainline2024b}. Throughout this paper, photometric magnitudes will be provided in the AB system \citep{oke1974}.  

\section{JADES and SMILES Observations and Data Reduction} \label{sec:observations}

In this section, we describe the near- and mid-IR photometric observations that form the basis for our selection and analysis of the brown dwarf candidates. We describe the JADES NIRCam observations, their reduction, and photometric measurements in Section \ref{subsec:jades_v1}. In GOODS-S, we can also use observations made by JWST/MIRI as part of both JADES as well as the Systematic Mid-infrared Instrument Legacy Extragalactic Survey \citep[SMILES, ][]{rieke2024} programs, which can be used in fitting the sources at longer wavelengths. These data will be briefly described in Section \ref{subsec:smiles_miri}. 

\subsection{JADES DR5 NIRCam Observations} \label{subsec:jades_v1}

JADES is a comprehensive observing program designed by the JWST/NIRCam and NIRSpec instrument teams, who combined Guaranteed Time Observations (GTO) with NIRCam, NIRSpec, and MIRI across the two well studied fields: GOODS-S (RA = 53.126 deg, DEC = -27.802 deg) and GOODS-N (RA = 189.229, DEC = +62.238 deg). The photometry that we use in this study comes from JADES DR5, where the collaboration has sought to combine JWST/NIRCam data from multiple JADES observational programs across GOODS-S and GOODS-N with those from multiple other GO programs, as described in the DR5 paper \citep{johnson2026}. The final survey footprint from which we selected our brown dwarfs and brown dwarf candidates has varying filter selection and observational depth. Using a 0.2$^{\prime\prime}$ diameter circular aperture, the F444W 10$\sigma$ photometric depth for a point source in the JADES GOODS-S observations is 28.96 AB mag across the wide ``medium'' depth region, 29.57 AB mag in the smaller area ``deep'' regions, and then 29.79 AB mag in the much smaller area ``deepest'' region. For GOODS-N, the depths are consistent, although slightly shallower in the largest area medium region. For this study, we searched for brown dwarf candidates across a total survey area of 408 square arcminutes, which includes 244 square arcminutes in GOODS-S (with 155 square arcminutes of JADES coverage) and 164 square arcminutes in GOODS-N (with 112 square arcminutes of JADES coverage). 

The raw observational data were assembled by the JADES team, and reduced, stacked, and mosaicked following the procedure outlined in \citet{robertson2023}, \citet{tacchella2023}, \citet{eisenstein2023}, and \citet{johnson2026}. For selecting brown dwarfs, we used fluxes measured using $0.2^{\prime\prime}$ diameter circular apertures, with aperture corrections derived from empirical JWST/NIRCam point spread functions, where we assumed point source morphologies for our sources \citep[see][for more details]{ji2023}. The measurement of photometry and its uncertainties for the JADES catalogs we used is discussed in detail in \citet{robertson2026}. For the JADES photometry, the underlying Poisson error from flux counts is combined with sky flux uncertainties estimated from random apertures measured across the mosaic to produce the final errors that we use. We refer the reader to Section 8 of \citet{robertson2026} for a full discussion. We selected brown dwarfs using the NIRCam filters F115W, F150W, F277W, F410M, and F444W, which have deep coverage across the majority of the GOODS-S and GOODS-N footprints. 

For fits to the sources, we used both the $0.2^{\prime\prime}$ diameter circular aperture fluxes, and for sources with MIRI coverage we utilized fluxes measured using a larger $0.5^{\prime\prime}$ diameter circular aperture to help mitigate the effects of the extended JWST/MIRI PSF as compared to NIRCam. For GOODS-S, we fit to the JWST/NIRCam filters F090W, F115W, F150W, F162M, F182M, F200W, F210M, F250M, F277W, F300M, F335M, F356W, F410M, and F444W. For GOODS-N, we fit to the JWST/NIRCam filters F090W, F115W, F150W, F162M, F182M, F200W, F210M, F277W, F300M, F335M, F356W, F410M, and F444W.

In order to explore whether these sources have proper motions from earlier observations we searched for detections in HST/ACS and WFC3 mosaics, which we discuss in Section \ref{sec:proper_motions}. We use the HST/ACS and HST/WFC3 mosaics from the Hubble Legacy Fields (HLF) v2.0 for GOODS-S and v2.5 for GOODS-N \citep[$25^{\prime} \times 25^{\prime}$ for GOODS-S, and $20.5^{\prime} \times 20.5^{\prime}$ for GOODS-N,][]{illingworth2013, whitaker2019}. We use data in the HST/ACS F435W, F606W, F775W, F814W, and F850LP filters, as well as HST/WFC3 F105W, F125W, F140W, and F160W. 

\subsection{JADES and SMILES MIRI Observations} \label{subsec:smiles_miri}

The JADES footprint in GOODS-S has some of the deepest extragalactic MIRI observations taken to date which we can include for aiding with fits to the brown dwarf candidate SEDs. We combined two programs, the JADES MIRI parallels, with observations with the F770W, F1280W and F1500W filters taken across relatively small areas (24.3, 19.4, and 24.0 square arcminutes, respectively), and SMILES, which covered the center of the JADES footprint in GOODS-S $(\sim$34 square arcminutes) with the F560W, F770W, F1000W, F1280W, F1500W, F1800W, F2100W, F2550W, although the data become shallower at longer wavelengths (5$\sigma$ flux sensitivities of 0.21 - 17$\mu$Jy from 5.6 - 25$\mu$m). For our fits we utilize MIRI data in the F560W, F770W, F1000W and F1280W filters, as our models only have predicted fluxes at $< 15\mu$m. Please see \citet{rieke2024}, \citet{alberts2024a}, and \citet{alberts2024b} for more information about the reduction and depth of the MIRI footprint in GOODS-S. Currently, this is the only comprehensive search for brown dwarfs that utilizes MIRI photometry, following work done exploring brown dwarfs in the mid-IR from the Spitzer Space Telescope and AKARI \citep{cushing2006, yamamura2010, tsuji2011}. Given the importance of the mid-IR for understanding the atmospheres of low-mass objects \citep{vos2023, miles2023, beiler2023, suarez2023, beiler2024, mccarthy2025}, the study of these sources offers a unique opportunity to constrain models out to $\sim 15\mu$m. 

\section{Selecting Brown Dwarfs} \label{sec:selecting_brown_dwarfs}

\begin{figure*}
  \centering
  \includegraphics[width=0.98\textwidth]{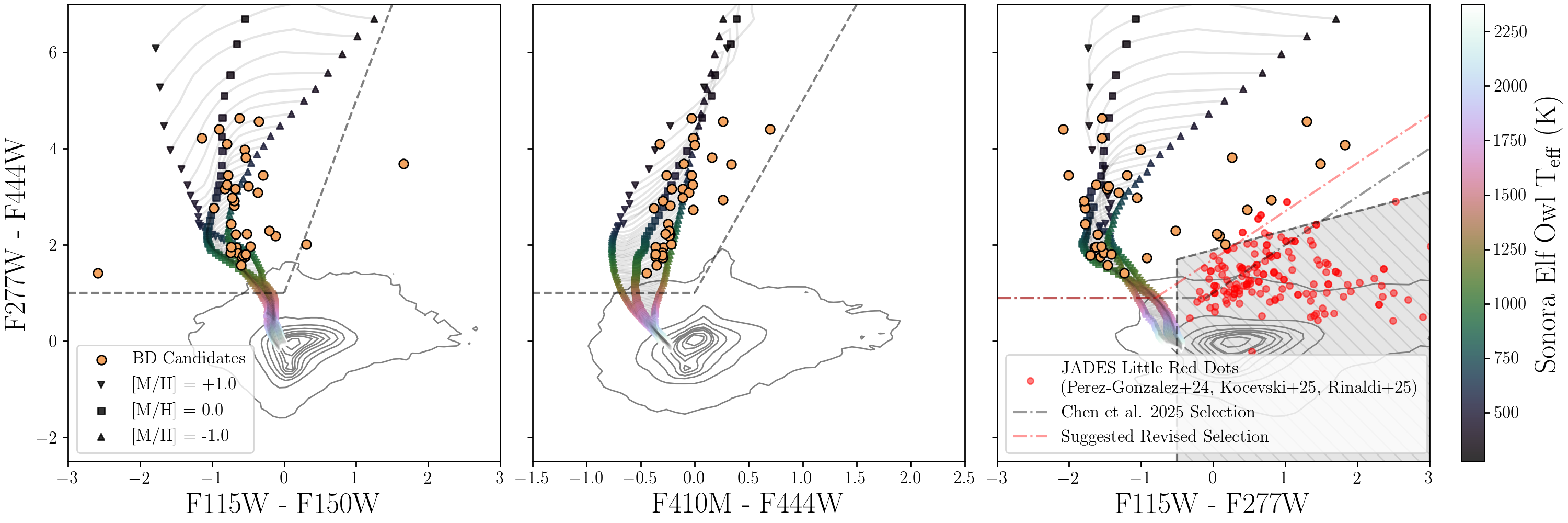}
  \caption{Color criteria used to select brown dwarf candidates in this work. In each panel, the colors of the underlying JADES GOODS-S population are represented with contours. Our 41 selected JADES brown dwarf candidates are shown using tan circles with black outlines. To show the colors spanned by brown dwarf models, we plot predicted color tracks derived from the Sonora Elf Owl models with downward triangles ([M/H] = +1.0), squares ([M/H] = 0.0), and upward triangles ([M/H] = -1.0), with the points colored by the T$_{\mathrm{eff}}$ values as shown in the color bar on the right. In each panel, we plot the color criteria used in the paper with black dashed lines. \textit{Left:} F277W - F444W color plotted against F115W - F150W color. \textit{Middle:} F277W - F444W color plotted against F410M - F444W color. There are sources outside the selection criteria in each panel as we only require selection using either the F277W - F444W vs F115W - F150W or the F277W - F444W vs F410M - F444W criteria. \textit{Right:} F277W - F444W color plotted against F115W - F277W color, where we show, in a grey hashed region, an additional color criterion designed to reject LRDs. We show a sample of GOODS-S and GOODS-N LRDs from \citet{perezgonzalez2024}, \citet{rinaldi2024}, and \citet{kocevski2025} as small red points. To help exclude these LRDs while including each of our candidates, we suggest a refined selection criteria which we plot as a red dashed line. 
  \label{fig:BD_Color_Color_All}}
\end{figure*}

In \KBD{}, the authors select brown dwarfs from JADES photometric data using a combination of common wide-band NIRCam colors. Specifically, they defined selection criteria that used F277W - F444W, F115W - F277W, F115W - F150W colors to select objects with a ``V-shaped'' SED that arises from brown dwarf atmospheric absorption at $1.5 - 3.5\mu$m by molecules such as \ce{NH3}, \ce{H2}, \ce{H2O}, and \ce{CH4}, leading to blue observed $1.5 - 2\mu$m colors and red $3 - 4.5\mu$m colors. The authors chose these specific filters as they were commonly used across the majority of the JWST extragalactic surveys including JADES, CEERS, and COSMOS-Web and many others. They also required that $m_{\mathrm{F444W,AB}} \leq 28.5$, but did not make cuts on SNR for any additional filters. The selection criteria from \KBD{} were refined using predicted colors derived from the Sonora Cholla brown dwarf atmospheric models from \cite{marley2021} and \cite{karalidi2021}. 

In this study, we have updated the selection criteria based on the growing evidence that brown dwarfs selected from extragalactic datasets are at large distances and have sub-solar metallicities \citep{burgasser2023, hainline2024a, hainline2024b, chen2025}, which will lead to significantly redder 1.0 - 1.5$\mu$m colors as compared to solar or supersolar metallicity sources. We introduce an updated selection based on both the newly-measured JADES DR5 colors for the brown dwarf candidates from \KBD{} and the predicted brown dwarf colors estimated across a range of metallicities from the Sonora Elf Owl models \citep{mukherjee2024, wogan2025, sonoraelfowl_doi}. Additionally, while in \KBD{} the authors restrict their sources to those with $m_{\mathrm{F444W,AB}} \leq 28.5$, we only require F444W flux signal-to-noise ratio (SNR) $ > 5$ (measured in a 0.2$^{\prime\prime}$ diameter circular aperture) to allow for fainter sources that may include farther and cooler dwarf candidates. We plot our color selection criteria in Figure \ref{fig:BD_Color_Color_All}. Our initial color criteria rely on finding sources with red 2.7$\mu$m - 4.4$\mu$m colors: 

\begin{equation}
(\mathrm{F277W} - \mathrm{F444W}) > 1.0
\end{equation}

and, then, depending on photometric availability, the sample assembled in this paper satisfies either:

\begin{equation}
(\mathrm{F277W} - \mathrm{F444W}) > 4.0 \times (\mathrm{F115W} - \mathrm{F150W})
\end{equation}

or 

\begin{equation}
(\mathrm{F277W} - \mathrm{F444W}) > 4.0 \times (\mathrm{F410M} - \mathrm{F444W})
\end{equation}

The F277W - F444W color limit is identical to that in \KBD{}, while we have updated our secondary color to allow for slightly redder F115W - F150W colors to select the low-metallicity Sonora Elf Owl models. We focus on F410M - F444W color since, as shown in the middle panel of Figure \ref{fig:BD_Color_Color_All}, the F410M - F444W colors for brown dwarfs change little as a function of metallicity for the Sonora Elf Owl models. However, a number of current JWST extragalactic surveys do not include deep F410M coverage, so we propose an alternate color selection without the need for F410M.

Our revised selection criteria do permit a large number of extragalactic sources. Specifically, one of the primary contaminants in samples of brown dwarfs are the previously mentioned LRDs, as well as galaxies with strong [OIII]$\lambda$4959,5007 line emission at $z \sim 7$. Many of these sources are compact and, in the case of the LRDs, can have visual morphologies similar to brown dwarfs: the \citet{langeroodi2023} and \citet{burgasser2023} UNCOVER sources were initially selected as interesting LRD candidates.

As a way of removing these extragalactic sources, we further require that: 
\begin{equation}
(\mathrm{F115W} - \mathrm{F277W}) < -0.5 \label{eq:lrd_1}
\end{equation}

and

\begin{equation}
(\mathrm{F277W} - \mathrm{F444W}) > 0.4 \times (\mathrm{F115W} - \mathrm{F277W}) + 1.9 \label{eq:lrd_2}
\end{equation}

In Figure \ref{fig:BD_Color_Color_All}, we show all three color spaces for our selection. We depict the underlying JADES GOODS-S color space with grey contours, and our final 41 brown dwarf candidates as tan circles with black outlines. We additionally plot the colors derived from the Sonora Elf Owl models with the [M/H] $= +1.0$ plotted as downward triangles, the those with [M/H] $= +0.0$ plotted as squares, and finally those with [M/H] $= -1.0$ plotted as upward triangles, to show the spread in color as a function of metallicity. The Sonora Elf Owl model points are colored by the effective temperature of the brown dwarf, as shown in the color bar on the right side of the figure. We show the color rejection from Equations \ref{eq:lrd_1} and \ref{eq:lrd_2} with a black hashed region in the rightmost panel, and the sources from which it was derived: a sample of LRDs in the literature assembled in GOODS-S and GOODS-N by \citet{perezgonzalez2024}, \citet{rinaldi2024}, and \citet{kocevski2025}, plotted with small red points.

In the rightmost panel of Figure \ref{fig:BD_Color_Color_All}, we show the F277W - F444W vs. F115W - F277W colors of our sample, and additionally plot the brown dwarf selection for the brown dwarfs chosen from COSMOS-Web in \citet{chen2025} with a black dot-dashed line. Their selection was based on model colors, and has a broad overlap with our own color selection region. While the relaxed F277W - F444W criteria from \citet{chen2025} ($> 0.9$ instead of $> 1.0$) as compared to our fiducial selection would have a minimal effect on the sources we find in JADES, we note that their secondary color limit would result in a large number of LRDs being chosen along with brown dwarfs. Given the large number of brown dwarf candidates with a spread of colors in our own sample, and the desire for a more clean selection, we suggest a revised color criteria from \citet{chen2025}:

\begin{equation}
(\mathrm{F277W} - \mathrm{F444W}) > 0.9 \label{eq:lrd_3}
\end{equation}

and 

\begin{equation}
(\mathrm{F277W} - \mathrm{F444W}) > (\mathrm{F115W} - \mathrm{F277W}) + 1.7 \label{eq:lrd_4}
\end{equation}

which we show with a red dot dashed line in the rightmost panel. This updated selection would encompass all of our candidates, while also removing the majority of the contaminating LRDs. 

Following the color selection of our sources, we performed visual inspection in order to remove obvious data reduction artifacts like bad pixels near the edges of mosaics, or sources that are obvious extragalactic sources (from their morphologies, or from fits to galaxy models, see Section \ref{subsec:tentative_sources}). After visual inspection, we ended up with 25 sources in GOODS-S, and 16 sources in GOODS-N. Each of the GOODS-S and GOODS-N brown dwarfs and brown dwarf candidates introduced in \KBD{} are also selected with our updated photometry and color criteria. 

Owing to their sizes, brown dwarfs will be unresolved in NIRCam imaging data, and we investigated the compactness of our color selected sources to help in differentiating brown dwarf candidates from extragalactic sources. To explore the sizes of the JADES catalog sources, we calculated a ``compactness ratio'' by comparing the fluxes measured using a 0.5$^{\prime\prime}$ diameter circular aperture to those measured using a 0.2$^{\prime\prime}$ diameter circular aperture. As these fluxes have had aperture corrections applied which are appropriate for point sources, the compactness ratio for unresolved sources should be near one, while extended objects will have compactness ratios significantly larger than one. For the 41 sources in our final brown dwarf candidate sample, we find that all but two have a compactness ratio consistent with one within the associated uncertainties. One of the final two sources (JADES-GN-BD-3) has a larger compactness ratio as it is smeared in the DR5 mosaic owing to its proper motion. The other source (JADES-GS-BD-17) appears to be a brown dwarf in front of a background galaxy. We will discuss both of these sources in more detail in Section \ref{subsec:individual_sources}.

\begin{figure*}
  \centering
  \includegraphics[width=0.475\textwidth]{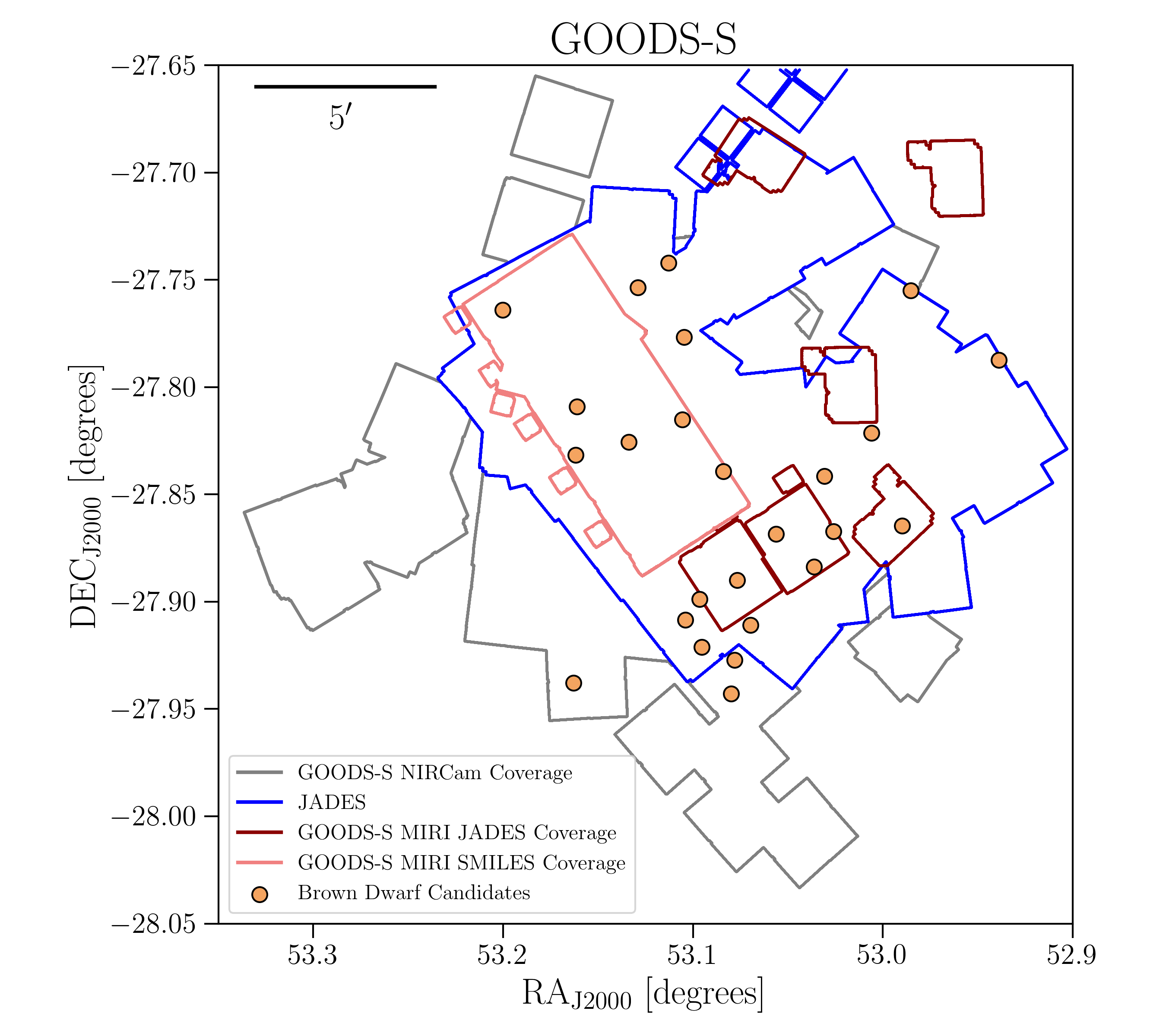}
  \includegraphics[width=0.512\textwidth]{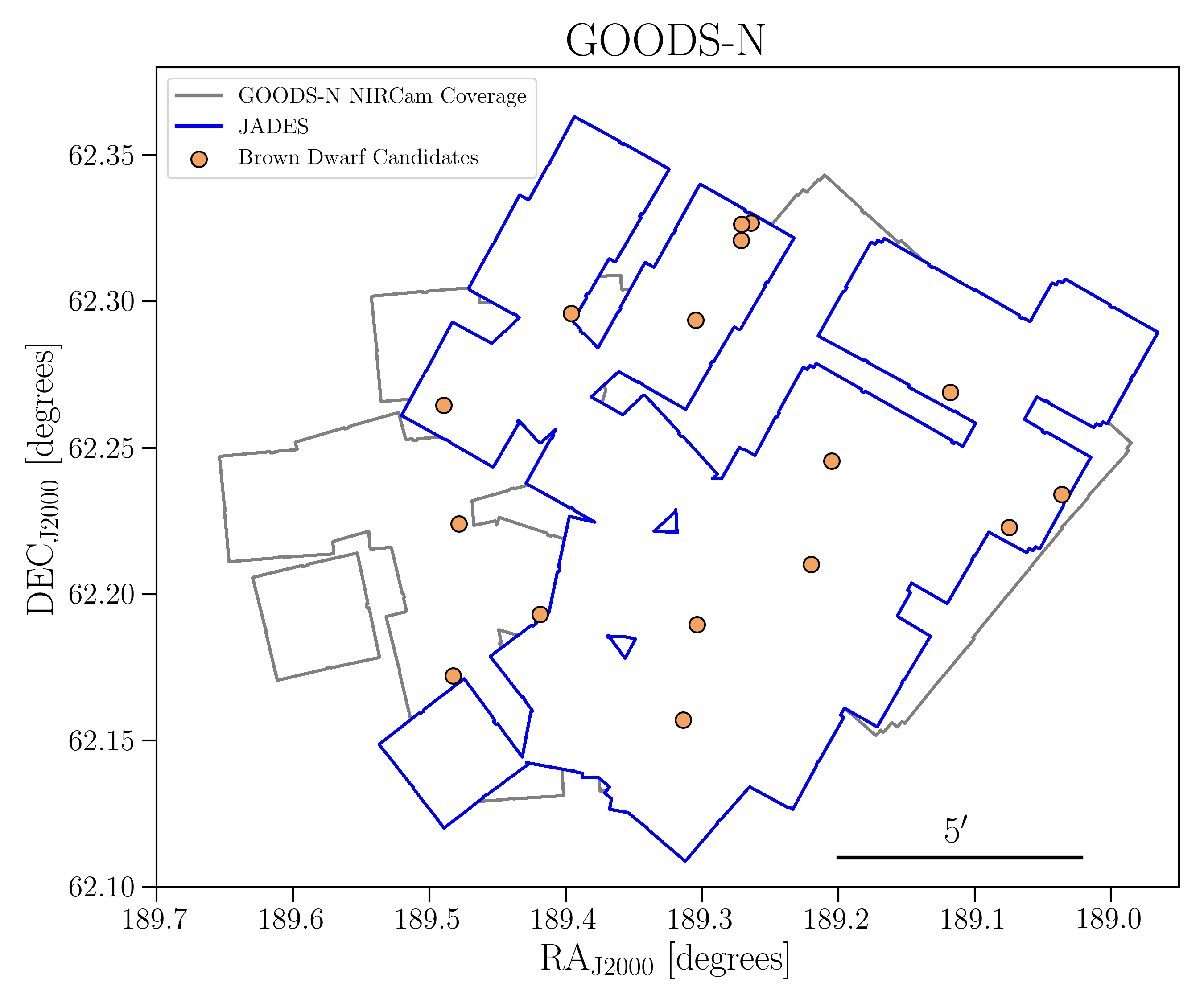}  
  \caption{JADES DR5 (Left) GOODS-S footprint and (Right) GOODS-N footprint over which we searched for brown dwarfs. We mark the full GOODS-S footprint including ancillary data with a grey line, and the JADES specific data with a blue line. Additionally, we plot the SMILES survey area with a light red line, and the additional area from the JADES MIRI parallels in F770W, F1280W, and F1500W with a dark red line. We plot the brown dwarfs and brown dwarf candidates recovered by our selection with tan points with black outlines. We note that the areas are not scaled proportionally between the GOODS-S and GOODS-N regions in each panel, and we indicate a 5$^{\prime}$ scale bar for reference in each panel.
  \label{fig:BD_GS_GN_footprint}}
\end{figure*}

We plot the positions of the final brown dwarfs and brown dwarf candidate sources across the GOODS-S and GOODS-N NIRCam footprints in Figure \ref{fig:BD_GS_GN_footprint}. In these figures we provide the JADES outlines in blue, and the larger area from DR5 in grey for both GOODS-S and GOODS-N. In the GOODS-S panel, we plot the SMILES MIRI survey area in light red,  and the ancillary F770W, F1280W, and F1500W MIRI observations from the JADES MIRI parallels in dark red. We present the IDs, positions, and fluxes for these sources in Tables \ref{tab:gs_bd_fluxes} and \ref{tab:gn_bd_fluxes} in Appendix \ref{sec:appendix_fluxes}. In this table, we provide either the fluxes measured in 0.2$^{\prime\prime}$ diameter circular apertures, or for the eleven GOODS-S sources with MIRI coverage, we instead include 0.5$^{\prime\prime}$ diameter circular apertures to more accurately account for the larger MIRI point spread function, indicated in a column in both tables. The area over which we searched for candidates was smaller than the total area of the survey given the need for filter coverage with JWST/NIRCam F115W, F277W, F444W and either F150W or F410M. When looking only at these areas, the total GOODS-S area we searched over was 204.5 square arcminutes, and the total GOODS-N area we searched over was 105.0 square arcminutes, for a total area of 309.5 square arcminutes. Excitingly, from Figure \ref{fig:BD_GS_GN_footprint}, it can be seen that in six cases (three in GOODS-S and three in GOODS-N), candidates were found outside of the JADES footprint, within the PANORAMIC parallel area that has been included as part of the DR5 JADES data release. We refer to these sources with the prefix ``PANO'' in subsequent tables and analyses to highlight them: PANO-GS-BD-18, PANO-GS-BD-19, PANO-GS-BD-25, PANO-GN-BD-13, PANO-GN-BD-14, and PANO-GN-BD-15.

\section{\texttt{NIFTY}: Near-Infrared Fitting for T and Y Dwarfs} \label{sec:nifty}

\begin{figure*}
  \centering
  \includegraphics[width=0.98\textwidth]{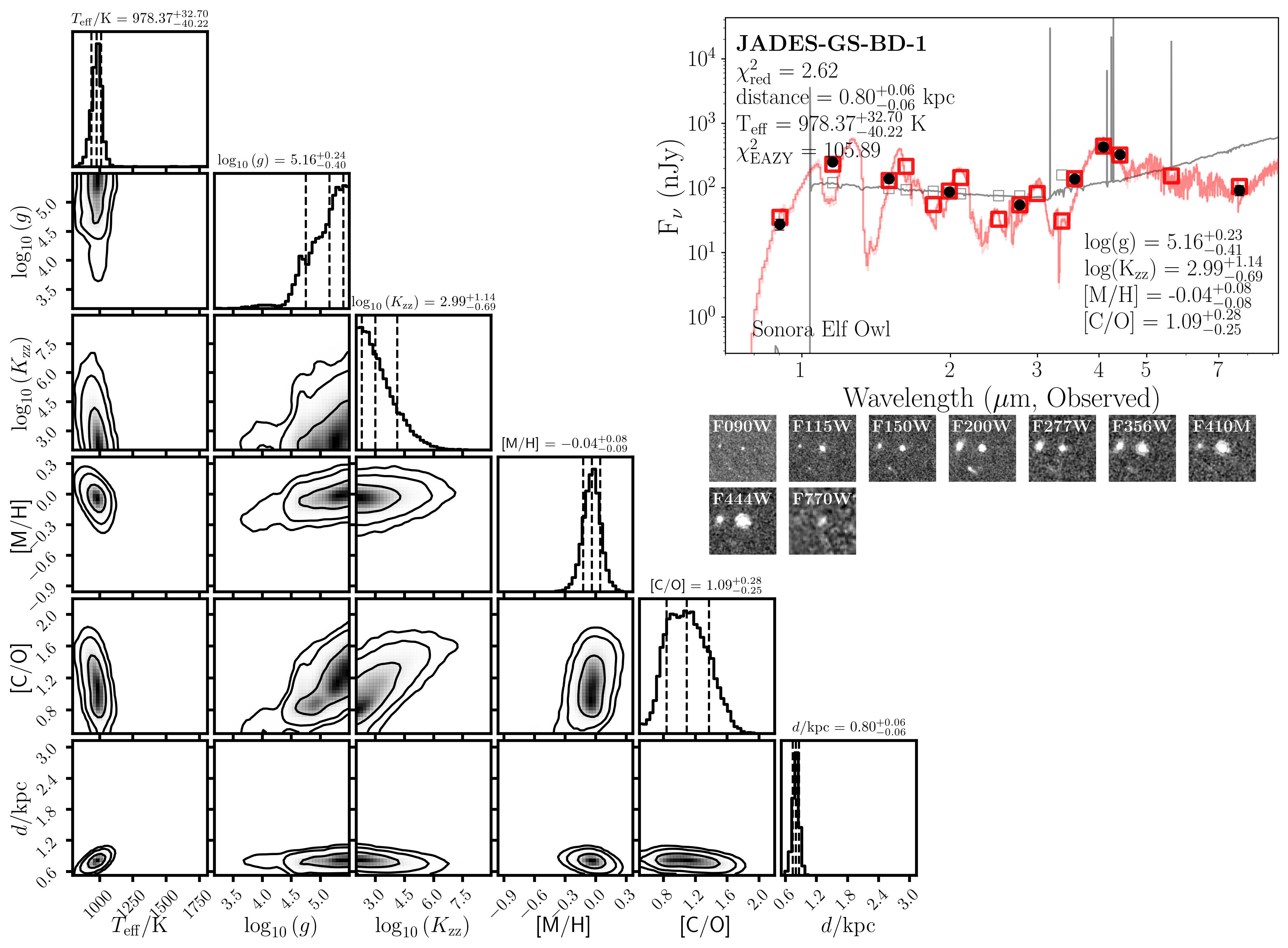}
  \caption{(Top right) Example \texttt{NIFTY} Fit to Source JADES-GS-BD-1. The JWST NIRCam and MIRI photometry is plotted in black, corresponding to the $2^{\prime\prime} \times 2^{\prime\prime}$ thumbnails below. The \texttt{NIFTY} median SED, and a 1$\sigma$ confidence interval, is plotted in red, with the median photometry in each filter shown with red squares. On the plot, we indicate the \texttt{NIFTY} parameters resulting from a fit with the Sonora Elf Owl models. We also plot a fit to the NIRCam photometry with galaxy templates using \texttt{EAZY} with a grey line and with grey squares indicating the model photometry, and provide the raw \texttt{EAZY} minimum $\chi^2$. (Left) The corner plot showing the posterior distributions for the free parameters from the Sonora Elf Owl fit to JADES-GS-BD-1.
  \label{fig:example_NIFTY_fit_44120}}
\end{figure*}

The large number of photometrically selected brown dwarfs and brown dwarf candidates discovered in JWST extragalactic surveys presents a new frontier for cold dwarf studies. In \KBD{} the authors fit their sources with a $\chi^2$ minimization scheme that utilized brown dwarf atmospheric models which covered a fairly sparse parameter space. While this technique would result in a ``best-fit solution,'' it is not optimal for providing robust error estimates on derived properties, or a way to explore the degeneracies between the inferred parameters. To that end, as a way for researchers to quickly fit JWST NIRCam and MIRI photometry for potential brown dwarfs, we developed a method similar to that outlined in \citet{chen2025}, utilizing a Bayesian framework for comparing the observed photometry to a suite of brown dwarf atmospheric models. We developed an open-source Python package,  \texttt{NIFTY} (Near-Infrared Fitting for T and Y Dwarfs), which is hosted on a public GitHub repository\footnote{\url{https://github.com/kevinhainline/NIFTY}}. \texttt{NIFTY} is designed to fit photometric observations from JWST/NIRCam and MIRI and makes use of the Markov-chain Monte Carlo sampler \textit{emcee} \citep{foremanmackey2013} to explore the posterior distributions for the model parameters. 


We have designed \texttt{NIFTY} to fit photometry with three suites of brown dwarf atmospheric models, Sonora Elf Owl \citep{mukherjee2024, wogan2025, sonoraelfowl_doi}, LOWZ \citep{meisner2021, lowz_doi}, and ATMO2020++ \citep{meisner2023}\footnote{\url{https://www.erc-atmo.eu/?page_id=322}}. These models were chosen to span a large range of L, T, and Y dwarf temperatures, while also allowing for an exploration of how metallicity can be retrieved from photometry alone, and whether, given the potential distances for these sources, low-metallicity models are preferred over those at solar metallicities. We specifically use the \citet{wogan2025} updated Sonora Elf Owl model spectra, which include the corrected \ce{CO2} concentrations and quench point. In addition, the models used by \texttt{NIFTY} are cloud-free, which will primarily impact fits to sources at $> 1200$K, where clouds have more of an impact on the observed SED. From evolutionary models \citep{phillips2020, marley2021, morley2024}, a source at 10 $M_{\mathrm{Jupiter}}$ will cool to 1200 K in less than 20 Myr, while a source at 50 $M_{\mathrm{Jupiter}}$ will take $\sim 2$ Gyr. For the distant, likely old, sources we are exploring in this study, cloud-free models are therefore appropriate. In the future, we will update \texttt{NIFTY} to include additional models, including the Sonora ``Diamondback'' cloudy substellar models \citep{morley2024, sonoradiamondback_doi} and the Sonora ``Flame Skimmer'' models, which extend the Elf Owl models to colder effective temperatures and lower surface gravities \citep[][J. Mang et al. 2025, in preparation, these models are not yet public]{crotts2025}.

\texttt{NIFTY} requires, first, that these models are interpolated across their free parameters in logarithmic space, and we provide scripts for this purpose in the \texttt{NIFTY} github repository. For each model, there are gaps where, for a given set of parameters, an atmospheric model was not produced (which is more common at the low-temperature, low-metallicity regions of parameter space), and we interpolate each model across these gaps (which includes extrapolation in a few specific edge gaps) as part of the creation of the final interpolation model. We use the resulting interpolation model within \texttt{NIFTY} to explore the parameter space and compare the models to the photometry. We present the free parameters used, and their ranges, for each model in Table \ref{tab:nifty_parameters}. We refer to the individual papers that present each model for further discussion of the underlying physics and chemistry involved in each model. Notably, the ATMO2020++ models do not vary the Eddy diffusion parameter K$_{\rm zz}$ or C/O, leading to two fewer free parameters as compared to Sonora Elf Owl and LOWZ. In this paper, while we will present and compare the output parameters for our brown dwarfs and brown dwarf candidates using all three model sets, we will adopt the Sonora Elf Owl models as our ``fiducial'' models for use in figures and in our discussion.

\begin{deluxetable}{lccc}[]
\tabletypesize{\footnotesize}
\tablecolumns{17}
\tablewidth{0pt}
\tablecaption{\texttt{NIFTY} Model Parameters and Ranges \label{tab:nifty_parameters}}
\tablehead{
\colhead{Parameter} & \colhead{Sonora Elf Owl} & \colhead{LOWZ} & \colhead{ATMO2020++} }
 \startdata
T$_{\mathrm{eff}}$ & 275 - 2400 K & 500 - 1600 K & 250 - 1200 K  \\
log(g) & 3 - 5 & 3.5 - 5.25  & 2.5 - 5.5  \\
$\left[ \mathrm{M/H} \right]$ & -1.0 - 1.0  & -2.5 - 1.0  & -1.0 - 0.3  \\
log(K$_{\mathrm{zz}}$) & 2.0 - 9.0 & -1.0 - 10.0  & -  \\
C/O & 0.5 - 2.5 & 0.1 - 0.85 & -  \\
\enddata
\end{deluxetable}

\texttt{NIFTY} assumes flat priors on each individual parameter and currently each brown dwarf is modeled at 1 Jupiter radius, which is used in scaling the normalization of the final model to estimate the distance. The normalization of each model is proportional to $(R/D)^2$, where R is the radius of the brown dwarf and D is its distance from the Sun. We ran \texttt{NIFTY} on the sample discussed in Section \ref{sec:selecting_brown_dwarfs}, using the 0.2$^{\prime\prime}$ and 0.5$^{\prime\prime}$ diameter circular aperture photometry presented in Tables \ref{tab:gs_bd_fluxes} and \ref{tab:gn_bd_fluxes}. We do not fit the HST photometry for the sources owing both to the fact that these objects are intrinsically faint at $<1\mu$m, and that any proper motion may result in sources that move outside of the aperture from the JADES NIRCam detection image. We have visually inspected each source in the HST/ACS images for F435W, F606W, or F775W filters shortward of 0.8$\mu$m to make sure that none are detected. For the GOODS-S sources, we include the MIRI observations if they are detected in a given MIRI filter out to F1280W owing to the fact that the models only extend out to 15$\mu$m. For each source, in order to avoid bright data points disproportionately influencing the fits, we institute a maximum SNR on individual fluxes of 20, which can be tuned within \texttt{NIFTY}. We fit each source with the Sonora Elf Owl, LOWZ, and ATMO2020++ models, and report the 16th, 50th, and 84th percentile output parameter from the fits in Tables \ref{tab:bd_properties_gs_gn},\ref{tab:bd_properties_gs_gn_pt2} and \ref{tab:bd_properties_gs_gn_pt3} in Appendix \ref{sec:appendix_parameters}. In these tables, we include the free parameters T$_{\mathrm{eff}}$, log(g), [M/H] for all three model sets, and log(K$_{\mathrm{zz}}$) and C/O for the Sonora Elf Owl and LOWZ models. We also include a best-derived distance in parsecs for each source from the fit, and a reduced $\chi^2$ value derived from comparing the observed fluxes to the 50th percentile fluxes in each filter. In a few instances, where there were more free parameters than data points, the reduced $\chi^2$ is unable to be calculated, and is instead reported as -9999. For those sources where the $\chi^2_{\mathrm{red}}$ can be calculated, the values are largely near one, except in a few notable cases we will discuss below. We will discuss the results from the fits in the next section. 

The uncertainties recovered by \texttt{NIFTY} for many of the sources in our sample, and especially the brightest sources, are small, as seen in Tables \ref{tab:bd_properties_gs_gn} - \ref{tab:bd_properties_gs_gn_pt3}. These uncertainties are driven largely by the small observed photometric errors from the JADES catalogs, and we consider them to be underpredicted, as they do not take into account uncertainties from the models themselves, as can be seen in the variance between fits to the same data from the LOWZ, ATMO2020 and Sonora Elf Owl fits. Across the sample, we find median inter-model differences (defined as half the range in inferred values for each source) of $\Delta T_{\mathrm{eff}, \mathrm{med}} = 46$ K, $\Delta \log(g)_{{\mathrm{med}}} = 0.4$, $\Delta \mathrm{[M/H]}_{\mathrm{med}} = 0.2$, and $\Delta \mathrm{distance}_{\mathrm{med}} = 170$ pc. Additional systematic uncertainty arises from our assumption of a fixed radius. Brown dwarf radii span approximately 0.6-1.2 R$_{\mathrm{Jupiter}}$ \citep{sorahana2013, tu2024}; within the \texttt{NIFTY} framework this primarily affects the inferred distances, introducing an additional uncertainty of $\sim$30\%.

We also use the reduced $\chi^2$ values reported in Tables \ref{tab:bd_properties_gs_gn}-\ref{tab:bd_properties_gs_gn_pt3} to estimate the degree to which the photometric uncertainties may be underestimated. Under the assumption that deviations from $\chi^2_{\mathrm{red}} = 1$ are dominated by these underestimated flux errors, the typical values of $\chi^2_{\mathrm{red}}$ imply that the uncertainties are too small by a factor of $\sim$1.5. To assess the impact of this effect, we refit the sample using the Sonora Elf Owl models with flux uncertainties scaled by this factor. The resulting parameter uncertainties increase accordingly: the uncertainties in $T_{\mathrm{eff}}$ increase by a factor of 1.45, those in log(g), log(K$_{\mathrm{zz}}$), [M/H], and C/O by a factor of $\sim$1.2, and those in distance by a factor of 1.41.

Similarly, there are likely additional sources of uncertainty from the use of linear interpolation on the model grids in \texttt{NIFTY}. To explore this latter point, \citet{tu2024} performed a set of ``leave-one-out'' validation tests on their own Bayesian fitting code using the Sonora Elf Owl models, finding that the process of model linear interpolation results in larger differences at 4 - 5$\mu$m than at shorter wavelengths, although they were fitting to full near-IR spectra, and so this effect is mitigated somewhat owing to the photometric fits in \texttt{NIFTY}. Overall, these effects indicate that systematic uncertainties, both from model assumptions and photometric error characterization, dominate over the formal statistical uncertainties returned by the MCMC fits. As such, the uncertainties we provide for the fits to the sources in this paper should be treated as strict lower limits. 

We plot a sample SED fit and corner plot for a \texttt{NIFTY} fit to JADES-GS-BD-1 using the Sonora Elf Owl models in Figure \ref{fig:example_NIFTY_fit_44120}. The fit is quite good, with $\chi^2_{\mathrm{red}} = 2.62$, largely owing to the bright fluxes and small uncertainties. This source is also detected at 7 microns with MIRI, helping to constrain the fit at longer wavelengths. The Sonora Elf Owl fit to this source returns a temperature T$_{\mathrm{eff}} = 978^{+33}_{-40}$ K, at a distance of $800 \pm 60$ pc. This source appears in \KBD{}, with a best-fit temperature using fits to the Sonora Cholla models (using a more simple $\chi^2$ minimization algorithm) of $1150$ K, but at a similar distance of 749 pc. The difference in temperature can also be attributed to the usage of MIRI F770W photometry, which was not included in the fits in \KBD{}, demonstrating the importance of mid-IR photometry to provide better constraints on the physical properties. The final SEDs and corner plots derived from the \texttt{NIFTY} fits are available in Zenodo via \dataset[DOI: 10.5281/zenodo.20127944]{https://doi.org/10.5281/zenodo.20127944}.

From Figure \ref{fig:example_NIFTY_fit_44120}, we can see that the specific gravity and eddy diffusion parameters are not well constrained, and run up against the edge of the Sonora Elf Owl ranges. While T$_{\mathrm{eff}}$, distance, and [M/H] are fairly well probed by broad-band photometry given the wavelength range spanned by our photometry, log(g) and log(K$_{\mathrm{zz}}$) have narrower, more degenerate spectral diagnostics. Specifically, changing log(g) alters the atmospheric pressure structure, which in turn affects pressure-broadened absorption wings, as well as collision-induced absorption from H$_2$ and the detailed shapes of molecular bands. In the broad JADES photometry, much of that information is averaged over and difficult to observe given the uncertainties, especially for lower temperature sources with fainter 1 - 3$\mu$m flux emission. Those changes are also highly degenerate with other parameters already present in the models, especially cloud properties, metallicity, and to some extent, T$_{\mathrm{eff}}$. The eddy diffusion parameter log(K$_{\mathrm{zz}}$) is more difficult to constrain from photometry, as this primarily acts through non-equilibrium chemistry rather than the full thermal continuum. Again, the subtle spectroscopic signatures that can be used to estimate eddy diffusion are degenerate with things like atmospheric metallicity [M/H], C/O, and the overall normalization of the continuum. The \texttt{NIFTY} fits may prefer one edge of the allowed log(g) or log(K$_{\mathrm{zz}}$) range simply because the likelihood is relatively flat across much of that parameter, not because the photometry demands an extreme surface gravity or mixing efficiency. As a result, we caution the interpretation of these results, especially when compared to a parameter like T$_{\mathrm{eff}}$, which is more robustly constrained. To test this interpretation, we would need to extend the boundaries on these parameters within atmospheric models, but also we could examine either additional photometry at wavelengths sensitive to these subtle changes, or more realistically, near-to-mid-IR spectroscopy for these sources.

\section{Results} \label{sec:results}

In this section we describe the results of our \texttt{NIFTY} fits to the brown dwarf and brown dwarf candidates and discuss individual sources. We also explore proper motions derived from data taken of GOODS-S and GOODS-N over the last several decades, and finally, we introduce an additional sample of candidates with weaker evidence for being brown dwarfs. 

\subsection{Brown Dwarf Properties from \texttt{NIFTY}}

\begin{figure*}
  \centering
  \includegraphics[width=0.32\textwidth]{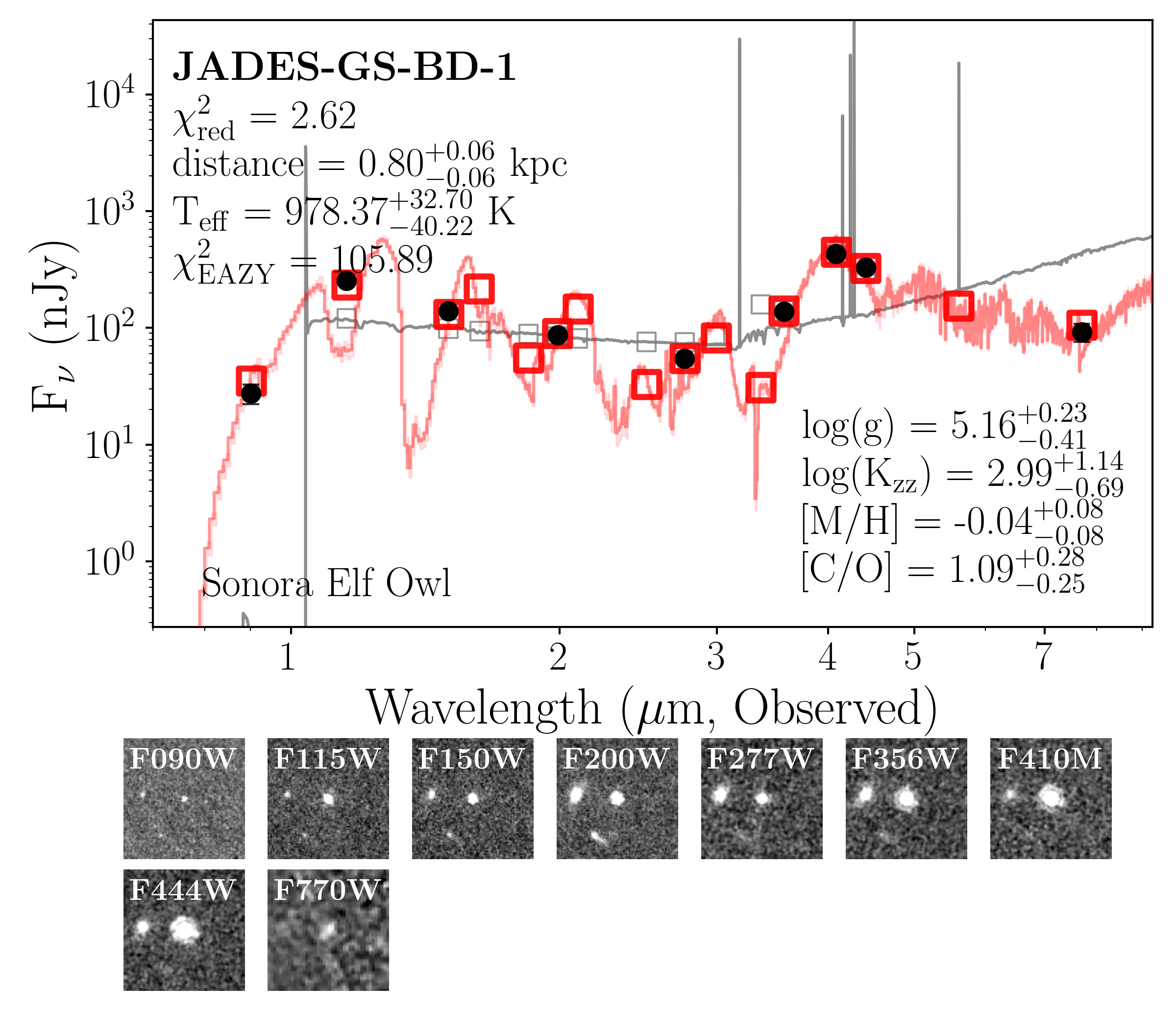}
  \includegraphics[width=0.32\textwidth]{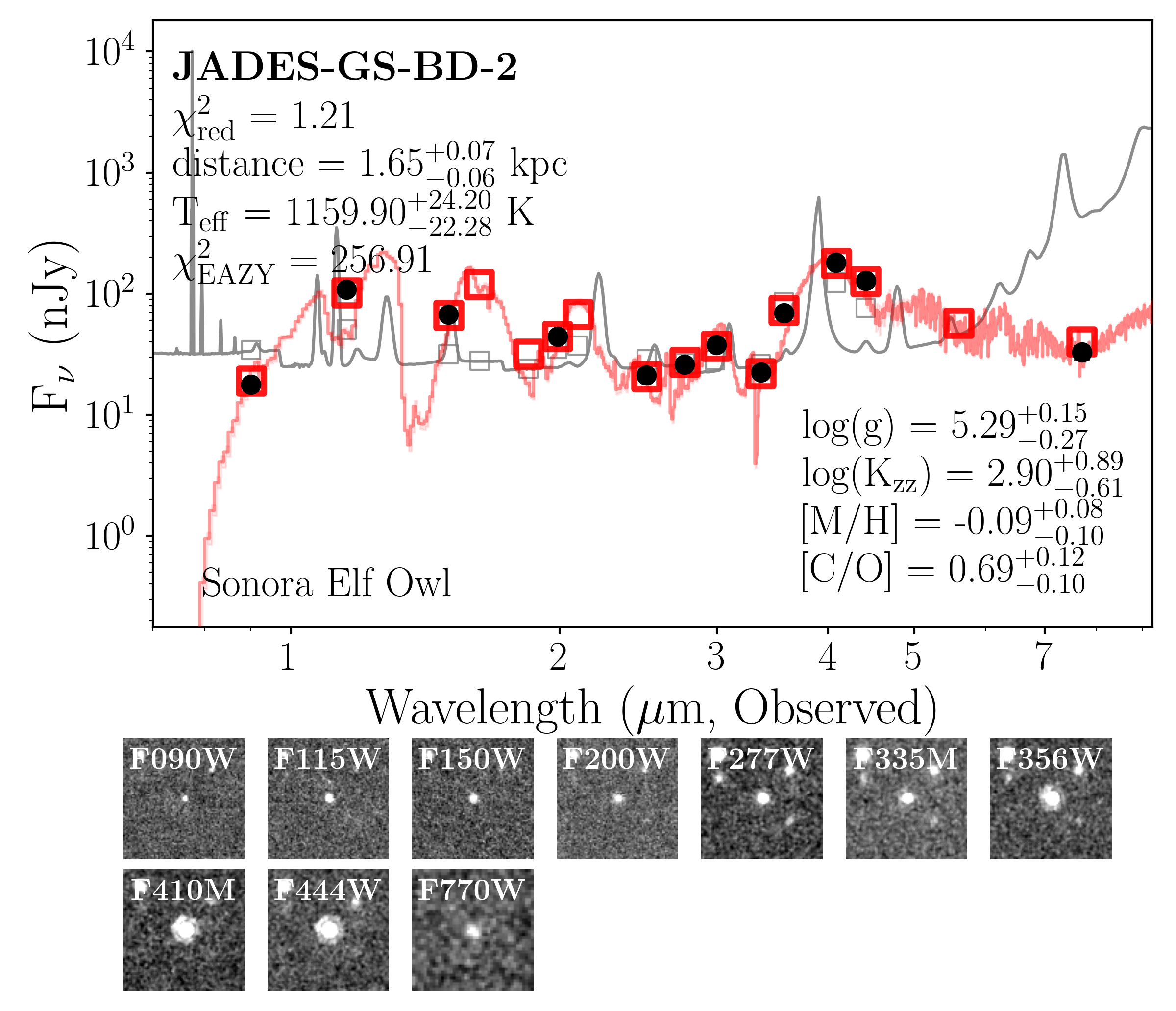}
  \includegraphics[width=0.32\textwidth]{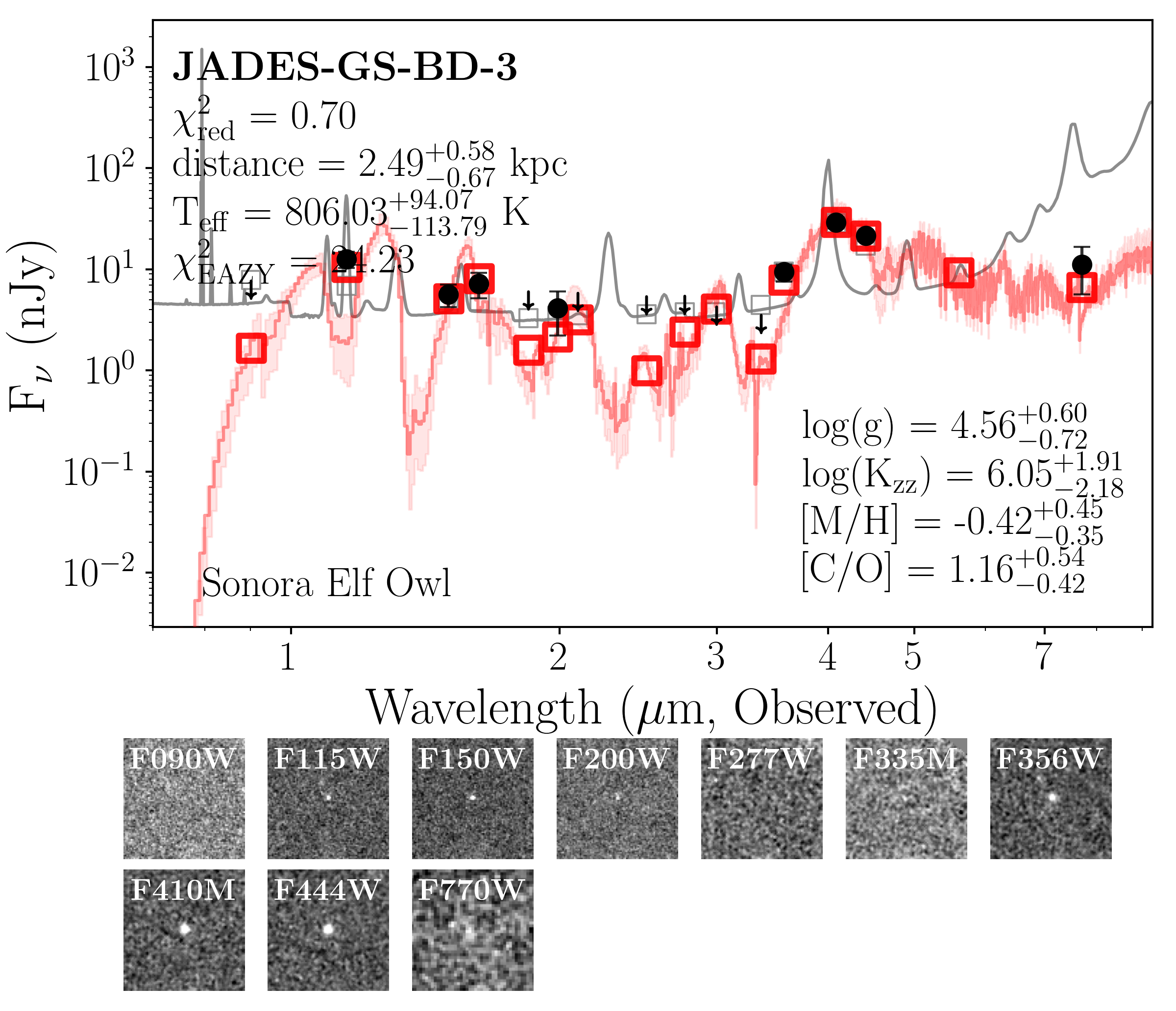}
  \includegraphics[width=0.32\textwidth]{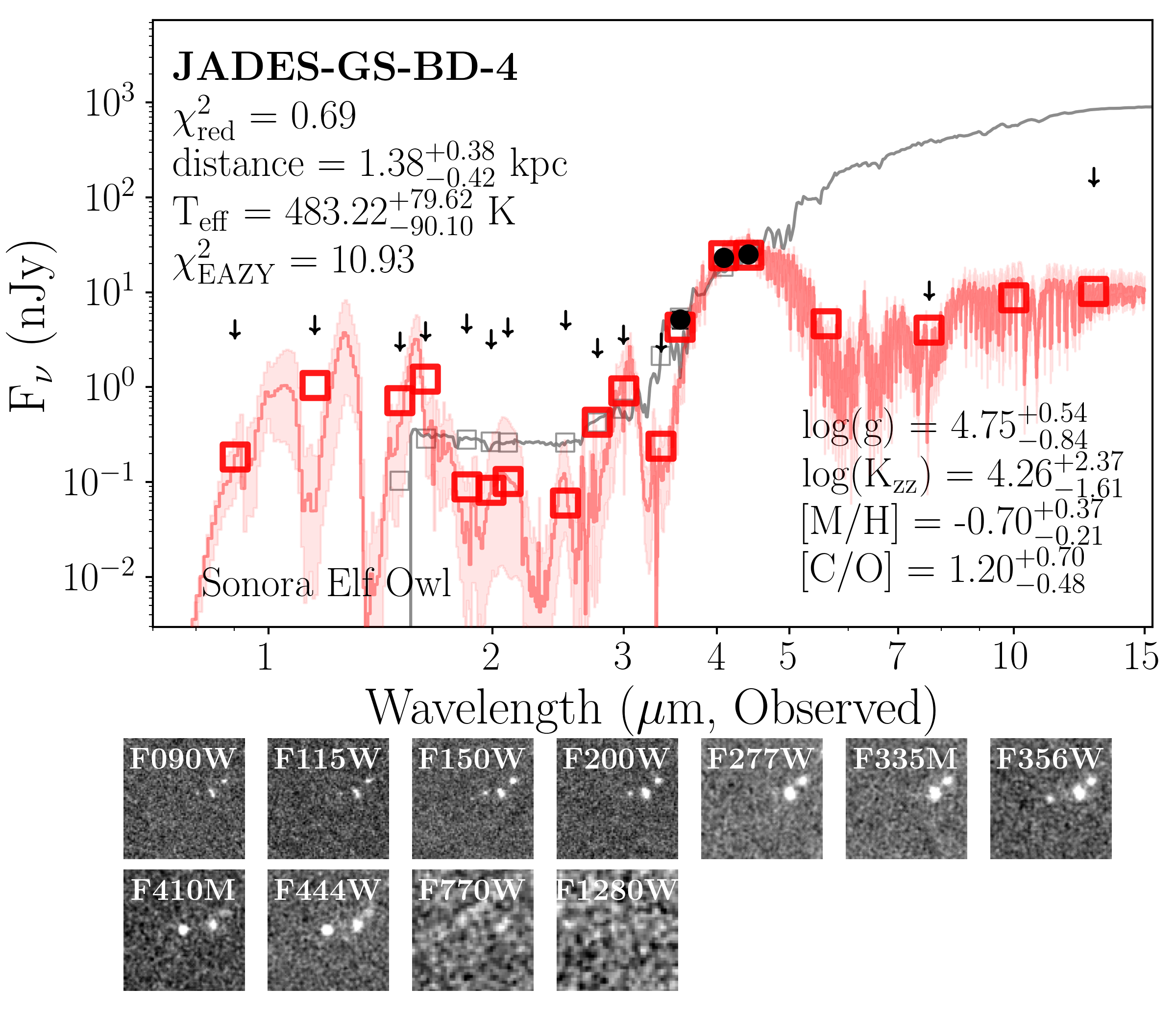}
  \includegraphics[width=0.32\textwidth]{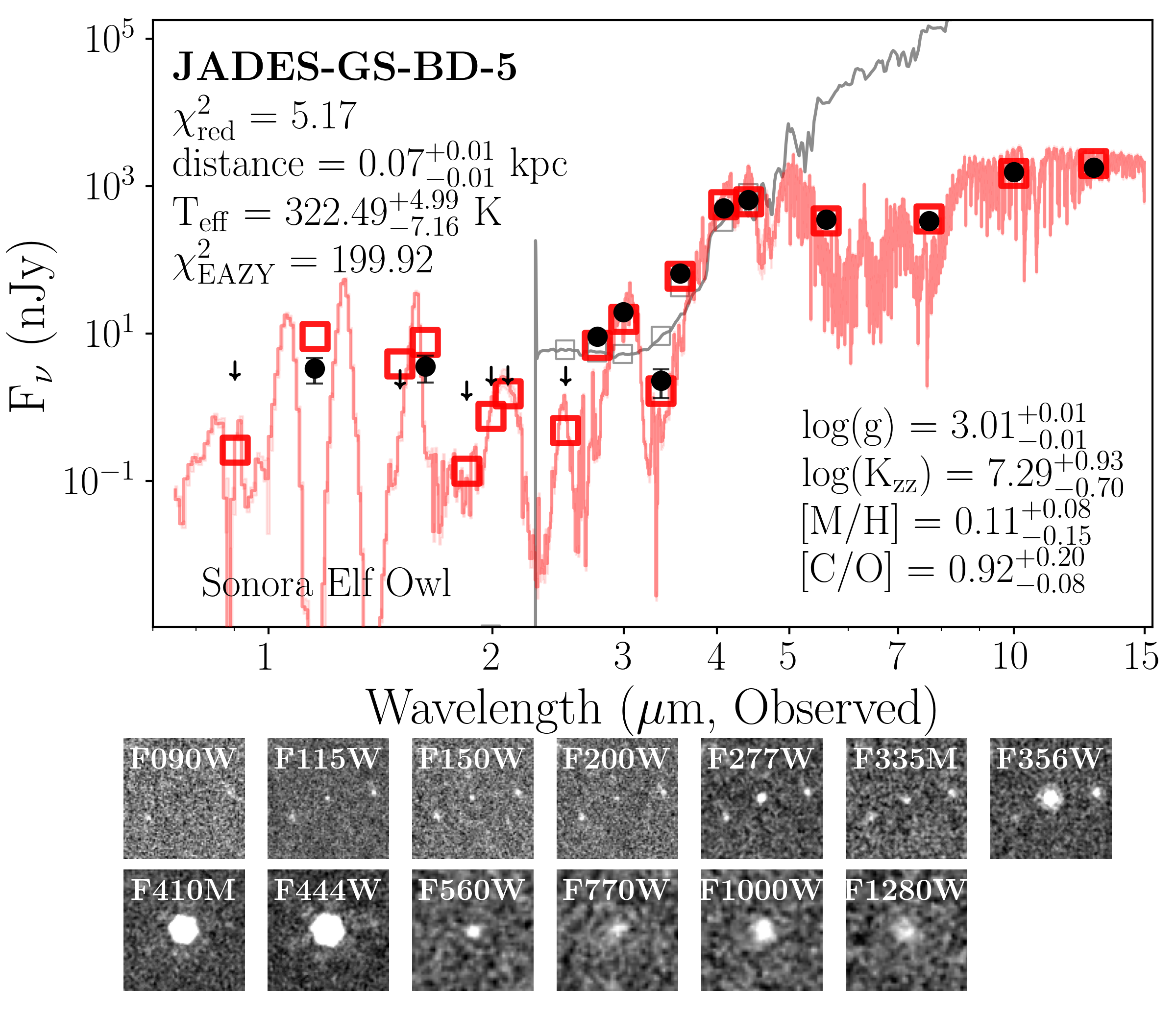}
  \includegraphics[width=0.32\textwidth]{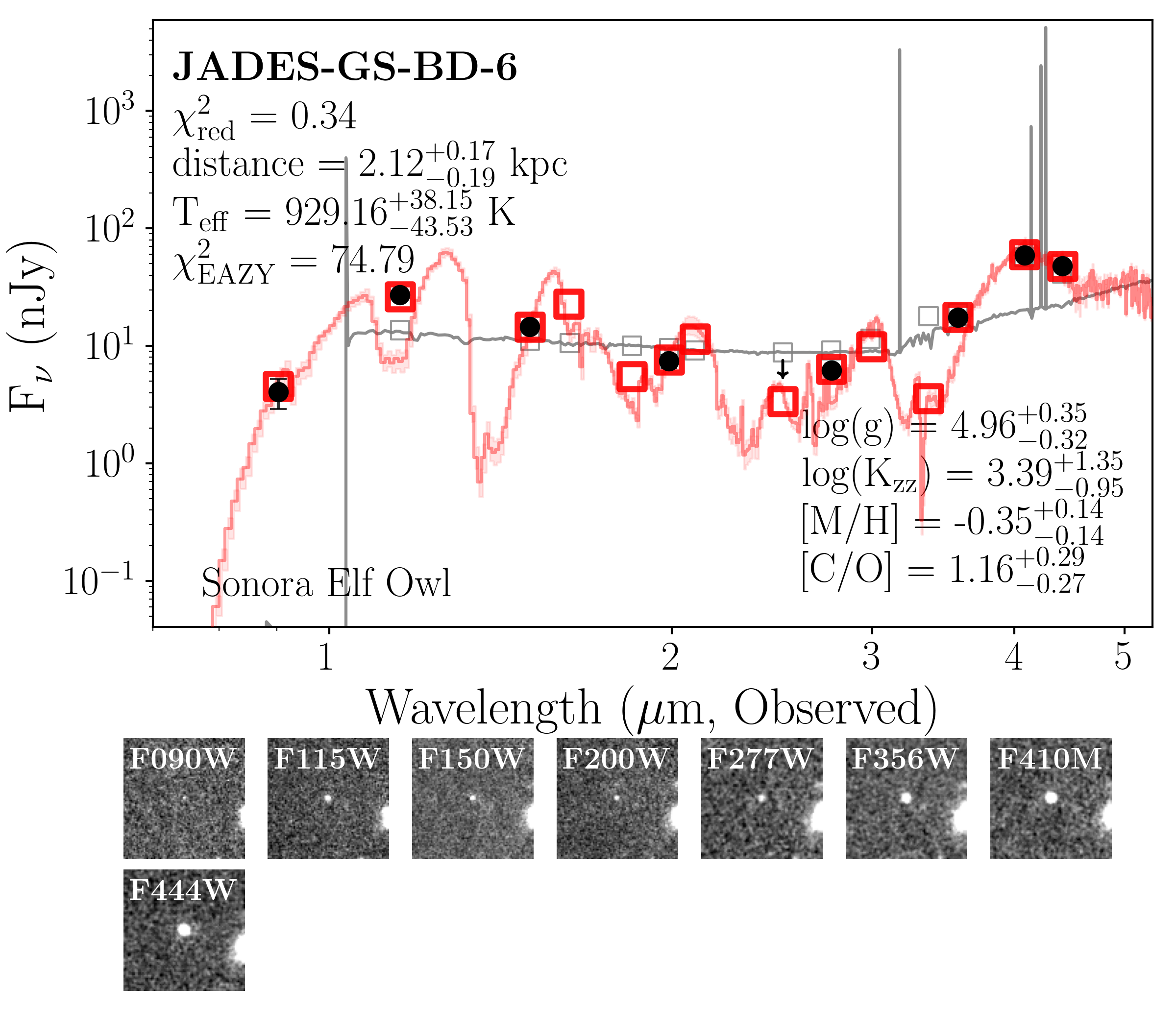}
  \includegraphics[width=0.32\textwidth]{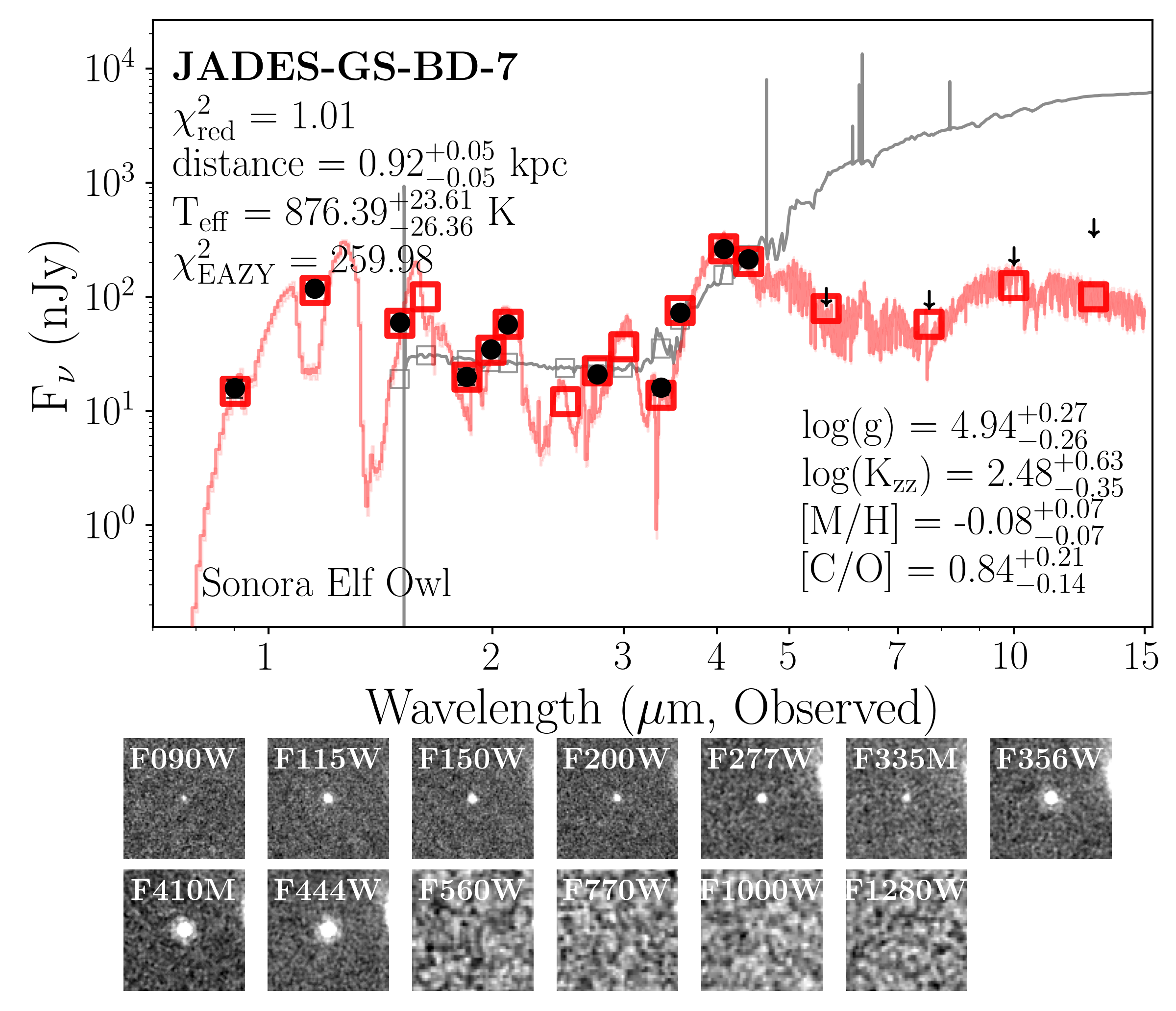}
  \includegraphics[width=0.32\textwidth]{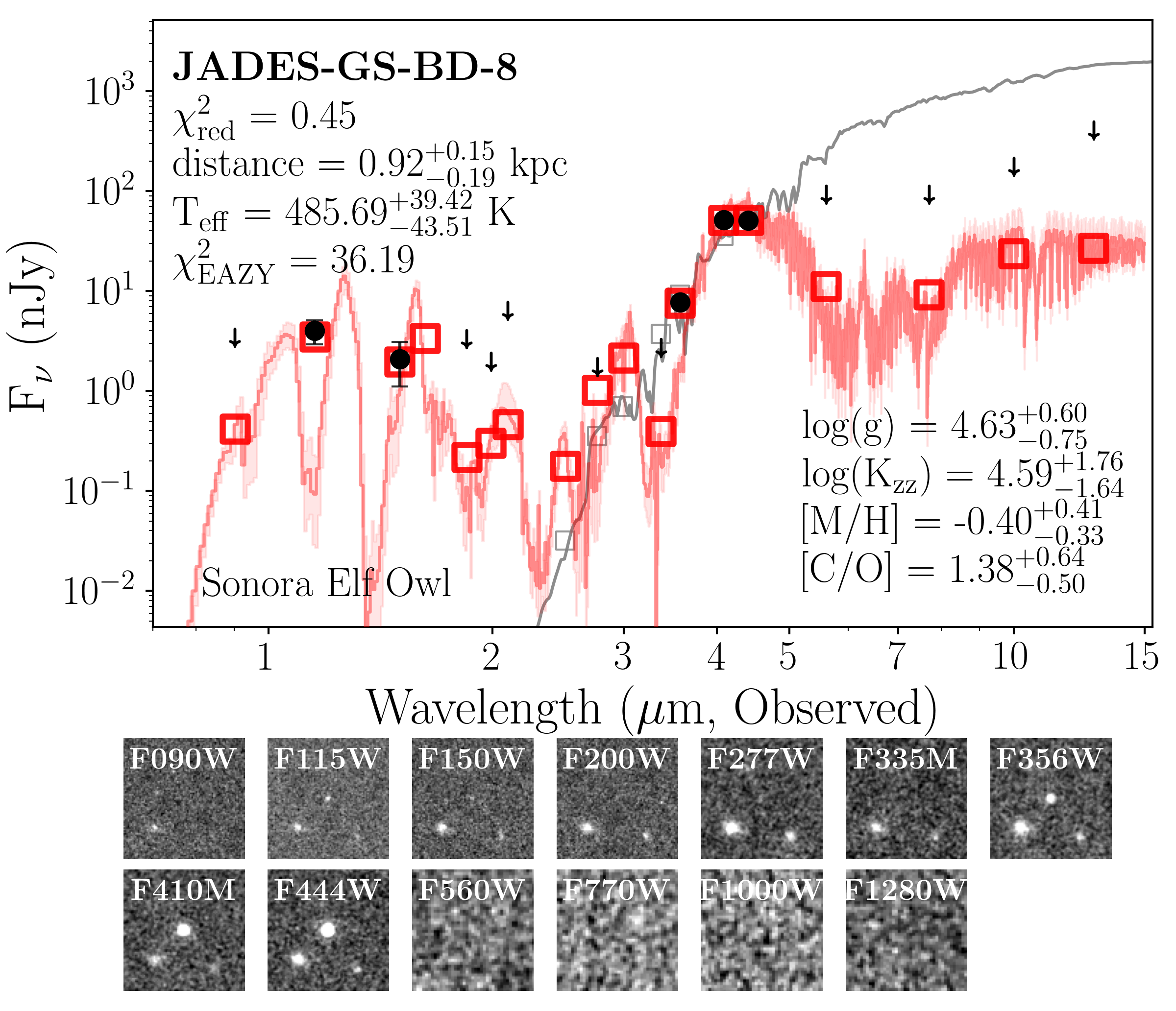}
  \includegraphics[width=0.32\textwidth]{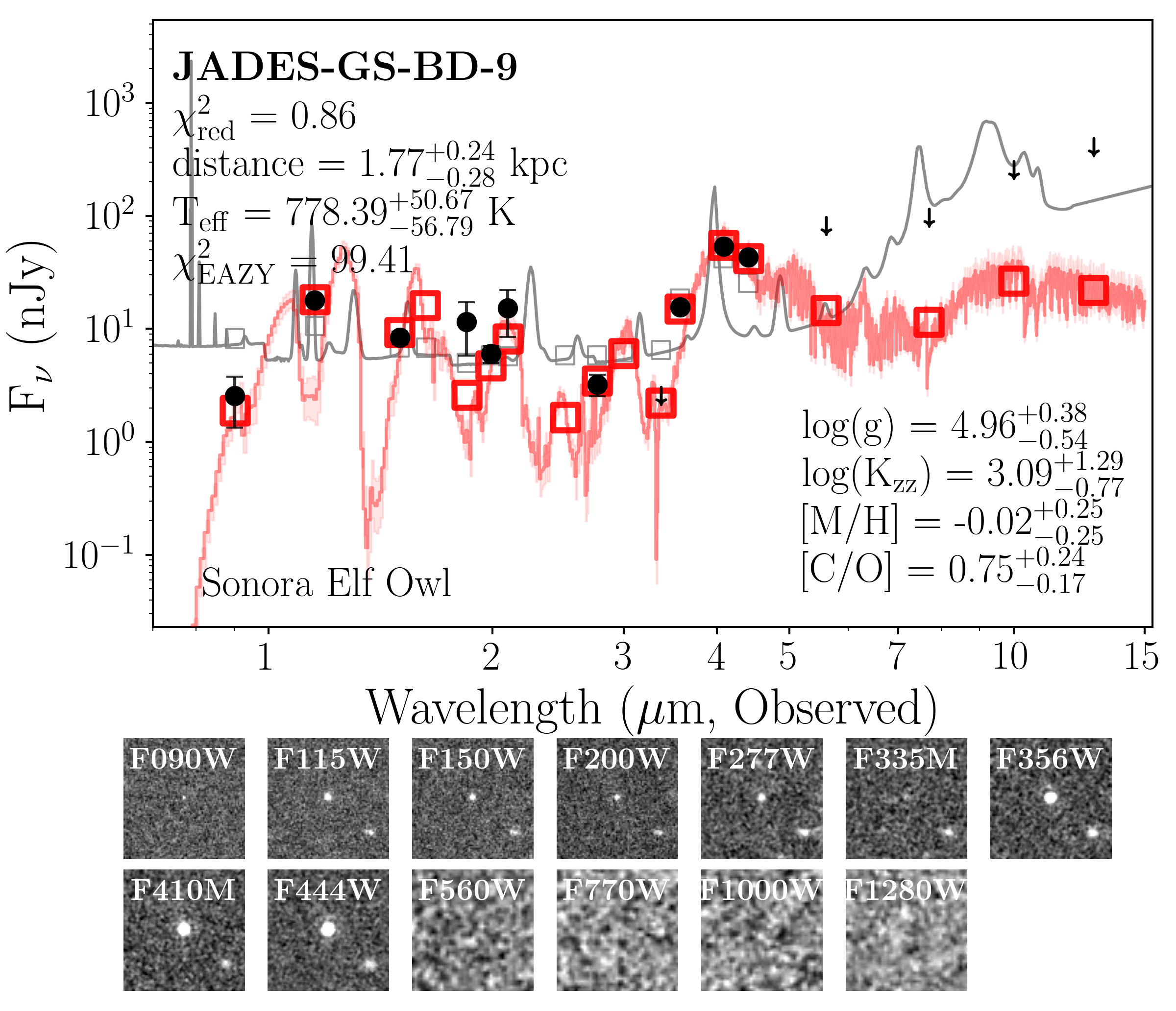}
  \includegraphics[width=0.32\textwidth]{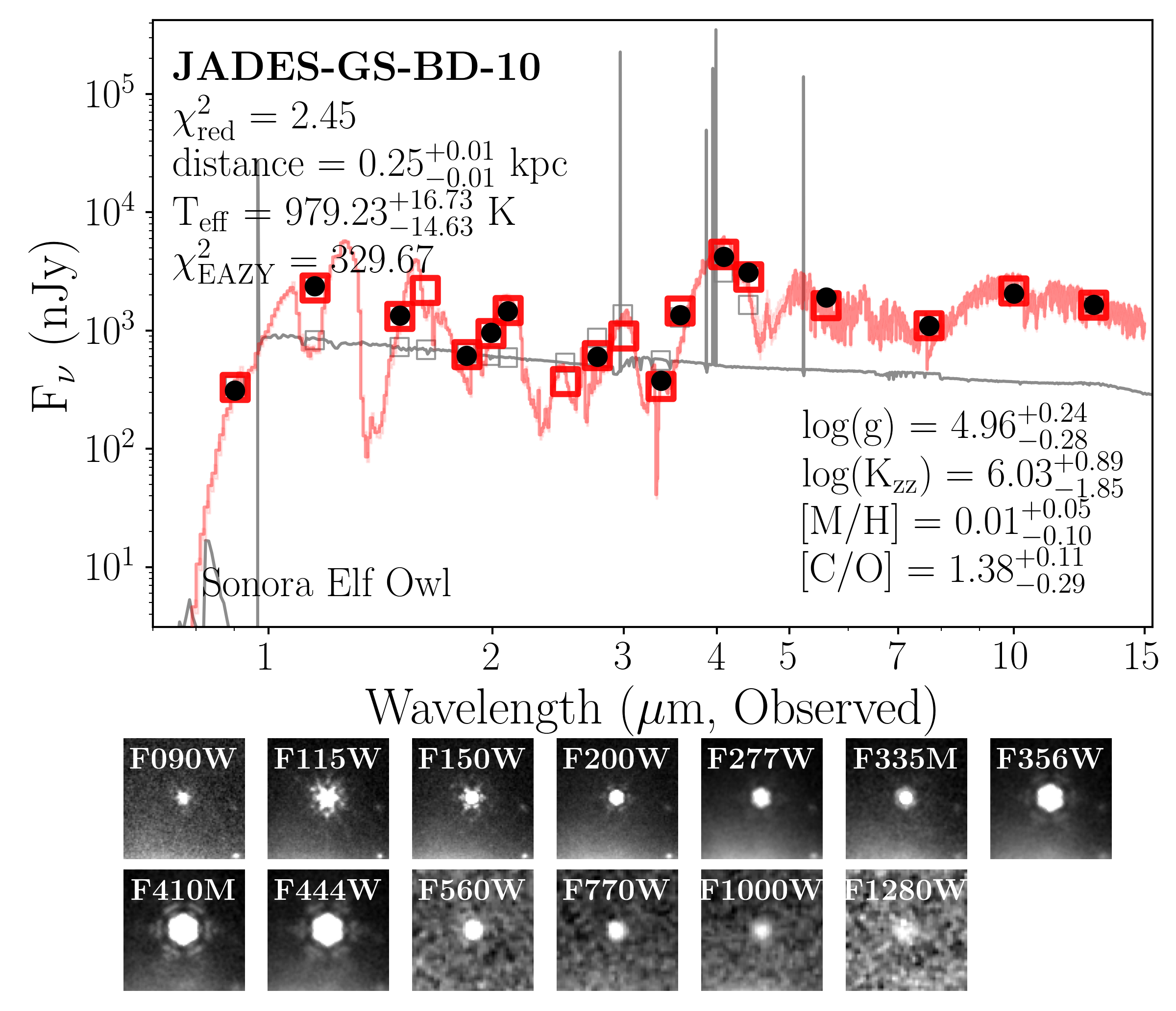}
  \includegraphics[width=0.32\textwidth]{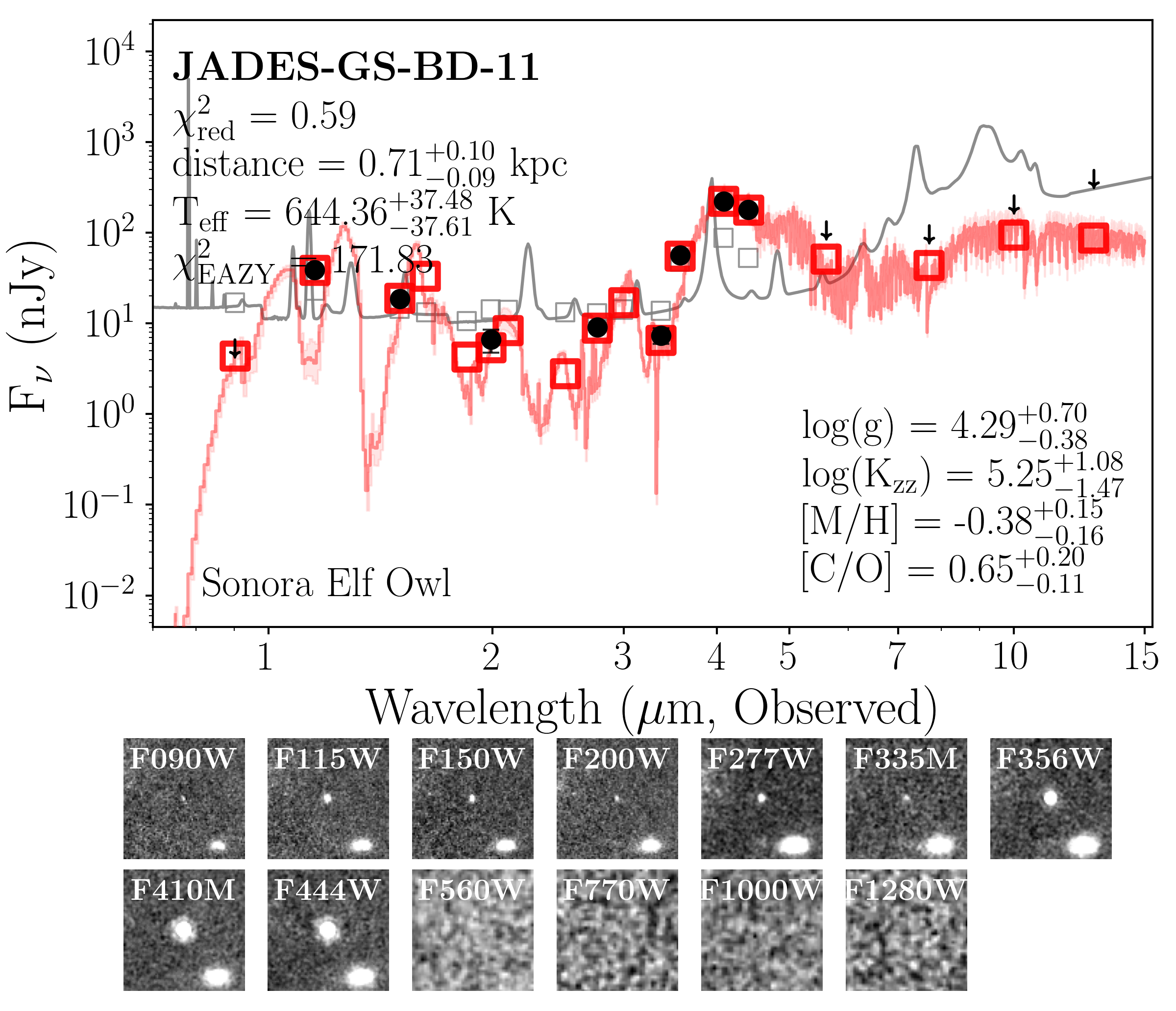}
  \includegraphics[width=0.32\textwidth]{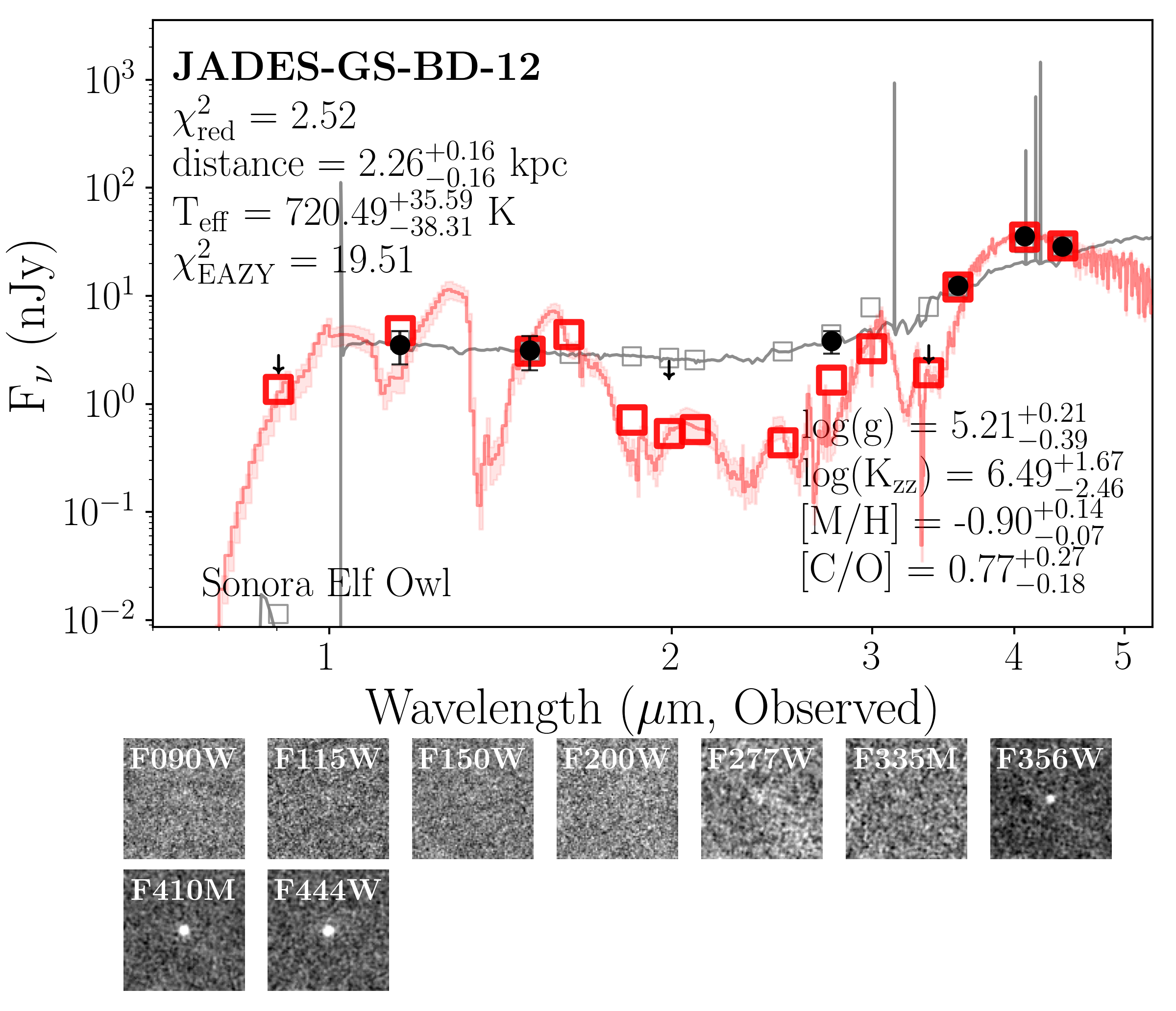}
  \caption{SED plots for the brown dwarf and brown dwarf candidates in our sample (black points in each panel), along with the Sonora Elf Owl fits and 1$\sigma$ confidence interval from \texttt{NIFTY} (red region), with the median \texttt{NIFTY} fit photometry plotted as red squares. In each panel, we indicate 2$\sigma$ upper limits with downward facing arrows. We also provide the Sonora Elf Owl model parameters in each panel, and below each SED we plot $2^{\prime\prime} \times 2^{\prime\prime}$ thumbnails for the majority of the filters that were used to observe these sources. This figure is continued in Figures \ref{fig:SED_plots_pt2}, \ref{fig:SED_plots_pt3}, and \ref{fig:SED_plots_pt4} in Appendix \ref{sec:appendix_parameters}.
  \label{fig:SED_plots_pt1}}
\end{figure*}

The \texttt{NIFTY} fits to our brown dwarf and brown dwarf candidates provide constraints on the effective temperature, specific gravity, metallicity, and distances. From the Sonora Elf Owl fits, we find ten (nearly a quarter of the entire sample) sources with effective temperatures consistent with them being Y dwarfs at T$_{\mathrm{eff}} < 500$K: JADES-GS-BD-4, JADES-GS-BD-5, JADES-GS-BD-8, JADES-GS-BD-18, JADES-GS-BD-21, JADES-GN-BD-5, JADES-GN-BD-8, JADES-GN-BD-9, JADES-GN-BD-11, and JADES-GN-BD-18. There are a few additional sources with fit effective temperatures within $1\sigma$ of 500K. The majority of the sources have effective temperatures from \texttt{NIFTY} fits with the Sonora Elf Owl models between 500 - 1200 K, indicating that they are T dwarfs. In Appendix \ref{sec:appendix_parameters}, we provide the output \texttt{NIFTY} properties for all of the sources in Tables \ref{tab:bd_properties_gs_gn}, \ref{tab:bd_properties_gs_gn_pt2}, and \ref{tab:bd_properties_gs_gn_pt3}, and plot the SEDs and thumbnails for the sources in Figures \ref{fig:SED_plots_pt1}, and then in \ref{fig:SED_plots_pt2}, \ref{fig:SED_plots_pt3}, and \ref{fig:SED_plots_pt4} in Appendix \ref{sec:appendix_fluxes}. We can see that many of the Y dwarf candidates are only detected at 4 - 5$\mu$m, with 2$\sigma$ upper limits at shorter wavelengths consistent with the best-fit model fluxes. 

\begin{figure*}[!ht]
  \centering
  \includegraphics[width=0.98\textwidth]{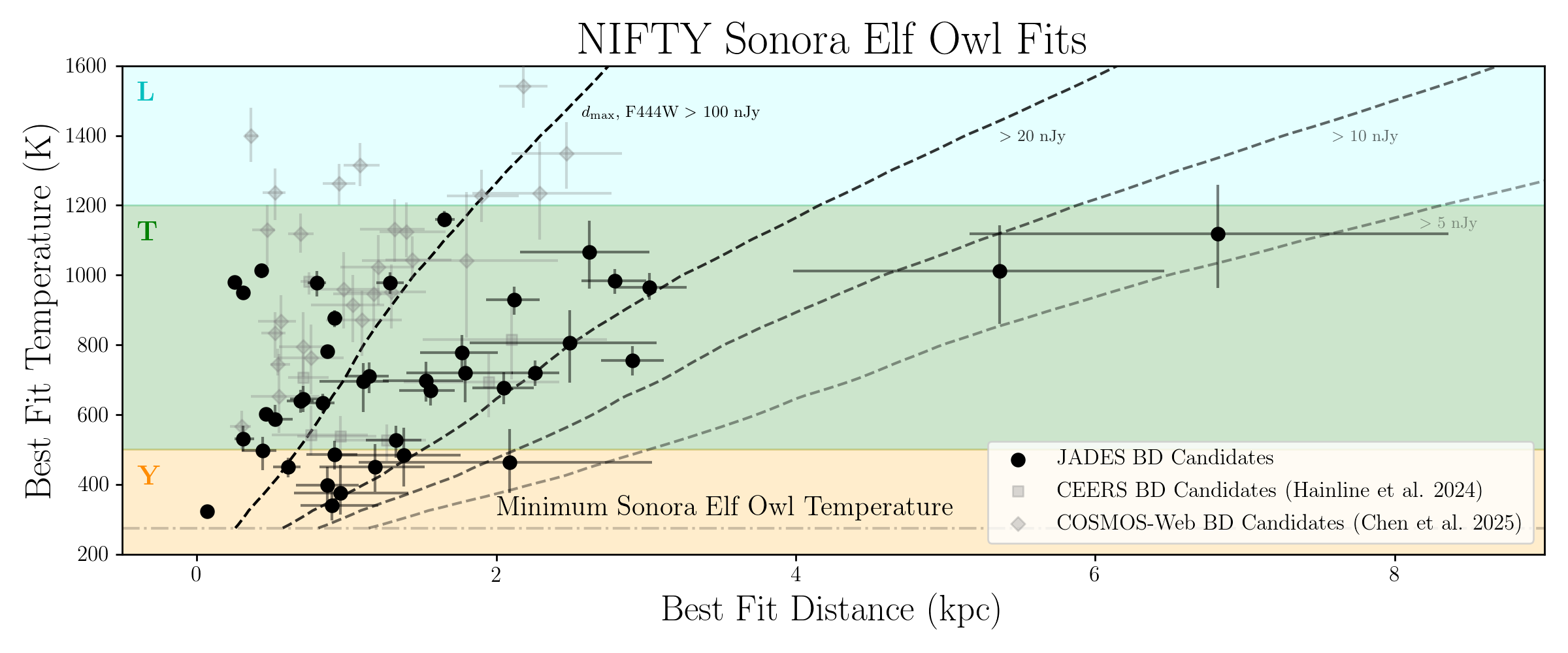}
  \caption{Best-fit effective temperature plotted against distance for the brown dwarf candidates selected in this work, derived using the Sonora Elf Owl models. The primary sample is plotted with black circles. In the background, we color the plot in three ranges, showing rough temperatures for L, T, and Y-dwarfs. Additionally, we plot the CEERS brown dwarf candidates presented in \KBD{} with grey squares, and the COSMOS-Web brown dwarf candidates presented in \citet{chen2025} with grey diamonds. For both of these samples, we re-fit the measured photometry from those studies with \texttt{NIFTY} using the Sonora Elf Owl models. To aid in understanding the distribution of sources, in dashed lines we plot the maximum distance allowable for Sonora Elf Owl models with F444W flux = 100 nJy, 20 nJy, 10 nJy, and 5 nJy. We are able to detect Y-dwarfs out to $\sim 2$kpc with JADES, demonstrating the power of deep NIRCam imaging.  
  \label{fig:temperature_vs_distance}}
\end{figure*}

In Figure \ref{fig:temperature_vs_distance} we plot the best-fit effective temperature from \texttt{NIFTY} fits to the Sonora Elf Owl models against the estimated distances of the sources. Confirming prior work in \citet{langeroodi2023}, \citet{burgasser2023}, \KBD{}, and \citet{chen2025}, deep extragalactic JWST imaging has opened the door for the detection of brown dwarf candidates out to many kiloparsecs. To help understand the distribution of sources in Figure \ref{fig:temperature_vs_distance}, we calculated the maximum distance a model brown dwarf could be recovered given a temperature and a NIRCam F444W detection flux, which we plot with dashed lines in Figure \ref{fig:temperature_vs_distance}. From left to right, we can see the maximum distance we can detect a source depends on the temperature such that, for surveys with $5 - 10$ nJy flux limits, like JADES, Y dwarfs can be observed out to $2 - 3$ kpc, T dwarfs out to $6 - 7$ kpc, and L dwarfs at even farther distances. The distribution of the recovered JADES brown dwarfs and brown dwarf candidates roughly follows these curves, demonstrating that potentially more distant brown dwarfs exist within the field, but are currently too faint to be observed. 

\begin{figure}[!h]
  \centering
  \includegraphics[width=0.51\textwidth]{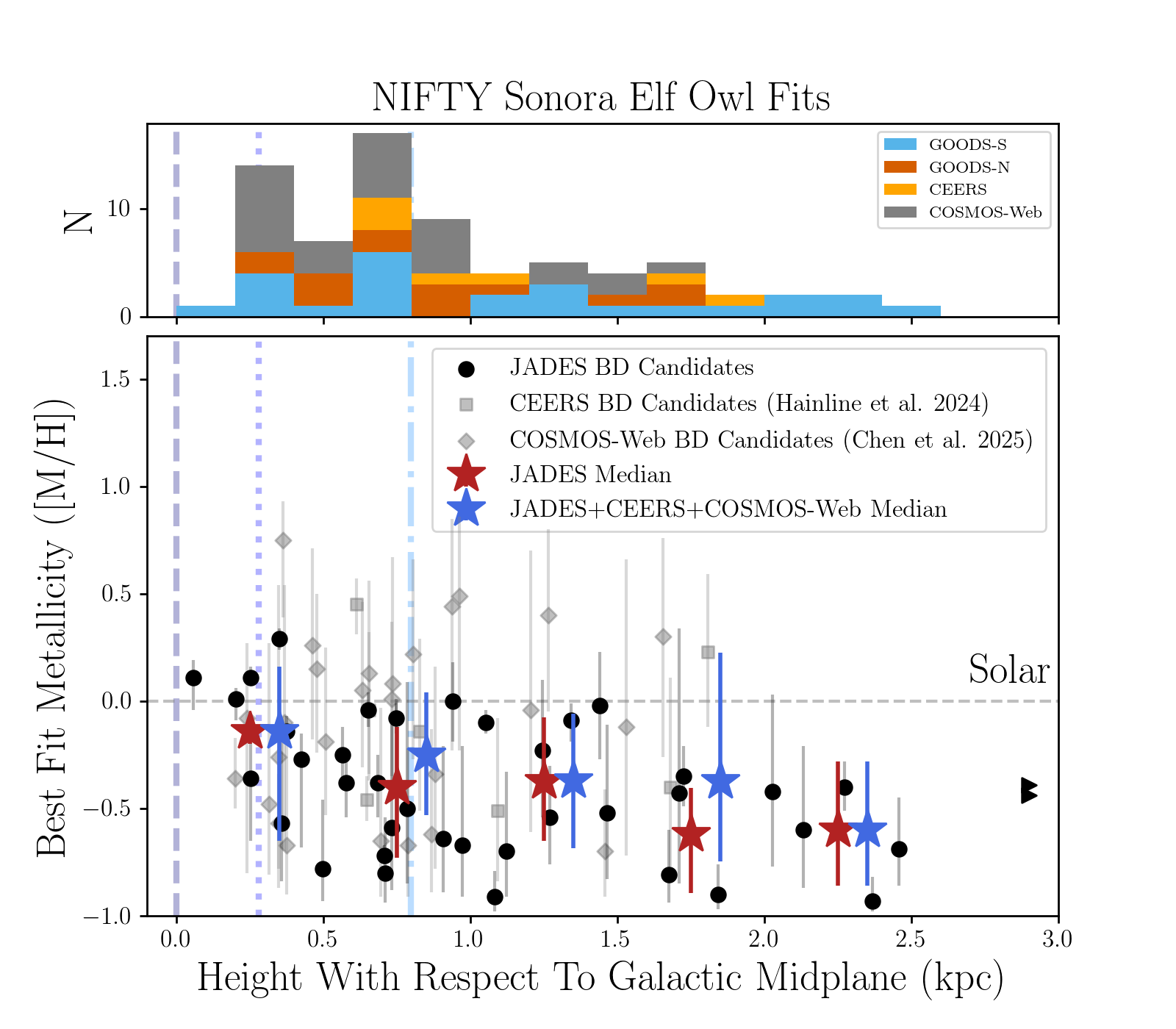}
  \caption{(Top) Stacked histogram of heights for the GOODS-S (blue), GOODS-N (red), CEERS (orange), and COSMOS-Web (grey) brown dwarfs and brown dwarf candidates above the Milky Way midplane. The bulk of the population has heights consistent with the thick disk, with a smaller number having heights that put them outside the thick disk, albeit with large uncertainties. (Bottom) Best-fit metallicity plotted against height for each source above the Milky Way midplane for the brown dwarf candidates selected in this work, derived using the Sonora Elf Owl models. The points are the same as what is shown in Figure \ref{fig:temperature_vs_distance}. We also plot median (and 1$\sigma$ ranges) metallicities in bins of height with the red (JADES alone) and blue (JADES combined with the CEERS and COSMOS-Web candidates) stars. At the right edge of the plot we show the two sources (JADES-GN-BD-6 and JADES-GN-BD-7) that lie at $> 4$ kpc in our modeling, indicating their metallicities. At larger heights above the midplane, the binned median of the metallicity declines. In both panels we plot with a dotted line the approximate scale height of the Milky Way thin disk (280 pc), and with a dot-dashed line the scale height of the thick disk (797 pc), from \citet{vieira2023}. 
  \label{fig:metallicity_vs_distance}}
\end{figure}

For comparison to literature samples, we also plot in Figure \ref{fig:temperature_vs_distance} brown dwarf candidates from both the CEERS and COSMOS-Web surveys. In order to accurately compare to the JADES sources, we re-derive temperatures and distances using \texttt{NIFTY} to fit the photometry (using the Sonora Elf Owl models) provided in \KBD{} (see Table 1 in that paper) for the CEERS sources and from \citet{chen2025} (see Table 2 in that paper) for the COSMOS-Web sources. The properties of these sources are plotted as grey symbols, and largely lie at higher temperatures, in line with the shallower observational depths of both the CEERS and COSMOS-Web surveys. The deep JADES photometry with multiple filters at $> 3\mu$m allows for the detection of faint Y-dwarf candidates, confirming the initial results from \KBD{}. In the top panel of Figure \ref{fig:metallicity_vs_distance}, we use the direction towards each extragalactic field and the \texttt{NIFTY} Sonora Elf Owl best-fit distances to plot the distribution of all of the sources of their heights with respect to the Milky Way midplane, with the thin disk and thick disk scale heights indicated, as adopted from \citet{vieira2023}. While the bulk of the sources have distances consistent with those in the Milky Way thick disk, there is a large distribution of distances associated with both the Milky Way thick disk, thin disk, and halo populations \citep{juric2008}, and we caution that distance alone cannot uniquely determine to which part of the Milky Way each source originates from. In \citet{burgasser2023}, the authors use a model for the Milky Way to compare their UNCOVER spectroscopically-confirmed brown dwarf distances to the thin disk, thick disk, and halo populations, finding that the fraction of thick disk sources exceeds thin disk sources at $\sim 1.4$ kpc, while the halo sources only outnumber thick disk sources at $\sim 6$ kpc. While we will discuss both metallicity below and proper motions for these sources in Section \ref{sec:proper_motions}, we caution that these objects, especially those that have smaller scale heights, could belong to the thin disk. 

It might be expected that, for more distant brown dwarfs, if they originate from the thick disk and halo of the Milky Way, they would have lower metallicities \citep{gilmore1995, cheng2012, burgasser2005, meisner2021}. To explore this, in the bottom panel of Figure \ref{fig:metallicity_vs_distance} we plot the same sources as in Figure \ref{fig:temperature_vs_distance}, showing the best-fit metallicity [M/H] against the height above the Milky Way midplane from \texttt{NIFTY} fits using the Sonora Elf Owl model. We bin these sources by height and plot median values and 1$\sigma$ distributions within the bins for both the JADES and JADES+CEERS+COSMOS-Web samples. From the points and the binned median values, we notice that the metallicities of the sources become more sub-solar (negative) at larger heights above the Milky Way midplane, although with considerable noise introduced by the difficulty in measuring precise metallicity values from photometric fits. Above the thick disk scale height ($\sim 1.4$ kpc), the JADES brown dwarf candidates are entirely sub-solar. 

Brown dwarf atmospheric metallicity is difficult to estimate from photometry alone, and there is a concern that perhaps the trend seen in Figure \ref{fig:metallicity_vs_distance} may be a result of an observational bias. To explore this, we created a sample of 100,000 simulated  brown dwarfs using the Sonora Elf Owl models where we uniformly sampled from the effective temperature, metallicity, log(g), [M/H], and C/O values (see Table \ref{tab:nifty_parameters}), and placed them between 0 and 4 kpc from the Earth, in line with the bulk of the sources in our sample. We then added noise to the resulting flux values in line with the observed noise properties in JADES, and explored which sources were selected using the criteria discussed in Section \ref{sec:selecting_brown_dwarfs}. When we looked at the fraction of sources selected as a function of simulated height above the galactic midplane and [M/H], we do see a bias towards selecting \textit{lower} metallicity sources at \textit{larger} heights above the midplane. Only 31\% of the full sample of simulated sources with heights of $> 2$ kpc above the midplane and super-solar metallicities are selected by our color and SNR criteria, compared to the 49\% of simulated sources at \textit{sub}-solar metallicity. At lower heights, we find that 52\% of simulated sources with super-solar metallicity at $< 2$ kpc are selected, and 70\% of simulated sources at sub-solar metallicity are selected. These results and conclusions do persist out to larger distances, but the fraction of sources selected is significantly lower due to them being quite faint. 

This observational bias is predominantly driven by the fact that, at larger simulated heights, sources with super-solar metallicity are more likely to have a faint F444W fluxes with SNR $ < 5$. At 4 - 5$\mu$m, an  opacity window primarily from \ce{H2O} allows a view deeper into the atmosphere of a brown dwarf. At super-solar metallicities, increased opacity due to the effects of \ce{H2O}, CO, and \ce{CH4} will lower the observed F444W flux. This can be seen in the grey lines in Figure \ref{fig:BD_Color_Color_All}, where the F277W - F444W color becomes significantly bluer for super-solar metallicity brown dwarfs at a given effective temperature. If there are distant brown dwarfs at super-solar metallicities in deep extragalactic datasets, they would require increased F444W observational depth to detect them in our sample.

As an additional test of whether fits to the near-to-mid infrared photometry with \texttt{NIFTY} are biased towards subsolar metallicities, we also adapted the code to fit a sample of local sources with observations compiled in the \textit{UltraCoolSheet}\footnote{\url{http://bit.ly/UltracoolSheet}, \citet{ultracoolsheet_doi}}. We selected a sample 25 sources with IR spectral types T0 to T8 (T$_{\mathrm{eff}} = 500 - 1400$K), and fit a combination of Pan-STARRS $z$ and $y$, 2MASS $J$, $H$, and $Ks$, Mauna Kea Observatory synthetic $Y$, $J$, $H$, and $K$, WISE $W1$, and $W2$, and Spitzer IRAC $Ch1$ and $Ch2$ photometry using the Sonora Elf Owl models. The resulting fits have similar effective temperatures to the ``teff\_atmo'' values derived in \citep{sanghi2023} (a median offset of $\Delta$T$_{\mathrm{eff}} = 54$K, with a scatter of 124 K) and distances as compared to the ``dist\_formula'' values reported in the \textit{UltraCoolSheet} (a median offset of $\Delta d  = 0$ pc, with a scatter of 8 pc). As these sources are all within 70 pc, they should primarily have solar or super-solar metallicities, and we find that eighteen of the twenty-five sources (72\%) have best-fit [M/H] $> 0$ from \texttt{NIFTY}, with only five sources in this subsample (20\%) having significant ($>1\sigma$) sub-solar metallicities, all with [M/H] $> -0.4$. These results support the observed subsolar metallicities for our primary sample as shown in Figure \ref{fig:metallicity_vs_distance}. 

\begin{figure*}
  \centering
  \includegraphics[width=0.98\textwidth]{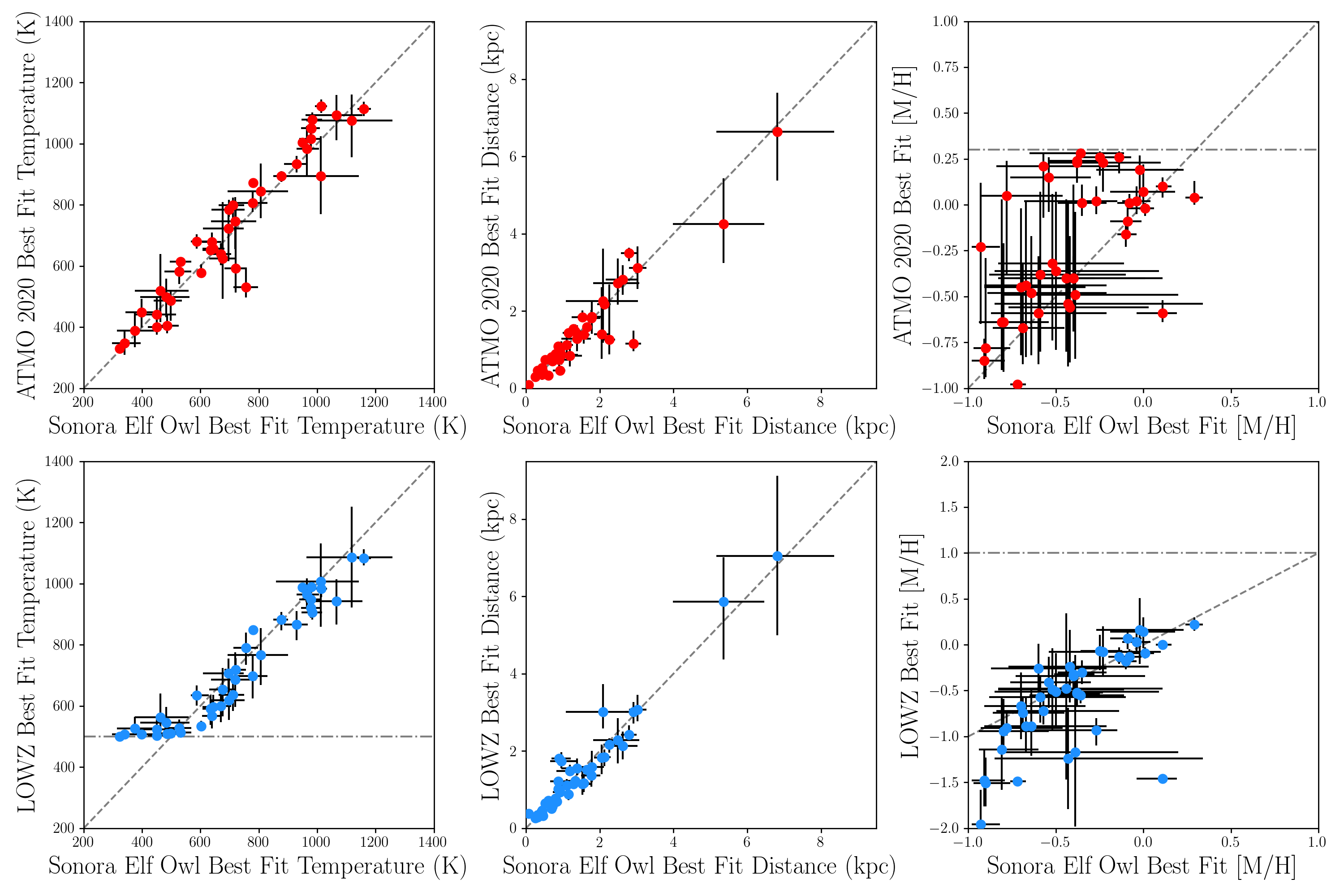}
  \caption{Comparison between \texttt{NIFTY} output parameters from fits to the brown dwarf and brown dwarf candidates. In the top row we plot, from left to right, the Temperature, Distance, and [M/H] comparison between the ATMO2020++ (y-axis) and Sonora Elf Owl (x-axis) fits, with red points. In the bottom row we plot the same comparisons, but for the LOWZ models (y-axis) and the Sonora Elf Owl (x-axis) models, in blue points. We use a dashed line to show a one-to-one relationship in each panel, and a dot-dashed line shows the minimum model temperature of 500 K for the LOWZ models in the bottom-left panel, and a maximum model metallicity of [M/H]$ = 0.3$ for the ATMO2020++ models in the top right, and [M/H]$ = 1.0$ for the LOWZ models in the bottom right. While the fit temperatures and distances are largely similar between the models, the metallicities have larger uncertainties and generally show a more scattered comparison between the model fits to the photometry. The distances derived from the ATMO2020++ models are slightly larger than what are derived from the fits to the Sonora Elf Owl models, but consistent within the uncertainties. The LOWZ models allow for lower [M/H] values, and while the Sonora Elf Owl and LOWZ models agree, there are a number that have LOWZ best-fit models at lower metallicities than those derived using the Sonora Elf Owl models. 
  \label{fig:model_comparison}}
\end{figure*}

To further explore how the \texttt{NIFTY} fits compare for the three different models we adopt for our brown dwarfs and brown dwarf candidates, we plot the temperature, distance, and metallicity [M/H] comparison both between the ATMO2020++ and Sonora Elf Owl, and the LOWZ and Sonora Elf Owl models in Figure \ref{fig:model_comparison}. In each panel we show a one-to-one relation with a dashed line, and in three of the panels we show the minimum or maximum values for that particular model (if they exist within the plot area, see Table \ref{tab:nifty_parameters}) with a dot-dashed line. The left column shows a comparison of the best-fit effective temperatures, and outside of how the LOWZ models are limited to T-dwarfs (T$_{\mathrm{eff}} > 500$ K), the fit values are very similar across all three model sets, with the majority being consistent within 1$\sigma$. There is a trend such that the ATMO2020++ model fits result in slightly larger effective temperatures (by 3\%), and the LOWZ model fits result in slightly smaller effective temperatures (by 1\%) than the Sonora Elf Owl fits. In the middle column we plot a comparison of the distances between the model fits, which are similar within 1$\sigma$ for all three model sets. Distance is derived from the normalization of each model to the photometry, 

In the right panel of Figure \ref{fig:model_comparison} we plot the \texttt{NIFTY} best-fit [M/H] parameter comparison between the models. Recovering accurate metallicity values from limited photometric data is difficult, and this is reflected in the large uncertainties and the large spread in the comparison between the values from each model. In general, the ATMO2020++ metallicities are higher (70\% have higher metallicity values), with more scatter, than those predicted from Sonora Elf Owl. The relationship between the LOWZ and Sonora Elf Owl models is significantly tighter, although the larger model range explored by the LOWZ models allows for individual sources to be fit at significantly sub-solar metallicities. From the fits, 61\% of our candidates have LOWZ metallicites below what is predicted from Sonora Elf Owl, and 8 sources (20\%) have LOWZ fits with [M/H] $< -1.0$, the lowest metallicity probed by the Sonora Elf Owl models.

\subsection{A Comparison of Brown Dwarf Candidate Photometry to Extragalactic Models}\label{subsec:extragalaactic_fits}

While Figure \ref{fig:BD_Color_Color_All} demonstrates that brown dwarfs have dissimilar colors from the bulk of the underlying population of galaxies within the JADES catalogs from which these sources were chosen, it would be helpful to explore whether the colors and fluxes for these sources can be replicated with galaxy templates. To that end, we re-fit the observed NIRCam photometry using \texttt{EAZY} \citep{brammer2008}, which uses $\chi^2$ minimization to find a best-fit combination of templates and redshift for sources. We follow the procedure outlined in \citet{hainline2023}, \citet{robertson2026}, and  \citet{hainline2026}, fitting each source with a suite of 16 templates that span a wide variety of colors and have been shown to accurately fit JADES galaxies, and we adopt the minimum $\chi^2$ fit to calculate the \texttt{EAZY} photometric redshift. Because the \texttt{EAZY} templates do not include a model for mid-IR dust emission, we do not include the MIRI photometry in the fit, and we re-fit our brown dwarf candidates without the MIRI photometry for an accurate comparison. As discussed in Section \ref{sec:nifty}, we do not use the HST fluxes in either fit. In Figures \ref{fig:example_NIFTY_fit_44120},  \ref{fig:SED_plots_pt1} and \ref{fig:SED_plots_pt2} - \ref{fig:SED_plots_pt4}, we include the minimum $\chi^2$ \texttt{EAZY} fit with a thin grey line, along with the model \texttt{EAZY} photometry plotted with grey squares. 

We compared both the resulting raw $\chi^2$ fluxes for both the \texttt{NIFTY} and \texttt{EAZY} fits to each source as well as the Bayesian Information Criteria (BIC), which folds in the number of free parameters and number of data points used in each fit. The majority of the GOODS-S and GOODS-N JADES candidates are better fit with \texttt{NIFTY}. The fraction of sources with a lower \texttt{NIFTY} $\chi^2$  is $90\%$, and $73\%$ of the sources have have $\Delta \chi^2 > 10$. Similarly, 80\% of the full brown dwarf sample have positive $\Delta$BIC $>2$ in favor of the \texttt{NIFTY} fit, 66\% of the sample has $\Delta$BIC $>6$, and 63\% has $\Delta$BIC $>10$. When we look at the sources with worse \texttt{NIFTY} fits than \texttt{EAZY}, they fall into two broad categories. The first are sources with \texttt{EAZY} $z_{\mathrm{phot}} < 0.3$ (two sources, JADES-GN-BD-7 and PANO-GN-BD-13), which we can reject owing to the non-detections in the observed HST bands for both, in contrast with the prediction from the \texttt{EAZY} fit. The second class are sources where $z_{\mathrm{phot}} > 10$ (three sources, JADES-GS-BD-4, JADES-GN-BD-11, and JADES-GN-BD-16). All three of these sources are only significantly detected at $> 3\mu$m, so the comparison between \texttt{NIFTY} and \texttt{EAZY} fits are only driven by three or four data points. Notably, as can be seen from the \texttt{EAZY} fits, the resulting combined-template fits are extremely red to the point of being unphysical. We discuss a similar source from the literature, Capotauro \citep{gandolfi2025b}, at length in Section \ref{subsec:capotauro}. The final sample consists of sources with $z_{\mathrm{phot}} > 7.5$ (two objects, JADES-GS-BD-20 and JADES-GN-BD-4), where the source is best fit as an LRD with strong optical emission from [\ion{O}{3}]$\lambda\lambda$4959,5007 and H$\beta$ boosting the F410M and F444W fluxes. In both of these sources, the flux at $< 2.5\mu$m is very faint, and the \texttt{NIFTY} fits underpredict the F277W flux. Fully disentangling which sources are brown dwarfs and which are galaxies, in the absence of proper motions, will require a deep spectroscopic campaign. In Section \ref{subsec:tentative_sources}, we will further discuss a subsample of additional sources that exist at the boundary between brown dwarf and galaxy colors. 

\subsection{Notes on Individual Sources}\label{subsec:individual_sources}

In this section, we discuss a subsample of eight notable brown dwarfs and brown dwarf candidates in more detail in order to explore source environment, indicate if spectra exist for any object, and describe sources at temperature and distance extrema. 

\textbf{JADES-GS-BD-5}: This brown dwarf, which was first described in \KBD{}, is the lowest-temperature source across the full sample, with an estimated \teff{}$ = 322^{+5}_{-7}$ K from the \texttt{NIFTY} fit with the Sonora Elf Owl models, and $ 330^{+7}_{-5}$ K from the ATMO2020++ fits (the LOWZ models do not go below 500 K). Owing to its brightness and distance (only 70 pc), it is significantly detected in all four MIRI photometric bands we use, with an excellent fit ($\chi^2_{\mathrm{red}} = 5.17$, slightly increased due to the high photometric SNR). This source was targeted with JWST/NIRSpec prism-resolution spectroscopy as part of JWST GO 5997 OASIS: Observing All phases of StochastIc Star formation. The reduction and analysis of this spectrum will be a part of an upcoming study (Hainline et al. in prep).

\textbf{JADES-GS-BD-7}: This source from \KBD{} has very consistent results from the three models used in \texttt{NIFTY}, although the predicted ATMO2020++ distance is $1.3^{+0.05}_{-0.04}$ kpc, significantly farther than the 0.9 kpc from the Sonora Elf Owl and LOWZ models. JADES-GS-BD-7 was initially found by \citet{oesch2012a} and \citet{lorenzoni2013}, the former study claimed it was likely a brown dwarf from the observed colors. It was a target for NIRSpec prism spectroscopy as part of the JADES ``Deep-HST'' campaign \citep{bunker2024, deugenio2025}, but it had moved off of the NIRSpec Multi-Shutter Array (MSA) slit placed at the location from the initial HST observation \citep[see Figure C3 in][who first reported the proper motion]{bunker2024}. However, the source is sufficiently bright enough that the reduced spectrum from the JADES data release does show flux in agreement with the observed photometry confirming that the object is a brown dwarf. Fits to the spectroscopic flux for this source will be part of a forthcoming paper. 

\textbf{JADES-GS-BD-9}: This source was spectroscopically confirmed as a 800-900 K (T5 - T6) dwarf at a distance of 1.8 - 2.3 kpc in \citet{hainline2024b}, where the authors also discovered proper motion for this source that helped support the idea that the object was a member of the Milky Way halo. The \texttt{NIFTY} photometric fits we performed are largely in agreement with the results from fits (using a $\chi^2$ minimization algorithm) to the NIRSpec spectrum in that paper, although the \texttt{NIFTY} distances (1.3 - 1.8 kpc depending on the model used) are lower than what was predicted for the source from the spectrum (1.8 - 2.2 kpc); this could be driven by the slit loss correction used in normalizing the NIRSpec spectrum. 

\textbf{JADES-GS-BD-10}: This source, explored in \KBD{}, was first identified as a brown dwarf from HST data in \citet{stanway2003} (as object SBM03\#5), and then discussed again in \citet{bunker2004} (as object 2140). It has an excellent Sonora Elf Owl \texttt{NIFTY} fit to the data ($\chi^2_{\mathrm{red}} = 2.45$) with high SNR detections in MIRI F560W, F770W, F1000W, and F1280W, indicating an effective temperature of \teff{} = $979^{+16}_{-15}$ K, and a distance of only $250 \pm 1$ pc. 

\textbf{JADES-GS-BD-14}: This source has the second-brightest F444W flux ($F_{\nu,\mathrm{F444W}} = 1994 \pm 2$ nJy) in the sample. While JADES-GS-BD-14 is not in the GOODS-S SMILES footprint, it sits in a JADES parallel region with deep MIRI F770W and F1280W coverage. The F1280W flux ($F_{\nu,\mathrm{F1280W}} = 3851 \pm 352$ nJy) is quite bright, and notably, it is almost four times brighter than the predicted \texttt{NIFTY} fit flux at 12.8$\mu$m. The F770W flux is consistent with the prediction, however, indicating that the excess of flux is only apparent at longer wavelengths. None of the other brown dwarfs and candidates with MIRI detections have such a discrepant MIRI flux between the observations and the model, and it is unclear what might be causing the observed flux excess. In some brown dwarfs, this excess could be a sign of warm dust emission from disks around the dwarf \citep{riaz2006, morrow2008}, an effect also seen in free-floating planetary mass objects \citep{damian2025}. This flux excess may reflect a limitation of the models to accurately account for the full range of colors for these sources at long wavelengths. Future MIRI spectroscopic observations of this and the other MIRI detected sources in our sample would be of interest in order to strengthen our models and understand the emission from these sources at longer wavelengths. 

\textbf{JADES-GS-BD-17}: This source has boosted F090W, F200W, F277W, and F335M flux as compared to the \texttt{NIFTY} Sonora Elf Owl fit, and we measure a compactness criterion value (see Section \ref{sec:selecting_brown_dwarfs}) for the source of 1.3. This object appears to be a brown dwarf candidate in front of a background galaxy that contaminates the source fluxes through our circular apertures (see the thumbnails in Figure \ref{fig:SED_plots_pt2}). Owing to the potential contamination, the photometry of this source is consistent with being a galaxy at $z \sim 7.5$, where [\ion{O}{3}]$\lambda\lambda 4959,5007$ nebular line emission would be boosting the flux in the F410M and F444W filters, and Lyman-$\alpha$ emission would be boosting F115W. If we assume this source were a distant galaxy, we can estimate a UV slope by fitting the F150W, F200W, and F277W $0.2^{\prime\prime}$ diameter circular aperture photometry of $\beta_{\mathrm{UV}} = -2.49 \pm 0.19$. This value is significantly bluer than the bulk of the sources in this redshift range within JADES \citep[$\beta_{\mathrm{UV,med}} = -2.16^{+0.06}_{-0.08}$ for sources at $z_{\mathrm{med}} = 7.57$, ][]{topping2023}. With this in mind, we choose to include JADES-GS-BD-17 in our final sample because of the quality of the fit to the \texttt{NIFTY} models. 

\textbf{JADES-GN-BD-3}: This source, which was first discovered in \KBD{}, was found to have a high proper motion between the HST/WFC3 and JWST/NIRCam imaging, which we re-measure from the JADES DR5 mosaics: $0.070^{\prime\prime}$/yr $\pm 0.005^{\prime\prime}$/yr. In JADES DR5, updated parallel imaging from the PANORAMIC survey \citep{williams2025} is included to produce the final stacked mosaic. Because this PANORAMIC data was taken 2 years after the original JADES imaging in GOODS-N, the source has moved, resulting in two peaks in the final mosaic. We measure a difference between the original JADES position (from F410M) and the PANORAMIC position (from F444W) of $0.084^{\prime\prime} \pm 0.004^{\prime\prime}$, which results in a proper motion of $0.044^{\prime\prime}$/yr $\pm 0.002^{\prime\prime}$/yr, smaller than the value from \KBD{}, but broadly consistent, given the extremely small separation between the JADES and PANORAMIC positions, and the uncertainty with which the exact date of the HST/WFC3 observation can be ascertained (see Section \ref{sec:proper_motions}). 

\textbf{JADES-GN-BD-6} and \textbf{JADES-GN-BD-7}: These two brown dwarf candidates are the farthest from the Sun in the sample, both with estimated \texttt{NIFTY} Sonora Elf Owl distances of $\sim 5-6$ kpc, consistent with each other within the large distance uncertainties. While the \texttt{NIFTY} fit for JADES-GN-BD-6 has $\chi^2_{\mathrm{red}} = 0.38$, the fit for JADES-GN-BD-7 is worse, with $\chi^2_{\mathrm{red}} = 3.25$, primarily due to the boosted F277W and F335M fluxes compared to the model. Notably, however, the two sources are only 11.8$^{\prime\prime}$ from each other on the sky, which, at 5 (6) kpc, corresponds to only 0.286 (0.343) pc, or 59,000 (70,748) AU. Additionally, JADES-GN-BD-6 is 3.8$^{\prime\prime}$ removed from a third source, JADES ID 1148239, also with fluxes and colors consistent with a brown dwarf (\texttt{NIFTY} Sonora Elf Owl distance of $7.0\pm1.2$ kpc), although this object is significantly fainter than the neighbors and we removed it from our primary brown dwarf candidate sample, which we will discuss in more detail in Section \ref{subsec:tentative_sources}. 

These two brown dwarf candidates are outliers compared to the rest of the sources in the sample, and there is some concern that they may be galaxies. Following the analysis we presented for JADES-GS-BD-17 previously, we can estimate a UV slope for JADES-GN-BD-6 by first assuming a photometric redshift ($z_{\mathrm{phot}}$ = 7.46), and then fitting the F150W, F200W, and F277W $0.2^{\prime\prime}$ diameter circular aperture photometry for the source. We estimate a slope of $\beta_{\mathrm{UV}} = -3.78 \pm 0.35$, which is significantly bluer than the bulk of the population of galaxies at $z = 7 - 8$ \citep{topping2023}. JADES-GN-BD-7, on the other hand, is too faint at $1 \sim 3\mu$m to properly make the same estimation. At such faint observed fluxes, spectroscopic confirmation for distant sources such as JADES-GN-BD-6 and JADES-GN-BD-7 will be challenging, but crucial for disentangling brown dwarfs and high-redshift galaxies.

\subsection{Proper Motions}\label{sec:proper_motions}

\begin{figure*}
  \centering
  \includegraphics[width=0.98\textwidth]{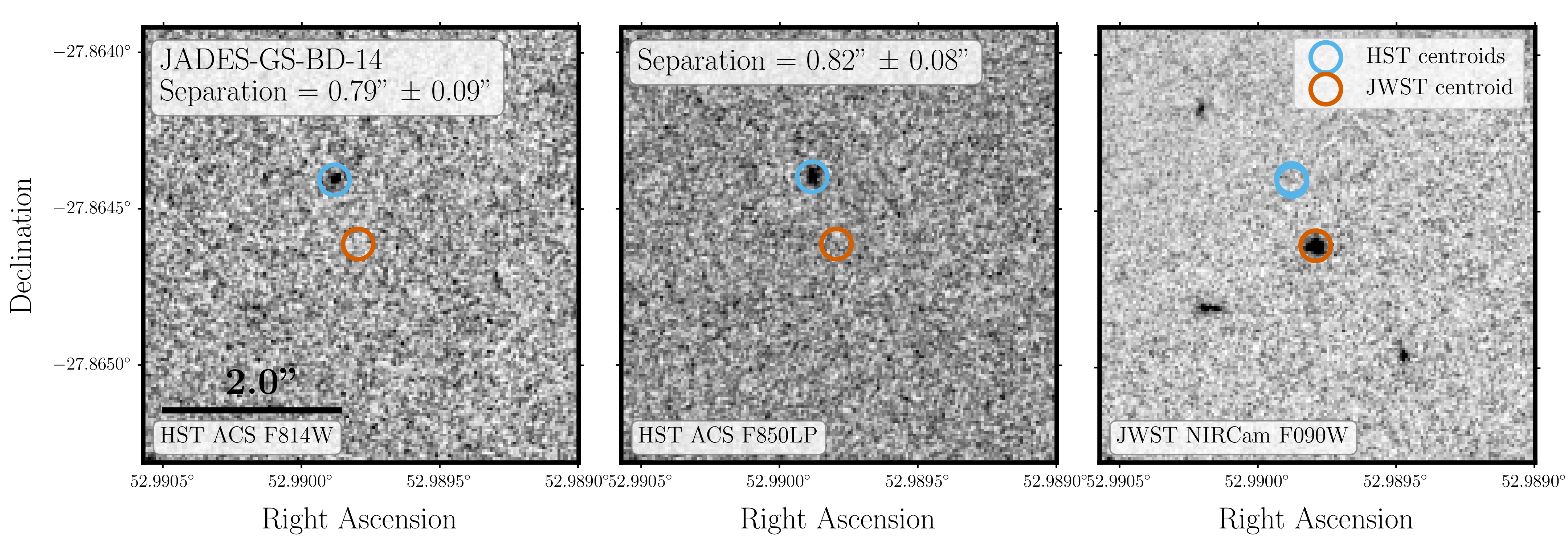}
  \includegraphics[width=0.49\textwidth]{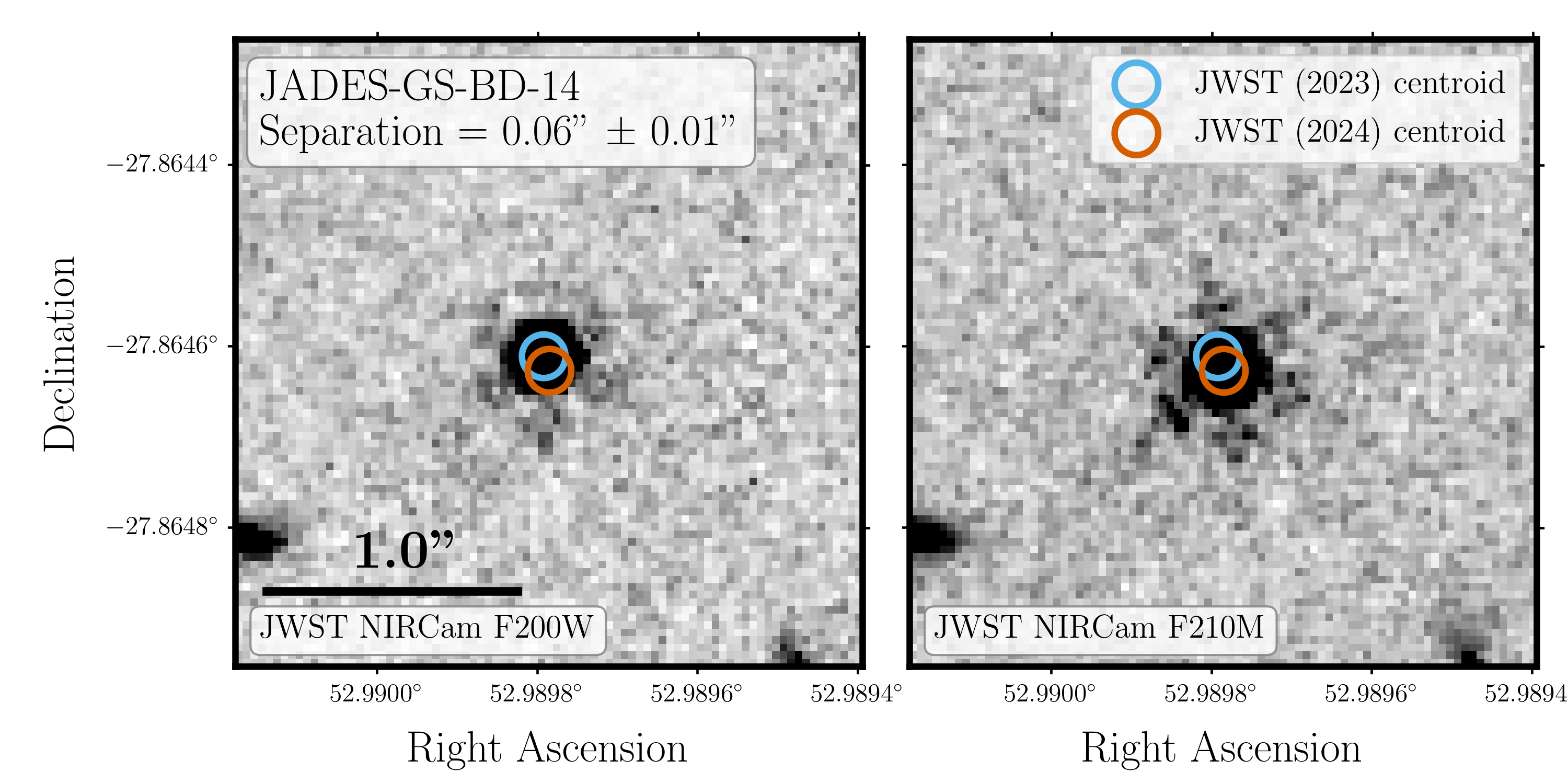}  
  \includegraphics[width=0.49\textwidth]{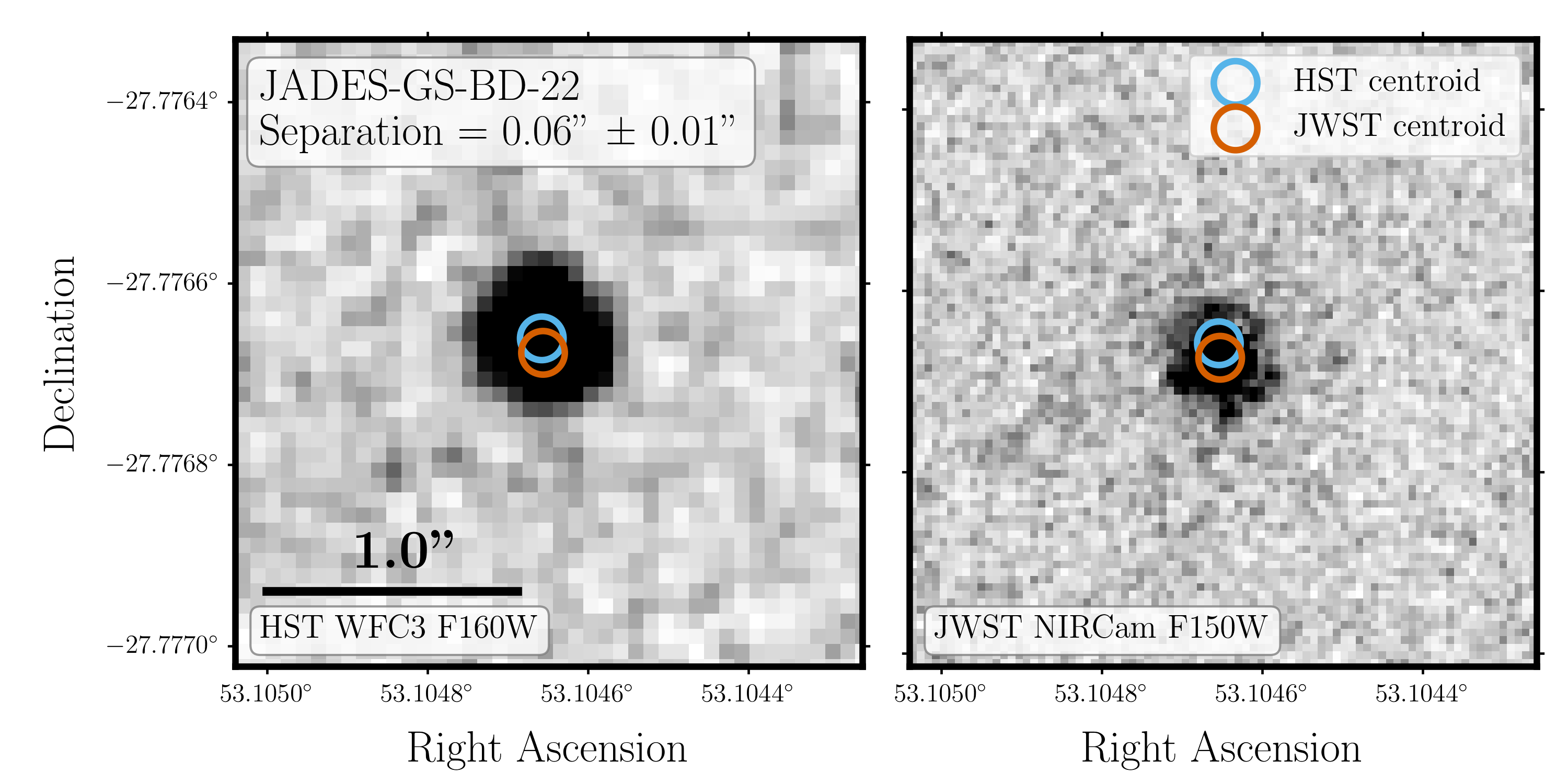}
  \includegraphics[width=0.49\textwidth]{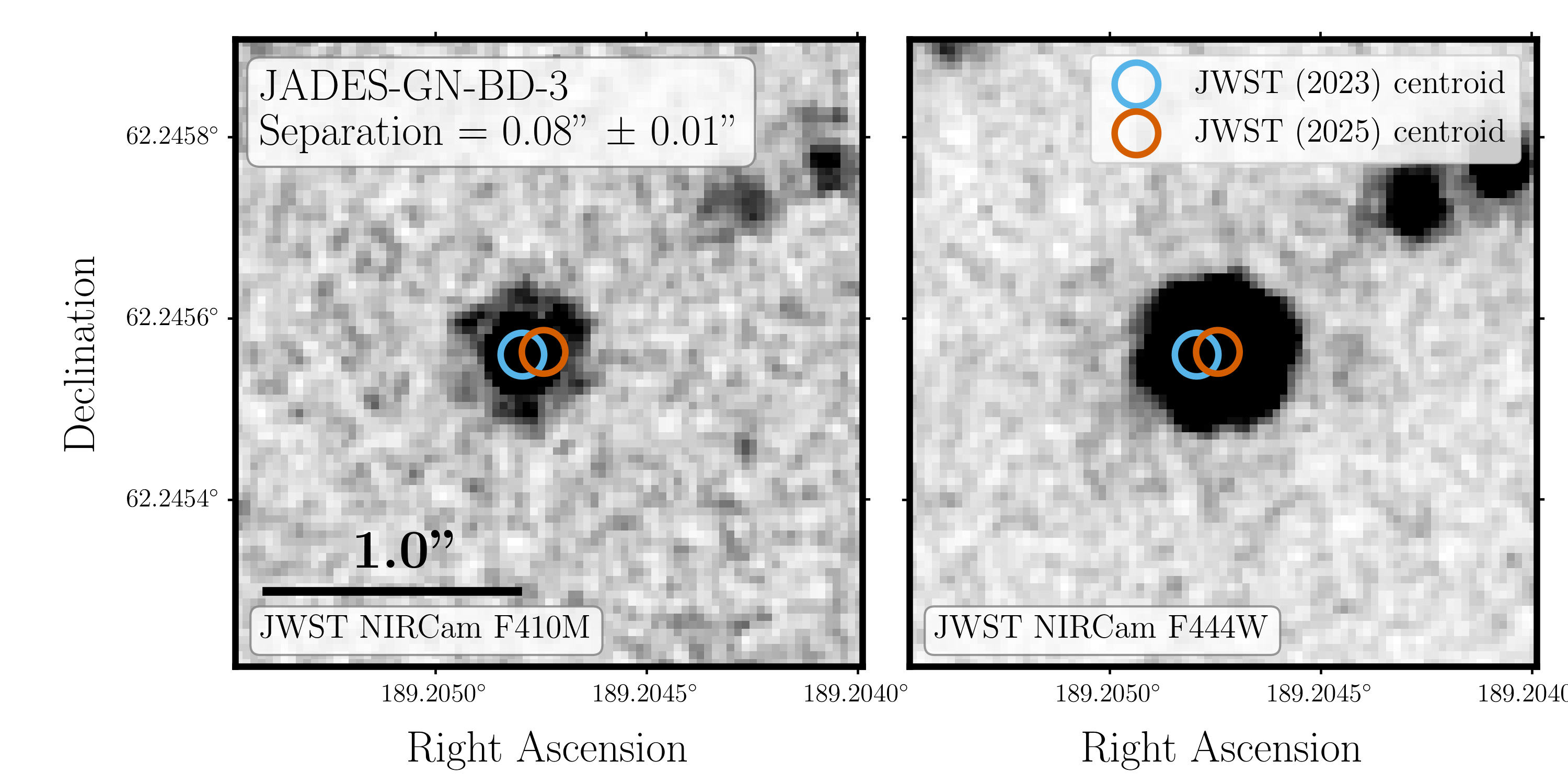}  
  \includegraphics[width=0.49\textwidth]{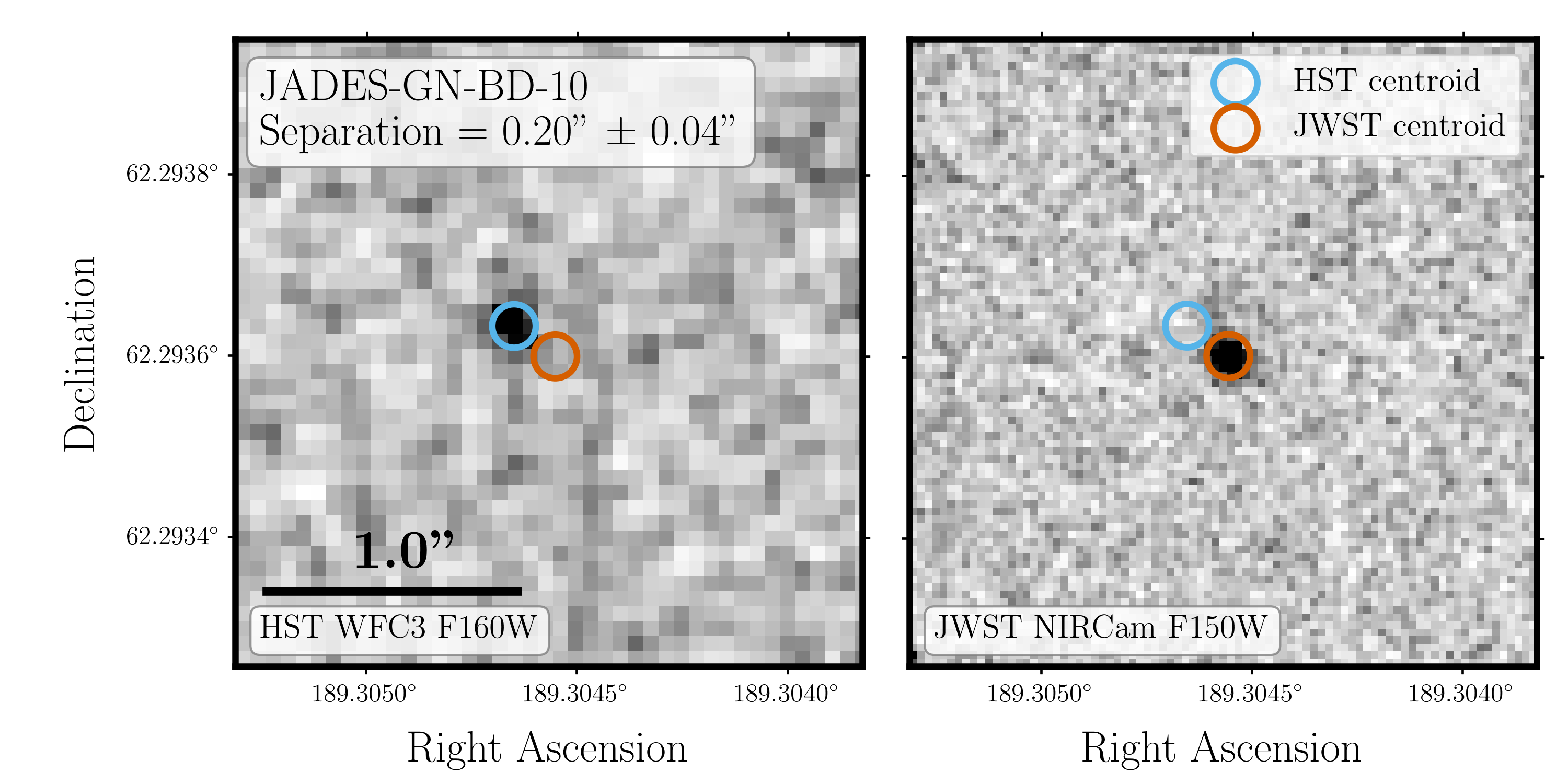}
  \caption{Newly uncovered proper motions for three brown dwarf candidates. In the top row we plot the HST/ACS F814W, F850LP, and JWST/NIRCam F090W thumbnails for JADES-GS-BD-14, where the measured position of the source in the HST images is indicated with a blue circle, and the position of the source in the NIRCam image is indicated with a red circle in each panel. The separation from both the F814W and F850LP images ($0.79^{\prime\prime} \pm 0.09^{\prime\prime}$ and $0.82^{\prime\prime} \pm 0.08^{\prime\prime}$ respectively) to the F090W image is consistent within the uncertainties. In the second row, on the left, we plot the same source, but with proper motion detected between the JWST/NIRCam F200W and F210M images. On the right we show the HST/WFC3 F160W and JWST/NIRCam F150W thumbnails for JADES-GS-BD-22, and in the bottom row we plot the proper motion for JADES-GN-BD-3 on the left and JADES-GN-BD-10 on the right. In each panel, we also provide a scale bar in the bottom left. 
  \label{fig:proper_motion}}
\end{figure*}

Given the matched astronomy between the JWST/NIRCam and HST images provided as part of JADES DR5 \citep{johnson2026}, there is an opportunity to search for proper motion for our brown dwarf candidates. The astrometric alignment for JADES was done using a large number of Gaia DR3 star positions observed across GOODS-S and GOODS-N, as discussed in Section 3.2.5 and 4.1 of \citet{johnson2026}. Both the HST and JWST images were both drizzled onto the same Gaia reference frame, and there is a $< 2$ mas median offset between the individual filter mosaics and the reference stars used. We searched through the full sample for proper motion between HST/ACS, HST/WFC3 and JWST/NIRCam observations. We also searched between multiple epochs of the NIRCam observations, as well. From the updated JADES DR5 reduction, all six of the JADES sources with proper motions from \KBD{} (JADES-GS-BD-1, JADES-GS-BD-5, JADES-GS-BD-7, JADES-GS-BD-10, JADES-GS-BD-11, and JADES-GN-BD-3) have proper motions consistent with the previous measurement, outside of JADES-GS-BD-1, where our updated measurement for the separation between the HST/WFC3 F160W and JWST/NIRCam F150W centroids is $0.07^{\prime\prime} \pm 0.02^{\prime\prime}$, below the value of 0.12$^{\prime\prime}$ in \KBD{}. A proper motion for JADES-GS-BD-9 was derived in \citet{hainline2024b}, and we confirm this proper motion with JADES DR5. As mentioned in the previous section, we also were able to make an independent confirmation of the proper motion for JADES-GN-BD-3. 

Because our color criteria allowed for selection of fainter sources than what was found in \KBD{}, it is not surprising that fewer were detected in HST/ACS or WFC3 for potential proper motion measurements. For the new candidates not discussed in \KBD{}, we also looked at Spitzer/IRAC observations at 3.6 and 4.5$\mu$m taken as part of the GOODS Spitzer Legacy program \citep{dickinson2003} to search for our sources, but the majority were too faint to be detected. The source that is potentially bright enough, JADES-GN-BD-15, is outside the GOODS Spitzer Legacy mosaic in GOODS-N. 

For this study, we measured proper motions for three additional sources not introduced in \KBD{} or \citet{hainline2024b}: JADES-GS-BD-14, JADES-GS-BD-22, and JADES-GN-BD-10. We found proper motions between the JWST/NIRCam F200W and F210M images, as well as the HST/ACS F814W and F850LP and the NIRCam positions for JADES-GS-BD-14, and the HST/WFC3 and the NIRCam positions for both JADES-GS-BD-22 and JADES-GN-BD-10. Across the whole sample, when we include the proper motions from \KBD{}, there is observed proper motions for 10 of the 41 brown dwarf candidates (24\%) providing strong evidence that they are members of our galaxy. 

To estimate proper motions for the new sources, we followed the procedure outlined in \KBD{}, where we used the \texttt{photutils} python package to measure centroids independently for each image, and then measured the offsets between the centroids. We plot the thumbnails for the sources with newly derived proper motions, along with the new estimate of proper motion from JWST for JADES-GN-BD-3, along with the measured separations, in Figure \ref{fig:proper_motion}. Here, for JADES-GS-BD-14 we plot the thumbnails and centroid positions for both F814W and F850LP, and find similar separations ($0.79^{\prime\prime} \pm 0.09^{\prime\prime}$ and $0.82^{\prime\prime} \pm 0.08^{\prime\prime}$ respectively) from the F090W position. We additionally plot the NIRCam F210M and F200W images, where we measure a smaller separation ($0.06^{\prime\prime} \pm 0.01^{\prime\prime}$) owing to the smaller time baseline between the observations. We note that the observed proper motion is in the same direction from the HST through to the NIRCam images (roughly southwest).

Using the separation between the HST and NIRCam positions for the three new sources, and the dates when the HST images were taken (derived from querying the STScI Mikulski Archive for Space Telescopes for the observations in these filters at the source positions), we estimated proper motions for these sources measured in arcseconds per year. We note that the HST mosaics we used are a combination of observations taken across multiple programs, and while we have attempted to ascertain the exact observational dates for the portions of the surveys that overlap with each source, there is an additional uncertainty when comparing the dates of the HST observations to the NIRCam observations which will effect the final proper motion estimates. This will primarily effect JADES-GS-BD-1, JADES-GS-BD-7, JADES-GS-BD-10, JADES-GS-BD-11, JADES-GS-BD-22, and JADES-GN-BD-10, while the other sources where proper motions were detected were confirmed with NIRCam to NIRCam observations. For JADES-GS-BD-14, we estimate a proper motion of $0.07^{\prime\prime}$/yr $\pm 0.01^{\prime\prime}$/yr from the HST/ACS F814W and F850LP to F090W NIRCam images, and $0.06^{\prime\prime}$/yr $\pm 0.01^{\prime\prime}$/yr from the NIRCam F210M to F200W images, consistent between the two methods within the uncertainties. For JADES-GS-BD-22, we estimate a proper motion of $0.004^{\prime\prime}$/yr $\pm 0.001^{\prime\prime}$/yr, and for JADES-GN-BD-10 we estimate a proper motion of $0.021^{\prime\prime}$/yr $\pm 0.0014^{\prime\prime}$/yr. We can use these proper motions to estimate transverse velocities assuming their Sonora Elf Owl distances and the equation: $v_T = 4.74 \mu d$, where $v_T$ is the transverse velocity in kilometers per second, $\mu$ is the proper motion in arcseconds per year, and $d$ is the distance in parsecs. We calculate a transverse velocity of $v_T = 102 \pm 15$ km s$^{-1}$ for JADES-GS-BD-14 (from the F200W to F210M proper motion), $v_T = 8 \pm 2$ km s$^{-1}$ for JADES-GS-BD-22, and $v_T = 128 \pm 15$ km s$^{-1}$ for JADES-GN-BD-10, which is among the highest transverse velocities across the entire sample, owing to the estimated distance ($1.29 \pm 0.09$ kpc from the \texttt{NIFTY} fits with the Sonora Elf Owl models). We report our observed proper motions and estimated transverse velocities assuming the Sonora Elf Owl distances (including the sources where we re-measured the proper motions from \KBD{}) in Table \ref{tab:proper_motions}. 

\begin{deluxetable}{cccc}[]
\tabletypesize{\footnotesize}
\tablecolumns{17}
\tablewidth{0pt}
\tablecaption{Measured Proper Motion and Transverse Velocities\label{tab:proper_motions}}
\tablehead{
\colhead{Object Name} & \colhead{filters} & \colhead{$\mu$ (mas/yr)} & \colhead{$v_T$ (km/s)$^{\tablenotemark{a}}$} }
 \startdata
JADES-GS-BD-1 & F160W$^{\tablenotemark{b}}$, F150W & $8 \pm 2$ & $30 \pm 24$ \\
JADES-GS-BD-5 & F277W, F300M & $50 \pm 20$ & $17 \pm 6$ \\
JADES-GS-BD-7 & F125W$^{\tablenotemark{b}}$, F115W & $18 \pm 2$ & $78 \pm 10$ \\
JADES-GS-BD-9 & F444W, F444W & $2 \pm 1$ & $20 \pm 4$ \\
JADES-GS-BD-10 & F160W$^{\tablenotemark{b}}$, F150W & $18 \pm 1$ & $21 \pm 1$ \\
JADES-GS-BD-11 & F125W$^{\tablenotemark{b}}$, F150W & $48 \pm 10$ & $162 \pm 40$ \\
JADES-GS-BD-14 & F814W$^{\tablenotemark{b}}$, F090W & $70 \pm 10$ & $102 \pm 15$ \\
'' & F850LP$^{\tablenotemark{b}}$, F090W & $70 \pm 10$ & $102 \pm 15$ \\
'' & F200W, F210M & $60 \pm 10$ & $88 \pm 15$ \\
JADES-GS-BD-22 & F160W$^{\tablenotemark{b}}$, F150W & $4 \pm 1$ & $8 \pm 2$ \\
JADES-GN-BD-3 & F410M, F444W & $44 \pm 2$ & $96 \pm 8$ \\
JADES-GN-BD-10 & F160W$^{\tablenotemark{b}}$, F150W & $21 \pm 1$ & $128 \pm 15$ \\
\enddata
\tablenotetext{a}{Transverse velocities ($v_T$) computed using estimated distances from \texttt{NIFTY} and the Sonora Elf Owl models.}
\tablenotetext{b}{The proper motions and transverse velocities presented for these sources will suffer from additional uncertainty owing to the difficulty in ascertaining the exact date of the HST observations from which the proper motions were measured, given the large range in dates of HST programs that observed GOODS-S and GOODS-N.}
\end{deluxetable}

\subsection{Additional Sources with Less Confidence}\label{subsec:tentative_sources}

The vast majority of the sources detected in JADES are extragalactic, with the brown dwarf candidates serving as ``contaminants'' to galaxy samples. In the right-most panel in Figure \ref{fig:BD_Color_Color_All}, it can be seen that a small number of the literature LRDs (five LRDs outside of our rejection criteria from Equations \ref{eq:lrd_1} and \ref{eq:lrd_2}, and only four above the revised color criteria we introduce and plot with a red dot-dashed line in the Figure) have NIRCam colors that are similar to our candidate brown dwarfs. Through visual inspection, we can reject these LRDs, which are otherwise compact and very red, but have fluxes at $< 2.5\mu$m that are significantly dissimilar from brown dwarfs. However, by only using two colors (derived from only three photometric datapoints), as in the F277W - F444W vs. F115W - F277W color-color plot, LRDs can have similar colors to brown dwarfs. However, even with multiple color criteria it is significantly more difficult to disentangle brown dwarf candidates and possible galaxies when they are more intrinsically faint.

\begin{figure*}
  \centering
  \includegraphics[width=0.32\textwidth]{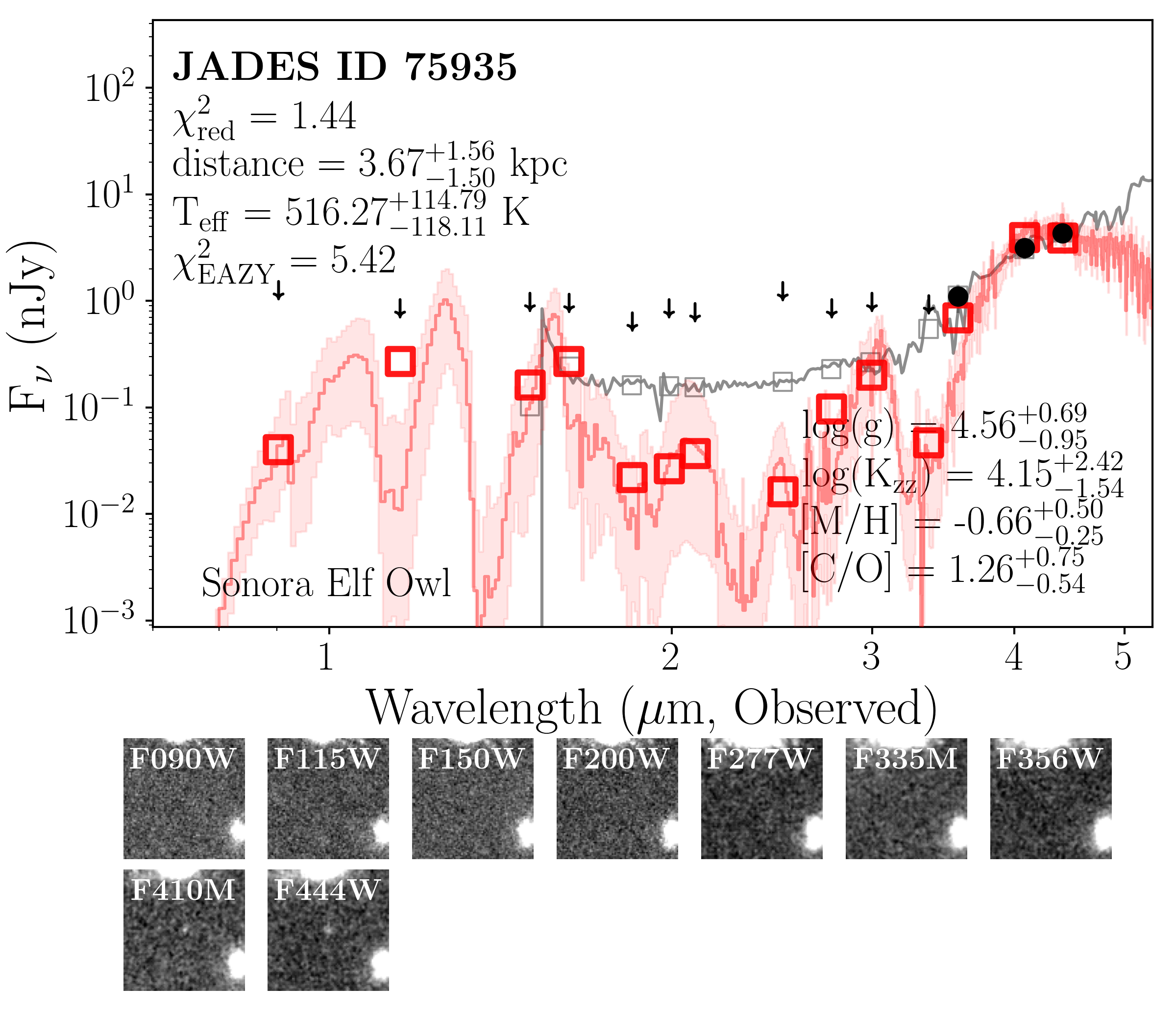}
  \includegraphics[width=0.32\textwidth]{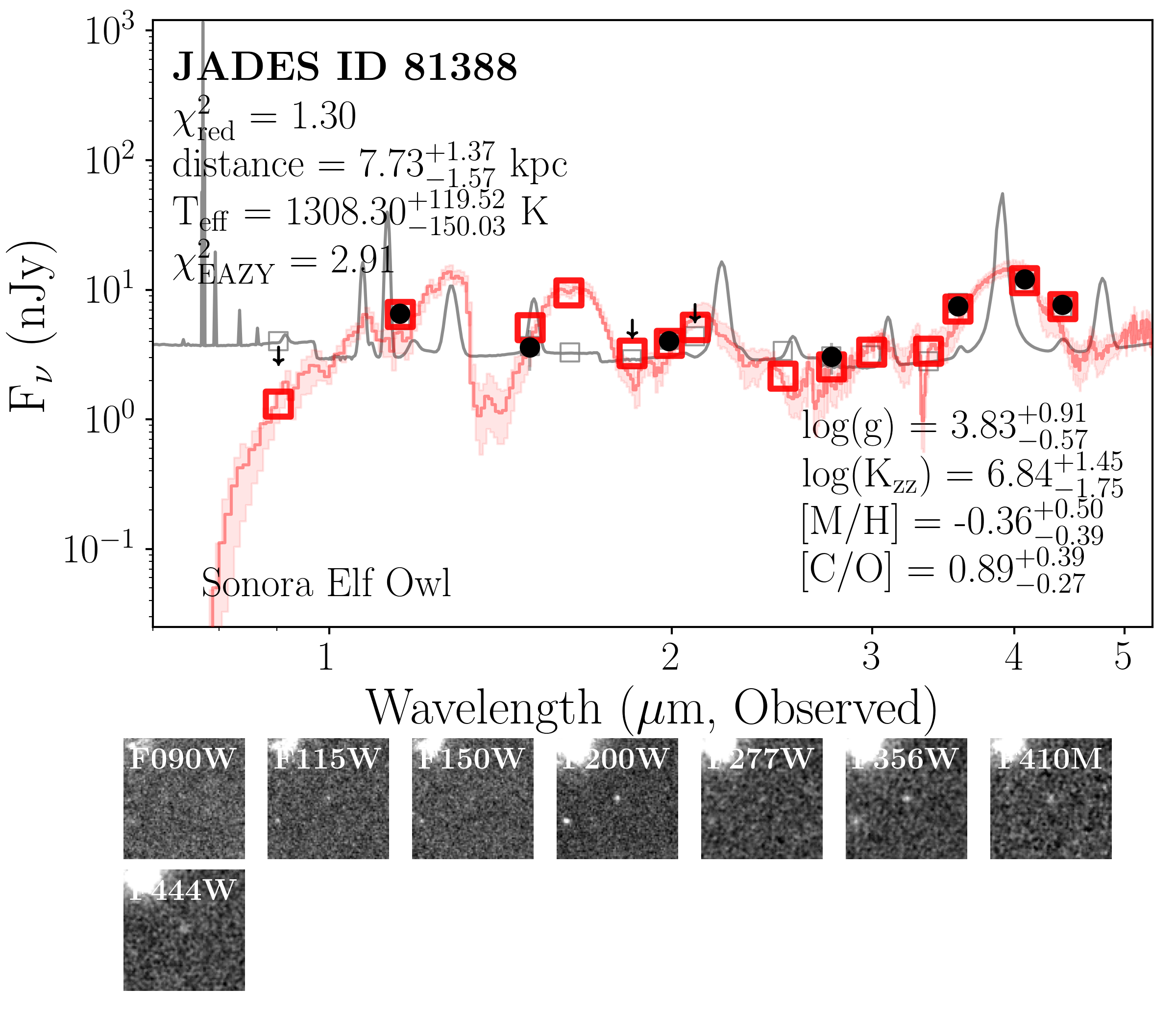}
  \includegraphics[width=0.32\textwidth]{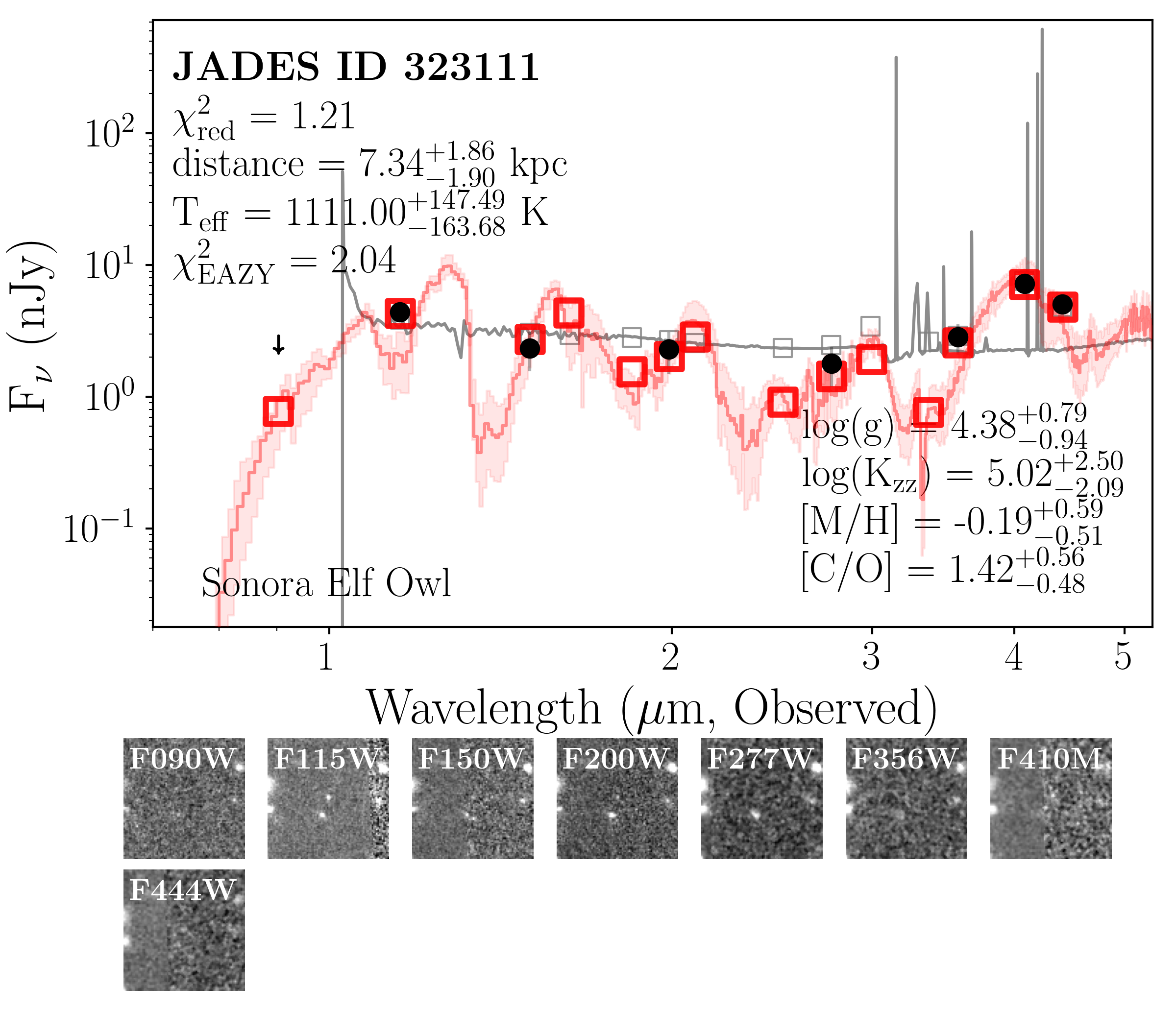}
  \includegraphics[width=0.32\textwidth]{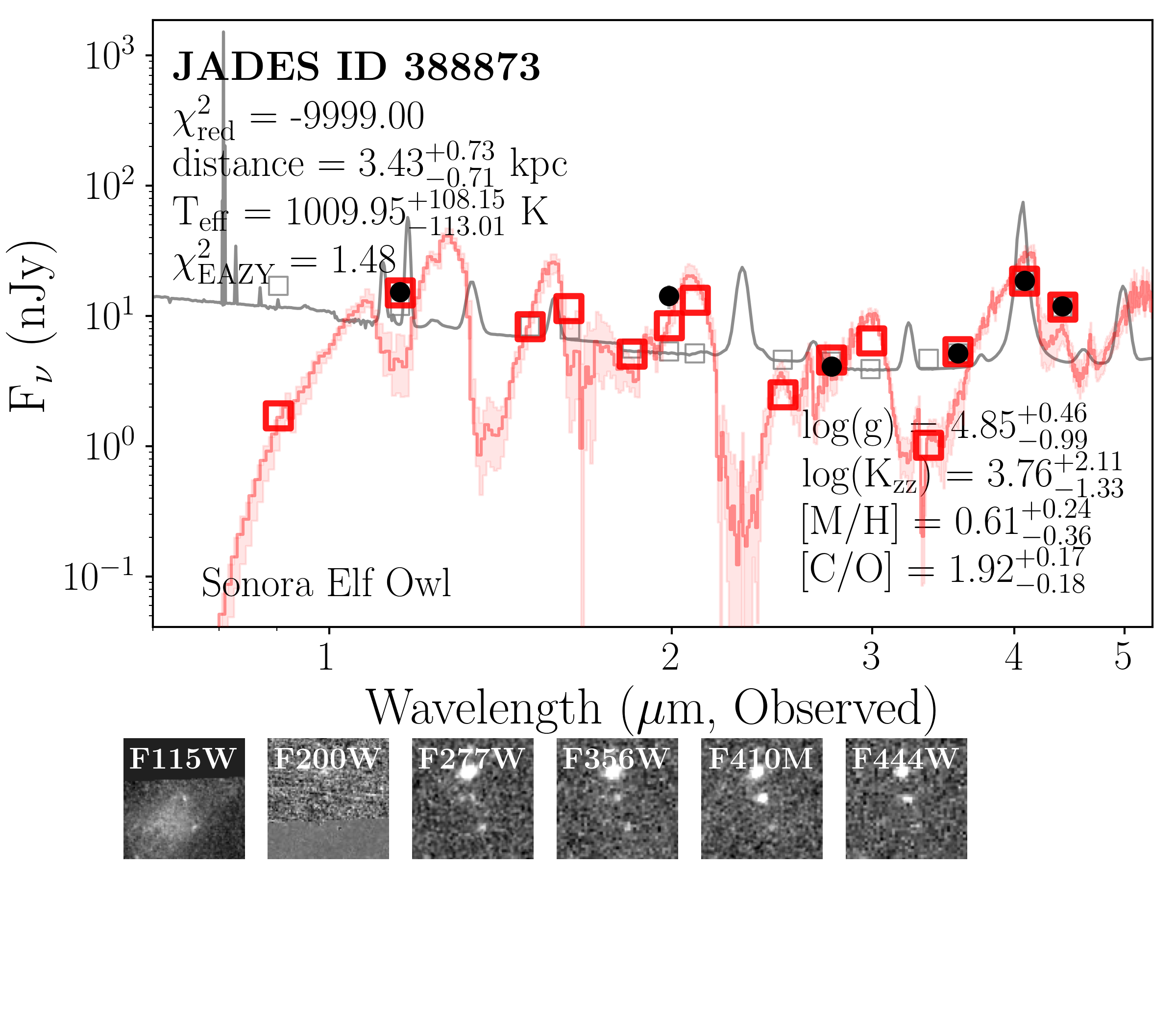}
  \includegraphics[width=0.32\textwidth]{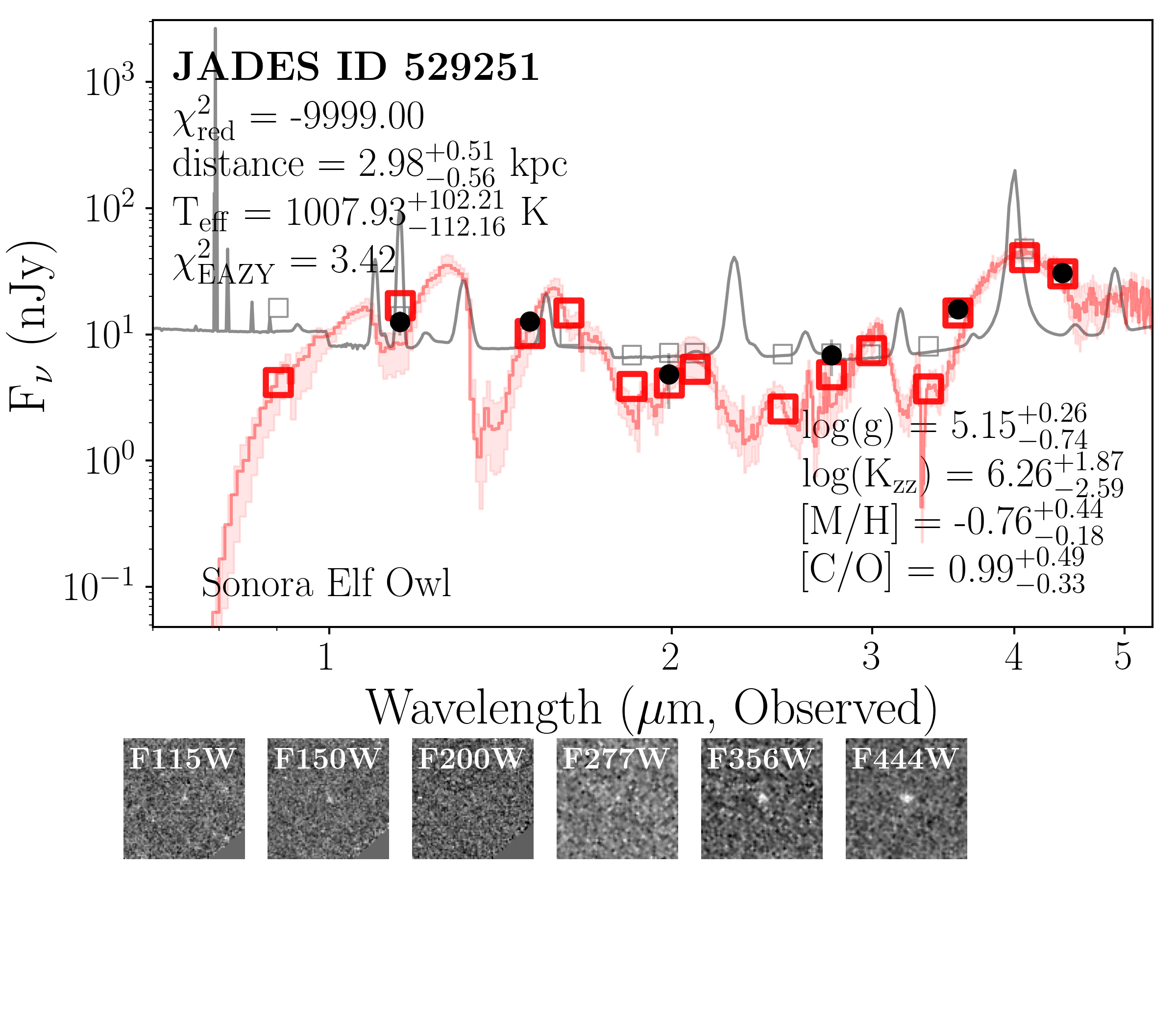}
  \includegraphics[width=0.32\textwidth]{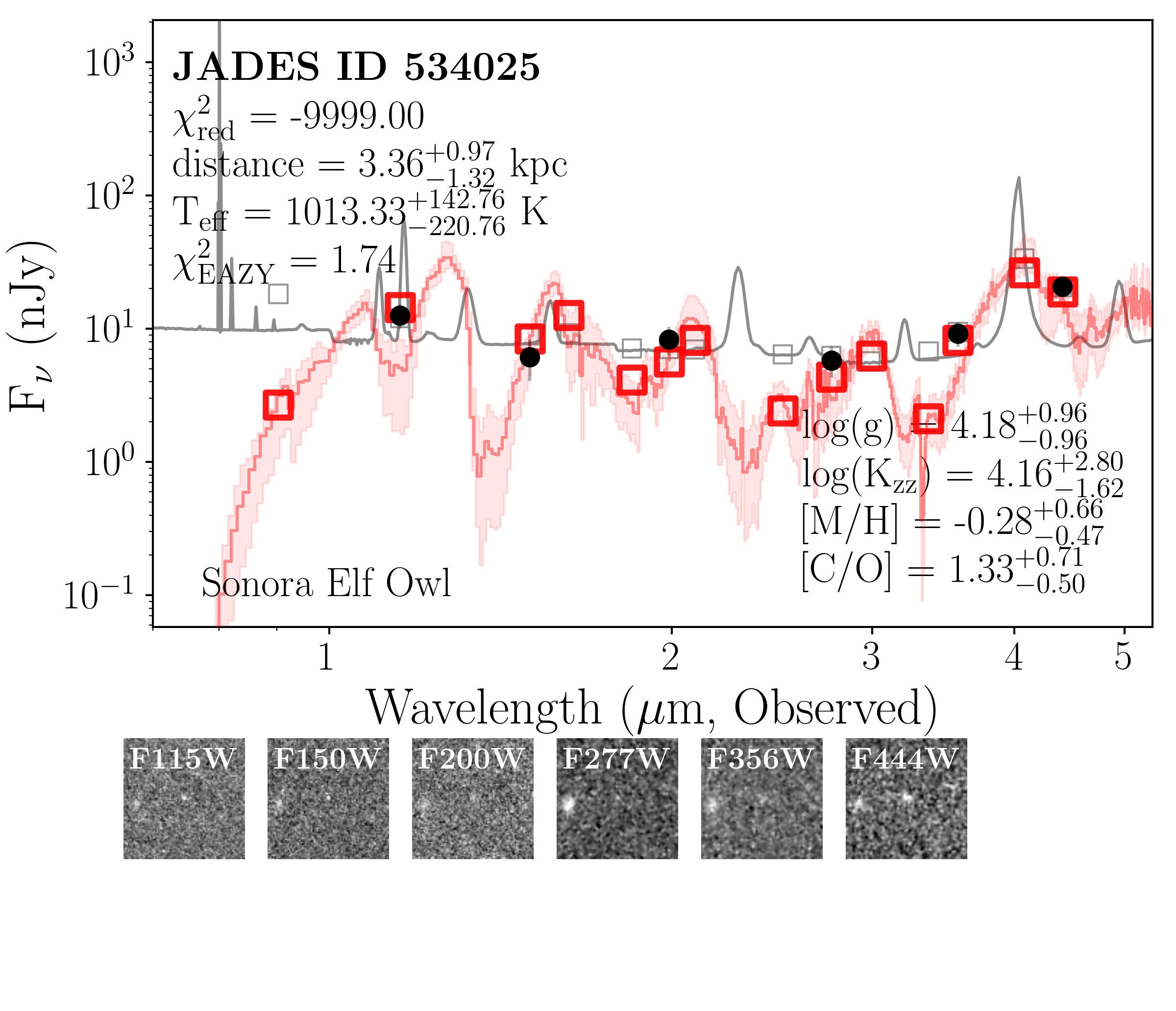}
  \includegraphics[width=0.32\textwidth]{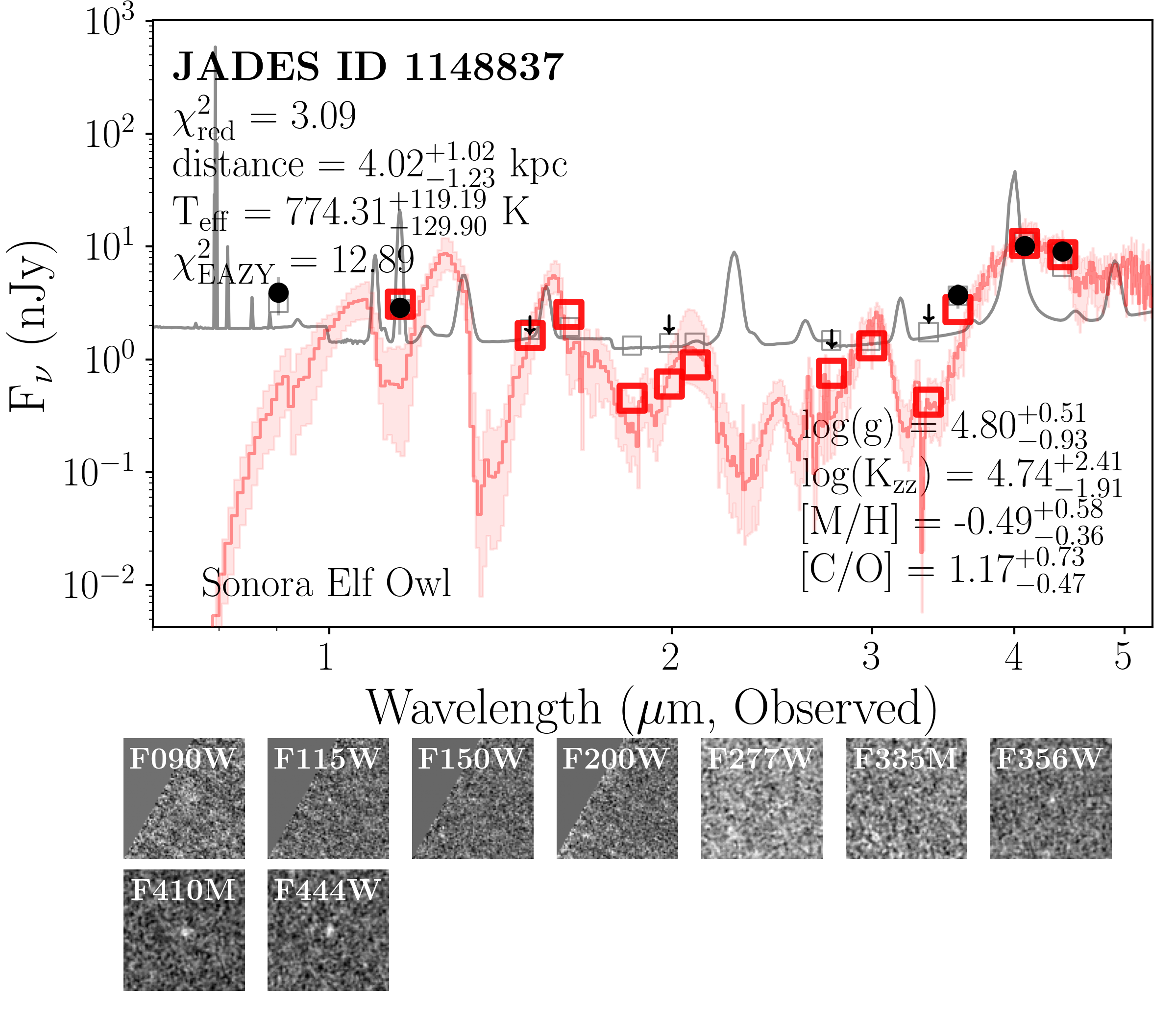}
  \includegraphics[width=0.32\textwidth]{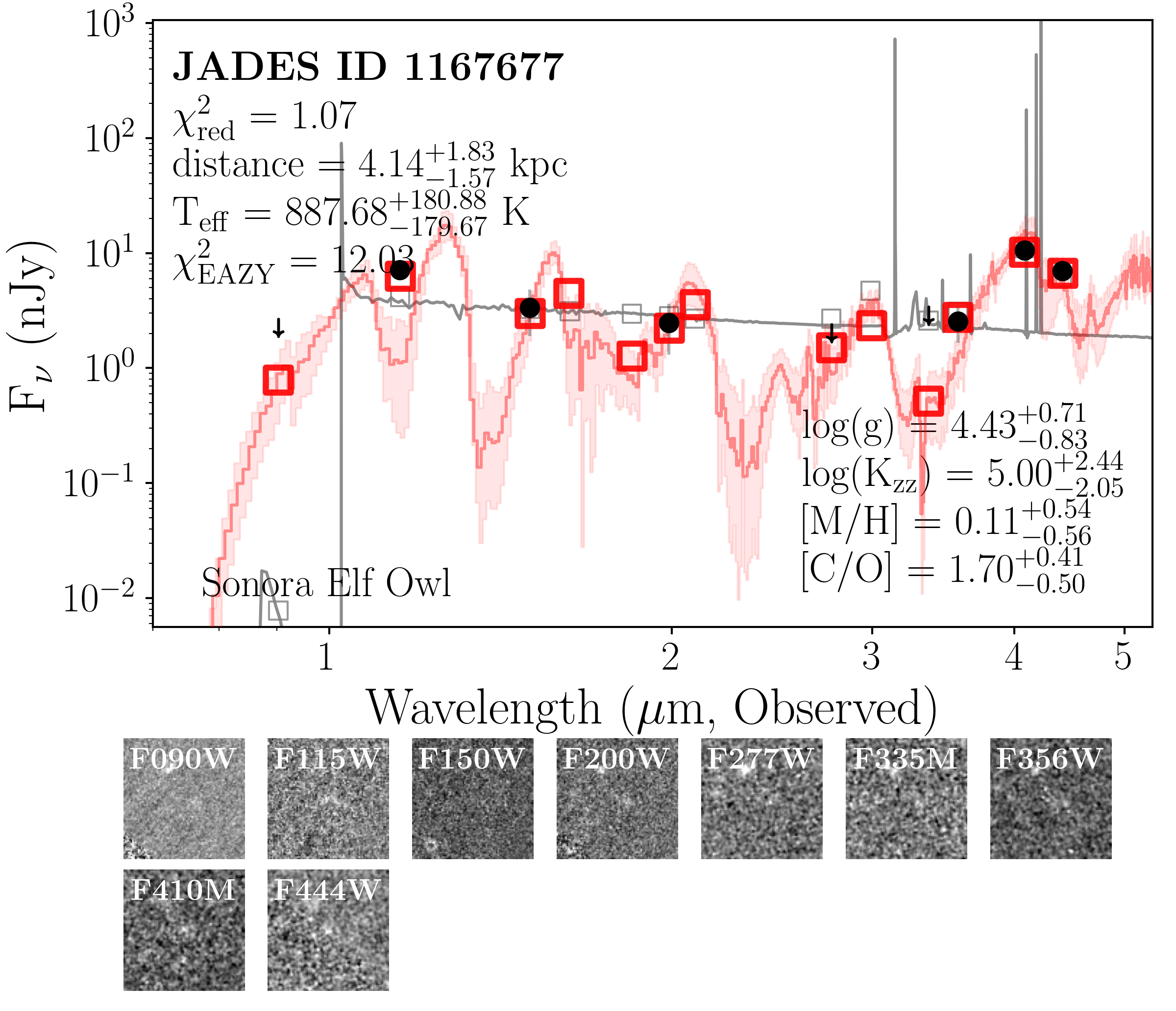}
  \includegraphics[width=0.32\textwidth]{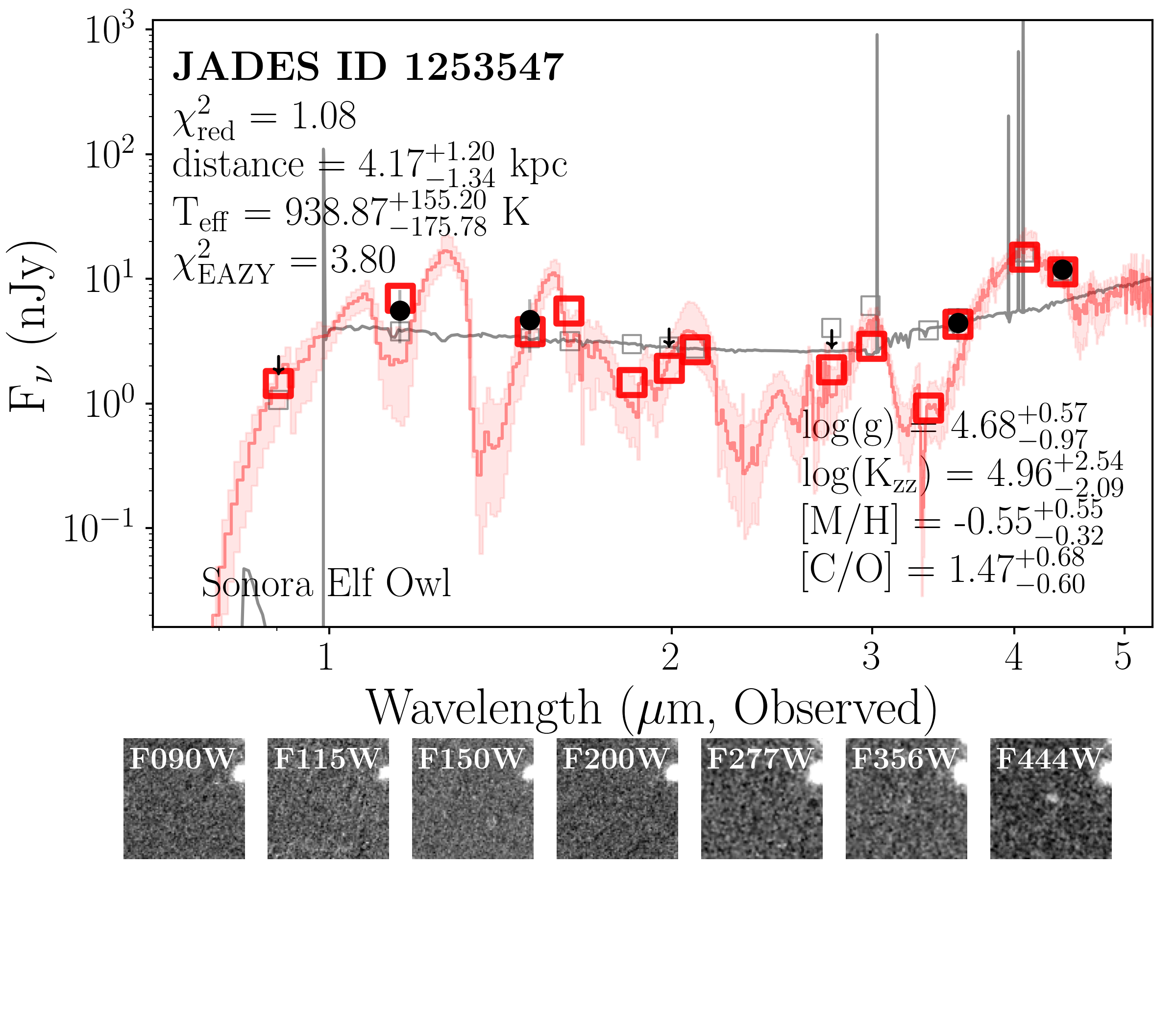}
  \includegraphics[width=0.32\textwidth]{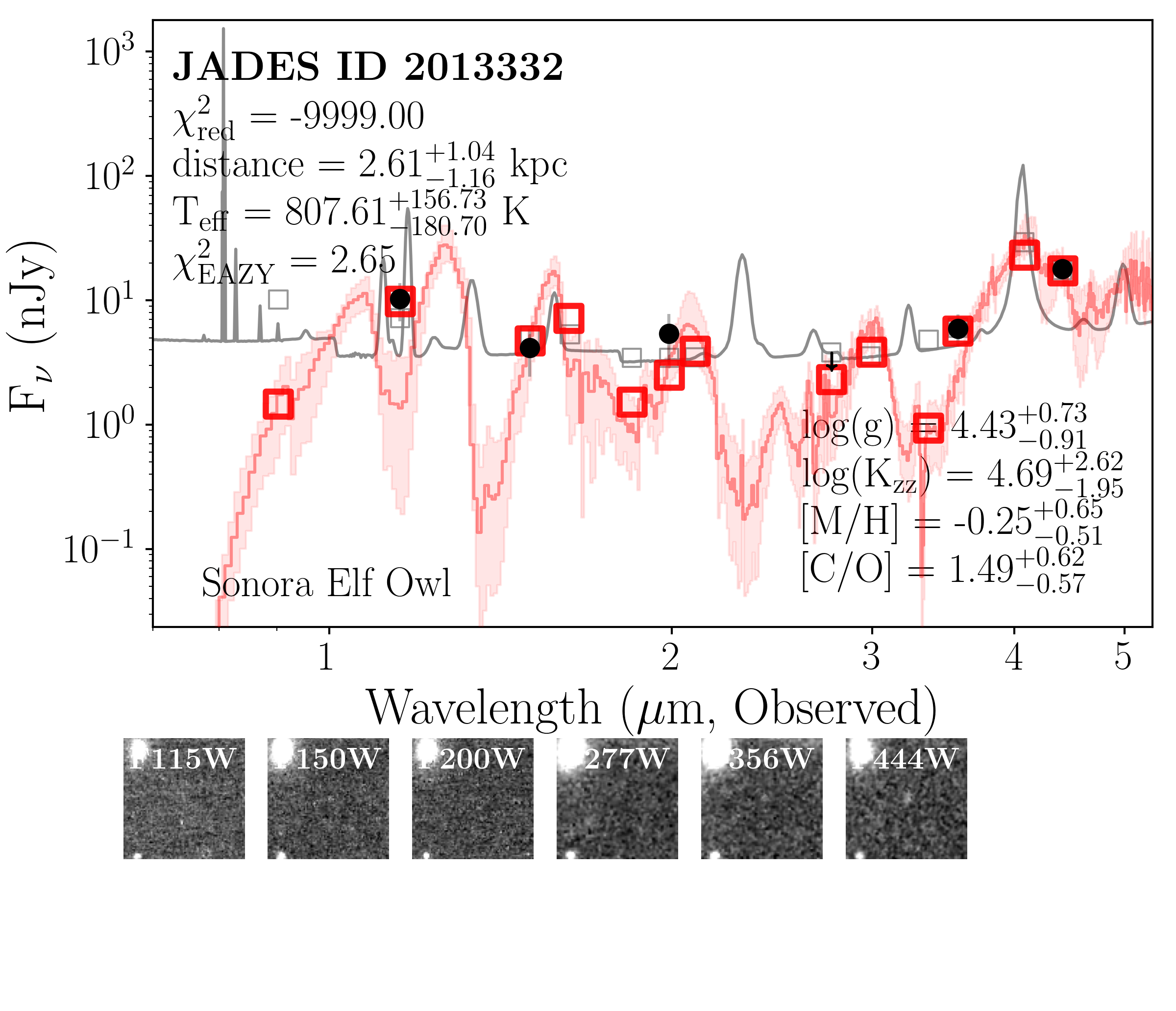}
  \includegraphics[width=0.32\textwidth]{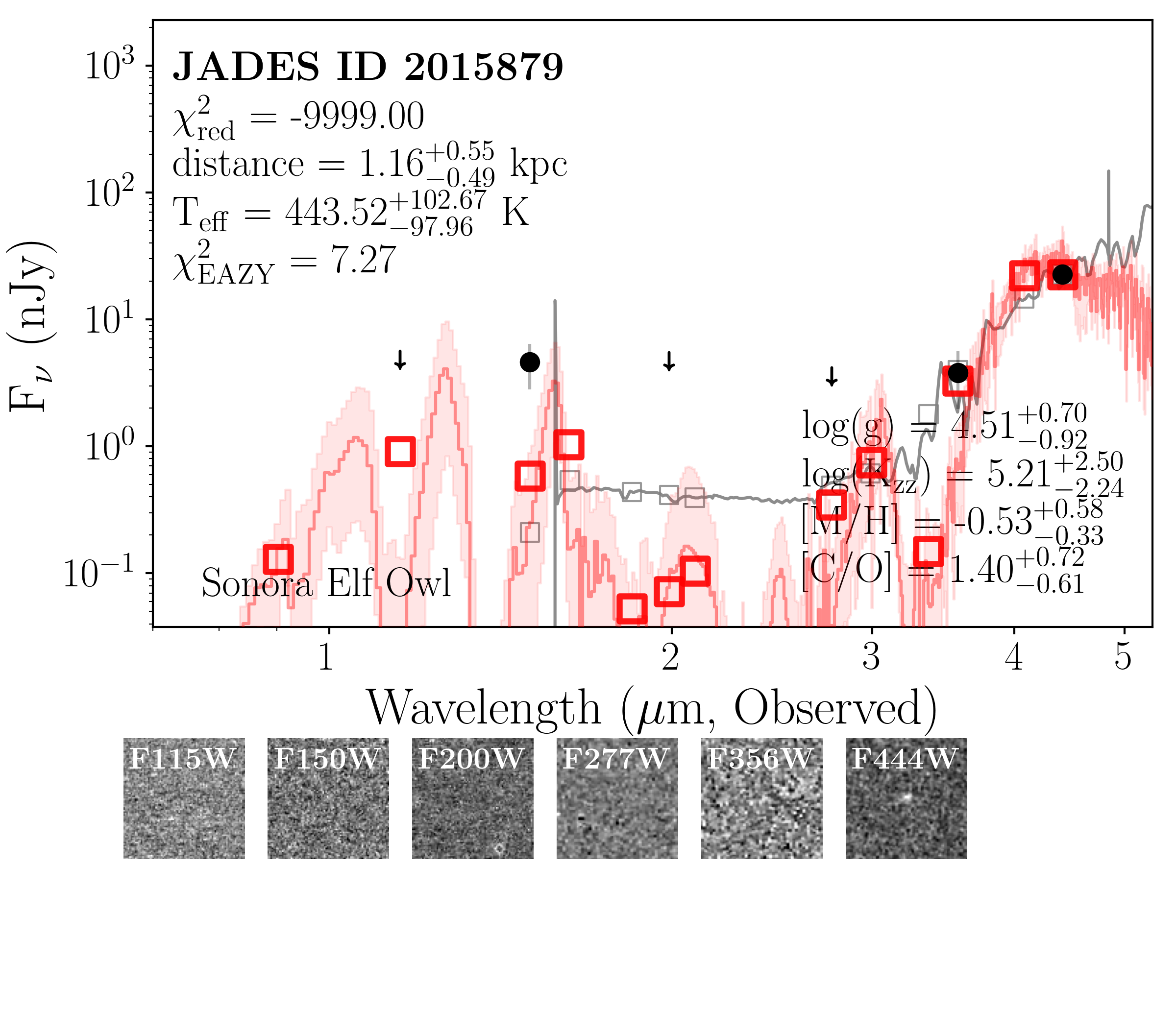}
  \caption{SED plots for the eleven sources in our sample with weaker evidence for being a brown dwarf, with similar points and \texttt{NIFTY} parameters as is provided in Figure \ref{fig:SED_plots_pt1}.
  \label{fig:tentative_source_seds}}
\end{figure*}

In this section, we will discuss a sample of eleven sources that were identified by the selection criteria discussed in Section \ref{sec:selecting_brown_dwarfs}, but due to their faint fluxes (none are detected with HST, and none are observed to have proper motions) and in a few cases the way in which their SED can be well-fit by galaxy models, we consider them to have weaker evidence of being true brown dwarf candidates. Each of these sources was identified during visual inspection, and fit with \texttt{NIFTY}, and we present the fluxes for these sources in Table \ref{tab:tentative_bd_fluxes} in Appendix \ref{sec:appendix_fluxes}, and plot the SEDs for this sample with fits to the Sonora Elf Owl models in Figure \ref{fig:tentative_source_seds}. 

Figure \ref{fig:tentative_source_seds} demonstrates the variety of sources with fluxes and colors consistent with brown dwarfs. Five out of the eleven sources have a F444W flux less than 10 nJy, and the maximum F444W flux across the whole subsample is 30.8 nJy for JADES ID 529251. In three cases, JADES ID 75935, 2013332 and 2015879, the source is only solidly detected at $ >4\mu$m, which would be expected for very faint brown dwarfs. Unsurprisingly, the Sonora Elf Owl fits for these three sources are solid, with $\chi^2_{\mathrm{red}} = 1.07 - 3.67$ for those with enough photometric detections to allow for a reduced $\chi^2$ to be calculated. Owing to the fact that JADES ID 1148837 is near the edge of the F090W mosaic, we did not use this flux in the fit, and the final Sonora Elf Owl fit $\chi^2_{\mathrm{red}} = 3.09$ is the largest value among these sources.

Notably, these sources, due to their being faint, have distances that range from $1.16 - 7.73$ kpc. The most distant source, JADES ID 81388, has an excellent \texttt{NIFTY} Sonora Elf Owl fit ($\chi^2_{\mathrm{red}} = 1.30$) with \teff{} $ = 1308^{+119}_{-150}$ K, with detections in 7 NIRCam filters. If we assume the source is at $z \sim 7$, we would measure a UV slope of only $\beta_{\mathrm{UV}} = -1.58 \pm 0.16$ (in a fit that doesn't include the F115W flux, which may be boosted by Lyman-$\alpha$ emission), which is similar to what would be expected for galaxies at this redshift. Similarly, for JADES ID 323111, also predicted to be at a distance above 7 kpc, we would estimate $\beta_{\mathrm{UV}} = -1.52 \pm 0.5$ if it were a galaxy at $z \sim 7$. For these faint sources, the lack of detection at F090W (or the lack of coverage in some cases, like in ID 534025) could either be because of the intrinsic faint predicted flux for a distant brown dwarf, or a Lyman-$\alpha$ break. 

More confusing are the sources, like JADES 75935, 1253547, and 2015879, where the source primarily appears at $3 - 5\mu$m. Only F444W is detected at a SNR $> 3$ for JADES ID 2015879, and while the fluxes are consistent with it being a 440K Y dwarf (with a highly uncertain F150W detection at a SNR = 2.6), it is difficult to say whether or not this object is a brown dwarf or some other strong nebular line emitting galaxy with a weak stellar continuum. Cold, distant Y-dwarfs would appear in our survey with fluxes and colors similar to what we find for these sources, but the lack of solid photometric detections promotes skepticism. 

\section{Discussion}\label{sec:discussion}

This sample of ultra-cool T- and Y-dwarfs and candidates offers a unique opportunity to explore the boundary between star and planet formation. In this section, we explore updated number counts and space densities based on our sample, as well as samples from the literature, and then we discuss how these faint, red sources can be mistaken for high-redshift galaxies. 

\subsection{Estimating Brown Dwarf Number Densities}\label{subsec:number_densities}

\begin{figure}[t]
  \centering
  \includegraphics[width=0.47\textwidth]{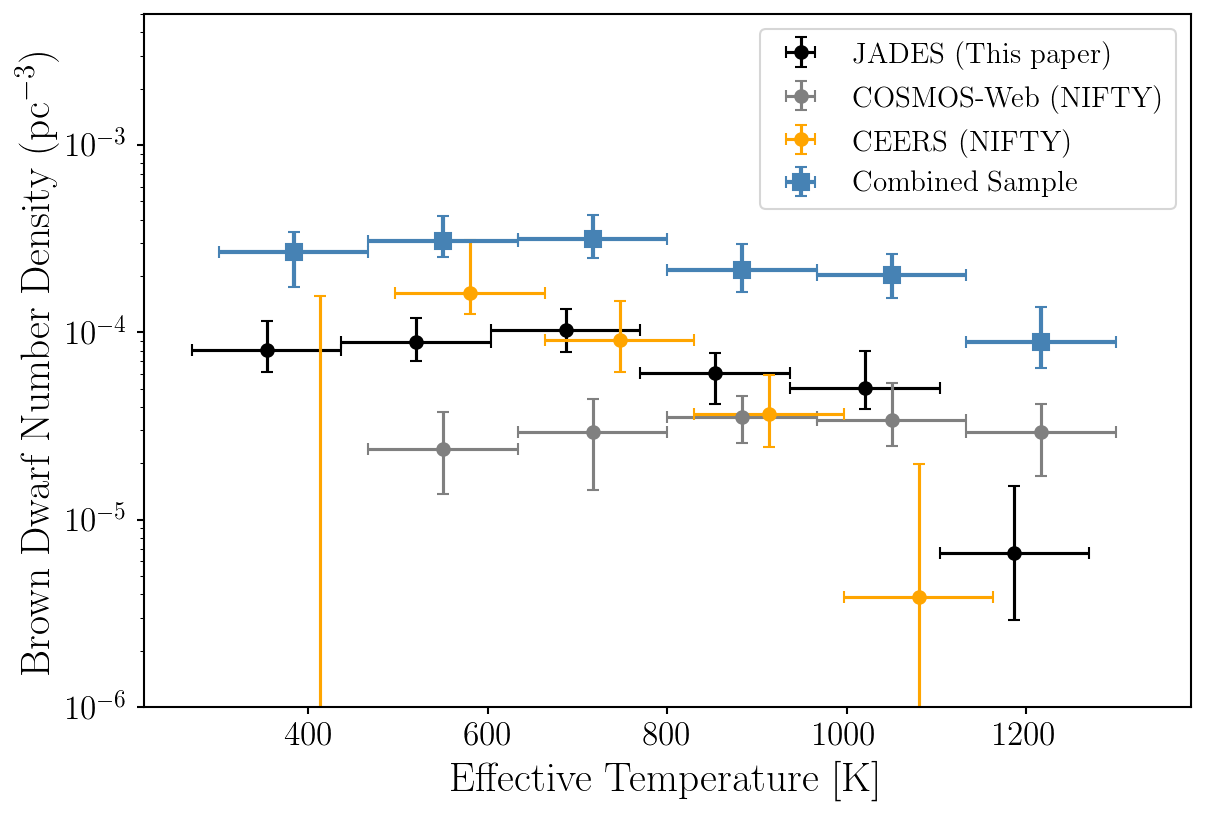}
  \includegraphics[width=0.47\textwidth]{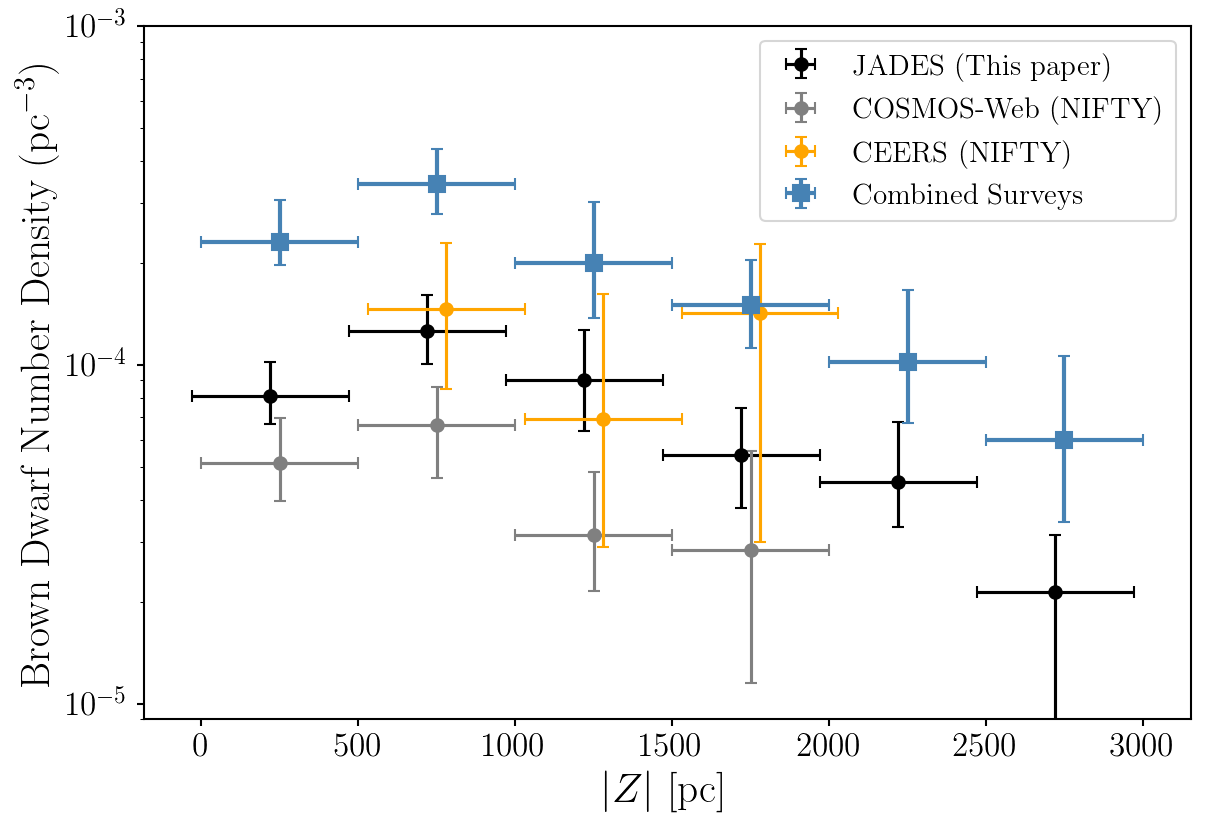}  
  \caption{Brown dwarf candidate number density (sources per pc$^{-3}$) plotted in bins of (top) effective temperature and (bottom) Milky Way disk scale height for multiple surveys. We use Lynden-Bell-Woodroofe-Wang (LBWW) nonparametric maximum-likelihood estimator, as described in the text, to account for the truncated data given the different flux limits and survey sizes from which the brown dwarf candidates were selected. The JADES GOODS-S and GOODS-N source number densities are plotted in black, and we also include number densities derived from \texttt{NIFTY} Sonora Elf Owl fits for the CEERS brown dwarf candidates (at a F444W limiting flux of 14.5 nJy, orange points) from \KBD{} and the COSMOS-Web sources (at a F444W limiting flux of 40.6 nJy, grey points) from \citet{chen2025}. We also show the number densities for the combined samples in each panel with blue points. The top panel indicates that the number density of sources flattens to lower effective temperatures, while the bottom panel demonstrates the fall-off with increasing height above the galactic midplane, in line with observations of stars from Gaia.
  \label{fig:number_densities}}
\end{figure}

Our final sample contains 41 brown dwarfs and brown dwarf candidates, sourced from across a total of 309.5 arcmin$^2$, for a total of 0.13 dwarfs at \teff{} $ < 1200 K$ per arcmin$^{2}$ (0.12 per arcmin$^{2}$ in GOODS-S, and 0.15 per arcmin$^{2}$ in GOODS-N). If we use the \texttt{NIFTY} Sonora Elf Owl fits, we find 0.03 Y-dwarfs per arcmin$^{2}$, and 0.10 T-dwarfs per arcmin$^{2}$. For comparison, \KBD{} published 14 brown dwarf candidates across a smaller survey area of 125 arcmin$^2$ in GOODS-S (67 arcmin$^2$) and GOODS-N (58 arcmin$^2$), yielding a total brown dwarf number density of 0.112 per arcmin$^{2}$, with 0.02 Y-dwarfs per arcmin$^{2}$, and 0.09 T-dwarfs per arcmin$^{2}$. While the previous selection criteria and ability to model temperatures below 500K differ from the updated approach presented in this work, the final values are broadly consistent.

In \citet{ryanreid2016}, the authors estimate the number density of T0-T5 \citep[roughly \teff{} = $1000 - 1300$K,][]{burgasser2002} brown dwarfs in multiple deep extragalactic fields using a star and brown dwarf luminosity function estimate and a model for the Milky Way. Summing across all observed magnitudes, and combining the thin and thick disk estimates, they report an estimate of 0.034 T0-T5 brown dwarfs per arcmin$^{2}$ in GOODS-S, and 0.033 T0-T5 brown dwarfs per arcmin$^{2}$ in GOODS-N. We only find 0.016 T0-T5 brown dwarfs per arcmin$^{2}$ across our entire sample, half the estimate from \citet{ryanreid2016}, likely owing to our color selection criteria which puts an upper limit on the sources we might detect of T$_{\mathrm{eff}} \lessapprox 1200$ K. 

To better compare our sample size with the literature, we also estimated the number densities of sources, taking into account the maximum distance we could discover a brown dwarf within JADES. Assuming a limiting F444W flux of 5 nJy for JADES in both GOODS-S and GOODS-N, we used the Lynden-Bell-Woodroofe-Wang (LBWW) nonparametric maximum-likelihood estimator \citep{lyndenbell1971, woodroofe1985, wang1986}. The LBWW estimator was designed to be used with randomly truncated data, here, for example, a source at a given distance and effective temperature is observable only if its predicted F444W flux, computed from our Sonora Elf Owl model interpolator, exceeds the survey flux limit at that distance. We explored the JADES sources presented in this study as well as the number densities for both the \citet{chen2025} COSMOS-Web sources and the \KBD{} CEERS sources using our \texttt{NIFTY} fits to their photometry. For the CEERS survey sources, we assumed a limiting flux density based on the minimum F444W magnitude used in \KBD{} of 14.5 nJy, across an area of 90.8 square arcminutes. For the COSMOS-Web sources, we assume a limiting flux density of 40.6 nJy, and an area of 874.8 square arcminutes, as reported by \citet{chen2025}. We repeated our calculation for each survey assuming the limiting flux densities for their selection and their survey volumes, as well as a combined sample. As part of this analysis, we verified the random truncation assumption required by LBWW using the Efron-Petrosian independence test \citep{efronpetrosian1992}, which estimates the Kendall rank correlation coefficient $\tau$ but with truncated data. A value of $\tau$ near zero indicates no correlation, supporting the LBWW method. We explored two distributions, number density vs. effective temperature ($\tau=0.27$ and $p=0.14$ for the combined sample), and number density vs. $\lvert Z \rvert$ ($\tau=0.13$, $p=0.53$). We find no statistically significant evidence for correlation between the variables and the truncation limit, supporting the applicability of the LBWW estimator. Uncertainties on the number densities were estimated by bootstrap resampling over the posterior parameter distributions from the \texttt{NIFTY} fits, recomputing the truncation limits at each draw.

In Figure \ref{fig:number_densities}, we plot the number densities for our brown dwarfs and brown dwarf candidates as a function of temperature (top panel), and absolute Milky Way vertical height (bottom panel). In both panels, the vertical error bars correspond to the Monte Carlo uncertainties on the number densities, and the horizontal error bars correspond to the bin sizes we use. Our results in bins of effective temperature demonstrate that the number density of sources only mildly declines to higher effective temperatures, seen both in the combined sample, but also in the JADES and CEERS sources. The COSMOS-Web sources, which span a larger area, but at shallower depth, do not show a significant change in number density across the temperature range we explore, although this is complicated by the different filter availability and selection criteria between the \citet{chen2025} sample and our own.  These results agree with predictions of the LFs for low-mass stellar and substellar objects from \citep{kirkpatrick2021}, which peak out to $400 - 500$K, with a steady decline to 1200K, with the exact slope dependent on underlying star formation rates and the IMF. 

In the bottom panel of Figure \ref{fig:number_densities}, we can see that all of the surveys indicate a drop-off in the number densities at absolute heights above 600 pc, as would be expected given results for the stellar population of the Milky Way as estimated with Gaia sources \citep{vieira2023}. The turnover at small heights is likely driven by the small area of the sky subtended by each survey, and their positions away from the plane of the Milky Way. At high Galactic latitude ($b \sim -54^{\circ}$ for GOODS-S, $b \sim 85^{\circ}$ for GOODS-N) even sources at modest distances of a few hundred pc project to $\lvert Z \rvert$ values of several hundred pc. As a result, the small $\lvert Z \rvert$ bins are populated only by the nearest sources, of which there are very few given the small survey solid angle. 

JWST has opened the door on a new, and abundant population of ultra-cool dwarfs with predicted distances where they may be probing multiple distinct regions of the Milky Way. As seen in Figure \ref{fig:metallicity_vs_distance} and the right panel of Figure \ref{fig:number_densities}, these brown dwarfs and brown dwarf candidates have distances that are consistent with the Milky Way thick disk, a conclusion that is supported by their high transverse velocities, and sub-solar metallicities. Follow-up spectroscopy for some of the distant brown dwarfs presented in UNCOVER in \citet{langeroodi2023} and \citet{burgasser2023}, as well as JADES-GS-BD-9 in \cite{hainline2024b} helps to confirm that a subsample are members of the halo. The estimate of accurate proper motions requires long time baselines for deep observations which is currently only attainable in a few extragalactic fields, like GOODS-S and GOODS-N, but with future surveys, like both LSST and the deep observations from Roman, more proper motions will be observed for these abundant brown dwarf candidates. The observed transverse velocity for JADES-GS-BD-9, $\sim 200$ km s$^{-1}$, is estimated for a source at around 2 kpc distance from the Earth, demonstrating that the current sensitivity and time baseline for GOODS-S would allow for stringent kinematic limits on even halo brown dwarf candidates. 

Our results also support the need for spectroscopic follow-up of the candidates. Beyond just confirming their origin as brown dwarfs, the degeneracies seen between parameters from the \texttt{NIFTY} fits points to a need for deep JWST/NIRSpec and MIRI spectroscopy, which can better constrain the atmospheric metallicity for T and Y dwarfs. Recent JWST/NIRSpec spectroscopy of T dwarf Wolf 1130C \citep{burgasser2025} and Y dwarf WISEA J153429.75-104303.3 \citep{faherty2025} have shown intriguing evidence for complex molecules (phosphine and silene respectively) in brown dwarfs only seen in Solar System objects Saturn and Jupiter, due in part to the low metallicity of these objects. As can be seen in Tables \ref{tab:bd_properties_gs_gn}, \ref{tab:bd_properties_gs_gn_pt2}, and \ref{tab:bd_properties_gs_gn_pt3}, from photometry alone, it is relatively difficult to estimate precise Eddy diffusion parameters, or C/O ratios. The Eddy diffusion parameter is degenerate with metallicity at $\sim 4-5\mu$m \citep{mukherjee2024}, which we observe for the coldest and faintest brown dwarf candidates in our sample. In addition, many of the colder brown dwarfs and brown dwarf candidates in our sample have very faint fluxes at 1 - 3$\mu$m, which may be interpreted as clouds that are absorbing the hotter emission from deeper in the brown dwarf \citep{morley2012}. Only deep spectroscopy will help to break these degeneracies, find evidence of these clouds, refine future cloud models, and understand the exact metallicity and the existence of molecules \citep[like phosphine, see][]{beiler2024a} in brown dwarf atmospheres for this new population of cold, distant subdwarfs. 

\subsection{Ultra-Cool Brown Dwarfs Masquerading as High-Redshift Galaxies: The Case of Capotauro}\label{subsec:capotauro}

As discussed in the introduction, historically, brown dwarfs were misidentified as possible high-redshift galaxies in HST surveys \citep[see ][]{bunker2004, oesch2012a, wilkins2014, finkelstein2010, finkelstein2015}, indeed, some of the sources in our brown dwarf sample were first discovered in searches for Lyman-$\alpha$ dropouts at 0.9$\mu$m. With the launch of JWST, the thermal emission and atmospheric absorption from ultra-cool brown dwarfs can be mistaken for Lyman-$\alpha$ break at $3-4\mu$m. Recently, \citet{gandolfi2025b} presented JWST/NIRCam and MIRI photometry, and NIRSpec prism spectroscopy of a source found in the CEERS survey, which they named ``Capotauro'' \citep[originally ``U-100588'' in][]{gandolfi2025a}. This source is only detected significantly in JWST F410M and F444W, and the spectrum, while at low SNR, agrees with the observed photometry. The authors explore multiple explanations for the origins of this source and favor the conclusion that Capotauro is a Lyman-$\alpha$ dropout galaxy at $z \sim 32$. Their analysis also indicates that the object may also be a Y-dwarf with \teff{} $\leq 300$K at d$ = 0.1 - 2$kpc, or even a free-floating exoplanet. The authors are skeptical of these stellar and planetary conclusions due to the lack of observed proper motion for the source (they derive an upper limit of $< 0.137^{\prime\prime}$/yr) and the predicted rarity of ultracool sources in the CEERS footprint ($\sim 0.01$) based on extrapolations of the number density from models presented in \citep{kirkpatrick2021}. 

We fit the JWST/NIRCam and MIRI photometry \citep[from Table A.1 of][]{gandolfi2025b} for Capotauro with \texttt{NIFTY}, and derive excellent fits to the photometry for the Sonora Elf Owl and ATMO2020++ models ($\chi^2_{\mathrm{red}} = 0.62$ and 0.42, respectively). The LOWZ models do not probe to the temperatures required to accurately fit the photometry for this source, and so overpredict the F356W flux. We derive temperatures that are slightly larger, but still in agreement within the uncertainties, with what is presented in \citet{gandolfi2025a}, \teff{}$ = 342^{+67}_{-46}$K for the Sonora Elf Owl model fits and \teff{}$ = 324^{+56}_{-44}$K for the ATMO2020++ model fits. Similarly, we derive distances of $d = 600^{+280}_{-180}$ pc for the Sonora Elf Owl model fits and $d = 450^{+190}_{-130}$ pc for the ATMO2020++ model fits. Both model fits indicate a sub-solar metallicity ([M/H] $\sim - 0.34 - -0.39$) for the object. The best-fit \texttt{NIFTY} Sonora Elf Owl model for the Capotauro fluxes agrees with the faint JWST/NIRSpec spectrum that the authors present in their study. 

To explore the origin of this source, we can compare it with the GOODS-S and GOODS-N brown dwarfs and brown dwarf candidates in our sample. Capotauro is very similar to JADES-GS-BD-21 and JADES-GN-BD-16 in that both of these sources are only significantly detected in F410M and F444W, and both have fits below \teff{}$ < 400$K, with best-fit distances of $\sim 0.9 - 1$ kpc. In addition, the source has many similarities to JADES-GS-BD-5, the coldest source in our sample, an object with both a detected proper motion and a spectrum confirming it as a brown dwarf (Hainline et al. in prep). If we scale the observed F444W fluxes for each of these objects to match what was presented in \citet{gandolfi2025b} for Capotauro (F444W flux $ = 30.7 \pm 2.4$ nJy), we find in each case that the fluxes short of F444W all agree with what was observed for Capotauro within the uncertainties. 

To reiterate, if JADES-GS-BD-5 were at $400 - 600$ pc and observed in the shallower CEERS field, it would look almost identical to Capotauro, with non-detections in the filters blueward of 4$\mu$m. The upper limit on the proper motion ($< 0.137^{\prime\prime}$/yr) for Capotauro is consistent with the fit distance \citep[see Figure 4 of ][]{gandolfi2025b}, the faint observed fluxes, and the relatively short time baseline (2.3 years) between observations. As a comparison, each of the proper motions we observe for our sample and report tn Table \ref{tab:proper_motions} are significantly below this upper limit.  

From the 10 Y-dwarf candidates in our sample we estimate that there would be $\sim 2.9$ Y-dwarfs across the 90.8 square arcminutes spanned by the CEERS survey, significantly in excess of the value of 0.01 given by \citet{gandolfi2025b}. If we only use the four objects in our sample with Sonora Elf Owl effective temperatures less than $400$ K to estimate a number density, we would still expect to find 1.2 sources below this temperature within CEERS. We note that only two of the sources (JADES-GS-BD-5 and JADES-GN-BD-5) are brighter than Capotauro at 4.4 $\mu$m while the other two sources, JADES-GS-BD-21 and JADES-GN-BD-16 have F444W flux above 13.4 nJy, the 5$\sigma$ flux limit presented in \citet{bagley2023} for the CEERS Epoch 1 NIRCam mosaics. As the CEERS field has only a slightly higher absolute galactic latitude (59$^{\circ}$) than GOODS-S (54$^{\circ}$) and GOODS-N (55$^{\circ}$), the number density estimate from JADES should be comparable to what would be expected in CEERS. It is very likely that Capotauro is an ultra-cool Y-dwarf, which is still of interest given its extreme temperature and the existence of a NIRSpec spectrum for the source. 

These results demonstrate how these ultra-cool brown dwarfs continue to be contaminants in Lyman-dropout samples, although now, with the NIRCam wavelength range, they will seemingly mimic $z \sim 30$ galaxies. If the sources discussed in Section \ref{subsec:tentative_sources} are brown dwarfs, there may even be larger predicted number densities of Y-dwarfs that will only appear in shallow surveys at $\sim 4\mu$m, as discussed at length in Section 4.3 of \citet{hainline2024a}. Further deep spectroscopy is crucial for confirming these sources and updating our models of distant brown dwarf number counts. 

\section{Conclusions}\label{sec:conclusions}

We searched through the 309.5 arcmin$^2$ JADES DR5 area in GOODS-S and GOODS-N to find brown dwarfs and brown dwarf candidates among the many hundreds of thousands of galaxies, expanding on the earlier work from \KBD{} in a prior data release. Deep extragalactic fields observed by NIRCam offer an exciting opportunity to find ultra-cool dwarfs at \teff{} $< 1200$ K out to multiple kiloparsecs. The primary conclusions of this study are:

\begin{itemize}
    \item With updated color selection criteria, we find a sample of 41 brown dwarfs and brown dwarf candidates across the survey. Our new criteria recovers the full sample of GOODS-S and GOODS-N sources presented in \KBD{}. Our new criteria are designed to utilize both common NIRCam filters seen in extragalactic surveys, and the less-commonly-used F410M filter if available, and we further tune our selection to reject red galaxies, like LRDs.
    \item We present \texttt{NIFTY}, Near-Infrared Fitting for T- and Y-dwarfs, an open source code designed to compare JWST NIRCam and MIRI photometry to cutting-edge brown dwarf atmospheric models in a Bayesian framework. With \texttt{NIFTY}, we fit each source in our selected sample and produced best-fit parameters and uncertainties, where we include MIRI 5 - 12.8$\mu$m fluxes for eleven of the sources in the sample.
    \item From the \texttt{NIFTY} fits, we find that 10 out of the 41 sources have photometry consistent with being at \teff{} $ < 500$ K (Y-dwarfs), and the remaining 31 sources are at 500 K $<$ \teff{} $ < 1200$ K (T-dwarfs), at distances out to 5 - 6 kpc from the Sun, although the majority are fit at $< 3$ kpc. In addition, while atmospheric metallicity is difficult to ascertain from photometry, the majority of the sample is best-fit at sub-solar metallicities, and we find that at larger heights above the galactic midplane our brown dwarf candidates have decreasing atmospheric metallicities, in line with expectations and observations of thick disk and halo stars. 
    \item We find broad agreement between the \texttt{NIFTY} derived temperatures and distances from the Sonora Elf Owl, LOWZ, and ATMO2020++ models, although metallicity has larger scatter. 
    \item Three of the sources that are new to this paper and not presented in \KBD{} have proper motions: JADES-GS-BD-14, JADES-GS-BD-22, and JADES-GN-BD-10. In total, 10 of the 41 sources (24\%) in our sample have observed proper motions.
    \item Caution must be taken in exploring ultra-faint brown dwarf candidates given the potential for selecting line-emitting galaxies at $z \sim 7.5$ where [\ion{O}{3}]$\lambda\lambda$4959,5007 may boost both F410M and F444W and mimic a brown dwarf $4 - 5\mu$m bump, along with faint or no F090W flux as would be observed in a Lyman-$\alpha$ dropout at this redshift.
    \item Taking into account the maximum distance we could find brown dwarf candidates in our survey, we estimate the number density of brown dwarfs as a function of temperature, demonstrating that brown dwarf number densities increase to lower temperatures ($\sim 3 \times 10^{-4}$ pc$^{-3}$ for ultra-cool Y-dwarf sources with \teff{} $ = 300 - 450$). In addition, our results demonstrate how ultracool dwarf number densities drop by almost an order of magnitude as a function of distance from the Milky Way midplane.
\end{itemize}

While deep near- to mid-IR spectra will be required to confirm these sources, early spectroscopic results for photometrically-selected candidates has already demonstrated the efficacy of this approach \citep{langeroodi2023, burgasser2023, hainline2024b}. A future paper will discuss the observed spectra for both JADES-GS-BD-7 and JADES-GS-BD-5, the latter of which has a \texttt{NIFTY} fit to both NIRCam and MIRI photometry with the Sonora Elf Owl models of only $322^{+5.0}_{-7.0}$ K. This work will also expand \texttt{NIFTY} to allow for spectroscopic fits. These sources, and their spectra, are fundamental for expanding our understanding of the atmospheres of brown dwarfs at low metallicities. 

\begin{acknowledgments}
The authors want to thank the anonymous referee whose review helped to strengthen this manuscript considerably. This work is based on observations made with the NASA/ESA/CSA James Webb Space Telescope, specifically JADES \citep{rieke2024, eisenstein2023, bunker2024, deugenio2025}. The data were obtained from the Mikulski Archive for Space Telescopes at the Space Telescope Science Institute, which is operated by the Association of Universities for Research in Astronomy, Inc., under NASA contract NAS5-03127 for JWST. KH, JH, JL, MR, BR, and CAW are supported by the JWST/NIRCam Science Team contract to the University of Arizona, NAS5-02015, and JWST Program 3215. We make use of data from JWST programs 1063, 1176, 1180, 1181, 1210, 1283, 1286, 1287, 1895, 1963, 2079, 2198, 2541, 2674, 3215, 3577, 3990, 4540, 5997, 6434, 6511, and 6541. The research of CCW is supported by NOIRLab, which is managed by the Association of Universities for Research in Astronomy (AURA) under a cooperative agreement with the National Science Foundation. AJB acknowledge funding from the ``FirstGalaxies'' Advanced Grant from the European Research Council (ERC) under the European Union’s Horizon 2020 research and innovation programme (Grant agreement No. 789056) RM acknowledges support by the Science and Technology Facilities Council (STFC), by the ERC through Advanced Grant 695671 “QUENCH”, and by the UKRI Frontier Research grant RISEandFALL. RM also acknowledges funding from a research professorship from the Royal Society. ST acknowledges support by the Royal Society Research Grant G125142. This work has benefitted from The UltracoolSheet at \url{http://bit.ly/UltracoolSheet}, maintained by Will Best, Trent Dupuy, Michael Liu, Aniket Sanghi, Rob Siverd, and Zhoujian Zhang, and developed from compilations by \citet{dupuy2012, dupuy2013, deacon2014, liu2016, best2018, best2021, sanghi2023, schneider2023}.
\end{acknowledgments}

\begin{contribution}

All authors contributed equally to the JADES collaboration.

\end{contribution}

\facilities{HST(ACS, WFC3), JWST(NIRCam, MIRI)}

\software{astropy \citep{astropy2013, astropy2018}, \texttt{NIFTY}}

\appendix

\section{Source Fluxes}\label{sec:appendix_fluxes}

In Table \ref{tab:gs_bd_fluxes} and \ref{tab:gn_bd_fluxes}, we provide the source IDs, positions, and $0.2^{\prime\prime}$ or $0.5^{\prime\prime}$ diameter circular aperture fluxes (as indicated) for both NIRCam and MIRI bands for our sample of brown dwarf and brown dwarf candidates. We indicate a lack of filter coverage with a dash in each table. Similarly, in Table \ref{tab:tentative_bd_fluxes}, we provide the source IDs, positions, and $0.2^{\prime\prime}$ diameter circular aperture NIRCam fluxes for the tentative candidates discussed in Section \ref{subsec:tentative_sources}. 

\begin{splitdeluxetable*}{lccc|crrrrrBl|rrrrrrrr}
\tabletypesize{\scriptsize}
\tablecolumns{12}
\tablewidth{0pt}
\tablecaption{GOODS-S Brown Dwarf Candidate Photometry (nJy) \label{tab:gs_bd_fluxes}}
\tablehead{
 \colhead{Object Name} & \colhead{JADES ID} & \colhead{RA} & \colhead{DEC} & \colhead{diam.} & \colhead{F090W} & \colhead{F115W} & \colhead{F150W} & \colhead{F200W}  & \colhead{F277W} & \colhead{Object ID} & \colhead{F335M} & \colhead{F356W} & \colhead{F410M} & \colhead{F444W} & \colhead{F560W} & \colhead{F770W} & \colhead{F1000W} & \colhead{F1280W}}
 \startdata
JADES-GS-BD-1 & 44120  & 53.026024 & -27.867167  & $0.5^{\prime\prime}$ & $27.62 \pm 5.26$ & $254.22 \pm 3.5$ & $138.91 \pm 3.46$ & $87.31 \pm 1.36$ & $54.47 \pm 2.04$ & JADES-GS-BD-1 & - & $138.2 \pm 1.77$ & $430.82 \pm 3.59$ & $330.15 \pm 1.68$ & - & $91.93 \pm 16.11$ & - & -  \\
JADES-GS-BD-2 & 20541  & 53.03629 & -27.883734  & $0.5^{\prime\prime}$ & $17.71 \pm 1.85$ & $108.01 \pm 2.22$ & $67.55 \pm 2.12$ & $44.49 \pm 1.68$ & $25.94 \pm 0.84$ & JADES-GS-BD-2 & $22.44 \pm 0.87$ & $69.3 \pm 0.9$ & $181.76 \pm 1.57$ & $128.17 \pm 0.95$ & - & $32.84 \pm 4.42$ & - & -  \\
JADES-GS-BD-3 & 42611  & 53.056305 & -27.868404  & $0.5^{\prime\prime}$ & $1.93 \pm 3.15$ & $12.62 \pm 2.33$ & $5.65 \pm 1.44$ & $4.14 \pm 1.93$ & $-0.07 \pm 2.24$ & JADES-GS-BD-3 & $0.29 \pm 1.44$ & $9.38 \pm 1.82$ & $29.0 \pm 3.43$ & $21.48 \pm 2.36$ & - & $11.17 \pm 5.53$ & - & -  \\
JADES-GS-BD-4 & 13985  & 53.076733 & -27.889942  & $0.5^{\prime\prime}$ & $-0.7 \pm 2.05$ & $3.76 \pm 2.28$ & $0.3 \pm 1.51$ & $1.83 \pm 1.6$ & $0.33 \pm 1.31$ & JADES-GS-BD-4 & $0.31 \pm 1.48$ & $5.19 \pm 1.08$ & $23.15 \pm 2.33$ & $25.18 \pm 1.44$ & - & $-1.53 \pm 5.21$ & - & $73.74 \pm 81.95$  \\
JADES-GS-BD-5 & 190413  & 53.084038 & -27.839346  & $0.5^{\prime\prime}$ & $-1.37 \pm 1.6$ & $3.38 \pm 1.27$ & $2.06 \pm 1.2$ & $1.65 \pm 1.34$ & $9.11 \pm 0.96$ & JADES-GS-BD-5 & $2.31 \pm 0.98$ & $65.78 \pm 0.88$ & $504.33 \pm 1.16$ & $655.23 \pm 1.08$ & $352.76 \pm 50.63$ & $335.91 \pm 60.6$ & $1552.62 \pm 135.99$ & $1786.92 \pm 221.54$  \\
JADES-GS-BD-6 & 2950  & 53.103905 & -27.908504  & $0.2^{\prime\prime}$ & $4.06 \pm 1.14$ & $27.12 \pm 0.99$ & $14.55 \pm 0.87$ & $7.41 \pm 0.92$ & $6.2 \pm 0.47$ & JADES-GS-BD-6 & - & $17.52 \pm 0.68$ & $59.17 \pm 1.4$ & $48.02 \pm 0.96$ & - & - & - & -  \\
JADES-GS-BD-7 & 101709  & 53.10554 & -27.815155  & $0.5^{\prime\prime}$ & $15.9 \pm 2.89$ & $118.06 \pm 1.81$ & $59.81 \pm 2.13$ & $34.87 \pm 2.07$ & $21.06 \pm 1.5$ & JADES-GS-BD-7 & $16.03 \pm 2.29$ & $72.63 \pm 1.28$ & $264.06 \pm 2.91$ & $213.47 \pm 1.61$ & $12.87 \pm 49.73$ & $-17.24 \pm 46.83$ & $116.5 \pm 113.48$ & $-16.56 \pm 200.5$  \\
JADES-GS-BD-8 & 92415  & 53.13376 & -27.825521  & $0.5^{\prime\prime}$ & $0.48 \pm 1.71$ & $4.02 \pm 1.07$ & $2.1 \pm 0.99$ & $0.72 \pm 0.99$ & $0.55 \pm 0.87$ & JADES-GS-BD-8 & $0.44 \pm 1.36$ & $7.71 \pm 0.7$ & $51.36 \pm 1.5$ & $51.2 \pm 1.2$ & $35.37 \pm 46.34$ & $-37.63 \pm 46.31$ & $-43.5 \pm 88.29$ & $-190.87 \pm 202.6$  \\
JADES-GS-BD-9 & 106126  & 53.16114 & -27.809166  & $0.5^{\prime\prime}$ & $2.56 \pm 1.22$ & $17.92 \pm 1.16$ & $8.36 \pm 1.08$ & $6.04 \pm 1.04$ & $3.23 \pm 0.71$ & JADES-GS-BD-9 & $2.31 \pm 1.28$ & $15.58 \pm 1.0$ & $53.84 \pm 1.4$ & $43.28 \pm 1.29$ & $-2.2 \pm 40.67$ & $26.43 \pm 48.49$ & $35.25 \pm 127.08$ & $-24.99 \pm 203.05$  \\
JADES-GS-BD-10 & 192282  & 53.1618 & -27.831615  & $0.5^{\prime\prime}$ & $311.83 \pm 2.16$ & $2384.35 \pm 2.04$ & $1342.19 \pm 1.67$ & $956.95 \pm 1.72$ & $603.82 \pm 0.93$ & JADES-GS-BD-10 & $377.85 \pm 1.51$ & $1356.21 \pm 1.12$ & $4212.34 \pm 2.3$ & $3106.51 \pm 1.36$ & $1905.18 \pm 51.36$ & $1093.37 \pm 63.14$ & $2057.29 \pm 155.43$ & $1649.16 \pm 203.41$  \\
JADES-GS-BD-11 & 140057  & 53.200314 & -27.764128  & $0.5^{\prime\prime}$ & $0.45 \pm 2.65$ & $38.78 \pm 2.13$ & $18.73 \pm 1.86$ & $6.64 \pm 1.89$ & $9.12 \pm 0.99$ & JADES-GS-BD-11 & $7.37 \pm 1.46$ & $56.23 \pm 1.36$ & $220.51 \pm 1.57$ & $180.22 \pm 1.34$ & $73.21 \pm 53.53$ & $33.13 \pm 48.12$ & $167.55 \pm 102.52$ & $49.14 \pm 196.53$  \\
JADES-GS-BD-12 & 486694  & 52.938843 & -27.78736  & $0.2^{\prime\prime}$ & $1.96 \pm 1.18$ & $3.52 \pm 1.19$ & $3.14 \pm 1.09$ & $0.86 \pm 1.03$ & $3.82 \pm 0.9$ & JADES-GS-BD-12 & $1.67 \pm 1.44$ & $12.4 \pm 0.81$ & $35.58 \pm 1.44$ & $28.62 \pm 1.1$ & - & - & - & -  \\
JADES-GS-BD-13 & 469315  & 52.98535 & -27.755032  & $0.2^{\prime\prime}$ & $3.34 \pm 0.69$ & $24.82 \pm 1.08$ & $11.58 \pm 0.82$ & $5.19 \pm 0.62$ & $6.4 \pm 0.47$ & JADES-GS-BD-13 & - & $28.08 \pm 0.47$ & $130.92 \pm 1.08$ & $118.19 \pm 0.99$ & - & - & - & -  \\
JADES-GS-BD-14 & 433690  & 52.98979 & -27.864613  & $0.5^{\prime\prime}$ & $160.45 \pm 2.59$ & $1480.76 \pm 4.1$ & $794.0 \pm 2.33$ & $563.45 \pm 3.24$ & $368.88 \pm 1.39$ & JADES-GS-BD-14 & $197.6 \pm 1.77$ & $784.43 \pm 1.35$ & $2645.97 \pm 2.8$ & $1994.52 \pm 1.96$ & - & $595.88 \pm 25.53$ & - & $3851.33 \pm 352.5$  \\
JADES-GS-BD-15 & 502828  & 53.005913 & -27.821442  & $0.2^{\prime\prime}$ & $3.07 \pm 0.49$ & $20.52 \pm 0.47$ & $10.39 \pm 0.58$ & $5.33 \pm 0.36$ & $4.6 \pm 0.41$ & JADES-GS-BD-15 & $4.17 \pm 0.8$ & $12.69 \pm 0.44$ & $39.23 \pm 0.8$ & $28.07 \pm 0.5$ & - & - & - & -  \\
JADES-GS-BD-16 & 495589  & 53.030766 & -27.841581  & $0.2^{\prime\prime}$ & $-0.09 \pm 1.16$ & $8.98 \pm 1.27$ & $6.42 \pm 1.14$ & $1.58 \pm 0.82$ & $2.67 \pm 0.77$ & JADES-GS-BD-16 & $2.92 \pm 1.43$ & $10.31 \pm 0.81$ & $48.56 \pm 1.6$ & $46.15 \pm 1.17$ & - & - & - & -  \\
JADES-GS-BD-17 & 2301  & 53.069588 & -27.910807  & $0.2^{\prime\prime}$ & $2.91 \pm 1.11$ & $12.03 \pm 0.94$ & $7.01 \pm 0.81$ & $5.43 \pm 0.69$ & $5.15 \pm 0.53$ & JADES-GS-BD-17 & $5.79 \pm 1.22$ & $13.0 \pm 0.65$ & $33.44 \pm 1.19$ & $25.43 \pm 0.84$ & - & - & - & $174.01 \pm 93.82$  \\
PANO-GS-BD-18 & 441205  & 53.078167 & -27.927135  & $0.2^{\prime\prime}$ & - & $15.15 \pm 2.48$ & $9.12 \pm 1.97$ & $5.96 \pm 2.35$ & $6.0 \pm 1.73$ & PANO-GS-BD-18 & - & $39.03 \pm 1.76$ & - & $235.65 \pm 2.57$ & - & - & - & -  \\
PANO-GS-BD-19 & 501052  & 53.07999 & -27.94288  & $0.2^{\prime\prime}$ & - & $28.5 \pm 2.88$ & $18.46 \pm 2.23$ & $8.33 \pm 1.86$ & $6.83 \pm 2.06$ & PANO-GS-BD-19 & - & $22.83 \pm 2.03$ & - & $41.98 \pm 2.85$ & - & - & - & -  \\
JADES-GS-BD-20 & 318023  & 53.09535 & -27.92108  & $0.2^{\prime\prime}$ & $-0.24 \pm 1.47$ & $2.59 \pm 0.89$ & $3.43 \pm 1.24$ & $0.9 \pm 0.98$ & $3.01 \pm 1.16$ & JADES-GS-BD-20 & - & $11.73 \pm 0.99$ & $23.55 \pm 1.66$ & $19.31 \pm 1.37$ & - & - & - & -  \\
JADES-GS-BD-21 & 7646  & 53.09658 & -27.89875  & $0.2^{\prime\prime}$ & $0.65 \pm 1.0$ & $-0.24 \pm 0.68$ & $-0.81 \pm 0.53$ & $0.36 \pm 0.6$ & $0.52 \pm 0.57$ & JADES-GS-BD-21 & $1.39 \pm 1.71$ & $1.5 \pm 0.41$ & $11.24 \pm 0.86$ & $15.33 \pm 0.79$ & - & - & - & -  \\
JADES-GS-BD-22 & 128632  & 53.104652 & -27.776674  & $0.2^{\prime\prime}$ & $118.15 \pm 1.22$ & $1021.31 \pm 2.11$ & $586.84 \pm 1.78$ & $466.8 \pm 1.71$ & $264.79 \pm 0.98$ & JADES-GS-BD-22 & $145.8 \pm 1.36$ & $504.29 \pm 0.98$ & $1564.37 \pm 2.36$ & $1130.0 \pm 1.35$ & - & - & - & -  \\
JADES-GS-BD-23 & 153326  & 53.112896 & -27.742144  & $0.2^{\prime\prime}$ & $7.78 \pm 0.81$ & $42.79 \pm 0.8$ & $35.26 \pm 1.03$ & $11.39 \pm 0.97$ & $26.44 \pm 0.9$ & JADES-GS-BD-23 & $32.56 \pm 1.41$ & $114.59 \pm 0.76$ & $273.96 \pm 1.6$ & $220.31 \pm 1.19$ & - & - & - & -  \\
JADES-GS-BD-24 & 233407  & 53.12898 & -27.753494  & $0.2^{\prime\prime}$ & $-1.76 \pm 1.34$ & $0.34 \pm 0.93$ & $1.55 \pm 0.98$ & $0.08 \pm 0.93$ & $1.32 \pm 0.77$ & JADES-GS-BD-24 & $2.22 \pm 1.76$ & $14.46 \pm 0.77$ & $43.38 \pm 1.53$ & $39.49 \pm 1.01$ & - & - & - & -  \\
PANO-GS-BD-25 & 484735  & 53.16294 & -27.937887  & $0.2^{\prime\prime}$ & - & $15.11 \pm 0.49$ & $6.16 \pm 0.88$ & $2.58 \pm 1.16$ & $2.93 \pm 0.59$ & PANO-GS-BD-25 & - & $13.5 \pm 0.68$ & $53.19 \pm 0.72$ & $37.54 \pm 1.23$ & - & - & - & -  \\
\enddata
\end{splitdeluxetable*}

\begin{splitdeluxetable*}{lccc|crrrrBl|rrrrr}
\tabletypesize{\scriptsize}
\tablecolumns{12}
\tablewidth{0pt}
\tablecaption{GOODS-N Brown Dwarf Candidate Photometry  (nJy) \label{tab:gn_bd_fluxes}}
\tablehead{
 \colhead{Object Name} & \colhead{ID} & \colhead{RA} & \colhead{DEC} & \colhead{diam.}  & \colhead{F090W} & \colhead{F115W} & \colhead{F150W} & \colhead{F200W} & \colhead{Object ID} & \colhead{F277W} & \colhead{F335M} & \colhead{F356W} & \colhead{F410M} & \colhead{F444W} }
 \startdata
JADES-GN-BD-1 & 1003262  & 189.03618 & 62.23416  & $0.2^{\prime\prime}$ & $4.23 \pm 1.76$ & $40.81 \pm 1.39$ & $19.62 \pm 3.13$ & $10.32 \pm 2.37$ & JADES-GN-BD-1 & $9.13 \pm 1.56$ & - & $35.46 \pm 1.39$ & $185.92 \pm 4.08$ & $182.25 \pm 2.2$  \\
JADES-GN-BD-2 & 1013184  & 189.118 & 62.26897  & $0.2^{\prime\prime}$ & $2.37 \pm 1.51$ & $27.21 \pm 1.18$ & $14.4 \pm 1.95$ & $7.42 \pm 0.95$ & JADES-GN-BD-2 & $5.22 \pm 1.51$ & $3.27 \pm 1.97$ & $22.34 \pm 1.11$ & $100.53 \pm 2.89$ & $76.76 \pm 1.6$  \\
JADES-GN-BD-3 & 1006775  & 189.20473 & 62.245564  & $0.2^{\prime\prime}$ & $6.16 \pm 0.81$ & $68.68 \pm 0.64$ & $29.77 \pm 1.05$ & $13.86 \pm 0.96$ & JADES-GN-BD-3 & $10.06 \pm 0.99$ & $6.31 \pm 2.13$ & $88.84 \pm 0.73$ & $307.2 \pm 3.11$ & $582.97 \pm 1.02$  \\
JADES-GN-BD-4 & 1000836  & 189.07454 & 62.222893  & $0.2^{\prime\prime}$ & $-0.91 \pm 1.64$ & $3.59 \pm 1.22$ & $2.22 \pm 2.57$ & $2.95 \pm 1.52$ & JADES-GN-BD-4 & $3.76 \pm 1.42$ & - & $12.55 \pm 1.03$ & $37.49 \pm 2.23$ & $29.25 \pm 1.75$  \\
JADES-GN-BD-5 & 1214440  & 189.22003 & 62.21018  & $0.2^{\prime\prime}$ & $-0.5 \pm 1.04$ & $0.15 \pm 0.75$ & $-1.35 \pm 0.91$ & $0.71 \pm 0.67$ & JADES-GN-BD-5 & $0.81 \pm 0.72$ & $1.15 \pm 1.01$ & $7.02 \pm 0.77$ & $34.81 \pm 1.5$ & $34.84 \pm 1.14$  \\
JADES-GN-BD-6 & 1148319  & 189.26414 & 62.32691  & $0.2^{\prime\prime}$ & $0.93 \pm 1.3$ & $6.17 \pm 1.65$ & $3.65 \pm 1.1$ & $1.82 \pm 0.93$ & JADES-GN-BD-6 & $1.28 \pm 0.63$ & $1.51 \pm 1.67$ & $3.82 \pm 0.92$ & $8.77 \pm 1.41$ & $6.64 \pm 1.06$  \\
JADES-GN-BD-7 & 1148179  & 189.27112 & 62.326435  & $0.2^{\prime\prime}$ & - & $6.95 \pm 2.07$ & $0.64 \pm 1.5$ & $0.98 \pm 1.14$ & JADES-GN-BD-7 & $2.23 \pm 0.8$ & $2.96 \pm 1.28$ & $6.26 \pm 1.01$ & $12.33 \pm 1.54$ & $8.18 \pm 0.86$  \\
JADES-GN-BD-8 & 1207397  & 189.27115 & 62.320843  & $0.2^{\prime\prime}$ & $-0.61 \pm 0.92$ & $2.26 \pm 1.15$ & $1.38 \pm 0.91$ & $1.62 \pm 0.94$ & JADES-GN-BD-8 & $2.88 \pm 0.7$ & $0.0 \pm 1.51$ & $13.17 \pm 0.89$ & $83.31 \pm 1.65$ & $96.61 \pm 1.53$  \\
JADES-GN-BD-9 & 1052050  & 189.3037 & 62.189587  & $0.2^{\prime\prime}$ & $0.48 \pm 0.63$ & $1.32 \pm 1.15$ & $-1.25 \pm 1.08$ & $-0.87 \pm 0.73$ & JADES-GN-BD-9 & $2.03 \pm 0.91$ & $0.21 \pm 1.79$ & $4.46 \pm 1.01$ & $25.57 \pm 2.02$ & $25.18 \pm 0.99$  \\
JADES-GN-BD-10 & 1139636  & 189.30457 & 62.2936  & $0.2^{\prime\prime}$ & $12.92 \pm 1.11$ & $87.98 \pm 1.14$ & $53.68 \pm 1.18$ & $32.64 \pm 1.07$ & JADES-GN-BD-10 & $23.84 \pm 0.94$ & $15.63 \pm 1.79$ & $56.78 \pm 1.03$ & $176.64 \pm 1.6$ & $126.31 \pm 1.29$  \\
JADES-GN-BD-11 & 1193778  & 189.31387 & 62.15705  & $0.2^{\prime\prime}$ & $-0.18 \pm 0.54$ & $-0.14 \pm 0.8$ & $-0.84 \pm 1.05$ & $0.55 \pm 0.72$ & JADES-GN-BD-11 & $-0.2 \pm 0.84$ & $1.08 \pm 1.41$ & $2.25 \pm 0.98$ & $11.21 \pm 1.77$ & $7.35 \pm 0.98$  \\
JADES-GN-BD-12 & 1160357  & 189.39603 & 62.296005  & $0.2^{\prime\prime}$ & $2.7 \pm 1.37$ & $34.12 \pm 1.28$ & $17.49 \pm 0.96$ & $10.7 \pm 1.15$ & JADES-GN-BD-12 & $12.84 \pm 0.82$ & $13.1 \pm 1.29$ & $52.1 \pm 1.22$ & $222.12 \pm 1.83$ & $200.3 \pm 1.37$  \\
PANO-GN-BD-13 & 1246477  & 189.41893 & 62.193085  & $0.2^{\prime\prime}$ & - & $9.49 \pm 2.84$ & $7.24 \pm 1.91$ & $1.74 \pm 2.43$ & PANO-GN-BD-13 & $1.49 \pm 2.15$ & - & $10.03 \pm 1.81$ & $45.56 \pm 2.76$ & $35.88 \pm 2.35$  \\
PANO-GN-BD-14 & 1240204  & 189.47832 & 62.224022  & $0.2^{\prime\prime}$ & - & $19.68 \pm 3.47$ & $12.46 \pm 1.74$ & $5.21 \pm 2.47$ & PANO-GN-BD-14 & $5.18 \pm 2.05$ & - & $34.61 \pm 2.25$ & - & $100.14 \pm 2.79$  \\
PANO-GN-BD-15 & 1223231  & 189.48245 & 62.172085  & $0.2^{\prime\prime}$ & - & $54.8 \pm 2.11$ & $26.65 \pm 1.9$ & $10.6 \pm 2.2$ & PANO-GN-BD-15 & $18.13 \pm 1.4$ & - & $74.84 \pm 1.74$ & $448.12 \pm 3.46$ & $435.49 \pm 4.98$  \\
JADES-GN-BD-16 & 1133501  & 189.48941 & 62.26455  & $0.2^{\prime\prime}$ & $0.6 \pm 1.28$ & $0.56 \pm 1.6$ & $-0.95 \pm 0.89$ & $1.95 \pm 1.32$ & JADES-GN-BD-16 & $1.18 \pm 0.88$ & $1.02 \pm 1.35$ & $1.99 \pm 0.97$ & $13.97 \pm 1.02$ & $17.74 \pm 1.16$  \\
\enddata
\end{splitdeluxetable*}

\begin{splitdeluxetable*}{lccc|rrrrBl|rrrrr}
\tabletypesize{\scriptsize}
\tablecolumns{12}
\tablewidth{0pt}
\tablecaption{Photometry for Tentative Brown Dwarf Candidates (nJy) \label{tab:tentative_bd_fluxes}}
\tablehead{
 \colhead{Object Name} & \colhead{RA} & \colhead{DEC} & \colhead{diam.}  & \colhead{F090W} & \colhead{F115W} & \colhead{F150W} & \colhead{F200W} & \colhead{Object ID} & \colhead{F277W} & \colhead{F335M} & \colhead{F356W} & \colhead{F410M} & \colhead{F444W} }
 \startdata
JADES ID 75935  & 53.069565 & -27.846165  & $0.2^{\prime\prime}$ & $0.09 \pm 0.63$ & $0.16 \pm 0.43$ & $0.18 \pm 0.49$ & $0.57 \pm 0.43$ & JADES ID 75935 & $0.34 \pm 0.43$ & $0.76 \pm 0.46$ & $1.1 \pm 0.33$ & $3.12 \pm 0.66$ & $4.3 \pm 0.49$  \\
JADES ID 81388  & 53.066303 & -27.840015  & $0.2^{\prime\prime}$ & $2.51 \pm 1.55$ & $6.59 \pm 1.03$ & $3.61 \pm 1.25$ & $4.05 \pm 0.47$ & JADES ID 81388 & $3.05 \pm 0.83$ & - & $7.49 \pm 0.89$ & $12.02 \pm 1.92$ & $7.68 \pm 0.88$  \\
JADES ID 323111  & 53.138393 & -27.892658  & $0.2^{\prime\prime}$ & $-0.72 \pm 1.26$ & $4.4 \pm 0.47$ & $2.33 \pm 0.78$ & $2.3 \pm 0.81$ & JADES ID 323111 & $1.79 \pm 0.61$ & - & $2.84 \pm 0.74$ & $7.23 \pm 0.98$ & $5.06 \pm 0.82$  \\
JADES ID 388873  & 53.169212 & -27.888872  & $0.2^{\prime\prime}$ & - & $15.26 \pm 0.71$ & - & $14.26 \pm 2.94$ & JADES ID 388873 & $4.11 \pm 0.65$ & - & $5.17 \pm 0.59$ & $18.64 \pm 0.69$ & $11.95 \pm 0.9$  \\
JADES ID 529251  & 53.11507 & -27.964632  & $0.2^{\prime\prime}$ & - & $12.56 \pm 2.73$ & $12.67 \pm 2.15$ & $4.83 \pm 2.26$ & JADES ID 529251 & $6.87 \pm 2.22$ & - & $15.9 \pm 1.93$ & - & $30.8 \pm 2.66$  \\
JADES ID 534025  & 53.043034 & -28.024458  & $0.2^{\prime\prime}$ & - & $12.56 \pm 2.57$ & $6.15 \pm 2.09$ & $8.33 \pm 2.06$ & JADES ID 534025 & $5.8 \pm 1.51$ & - & $9.23 \pm 1.9$ & - & $20.69 \pm 2.39$  \\
JADES ID 1148837  & 189.38713 & 62.329243  & $0.2^{\prime\prime}$ & $3.92 \pm 1.46$ & $2.86 \pm 1.2$ & $1.91 \pm 1.0$ & $1.96 \pm 1.03$ & JADES ID 1148837 & $0.43 \pm 0.76$ & $-1.66 \pm 1.28$ & $3.73 \pm 0.92$ & $10.14 \pm 1.27$ & $9.09 \pm 1.12$  \\
JADES ID 1167677  & 189.25372 & 62.149643  & $0.2^{\prime\prime}$ & $-0.33 \pm 1.13$ & $7.17 \pm 1.13$ & $3.36 \pm 1.47$ & $2.47 \pm 1.16$ & JADES ID 1167677 & $0.56 \pm 1.0$ & $1.52 \pm 1.43$ & $2.54 \pm 0.89$ & $10.61 \pm 1.89$ & $7.01 \pm 1.26$  \\
JADES ID 1253547  & 189.44133 & 62.138565  & $0.2^{\prime\prime}$ & $1.91 \pm 1.02$ & $5.57 \pm 2.54$ & $4.7 \pm 2.15$ & $2.78 \pm 1.7$ & JADES ID 1253547 & $1.62 \pm 1.64$ & - & $4.45 \pm 1.36$ & - & $11.88 \pm 1.33$  \\
JADES ID 2013332  & 189.64673 & 62.220184  & $0.2^{\prime\prime}$ & - & $10.26 \pm 3.48$ & $4.15 \pm 1.72$ & $5.41 \pm 2.36$ & JADES ID 2013332 & $2.37 \pm 1.61$ & - & $5.94 \pm 1.72$ & - & $17.88 \pm 1.86$  \\
JADES ID 2015879  & 189.62143 & 62.2367  & $0.2^{\prime\prime}$ & - & $-1.78 \pm 2.41$ & $4.62 \pm 1.8$ & $-0.76 \pm 2.34$ & JADES ID 2015879 & $-1.08 \pm 1.78$ & - & $3.79 \pm 1.81$ & - & $22.72 \pm 1.96$  \\
\enddata
\end{splitdeluxetable*}

\section{\texttt{NIFTY} SED Plots and Source Properties}\label{sec:appendix_parameters}

In Tables \ref{tab:bd_properties_gs_gn} - \ref{tab:bd_properties_gs_gn_pt3}, we provide the source IDs and resulting \texttt{NIFTY} fitted model parameters from fits to the Sonora Elf Owl, LOWZ, and ATMO2020++ models. For each source we include the free parameters T$_{\mathrm{eff}}$, log(g), [M/H] for all three model sets, and log(K$_{\mathrm{zz}}$) and C/O for the Sonora Elf Owl and LOWZ models. We also include the distances in parsecs from the fit, and the reduced $\chi^2$ values derived from comparing the observed fluxes to the 50th percentile fluxes in each filter. In a few instance where there were more free parameters than data points (for very faint sources, and sources at temperatures where only the 4 - 5$\mu$m fluxes were observed) the reduced $\chi^2$ is unable to be calculated, which we present as -9999. Additionally, in Figures \ref{fig:SED_plots_pt2} - \ref{fig:SED_plots_pt4}, we continue to plot the SEDs and \texttt{NIFTY} Sonora Elf Owl fits for each of our sources, following Figure \ref{fig:SED_plots_pt1}. For each object, we plot the observed fluxes with black points, and we indicate 2$\sigma$ upper limits with downward facing arrows. The median \texttt{NIFTY} SED is provided in red, with a 1$\sigma$ error region shown with a solid light red region, and the model fluxes in each filter are shown with red squares. 

\begin{deluxetable*}{lcccccccc}
\tabletypesize{\scriptsize}
\tablecolumns{12}
\tablewidth{0pt}
\tablecaption{\texttt{NIFTY} Fitting Output Properties \label{tab:bd_properties_gs_gn}}
\tablehead{
 \colhead{Object ID} & \colhead{Model} & \colhead{T$_{\mathrm{eff}}$ (K)} & \colhead{log(g) (cgs)} & \colhead{log(K$_{\mathrm{zz}}$)} & \colhead{[M/H]} & \colhead{C/O}  & \colhead{distance (pc)}  & \colhead{$\chi^2_{\mathrm{red}}$} }
 \startdata
 & Sonora Elf Owl & $978.0^{+33.0}_{-40.0}$ & $5.2^{+0.2}_{-0.4}$ & $3.0^{+1.1}_{-0.7}$ & $-0.0^{+0.1}_{-0.1}$ & $1.1^{+0.3}_{-0.0}$ & $800.0^{+60.0}_{-60.0}$ & 2.62\\
JADES-GS-BD-1 & LOWZ & $948.0^{+38.0}_{-34.0}$ & $5.1^{+0.1}_{-0.3}$ & $5.2^{+1.1}_{-1.2}$ & $0.0^{+0.1}_{-0.1}$ & $0.8^{+0.1}_{-0.0}$ & $770.0^{+60.0}_{-60.0}$ & 2.08\\
 & ATMO2020 & $1015.0^{+25.0}_{-27.0}$ & $5.5^{+0.1}_{-0.1}$ & -  & $0.0^{+0.1}_{-0.1}$ & -  & $890.0^{+50.0}_{-50.0}$ & 12.82\\
\hline
 & Sonora Elf Owl & $1160.0^{+24.0}_{-22.0}$ & $5.3^{+0.2}_{-0.3}$ & $2.9^{+0.9}_{-0.6}$ & $-0.1^{+0.1}_{-0.1}$ & $0.7^{+0.1}_{-0.0}$ & $1650.0^{+70.0}_{-60.0}$ & 1.21\\
JADES-GS-BD-2 & LOWZ & $1083.0^{+29.0}_{-23.0}$ & $5.2^{+0.1}_{-0.1}$ & $4.8^{+1.2}_{-1.3}$ & $0.1^{+0.1}_{-0.1}$ & $0.6^{+0.1}_{-0.0}$ & $1510.0^{+90.0}_{-70.0}$ & 4.64\\
 & ATMO2020 & $1114.0^{+23.0}_{-25.0}$ & $5.5^{+0.1}_{-0.1}$ & -  & $-0.1^{+0.1}_{-0.1}$ & -  & $1590.0^{+80.0}_{-80.0}$ & 10.69\\
\hline
 & Sonora Elf Owl & $806.0^{+94.0}_{-114.0}$ & $4.6^{+0.6}_{-0.7}$ & $6.0^{+1.9}_{-2.2}$ & $-0.4^{+0.4}_{-0.4}$ & $1.2^{+0.5}_{-0.0}$ & $2490.0^{+580.0}_{-670.0}$ & 0.7\\
JADES-GS-BD-3 & LOWZ & $766.0^{+88.0}_{-98.0}$ & $4.6^{+0.5}_{-0.6}$ & $6.6^{+2.2}_{-3.6}$ & $-0.2^{+0.4}_{-0.4}$ & $0.6^{+0.2}_{-0.0}$ & $2280.0^{+570.0}_{-600.0}$ & 0.66\\
 & ATMO2020 & $844.0^{+91.0}_{-88.0}$ & $4.3^{+1.0}_{-0.8}$ & -  & $-0.6^{+0.4}_{-0.3}$ & -  & $2720.0^{+640.0}_{-560.0}$ & 1.29\\
\hline
 & Sonora Elf Owl & $483.0^{+80.0}_{-90.0}$ & $4.8^{+0.5}_{-0.8}$ & $4.3^{+2.4}_{-1.6}$ & $-0.7^{+0.4}_{-0.2}$ & $1.2^{+0.7}_{-0.0}$ & $1380.0^{+380.0}_{-420.0}$ & 0.69\\
JADES-GS-BD-4 & LOWZ & $545.0^{+51.0}_{-32.0}$ & $4.4^{+0.6}_{-0.6}$ & $1.9^{+2.6}_{-2.0}$ & $-0.7^{+0.4}_{-0.4}$ & $0.7^{+0.1}_{-0.0}$ & $1550.0^{+260.0}_{-170.0}$ & 0.65\\
 & ATMO2020 & $498.0^{+60.0}_{-66.0}$ & $5.0^{+0.4}_{-0.7}$ & -  & $-0.4^{+0.4}_{-0.4}$ & -  & $1280.0^{+350.0}_{-320.0}$ & 0.66\\
\hline
 & Sonora Elf Owl & $322.0^{+5.0}_{-7.0}$ & $3.0^{+0.1}_{-0.1}$ & $7.3^{+0.9}_{-0.7}$ & $0.1^{+0.1}_{-0.2}$ & $0.9^{+0.2}_{-0.0}$ & $70.0^{+1.0}_{-1.0}$ & 5.17\\
JADES-GS-BD-5 & LOWZ & $500.0^{+1.0}_{-1.0}$ & $3.5^{+0.1}_{-0.1}$ & $2.0^{+0.5}_{-0.5}$ & $-1.5^{+0.1}_{-0.1}$ & $0.8^{+0.1}_{-0.0}$ & $380.0^{+10.0}_{-1.0}$ & 43.79\\
 & ATMO2020 & $330.0^{+7.0}_{-5.0}$ & $2.5^{+0.1}_{-0.1}$ & -  & $-0.6^{+0.1}_{-0.1}$ & -  & $90.0^{+10.0}_{-1.0}$ & 3.19\\
\hline
 & Sonora Elf Owl & $929.0^{+38.0}_{-44.0}$ & $5.0^{+0.3}_{-0.3}$ & $3.4^{+1.4}_{-1.0}$ & $-0.4^{+0.1}_{-0.1}$ & $1.2^{+0.3}_{-0.0}$ & $2120.0^{+170.0}_{-190.0}$ & 0.34\\
JADES-GS-BD-6 & LOWZ & $866.0^{+44.0}_{-51.0}$ & $4.6^{+0.4}_{-0.4}$ & $5.6^{+1.3}_{-1.4}$ & $-0.3^{+0.1}_{-0.1}$ & $0.7^{+0.1}_{-0.0}$ & $1840.0^{+200.0}_{-230.0}$ & 0.45\\
 & ATMO2020 & $933.0^{+27.0}_{-28.0}$ & $5.5^{+0.1}_{-0.1}$ & -  & $0.0^{+0.1}_{-0.1}$ & -  & $2180.0^{+130.0}_{-140.0}$ & 10.18\\
\hline
 & Sonora Elf Owl & $876.0^{+24.0}_{-26.0}$ & $4.9^{+0.3}_{-0.3}$ & $2.5^{+0.6}_{-0.4}$ & $-0.1^{+0.1}_{-0.1}$ & $0.8^{+0.2}_{-0.0}$ & $920.0^{+50.0}_{-50.0}$ & 1.01\\
JADES-GS-BD-7 & LOWZ & $882.0^{+26.0}_{-34.0}$ & $5.0^{+0.2}_{-0.3}$ & $3.2^{+1.6}_{-2.1}$ & $-0.1^{+0.1}_{-0.1}$ & $0.6^{+0.1}_{-0.0}$ & $930.0^{+60.0}_{-80.0}$ & 1.13\\
 & ATMO2020 & $894.0^{+16.0}_{-19.0}$ & $5.5^{+0.1}_{-0.1}$ & -  & $0.0^{+0.1}_{-0.1}$ & -  & $1000.0^{+40.0}_{-40.0}$ & 10.56\\
\hline
 & Sonora Elf Owl & $486.0^{+39.0}_{-44.0}$ & $4.6^{+0.6}_{-0.8}$ & $4.6^{+1.8}_{-1.6}$ & $-0.4^{+0.4}_{-0.3}$ & $1.4^{+0.6}_{-0.0}$ & $920.0^{+150.0}_{-190.0}$ & 0.45\\
JADES-GS-BD-8 & LOWZ & $508.0^{+14.0}_{-6.0}$ & $4.8^{+0.3}_{-0.5}$ & $3.4^{+1.5}_{-1.5}$ & $-0.3^{+0.2}_{-0.2}$ & $0.8^{+0.1}_{-0.0}$ & $940.0^{+60.0}_{-40.0}$ & 1.91\\
 & ATMO2020 & $404.0^{+38.0}_{-25.0}$ & $3.3^{+1.0}_{-0.4}$ & -  & $-0.4^{+0.5}_{-0.3}$ & -  & $460.0^{+130.0}_{-70.0}$ & 1.69\\
\hline
 & Sonora Elf Owl & $778.0^{+51.0}_{-57.0}$ & $5.0^{+0.4}_{-0.5}$ & $3.1^{+1.3}_{-0.8}$ & $-0.0^{+0.2}_{-0.2}$ & $0.8^{+0.2}_{-0.0}$ & $1770.0^{+240.0}_{-280.0}$ & 0.86\\
JADES-GS-BD-9 & LOWZ & $697.0^{+68.0}_{-73.0}$ & $4.8^{+0.3}_{-0.6}$ & $5.2^{+2.0}_{-2.8}$ & $0.2^{+0.4}_{-0.4}$ & $0.5^{+0.1}_{-0.0}$ & $1370.0^{+300.0}_{-300.0}$ & 0.77\\
 & ATMO2020 & $806.0^{+38.0}_{-30.0}$ & $5.4^{+0.1}_{-0.1}$ & -  & $0.2^{+0.1}_{-0.2}$ & -  & $1820.0^{+200.0}_{-120.0}$ & 2.12\\
\hline
 & Sonora Elf Owl & $979.0^{+17.0}_{-15.0}$ & $5.0^{+0.2}_{-0.3}$ & $6.0^{+0.9}_{-1.9}$ & $0.0^{+0.1}_{-0.1}$ & $1.4^{+0.1}_{-0.0}$ & $250.0^{+10.0}_{-1.0}$ & 2.45\\
JADES-GS-BD-10 & LOWZ & $989.0^{+14.0}_{-18.0}$ & $4.8^{+0.2}_{-0.2}$ & $5.5^{+0.7}_{-0.8}$ & $-0.1^{+0.1}_{-0.1}$ & $0.8^{+0.1}_{-0.0}$ & $260.0^{+10.0}_{-1.0}$ & 1.56\\
 & ATMO2020 & $1050.0^{+15.0}_{-13.0}$ & $5.5^{+0.1}_{-0.1}$ & -  & $-0.0^{+0.1}_{-0.1}$ & -  & $300.0^{+10.0}_{-10.0}$ & 11.84\\
\hline
 & Sonora Elf Owl & $644.0^{+37.0}_{-38.0}$ & $4.3^{+0.7}_{-0.4}$ & $5.2^{+1.1}_{-1.5}$ & $-0.4^{+0.2}_{-0.2}$ & $0.6^{+0.2}_{-0.0}$ & $710.0^{+100.0}_{-90.0}$ & 0.59\\
JADES-GS-BD-11 & LOWZ & $596.0^{+39.0}_{-43.0}$ & $4.2^{+0.5}_{-0.4}$ & $7.4^{+1.4}_{-1.6}$ & $-0.5^{+0.2}_{-0.2}$ & $0.6^{+0.1}_{-0.0}$ & $570.0^{+100.0}_{-90.0}$ & 0.7\\
 & ATMO2020 & $656.0^{+19.0}_{-18.0}$ & $5.5^{+0.1}_{-0.1}$ & -  & $0.2^{+0.1}_{-0.1}$ & -  & $700.0^{+40.0}_{-50.0}$ & 6.14\\
\hline
 & Sonora Elf Owl & $720.0^{+36.0}_{-38.0}$ & $5.2^{+0.2}_{-0.4}$ & $6.5^{+1.7}_{-2.5}$ & $-0.9^{+0.1}_{-0.1}$ & $0.8^{+0.3}_{-0.0}$ & $2260.0^{+160.0}_{-160.0}$ & 2.52\\
JADES-GS-BD-12 & LOWZ & $718.0^{+38.0}_{-43.0}$ & $4.2^{+0.6}_{-0.5}$ & $7.3^{+2.0}_{-3.5}$ & $-1.5^{+0.3}_{-0.2}$ & $0.7^{+0.1}_{-0.0}$ & $2160.0^{+170.0}_{-200.0}$ & 1.89\\
 & ATMO2020 & $592.0^{+219.0}_{-79.0}$ & $3.4^{+1.9}_{-0.7}$ & -  & $-0.8^{+0.5}_{-0.2}$ & -  & $1260.0^{+1430.0}_{-380.0}$ & 1.23\\
\hline
 & Sonora Elf Owl & $634.0^{+27.0}_{-27.0}$ & $3.8^{+0.3}_{-0.3}$ & $3.5^{+1.0}_{-0.9}$ & $-0.4^{+0.1}_{-0.2}$ & $0.7^{+0.2}_{-0.0}$ & $840.0^{+80.0}_{-90.0}$ & 2.12\\
JADES-GS-BD-13 & LOWZ & $590.0^{+44.0}_{-52.0}$ & $3.8^{+0.3}_{-0.2}$ & $5.9^{+1.4}_{-1.8}$ & $-0.5^{+0.2}_{-0.2}$ & $0.7^{+0.1}_{-0.0}$ & $690.0^{+130.0}_{-140.0}$ & 5.8\\
 & ATMO2020 & $652.0^{+19.0}_{-12.0}$ & $5.5^{+0.1}_{-0.1}$ & -  & $0.2^{+0.1}_{-0.1}$ & -  & $900.0^{+60.0}_{-40.0}$ & 24.05\\
\hline
 & Sonora Elf Owl & $950.0^{+13.0}_{-14.0}$ & $5.4^{+0.1}_{-0.1}$ & $6.2^{+0.8}_{-0.8}$ & $0.1^{+0.1}_{-0.1}$ & $1.5^{+0.1}_{-0.0}$ & $310.0^{+10.0}_{-10.0}$ & 9.13\\
JADES-GS-BD-14 & LOWZ & $987.0^{+12.0}_{-17.0}$ & $5.2^{+0.1}_{-0.1}$ & $5.1^{+0.9}_{-0.9}$ & $-0.0^{+0.1}_{-0.1}$ & $0.8^{+0.1}_{-0.0}$ & $340.0^{+10.0}_{-10.0}$ & 8.65\\
 & ATMO2020 & $1004.0^{+14.0}_{-16.0}$ & $5.5^{+0.1}_{-0.1}$ & -  & $0.1^{+0.1}_{-0.1}$ & -  & $360.0^{+10.0}_{-10.0}$ & 19.79\\
\hline
 & Sonora Elf Owl & $983.0^{+34.0}_{-37.0}$ & $4.4^{+0.3}_{-0.4}$ & $4.8^{+1.5}_{-1.2}$ & $-0.4^{+0.1}_{-0.1}$ & $0.8^{+0.3}_{-0.0}$ & $2790.0^{+210.0}_{-220.0}$ & 1.67\\
JADES-GS-BD-15 & LOWZ & $906.0^{+41.0}_{-24.0}$ & $4.5^{+0.5}_{-0.3}$ & $7.6^{+1.1}_{-1.2}$ & $-0.3^{+0.1}_{-0.1}$ & $0.7^{+0.1}_{-0.0}$ & $2420.0^{+240.0}_{-140.0}$ & 3.35\\
 & ATMO2020 & $1079.0^{+23.0}_{-34.0}$ & $5.5^{+0.1}_{-0.1}$ & -  & $-0.4^{+0.1}_{-0.1}$ & -  & $3500.0^{+140.0}_{-210.0}$ & 11.88\\
\hline
 & Sonora Elf Owl & $670.0^{+36.0}_{-43.0}$ & $4.9^{+0.5}_{-1.0}$ & $3.7^{+1.5}_{-1.2}$ & $-0.5^{+0.2}_{-0.2}$ & $1.5^{+0.6}_{-1.0}$ & $1560.0^{+160.0}_{-210.0}$ & 1.83\\
JADES-GS-BD-16 & LOWZ & $600.0^{+52.0}_{-54.0}$ & $4.2^{+0.6}_{-0.5}$ & $5.3^{+1.6}_{-1.9}$ & $-0.4^{+0.3}_{-0.3}$ & $0.8^{+0.1}_{-0.0}$ & $1170.0^{+230.0}_{-230.0}$ & 2.36\\
 & ATMO2020 & $640.0^{+35.0}_{-37.0}$ & $5.3^{+0.2}_{-0.5}$ & -  & $0.2^{+0.1}_{-0.2}$ & -  & $1390.0^{+150.0}_{-230.0}$ & 4.5\\
\hline
 & Sonora Elf Owl & $965.0^{+42.0}_{-36.0}$ & $3.1^{+0.1}_{-0.1}$ & $6.1^{+0.9}_{-0.8}$ & $-0.7^{+0.2}_{-0.2}$ & $0.6^{+0.1}_{-0.0}$ & $3020.0^{+250.0}_{-230.0}$ & 3.98\\
JADES-GS-BD-17 & LOWZ & $964.0^{+54.0}_{-46.0}$ & $3.7^{+0.3}_{-0.1}$ & $4.9^{+2.0}_{-2.6}$ & $-0.7^{+0.2}_{-0.2}$ & $0.4^{+0.1}_{-0.0}$ & $3070.0^{+380.0}_{-300.0}$ & 4.64\\
 & ATMO2020 & $984.0^{+75.0}_{-88.0}$ & $4.6^{+0.6}_{-0.4}$ & -  & $-0.7^{+0.4}_{-0.2}$ & -  & $3120.0^{+550.0}_{-550.0}$ & 1.35\\
\hline
 & Sonora Elf Owl & $498.0^{+38.0}_{-58.0}$ & $4.0^{+0.9}_{-0.6}$ & $5.0^{+2.5}_{-2.0}$ & $-0.6^{+0.4}_{-0.3}$ & $1.2^{+0.7}_{-0.0}$ & $440.0^{+90.0}_{-140.0}$ & -9999.0\\
PANO-GS-BD-18 & LOWZ & $511.0^{+15.0}_{-8.0}$ & $3.9^{+0.6}_{-0.3}$ & $1.6^{+1.1}_{-1.3}$ & $-0.7^{+0.2}_{-0.2}$ & $0.8^{+0.1}_{-0.0}$ & $460.0^{+30.0}_{-20.0}$ & -9999.0\\
 & ATMO2020 & $487.0^{+30.0}_{-67.0}$ & $4.8^{+0.4}_{-1.1}$ & -  & $0.2^{+0.1}_{-0.3}$ & -  & $360.0^{+90.0}_{-140.0}$ & 7.15\\
\hline
 & Sonora Elf Owl & $1066.0^{+89.0}_{-105.0}$ & $5.0^{+0.4}_{-0.7}$ & $6.1^{+1.9}_{-2.2}$ & $-0.6^{+0.4}_{-0.3}$ & $1.0^{+0.4}_{-0.0}$ & $2620.0^{+400.0}_{-460.0}$ & -9999.0\\
PANO-GS-BD-19 & LOWZ & $942.0^{+71.0}_{-75.0}$ & $4.8^{+0.4}_{-0.6}$ & $7.9^{+1.5}_{-3.0}$ & $-0.3^{+0.3}_{-0.3}$ & $0.6^{+0.1}_{-0.0}$ & $2130.0^{+370.0}_{-350.0}$ & -9999.0\\
 & ATMO2020 & $1093.0^{+66.0}_{-81.0}$ & $5.1^{+0.3}_{-0.9}$ & -  & $-0.6^{+0.3}_{-0.3}$ & -  & $2820.0^{+360.0}_{-400.0}$ & 4.73\\
\enddata
\end{deluxetable*}

\begin{deluxetable*}{lcccccccc}
\tabletypesize{\scriptsize}
\tablecolumns{12}
\tablewidth{0pt}
\tablecaption{\texttt{NIFTY} Fitting Output Properties (continued) \label{tab:bd_properties_gs_gn_pt2}}
\tablehead{
 \colhead{Object ID} & \colhead{Model} & \colhead{T$_{\mathrm{eff}}$ (K)} & \colhead{log(g) (cgs)} & \colhead{log(K$_{\mathrm{zz}}$)} & \colhead{[M/H]} & \colhead{C/O}  & \colhead{distance (pc)}  & \colhead{$\chi^2_{\mathrm{red}}$} }
 \startdata
 & Sonora Elf Owl & $756.0^{+41.0}_{-43.0}$ & $5.4^{+0.1}_{-0.2}$ & $6.8^{+1.7}_{-2.8}$ & $-0.9^{+0.1}_{-0.1}$ & $0.6^{+0.1}_{-0.0}$ & $2910.0^{+210.0}_{-210.0}$ & 4.69\\
JADES-GS-BD-20 & LOWZ & $790.0^{+49.0}_{-56.0}$ & $4.3^{+0.6}_{-0.5}$ & $5.0^{+3.4}_{-3.9}$ & $-2.0^{+0.4}_{-0.3}$ & $0.5^{+0.2}_{-0.0}$ & $3010.0^{+260.0}_{-300.0}$ & 1.52\\
 & ATMO2020 & $531.0^{+64.0}_{-34.0}$ & $2.7^{+0.3}_{-0.1}$ & -  & $-0.2^{+0.4}_{-0.4}$ & -  & $1160.0^{+330.0}_{-200.0}$ & 1.55\\
\hline
 & Sonora Elf Owl & $340.0^{+56.0}_{-43.0}$ & $4.1^{+0.8}_{-0.8}$ & $4.4^{+2.3}_{-1.7}$ & $-0.6^{+0.5}_{-0.3}$ & $1.4^{+0.7}_{-1.0}$ & $900.0^{+310.0}_{-210.0}$ & 1.47\\
JADES-GS-BD-21 & LOWZ & $507.0^{+13.0}_{-6.0}$ & $4.0^{+0.7}_{-0.4}$ & $0.8^{+1.3}_{-1.2}$ & $-0.6^{+0.2}_{-0.3}$ & $0.8^{+0.1}_{-0.0}$ & $1800.0^{+120.0}_{-100.0}$ & 4.84\\
 & ATMO2020 & $347.0^{+39.0}_{-38.0}$ & $4.5^{+0.7}_{-1.0}$ & -  & $-0.4^{+0.4}_{-0.4}$ & -  & $740.0^{+200.0}_{-160.0}$ & 0.84\\
\hline
 & Sonora Elf Owl & $1014.0^{+19.0}_{-22.0}$ & $5.3^{+0.2}_{-0.3}$ & $5.4^{+0.8}_{-1.0}$ & $0.3^{+0.1}_{-0.1}$ & $1.6^{+0.1}_{-0.0}$ & $430.0^{+20.0}_{-30.0}$ & 3.13\\
JADES-GS-BD-22 & LOWZ & $983.0^{+41.0}_{-27.0}$ & $4.8^{+0.4}_{-0.3}$ & $4.8^{+1.0}_{-0.9}$ & $0.2^{+0.1}_{-0.1}$ & $0.8^{+0.1}_{-0.0}$ & $410.0^{+40.0}_{-30.0}$ & 2.23\\
 & ATMO2020 & $1122.0^{+22.0}_{-23.0}$ & $5.5^{+0.1}_{-0.1}$ & -  & $0.0^{+0.1}_{-0.1}$ & -  & $530.0^{+20.0}_{-30.0}$ & 16.11\\
\hline
 & Sonora Elf Owl & $781.0^{+14.0}_{-13.0}$ & $3.0^{+2.4}_{-0.1}$ & $9.0^{+0.1}_{-1.6}$ & $-0.7^{+0.1}_{-0.1}$ & $0.5^{+0.1}_{-0.0}$ & $870.0^{+30.0}_{-40.0}$ & 25.82\\
JADES-GS-BD-23 & LOWZ & $849.0^{+3.0}_{-4.0}$ & $3.5^{+0.1}_{-0.1}$ & $3.3^{+1.9}_{-1.1}$ & $-1.5^{+0.1}_{-0.1}$ & $0.2^{+0.1}_{-0.0}$ & $1020.0^{+20.0}_{-10.0}$ & 8.61\\
 & ATMO2020 & $872.0^{+13.0}_{-13.0}$ & $5.4^{+0.1}_{-0.1}$ & -  & $-1.0^{+0.1}_{-0.1}$ & -  & $1090.0^{+30.0}_{-40.0}$ & 8.8\\
\hline
 & Sonora Elf Owl & $527.0^{+41.0}_{-51.0}$ & $5.3^{+0.2}_{-0.3}$ & $5.3^{+2.3}_{-2.3}$ & $-0.9^{+0.1}_{-0.1}$ & $0.6^{+0.2}_{-0.0}$ & $1330.0^{+170.0}_{-200.0}$ & 0.93\\
JADES-GS-BD-24 & LOWZ & $527.0^{+28.0}_{-19.0}$ & $4.2^{+0.6}_{-0.5}$ & $5.0^{+3.4}_{-3.8}$ & $-1.5^{+0.3}_{-0.3}$ & $0.6^{+0.2}_{-0.0}$ & $1220.0^{+110.0}_{-80.0}$ & 1.2\\
 & ATMO2020 & $582.0^{+39.0}_{-41.0}$ & $5.3^{+0.1}_{-0.2}$ & -  & $-0.8^{+0.1}_{-0.1}$ & -  & $1470.0^{+180.0}_{-150.0}$ & 0.45\\
\hline
 & Sonora Elf Owl & $697.0^{+54.0}_{-59.0}$ & $4.4^{+0.5}_{-0.5}$ & $6.1^{+1.6}_{-1.7}$ & $-0.2^{+0.3}_{-0.3}$ & $0.9^{+0.3}_{-0.0}$ & $1530.0^{+250.0}_{-290.0}$ & 0.77\\
PANO-GS-BD-25 & LOWZ & $618.0^{+75.0}_{-64.0}$ & $4.4^{+0.5}_{-0.5}$ & $8.8^{+0.8}_{-1.0}$ & $-0.1^{+0.3}_{-0.3}$ & $0.6^{+0.1}_{-0.0}$ & $1140.0^{+330.0}_{-270.0}$ & 0.86\\
 & ATMO2020 & $784.0^{+32.0}_{-23.0}$ & $5.4^{+0.1}_{-0.1}$ & -  & $0.2^{+0.1}_{-0.2}$ & -  & $1840.0^{+170.0}_{-100.0}$ & 12.63\\
\hline
 & Sonora Elf Owl & $640.0^{+31.0}_{-34.0}$ & $4.6^{+0.7}_{-0.7}$ & $2.8^{+1.0}_{-0.6}$ & $-0.2^{+0.1}_{-0.1}$ & $1.2^{+0.6}_{-0.0}$ & $690.0^{+90.0}_{-90.0}$ & 1.29\\
JADES-GN-BD-1 & LOWZ & $568.0^{+48.0}_{-41.0}$ & $4.4^{+0.6}_{-0.6}$ & $4.7^{+1.2}_{-1.5}$ & $-0.1^{+0.2}_{-0.2}$ & $0.8^{+0.1}_{-0.0}$ & $510.0^{+110.0}_{-90.0}$ & 2.97\\
 & ATMO2020 & $678.0^{+31.0}_{-18.0}$ & $5.5^{+0.1}_{-0.1}$ & -  & $0.3^{+0.1}_{-0.1}$ & -  & $810.0^{+80.0}_{-40.0}$ & 18.43\\
\hline
 & Sonora Elf Owl & $711.0^{+39.0}_{-49.0}$ & $5.2^{+0.2}_{-0.4}$ & $4.0^{+1.5}_{-1.2}$ & $0.0^{+0.2}_{-0.2}$ & $1.2^{+0.3}_{-0.0}$ & $1150.0^{+130.0}_{-160.0}$ & 0.34\\
JADES-GN-BD-2 & LOWZ & $635.0^{+38.0}_{-52.0}$ & $5.0^{+0.2}_{-0.4}$ & $7.5^{+1.1}_{-1.2}$ & $0.1^{+0.2}_{-0.2}$ & $0.8^{+0.1}_{-0.0}$ & $880.0^{+120.0}_{-150.0}$ & 0.97\\
 & ATMO2020 & $799.0^{+25.0}_{-46.0}$ & $5.5^{+0.1}_{-0.1}$ & -  & $0.1^{+0.2}_{-0.1}$ & -  & $1440.0^{+100.0}_{-150.0}$ & 8.17\\
\hline
 & Sonora Elf Owl & $603.0^{+14.0}_{-15.0}$ & $4.5^{+0.2}_{-0.1}$ & $2.1^{+0.2}_{-0.1}$ & $-0.1^{+0.1}_{-0.1}$ & $0.8^{+0.1}_{-0.0}$ & $460.0^{+30.0}_{-30.0}$ & 25.43\\
JADES-GN-BD-3 & LOWZ & $533.0^{+18.0}_{-16.0}$ & $4.4^{+0.3}_{-0.3}$ & $0.3^{+0.6}_{-0.6}$ & $-0.1^{+0.1}_{-0.1}$ & $0.8^{+0.1}_{-0.0}$ & $320.0^{+20.0}_{-30.0}$ & 13.15\\
 & ATMO2020 & $577.0^{+28.0}_{-15.0}$ & $5.5^{+0.1}_{-0.1}$ & -  & $0.3^{+0.1}_{-0.1}$ & -  & $420.0^{+50.0}_{-20.0}$ & 52.51\\
\hline
 & Sonora Elf Owl & $676.0^{+44.0}_{-47.0}$ & $5.2^{+0.2}_{-0.5}$ & $6.9^{+1.5}_{-2.5}$ & $-0.8^{+0.2}_{-0.1}$ & $0.7^{+0.2}_{-0.0}$ & $2050.0^{+200.0}_{-210.0}$ & 2.0\\
JADES-GN-BD-4 & LOWZ & $653.0^{+71.0}_{-88.0}$ & $4.2^{+0.6}_{-0.5}$ & $7.0^{+2.1}_{-3.3}$ & $-1.1^{+0.6}_{-0.4}$ & $0.4^{+0.2}_{-0.0}$ & $1830.0^{+330.0}_{-430.0}$ & 2.15\\
 & ATMO2020 & $624.0^{+185.0}_{-132.0}$ & $4.1^{+1.2}_{-1.4}$ & -  & $-0.6^{+0.7}_{-0.3}$ & -  & $1400.0^{+1220.0}_{-640.0}$ & 1.12\\
\hline
 & Sonora Elf Owl & $398.0^{+52.0}_{-55.0}$ & $4.9^{+0.4}_{-0.7}$ & $5.5^{+2.3}_{-2.3}$ & $-0.8^{+0.3}_{-0.1}$ & $0.9^{+0.5}_{-0.0}$ & $870.0^{+210.0}_{-210.0}$ & 1.33\\
JADES-GN-BD-5 & LOWZ & $507.0^{+11.0}_{-5.0}$ & $4.6^{+0.5}_{-0.7}$ & $2.0^{+3.4}_{-2.2}$ & $-0.9^{+0.4}_{-0.2}$ & $0.7^{+0.1}_{-0.0}$ & $1210.0^{+60.0}_{-50.0}$ & 3.35\\
 & ATMO2020 & $448.0^{+45.0}_{-49.0}$ & $4.5^{+0.8}_{-1.0}$ & -  & $-0.6^{+0.4}_{-0.3}$ & -  & $830.0^{+230.0}_{-190.0}$ & 1.22\\
\hline
 & Sonora Elf Owl & $1118.0^{+140.0}_{-154.0}$ & $4.6^{+0.6}_{-0.9}$ & $5.5^{+2.2}_{-2.2}$ & $-0.4^{+0.6}_{-0.4}$ & $1.2^{+0.5}_{-0.0}$ & $6820.0^{+1540.0}_{-1660.0}$ & 0.38\\
JADES-GN-BD-6 & LOWZ & $1085.0^{+166.0}_{-163.0}$ & $4.5^{+0.5}_{-0.6}$ & $3.9^{+3.5}_{-3.2}$ & $-0.5^{+0.8}_{-1.0}$ & $0.5^{+0.2}_{-0.0}$ & $7050.0^{+2070.0}_{-2060.0}$ & 0.86\\
 & ATMO2020 & $1075.0^{+85.0}_{-120.0}$ & $4.7^{+0.5}_{-0.8}$ & -  & $-0.4^{+0.4}_{-0.4}$ & -  & $6640.0^{+1010.0}_{-1260.0}$ & 0.65\\
\hline
 & Sonora Elf Owl & $1012.0^{+130.0}_{-152.0}$ & $4.1^{+1.0}_{-0.8}$ & $7.4^{+1.1}_{-2.0}$ & $-0.4^{+0.6}_{-0.4}$ & $0.8^{+0.4}_{-0.0}$ & $5360.0^{+1100.0}_{-1380.0}$ & 3.25\\
JADES-GN-BD-7 & LOWZ & $1006.0^{+125.0}_{-147.0}$ & $4.3^{+0.6}_{-0.5}$ & $6.4^{+2.5}_{-3.6}$ & $-1.2^{+1.1}_{-0.8}$ & $0.3^{+0.2}_{-0.0}$ & $5860.0^{+1150.0}_{-1490.0}$ & 3.71\\
 & ATMO2020 & $894.0^{+130.0}_{-125.0}$ & $3.5^{+0.5}_{-0.4}$ & -  & $-0.5^{+0.4}_{-0.4}$ & -  & $4250.0^{+1190.0}_{-1010.0}$ & 1.07\\
\hline
 & Sonora Elf Owl & $450.0^{+26.0}_{-30.0}$ & $3.6^{+1.1}_{-0.4}$ & $4.2^{+1.8}_{-1.5}$ & $-0.8^{+0.3}_{-0.2}$ & $1.2^{+0.7}_{-0.0}$ & $610.0^{+80.0}_{-100.0}$ & 2.86\\
JADES-GN-BD-8 & LOWZ & $502.0^{+4.0}_{-2.0}$ & $3.7^{+0.5}_{-0.2}$ & $2.0^{+0.8}_{-0.9}$ & $-0.9^{+0.2}_{-0.1}$ & $0.8^{+0.1}_{-0.0}$ & $720.0^{+20.0}_{-20.0}$ & 11.76\\
 & ATMO2020 & $400.0^{+27.0}_{-25.0}$ & $4.0^{+0.6}_{-0.6}$ & -  & $0.0^{+0.2}_{-0.4}$ & -  & $330.0^{+90.0}_{-60.0}$ & 1.4\\
\hline
 & Sonora Elf Owl & $450.0^{+66.0}_{-72.0}$ & $4.5^{+0.7}_{-1.0}$ & $5.6^{+2.1}_{-2.3}$ & $-0.7^{+0.5}_{-0.2}$ & $1.2^{+0.7}_{-0.0}$ & $1190.0^{+330.0}_{-370.0}$ & 2.8\\
JADES-GN-BD-9 & LOWZ & $524.0^{+31.0}_{-18.0}$ & $4.3^{+0.6}_{-0.5}$ & $3.4^{+2.9}_{-2.7}$ & $-0.9^{+0.3}_{-0.3}$ & $0.7^{+0.1}_{-0.0}$ & $1480.0^{+160.0}_{-110.0}$ & 4.02\\
 & ATMO2020 & $441.0^{+60.0}_{-56.0}$ & $3.7^{+1.0}_{-0.8}$ & -  & $-0.4^{+0.5}_{-0.4}$ & -  & $840.0^{+390.0}_{-280.0}$ & 1.63\\
\hline
 & Sonora Elf Owl & $978.0^{+31.0}_{-32.0}$ & $5.2^{+0.2}_{-0.3}$ & $4.9^{+1.2}_{-1.3}$ & $-0.1^{+0.1}_{-0.1}$ & $1.2^{+0.2}_{-0.0}$ & $1290.0^{+90.0}_{-90.0}$ & 1.75\\
JADES-GN-BD-10 & LOWZ & $921.0^{+37.0}_{-33.0}$ & $4.3^{+0.4}_{-0.3}$ & $6.3^{+0.9}_{-0.9}$ & $-0.2^{+0.1}_{-0.1}$ & $0.7^{+0.1}_{-0.0}$ & $1150.0^{+110.0}_{-90.0}$ & 1.14\\
 & ATMO2020 & $1051.0^{+28.0}_{-26.0}$ & $5.5^{+0.1}_{-0.1}$ & -  & $-0.2^{+0.1}_{-0.1}$ & -  & $1540.0^{+80.0}_{-80.0}$ & 10.33\\
\hline
 & Sonora Elf Owl & $463.0^{+96.0}_{-88.0}$ & $4.7^{+0.6}_{-1.0}$ & $6.8^{+1.6}_{-2.7}$ & $-0.4^{+0.8}_{-0.4}$ & $0.9^{+0.6}_{-0.0}$ & $2090.0^{+950.0}_{-1010.0}$ & 1.74\\
JADES-GN-BD-11 & LOWZ & $562.0^{+78.0}_{-45.0}$ & $4.4^{+0.6}_{-0.6}$ & $5.2^{+3.4}_{-4.0}$ & $-1.2^{+0.6}_{-0.6}$ & $0.5^{+0.2}_{-0.0}$ & $3010.0^{+720.0}_{-430.0}$ & 2.32\\
 & ATMO2020 & $519.0^{+120.0}_{-102.0}$ & $4.5^{+0.8}_{-1.5}$ & -  & $-0.5^{+0.6}_{-0.3}$ & -  & $2260.0^{+1360.0}_{-1080.0}$ & 1.27\\
\hline
 & Sonora Elf Owl & $588.0^{+41.0}_{-20.0}$ & $3.1^{+0.2}_{-0.1}$ & $5.3^{+0.6}_{-0.6}$ & $-0.3^{+0.1}_{-0.4}$ & $0.6^{+0.1}_{-0.0}$ & $520.0^{+120.0}_{-60.0}$ & 14.76\\
JADES-GN-BD-12 & LOWZ & $635.0^{+32.0}_{-34.0}$ & $3.6^{+0.1}_{-0.1}$ & $4.8^{+2.1}_{-2.0}$ & $-0.9^{+0.1}_{-0.2}$ & $0.6^{+0.1}_{-0.0}$ & $650.0^{+70.0}_{-90.0}$ & 21.03\\
 & ATMO2020 & $680.0^{+23.0}_{-23.0}$ & $5.3^{+0.1}_{-0.2}$ & -  & $0.0^{+0.1}_{-0.1}$ & -  & $740.0^{+60.0}_{-50.0}$ & 7.73\\
\hline
 & Sonora Elf Owl & $720.0^{+71.0}_{-84.0}$ & $4.9^{+0.4}_{-0.8}$ & $5.8^{+1.9}_{-2.2}$ & $-0.5^{+0.4}_{-0.3}$ & $1.5^{+0.7}_{-1.0}$ & $1790.0^{+320.0}_{-390.0}$ & 2.29\\
PANO-GN-BD-13 & LOWZ & $685.0^{+90.0}_{-89.0}$ & $4.5^{+0.5}_{-0.7}$ & $7.0^{+1.7}_{-2.3}$ & $-0.5^{+0.4}_{-0.4}$ & $0.7^{+0.1}_{-0.0}$ & $1590.0^{+410.0}_{-400.0}$ & 3.12\\
 & ATMO2020 & $746.0^{+79.0}_{-91.0}$ & $4.8^{+0.5}_{-0.9}$ & -  & $-0.3^{+0.4}_{-0.4}$ & -  & $1850.0^{+390.0}_{-440.0}$ & 2.69\\
\enddata
\end{deluxetable*}

\begin{deluxetable*}{lcccccccc}
\tabletypesize{\scriptsize}
\tablecolumns{12}
\tablewidth{0pt}
\tablecaption{\texttt{NIFTY} Fitting Output Properties (continued) \label{tab:bd_properties_gs_gn_pt3}}
\tablehead{
 \colhead{Object ID} & \colhead{Model} & \colhead{T$_{\mathrm{eff}}$ (K)} & \colhead{log(g) (cgs)} & \colhead{log(K$_{\mathrm{zz}}$)} & \colhead{[M/H]} & \colhead{C/O}  & \colhead{distance (pc)}  & \colhead{$\chi^2_{\mathrm{red}}$} }
 \startdata
 & Sonora Elf Owl & $696.0^{+51.0}_{-88.0}$ & $4.9^{+0.4}_{-0.8}$ & $6.0^{+2.0}_{-2.5}$ & $-0.6^{+0.4}_{-0.2}$ & $0.9^{+0.4}_{-0.0}$ & $1110.0^{+170.0}_{-290.0}$ & -9999.0\\
PANO-GN-BD-14 & LOWZ & $706.0^{+49.0}_{-74.0}$ & $4.4^{+0.6}_{-0.6}$ & $3.8^{+3.8}_{-3.2}$ & $-0.9^{+0.4}_{-0.3}$ & $0.6^{+0.2}_{-0.0}$ & $1110.0^{+160.0}_{-230.0}$ & -9999.0\\
 & ATMO2020 & $723.0^{+60.0}_{-107.0}$ & $4.8^{+0.5}_{-1.3}$ & -  & $-0.5^{+0.3}_{-0.3}$ & -  & $1120.0^{+230.0}_{-370.0}$ & 3.51\\
\hline
 & Sonora Elf Owl & $531.0^{+38.0}_{-36.0}$ & $3.5^{+0.3}_{-0.3}$ & $4.4^{+1.0}_{-1.1}$ & $-0.4^{+0.2}_{-0.3}$ & $1.0^{+0.3}_{-0.0}$ & $310.0^{+70.0}_{-60.0}$ & 1.9\\
PANO-GN-BD-15 & LOWZ & $513.0^{+24.0}_{-10.0}$ & $3.6^{+0.2}_{-0.1}$ & $5.3^{+1.0}_{-1.2}$ & $-0.6^{+0.1}_{-0.1}$ & $0.8^{+0.1}_{-0.0}$ & $290.0^{+40.0}_{-20.0}$ & 15.83\\
 & ATMO2020 & $614.0^{+6.0}_{-6.0}$ & $5.5^{+0.1}_{-0.1}$ & -  & $0.3^{+0.1}_{-0.1}$ & -  & $460.0^{+10.0}_{-20.0}$ & 29.75\\
\hline
 & Sonora Elf Owl & $375.0^{+80.0}_{-62.0}$ & $4.1^{+0.9}_{-0.8}$ & $4.3^{+2.4}_{-1.6}$ & $-0.5^{+0.6}_{-0.4}$ & $1.5^{+0.7}_{-1.0}$ & $960.0^{+450.0}_{-310.0}$ & 1.58\\
JADES-GN-BD-16 & LOWZ & $526.0^{+38.0}_{-20.0}$ & $4.1^{+0.7}_{-0.5}$ & $1.2^{+1.7}_{-1.5}$ & $-0.5^{+0.4}_{-0.4}$ & $0.7^{+0.1}_{-0.0}$ & $1740.0^{+230.0}_{-190.0}$ & 2.43\\
 & ATMO2020 & $388.0^{+60.0}_{-55.0}$ & $4.5^{+0.7}_{-1.0}$ & -  & $-0.4^{+0.4}_{-0.4}$ & -  & $870.0^{+340.0}_{-260.0}$ & 0.99\\
\enddata
\end{deluxetable*}

\begin{figure*}
  \centering
  \includegraphics[width=0.32\textwidth]{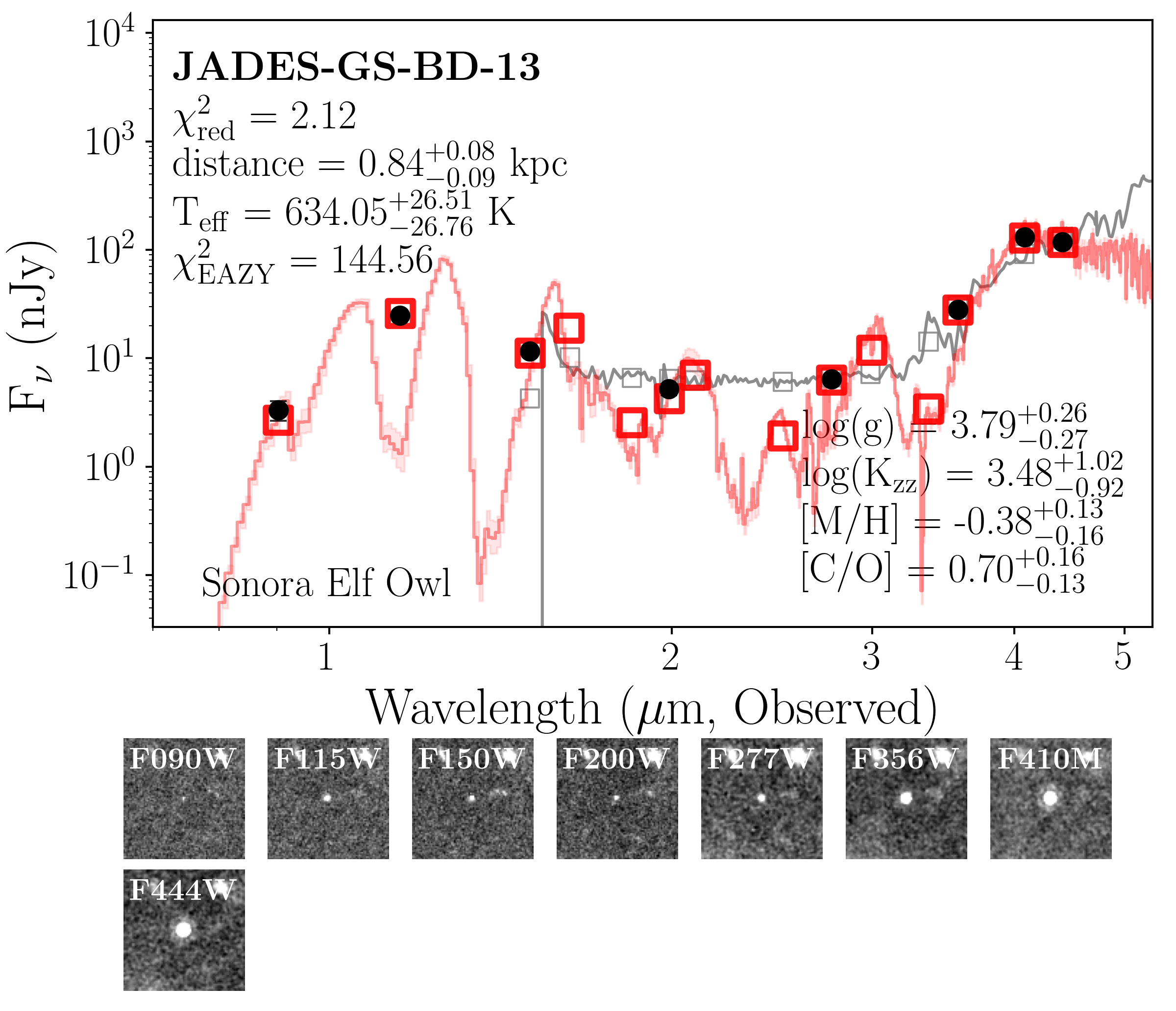}
  \includegraphics[width=0.32\textwidth]{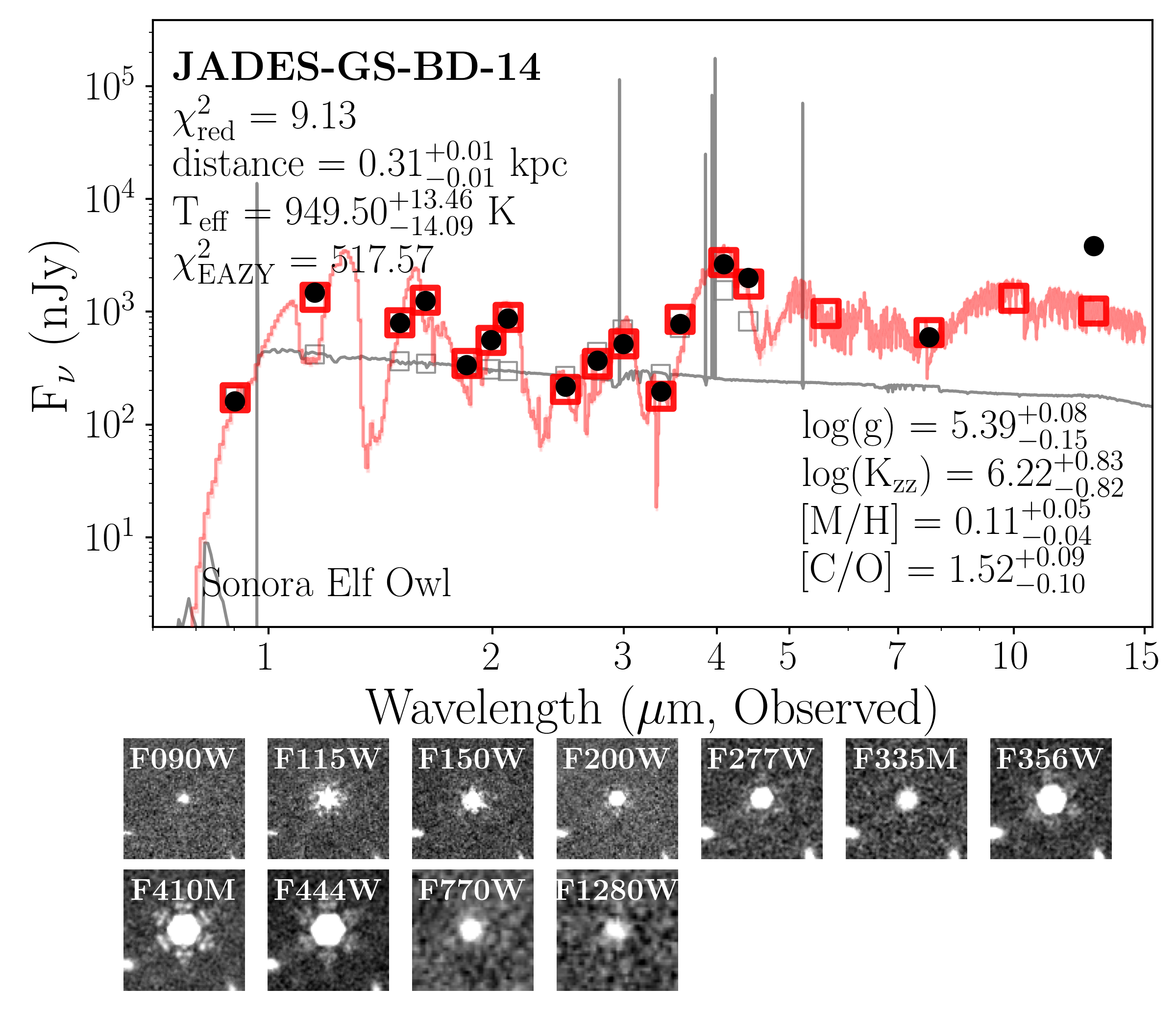}
  \includegraphics[width=0.32\textwidth]{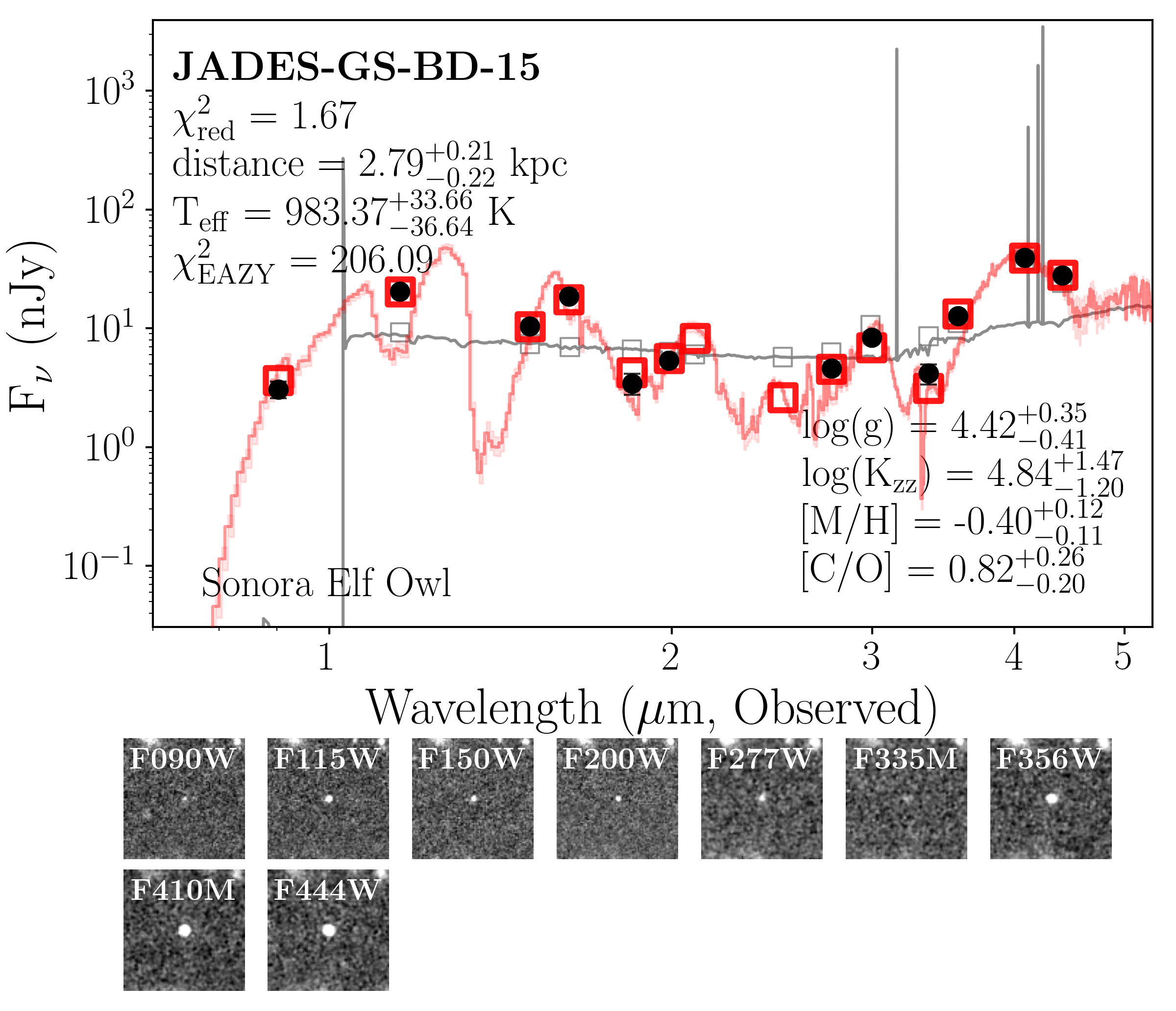}
  \includegraphics[width=0.32\textwidth]{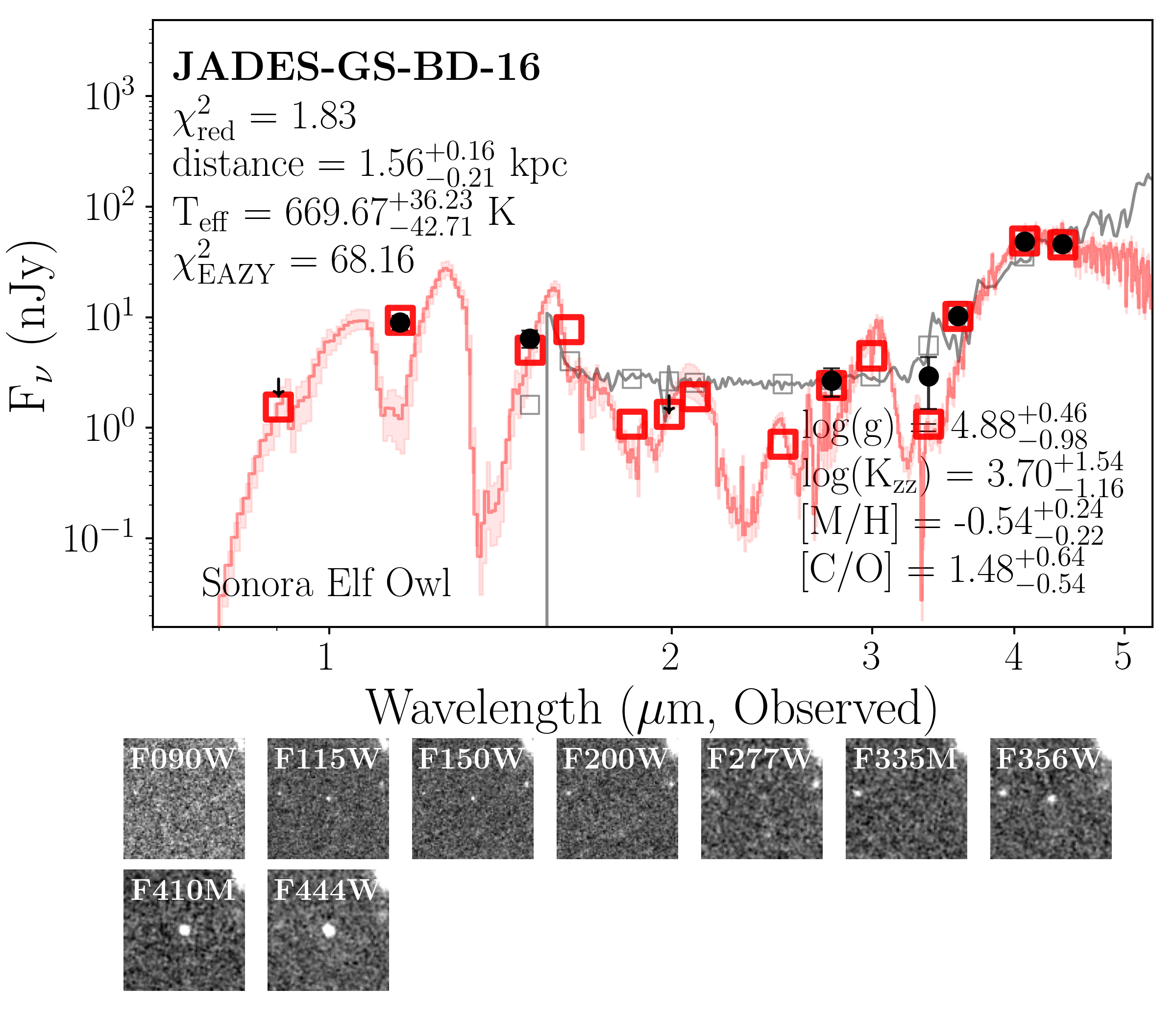}
  \includegraphics[width=0.32\textwidth]{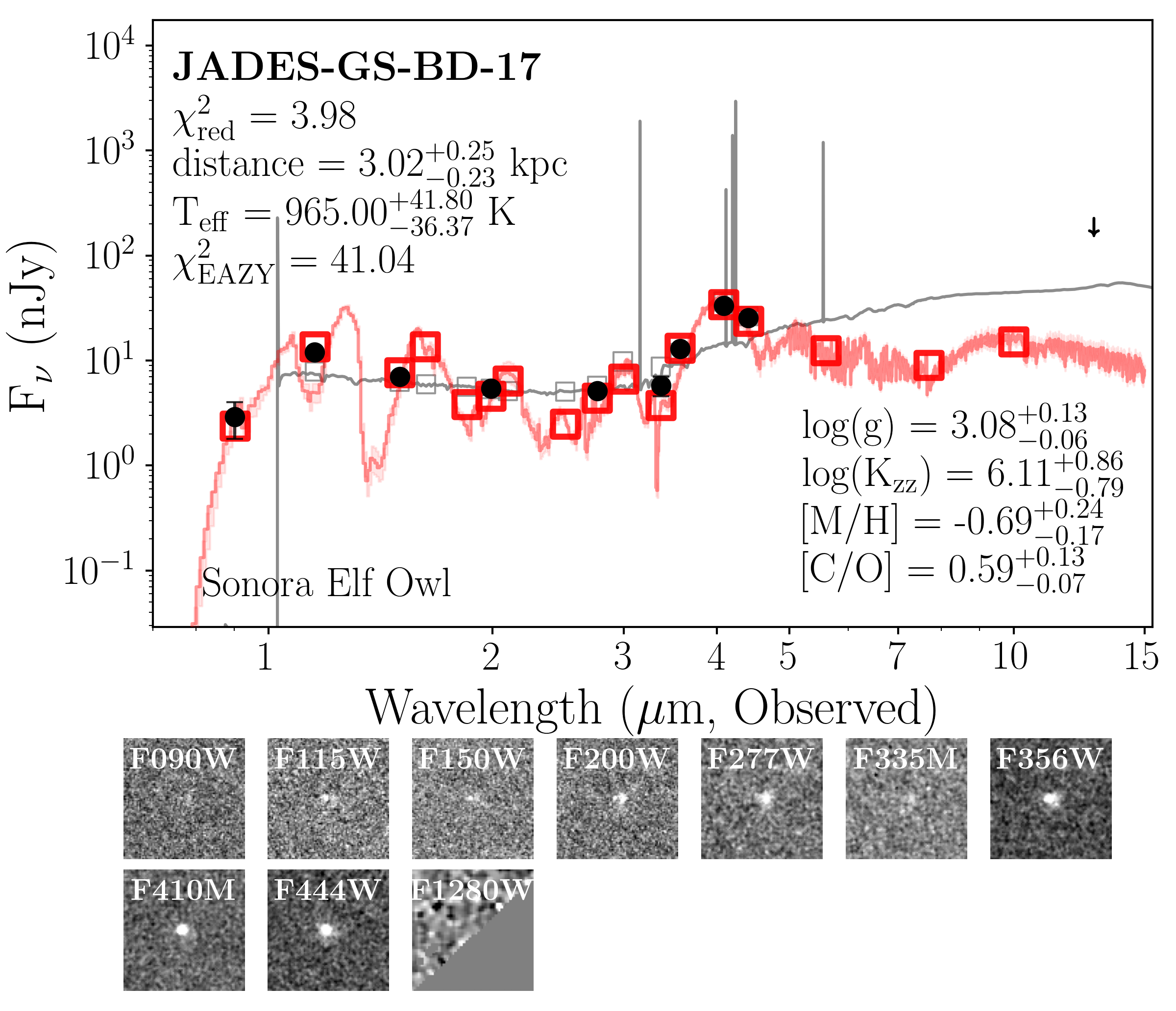}
  \includegraphics[width=0.32\textwidth]{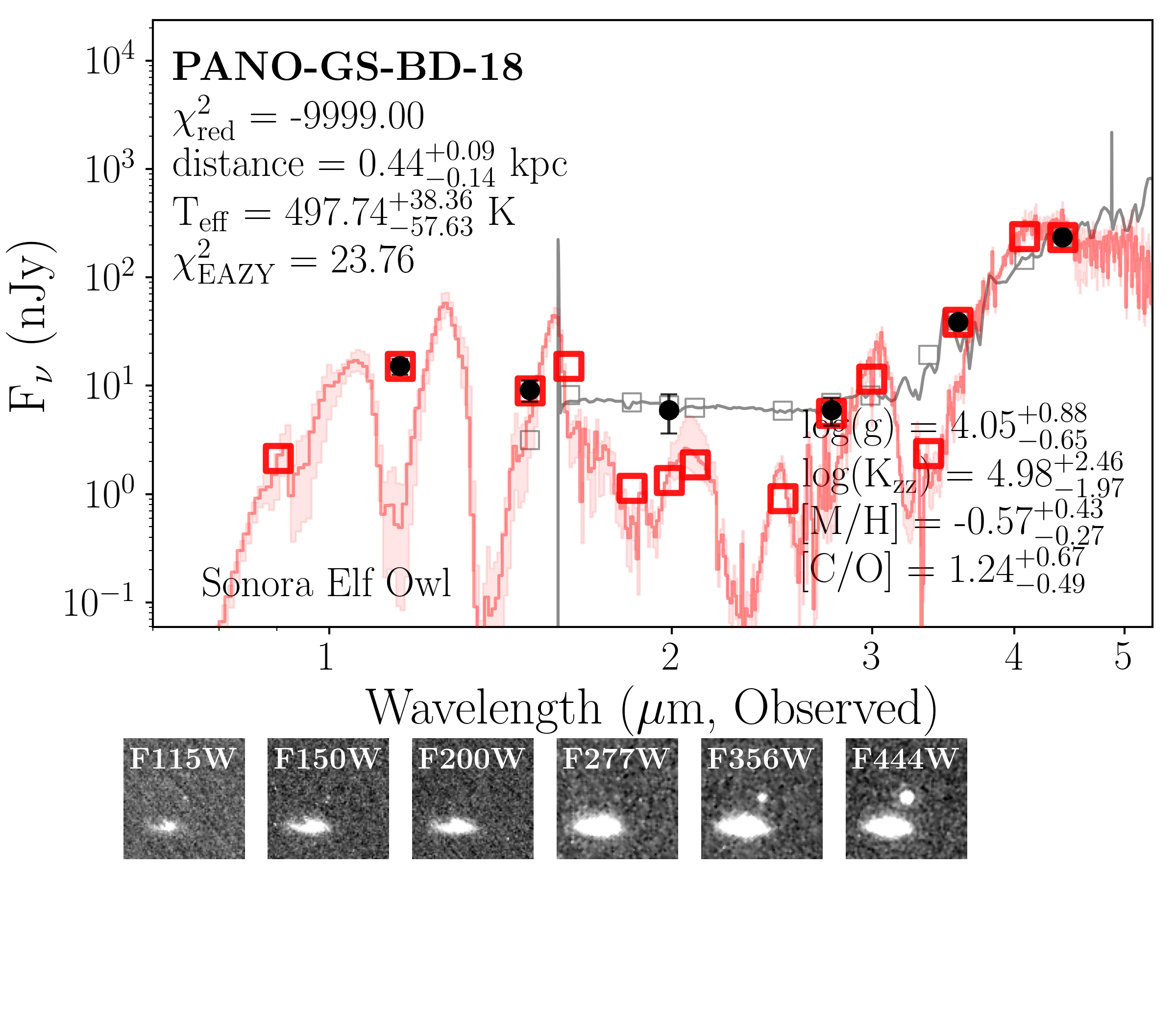}
  \includegraphics[width=0.32\textwidth]{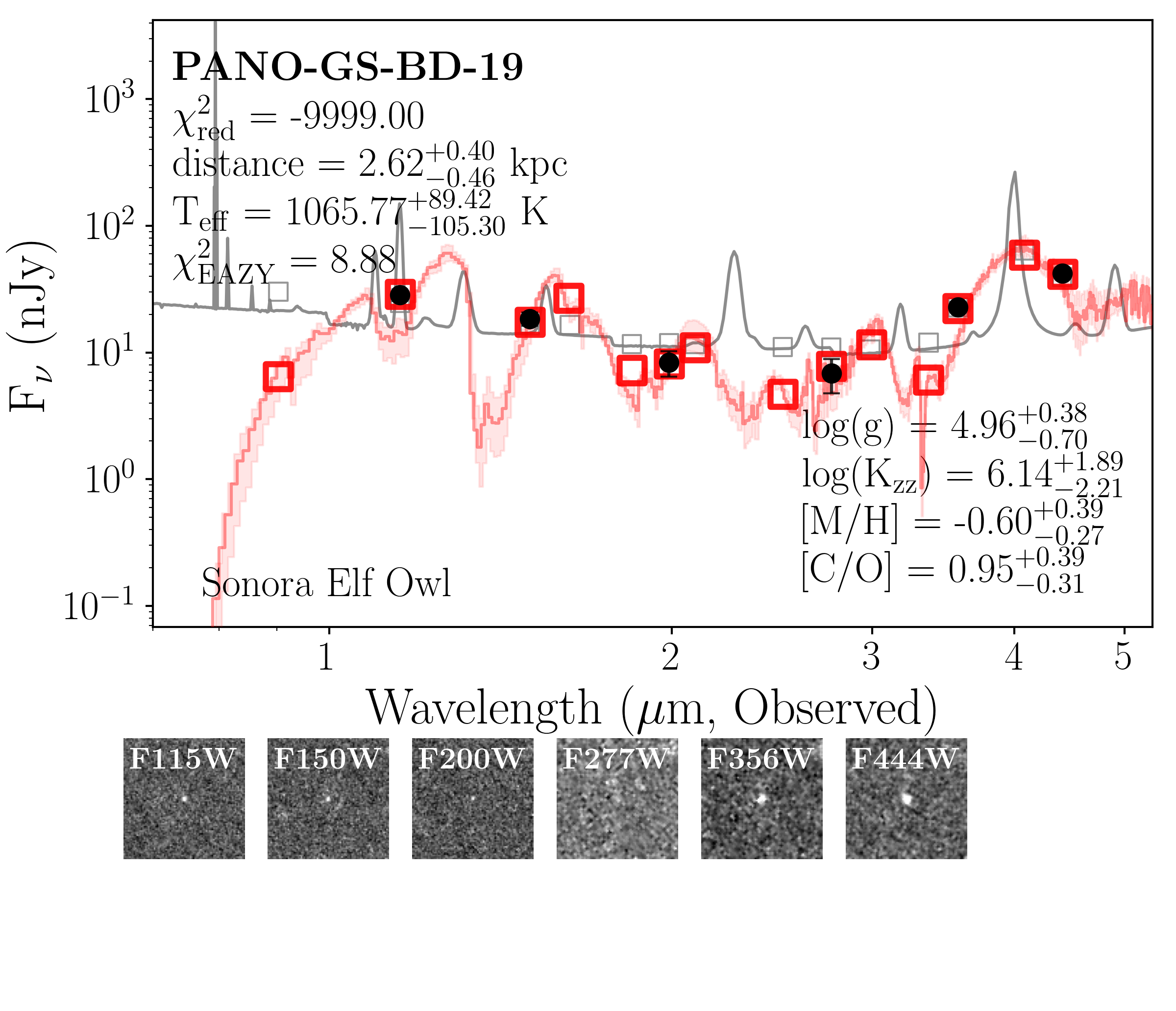}
  \includegraphics[width=0.32\textwidth]{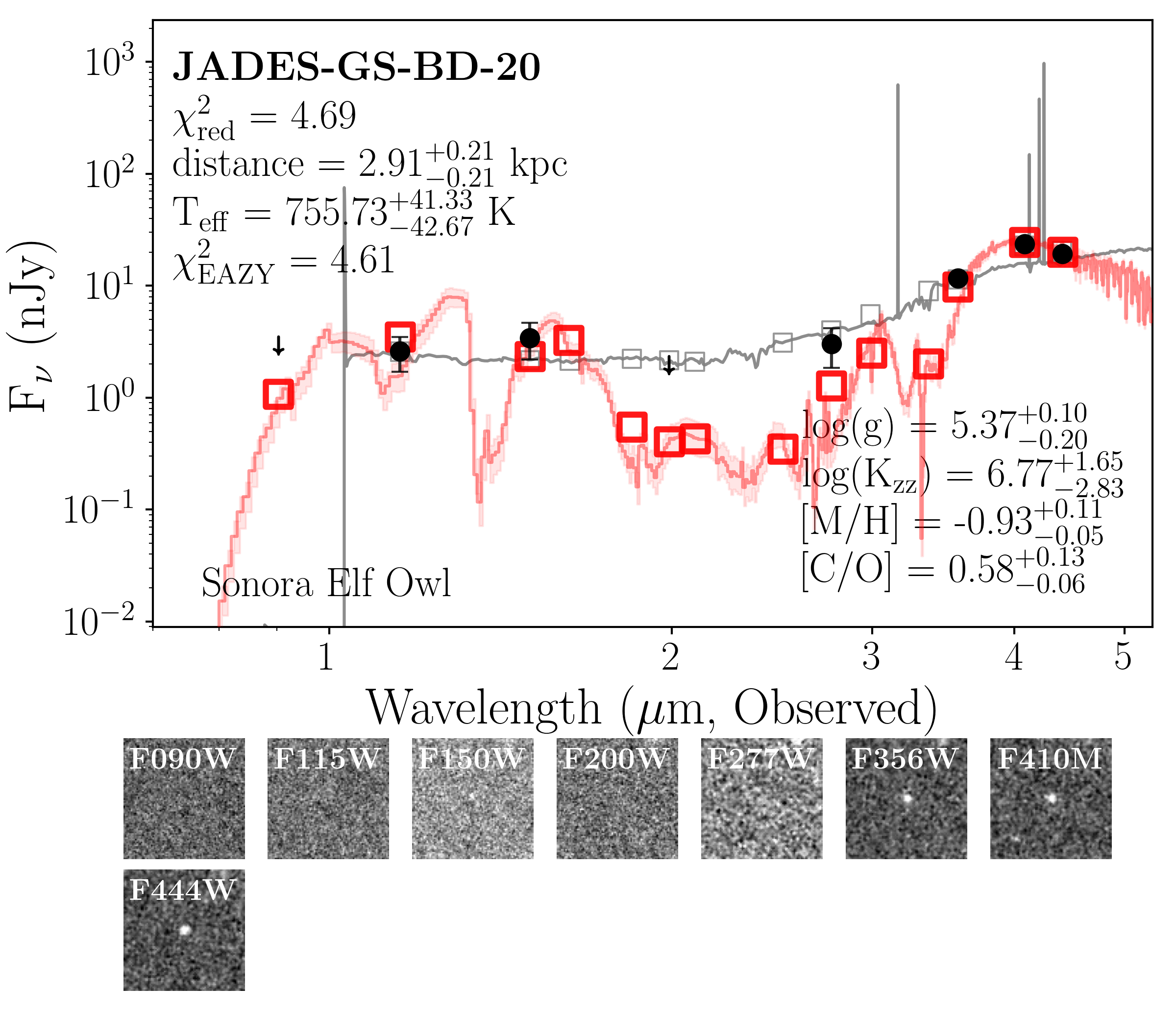}
  \includegraphics[width=0.32\textwidth]{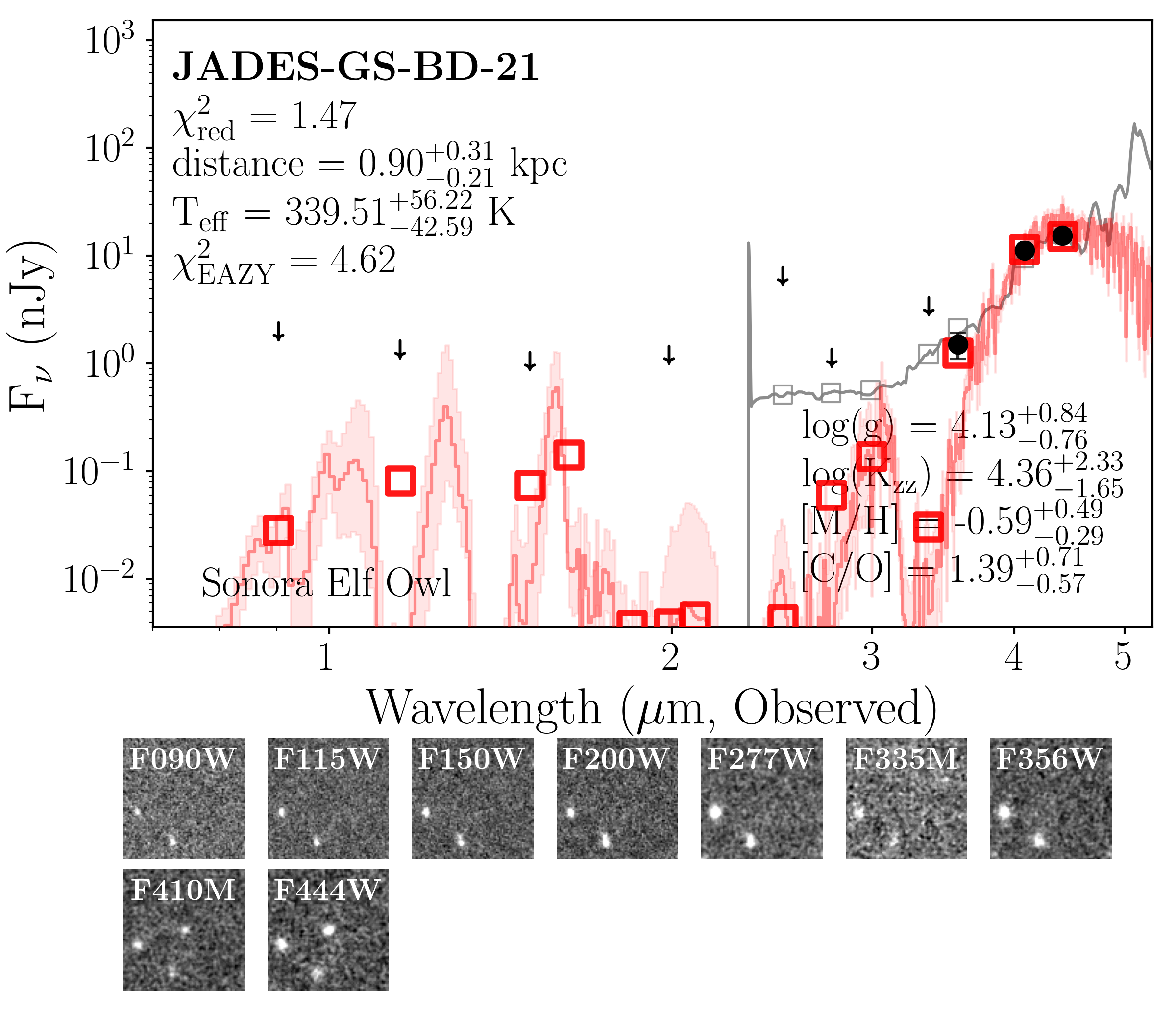}
  \includegraphics[width=0.32\textwidth]{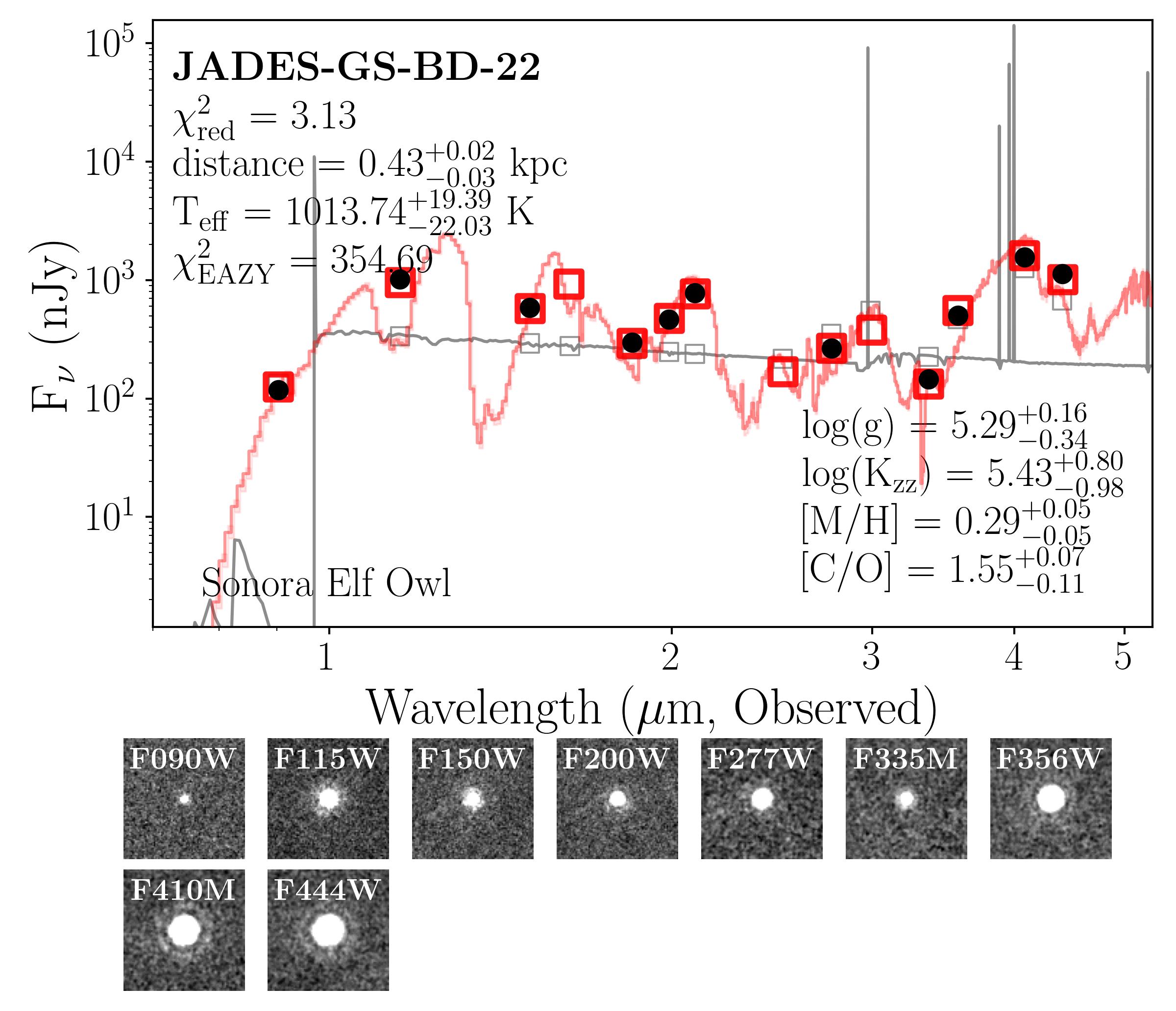}
  \includegraphics[width=0.32\textwidth]{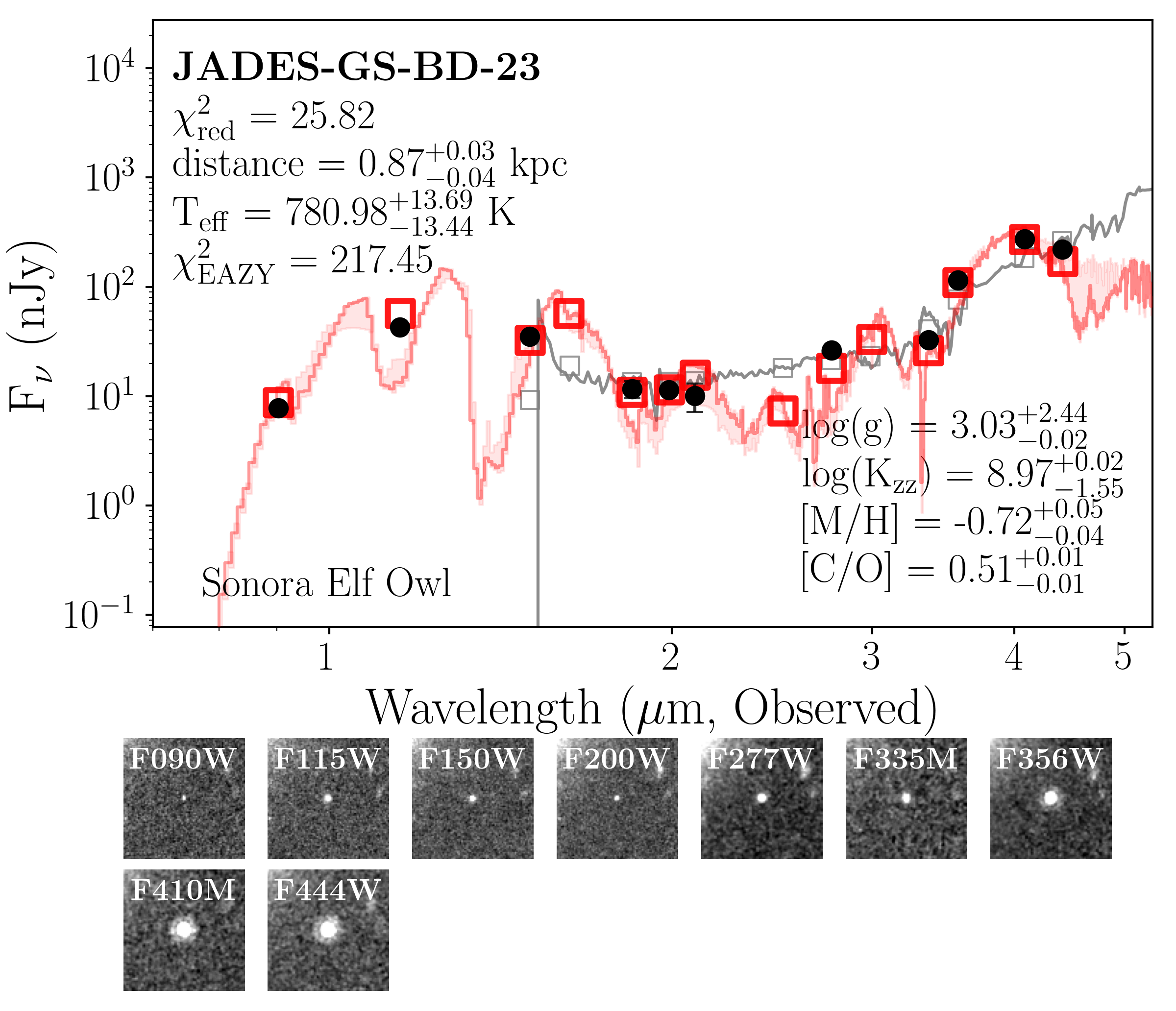}
  \includegraphics[width=0.32\textwidth]{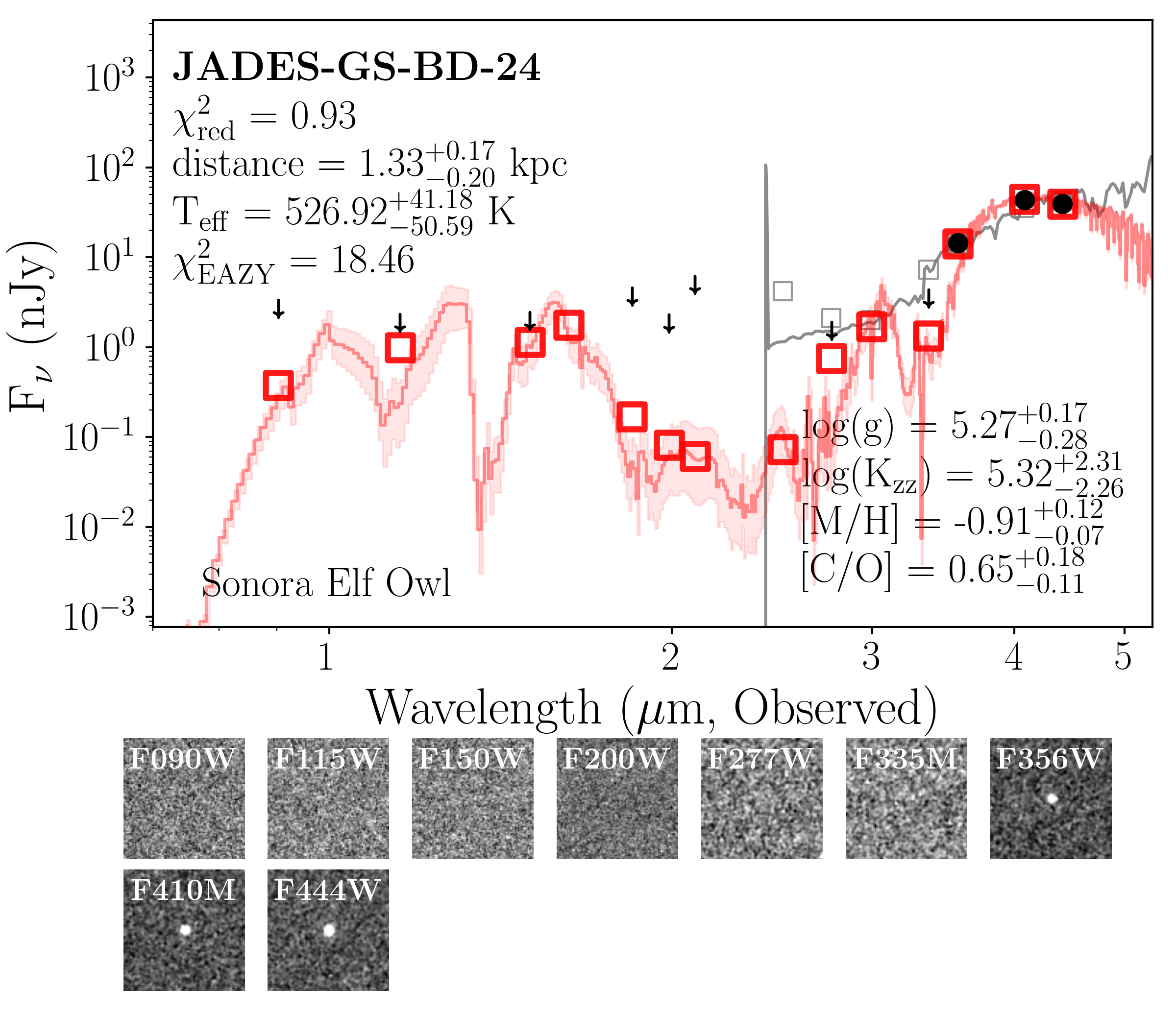}
  \caption{Continued from Figure \ref{fig:SED_plots_pt1}.
  \label{fig:SED_plots_pt2}}
\end{figure*}

\begin{figure*}
  \centering
  \includegraphics[width=0.32\textwidth]{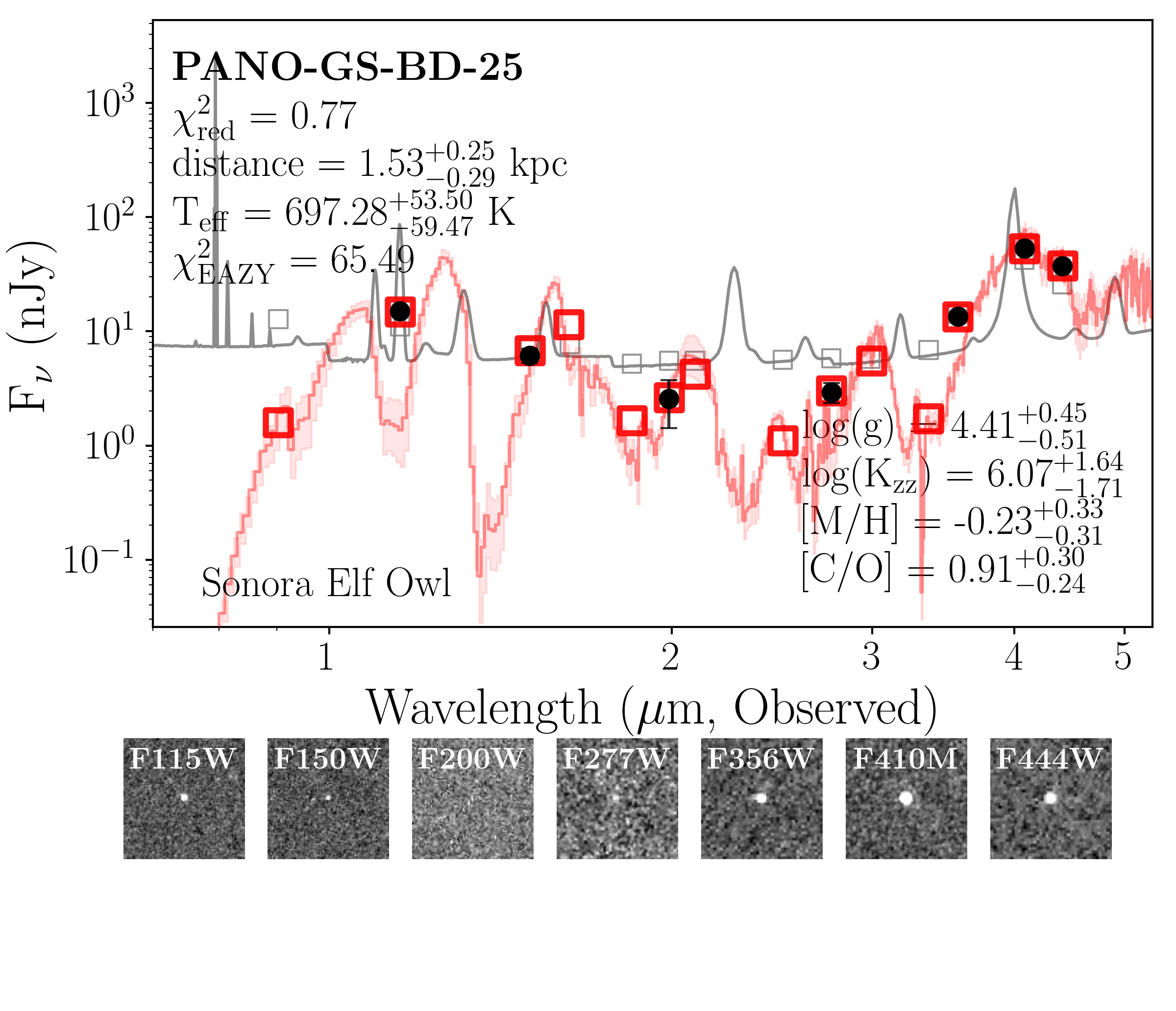}
  \includegraphics[width=0.32\textwidth]{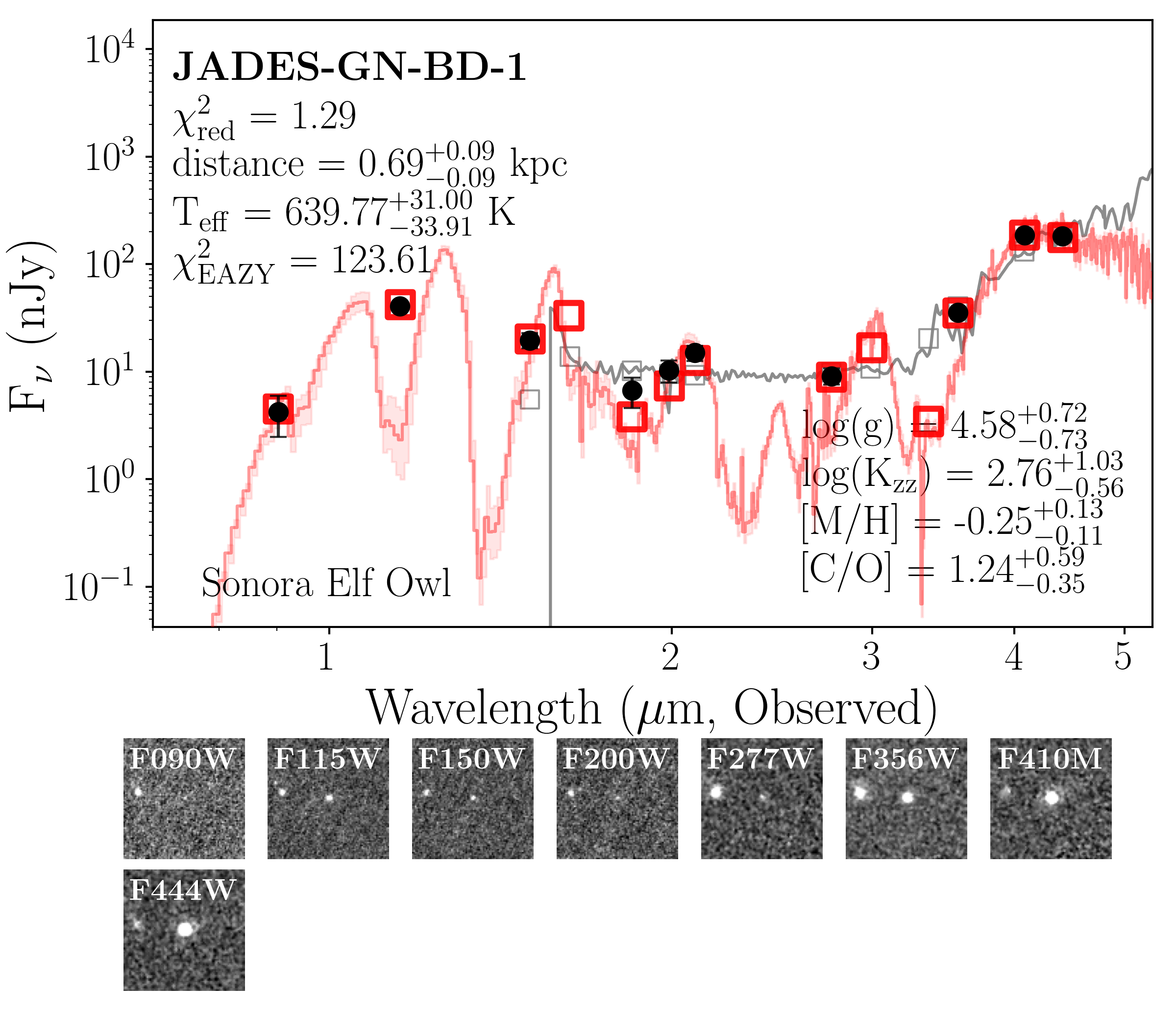}
  \includegraphics[width=0.32\textwidth]{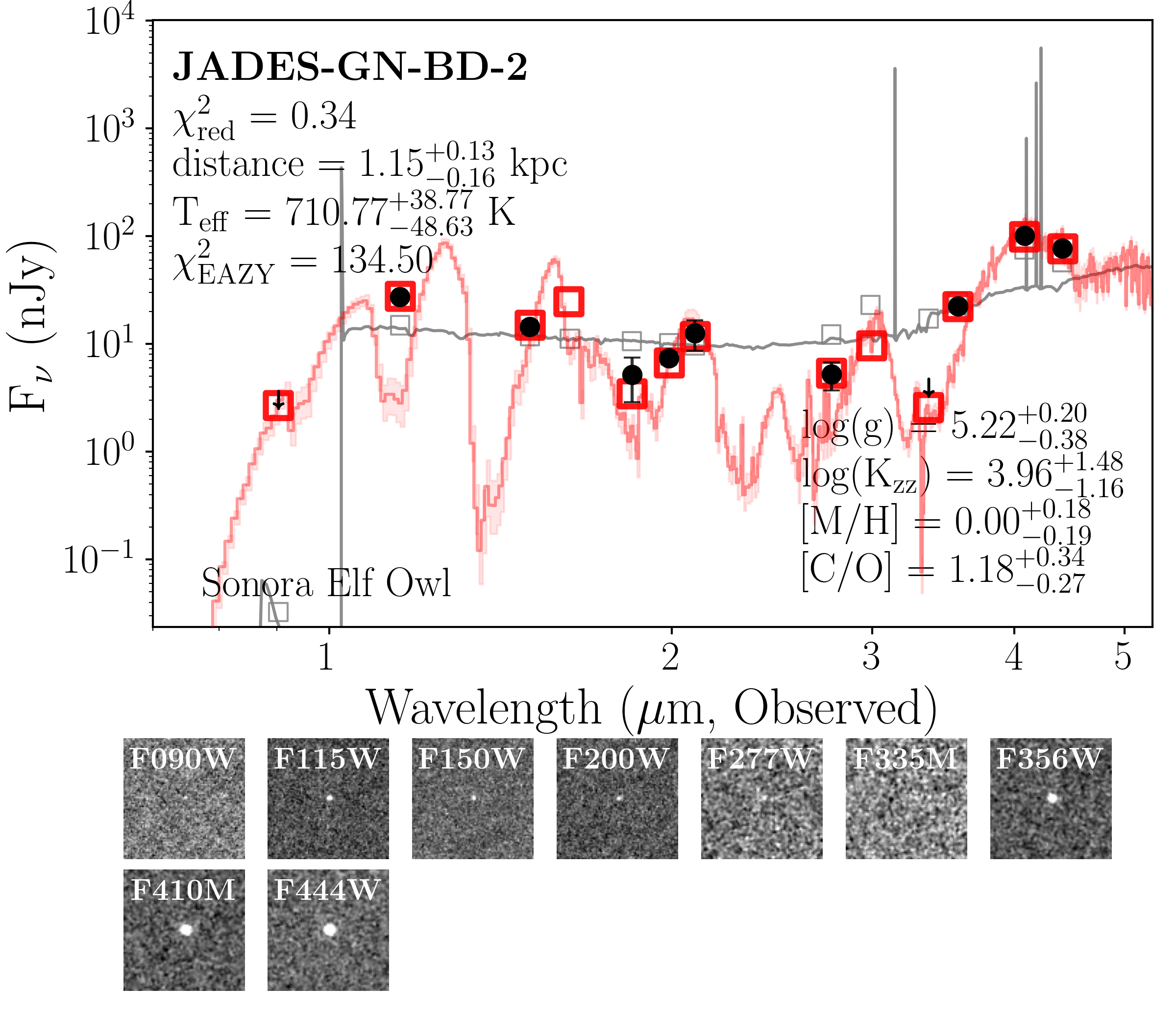}
  \includegraphics[width=0.32\textwidth]{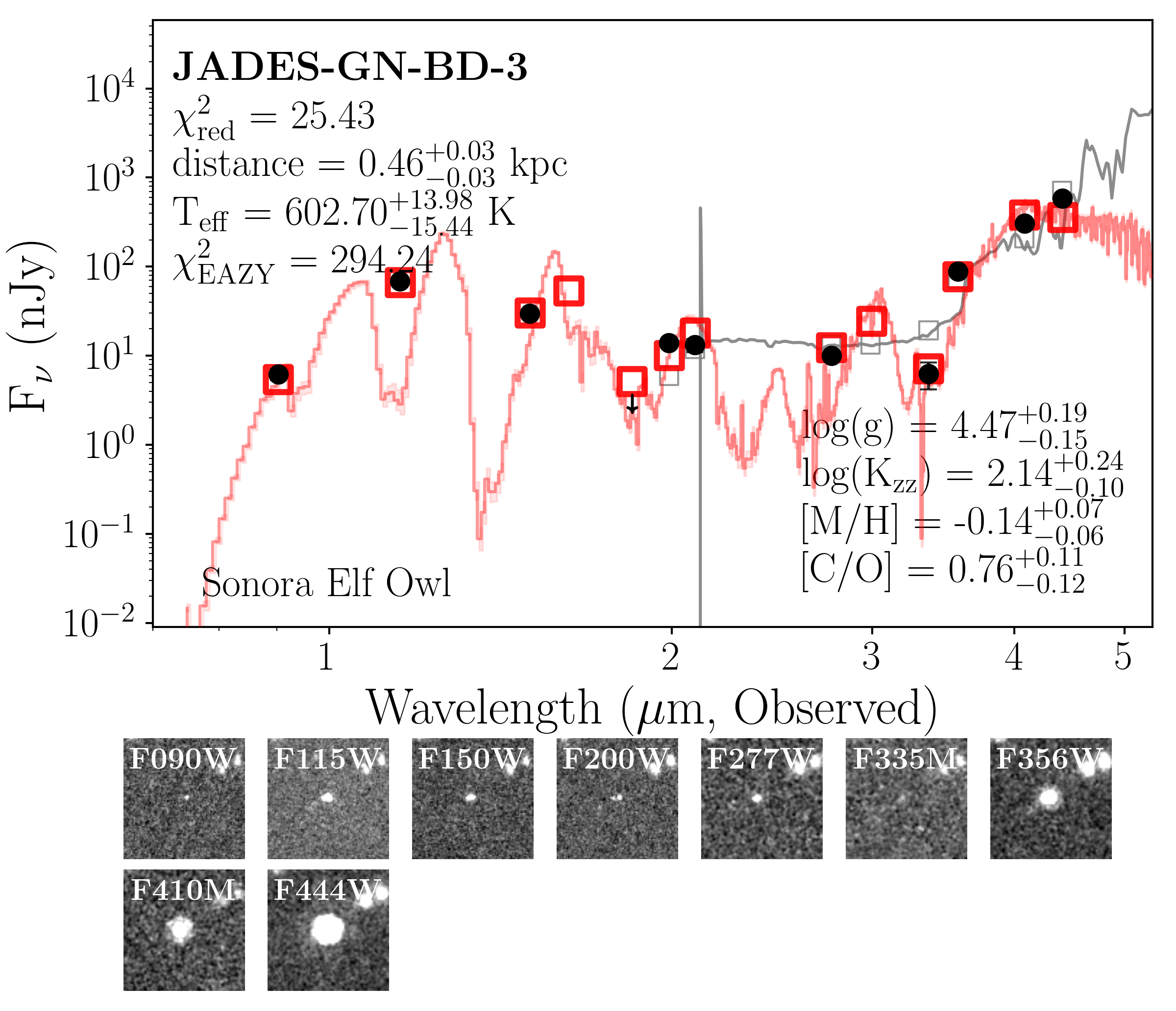}
  \includegraphics[width=0.32\textwidth]{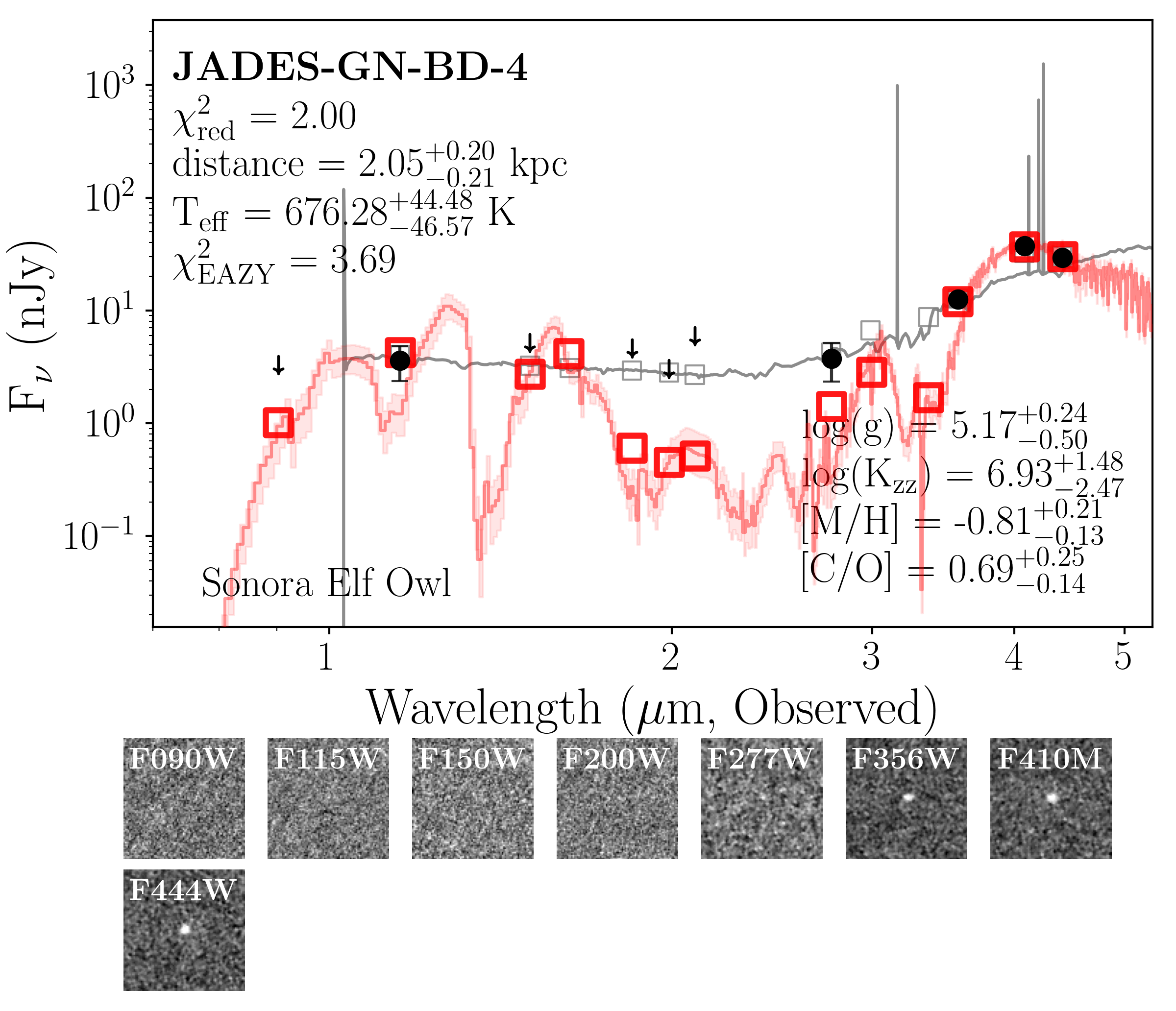}
  \includegraphics[width=0.32\textwidth]{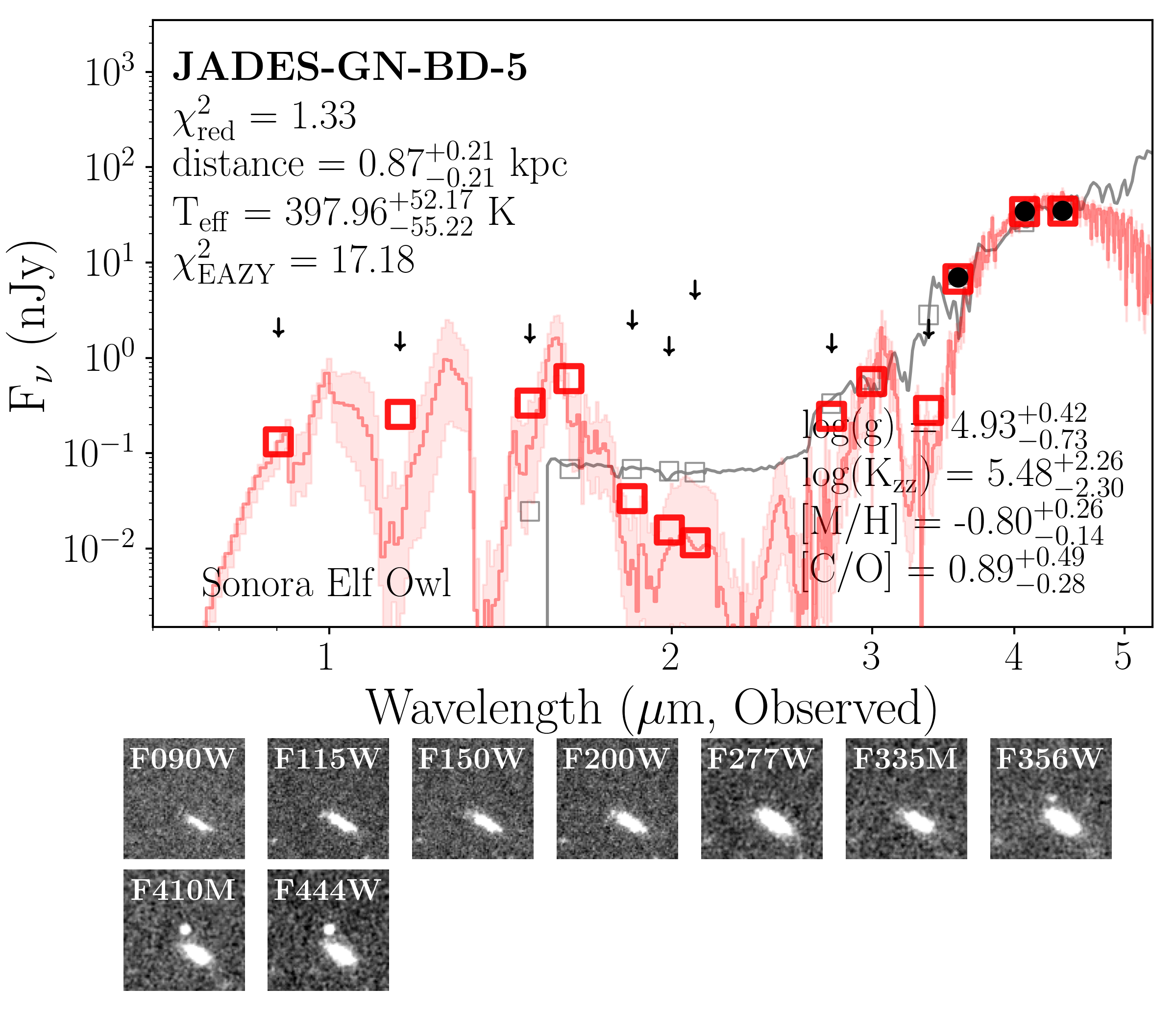}
  \includegraphics[width=0.32\textwidth]{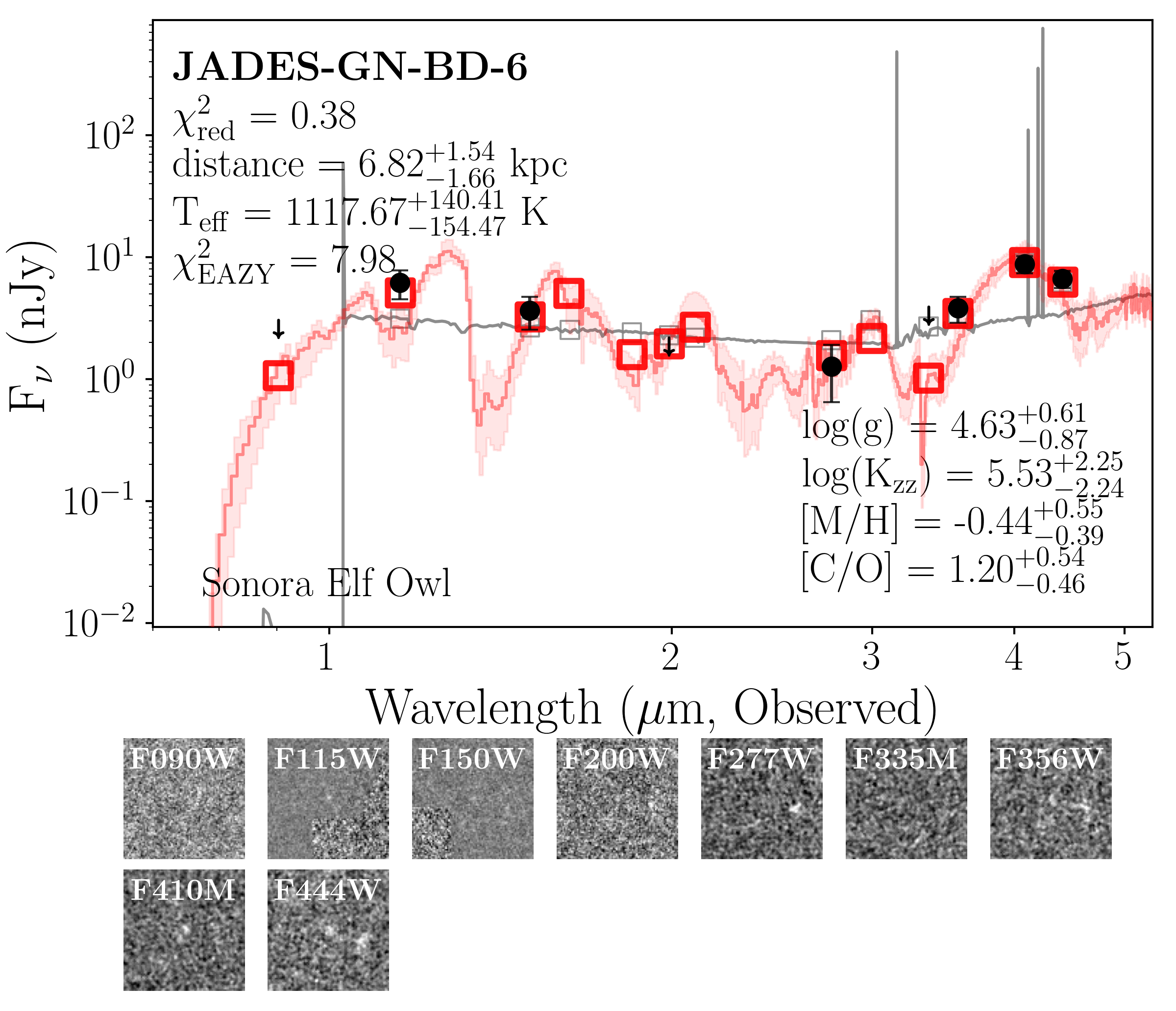}
  \includegraphics[width=0.32\textwidth]{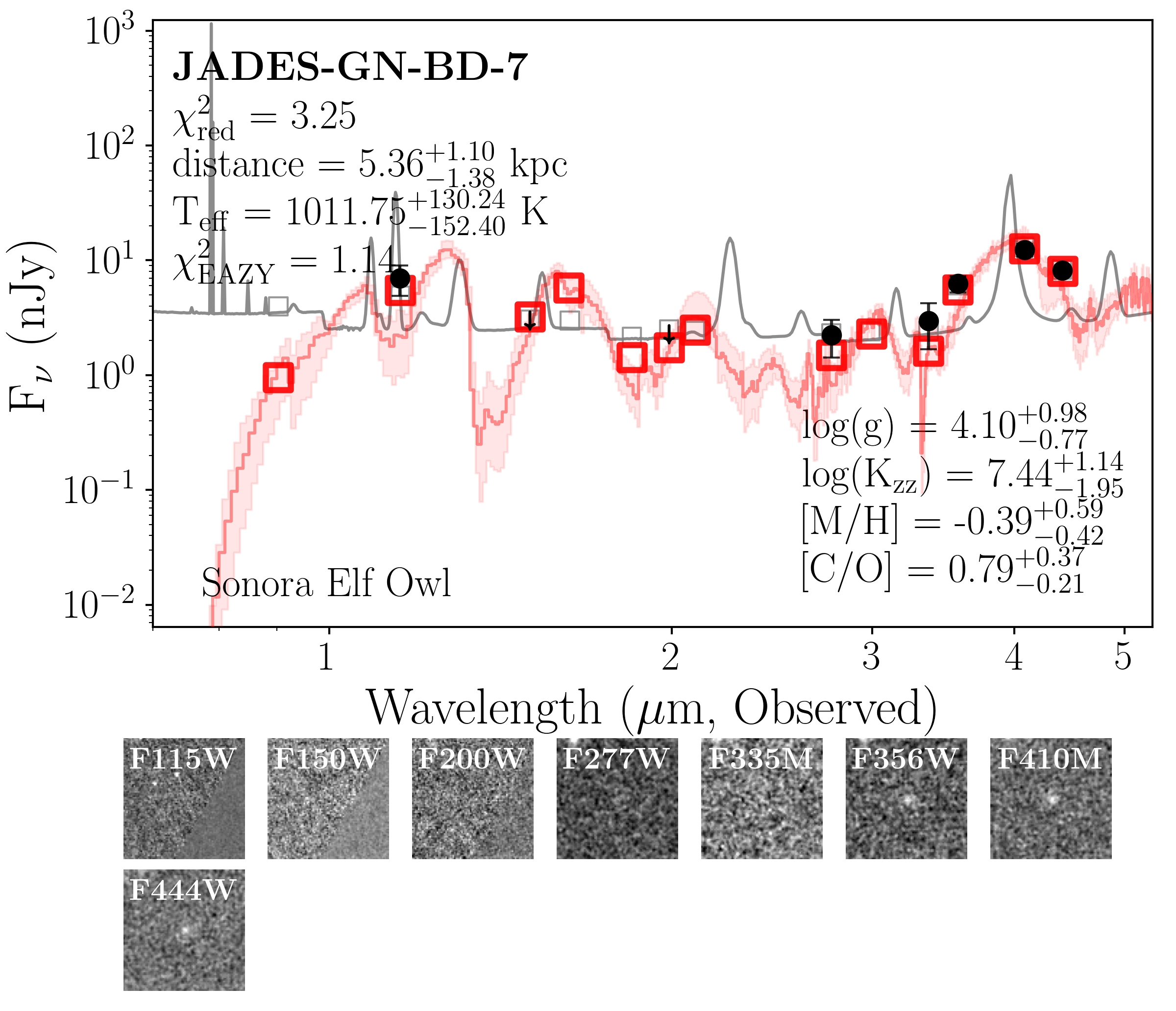}
  \includegraphics[width=0.32\textwidth]{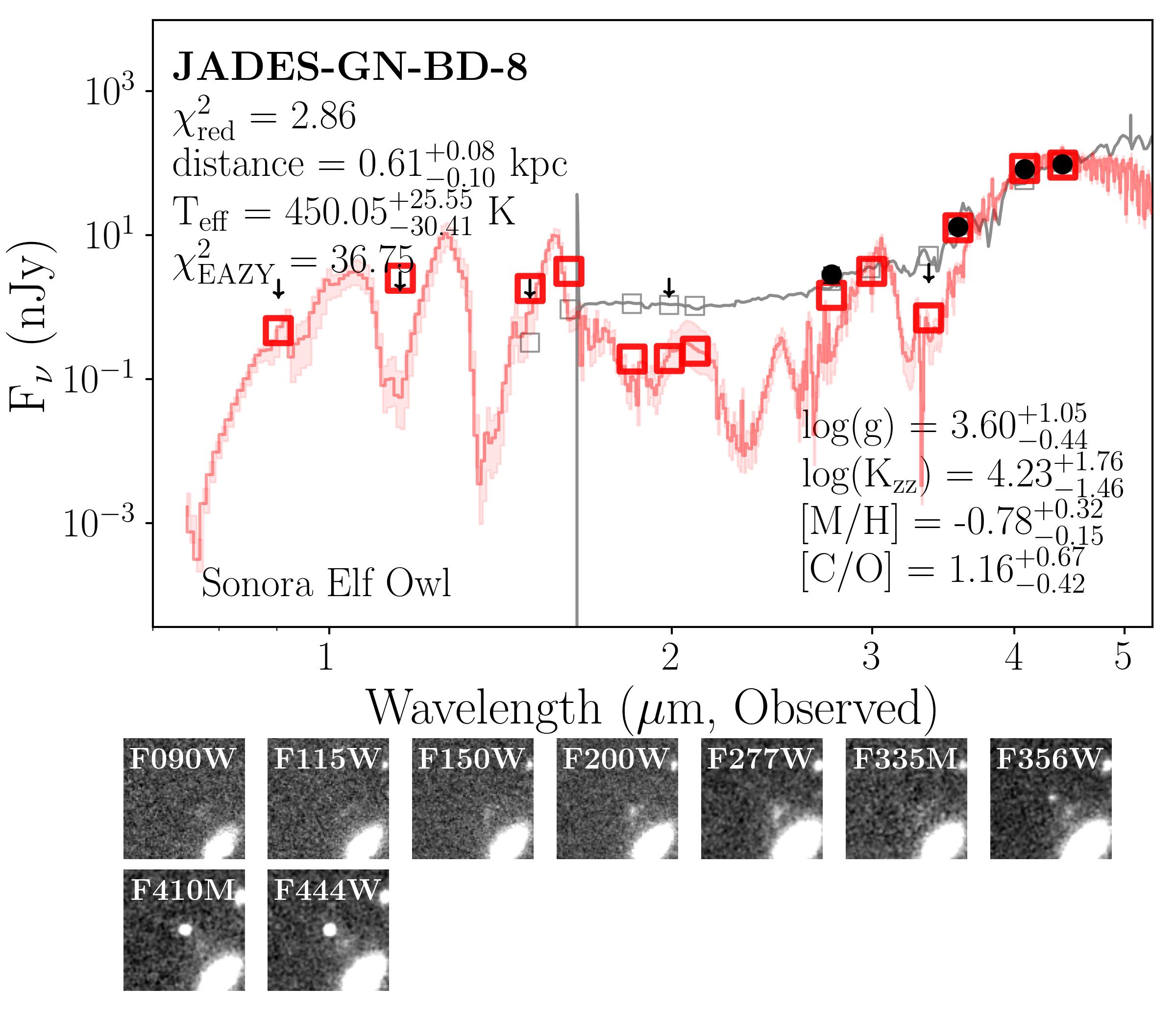}
  \includegraphics[width=0.32\textwidth]{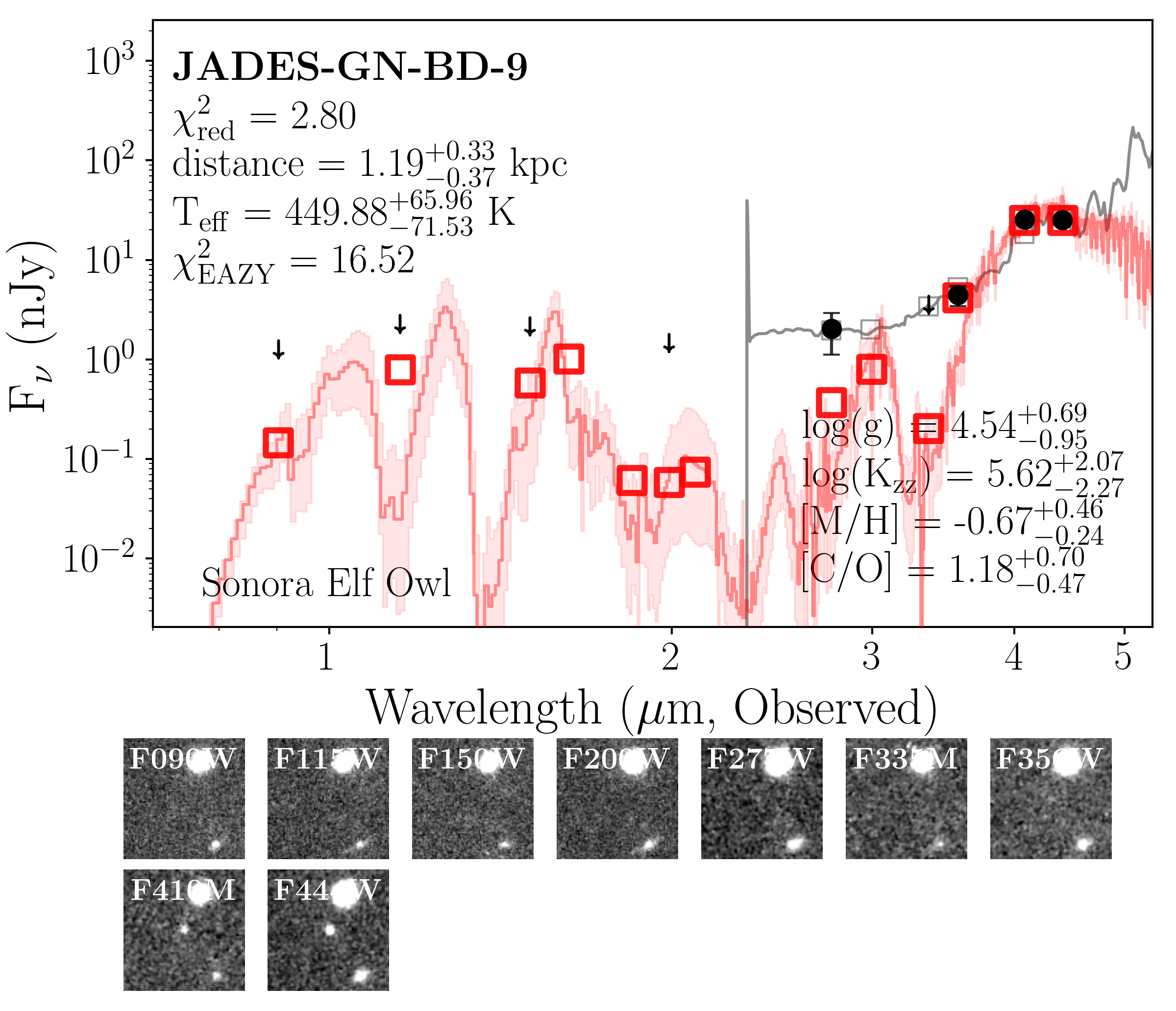}
  \includegraphics[width=0.32\textwidth]{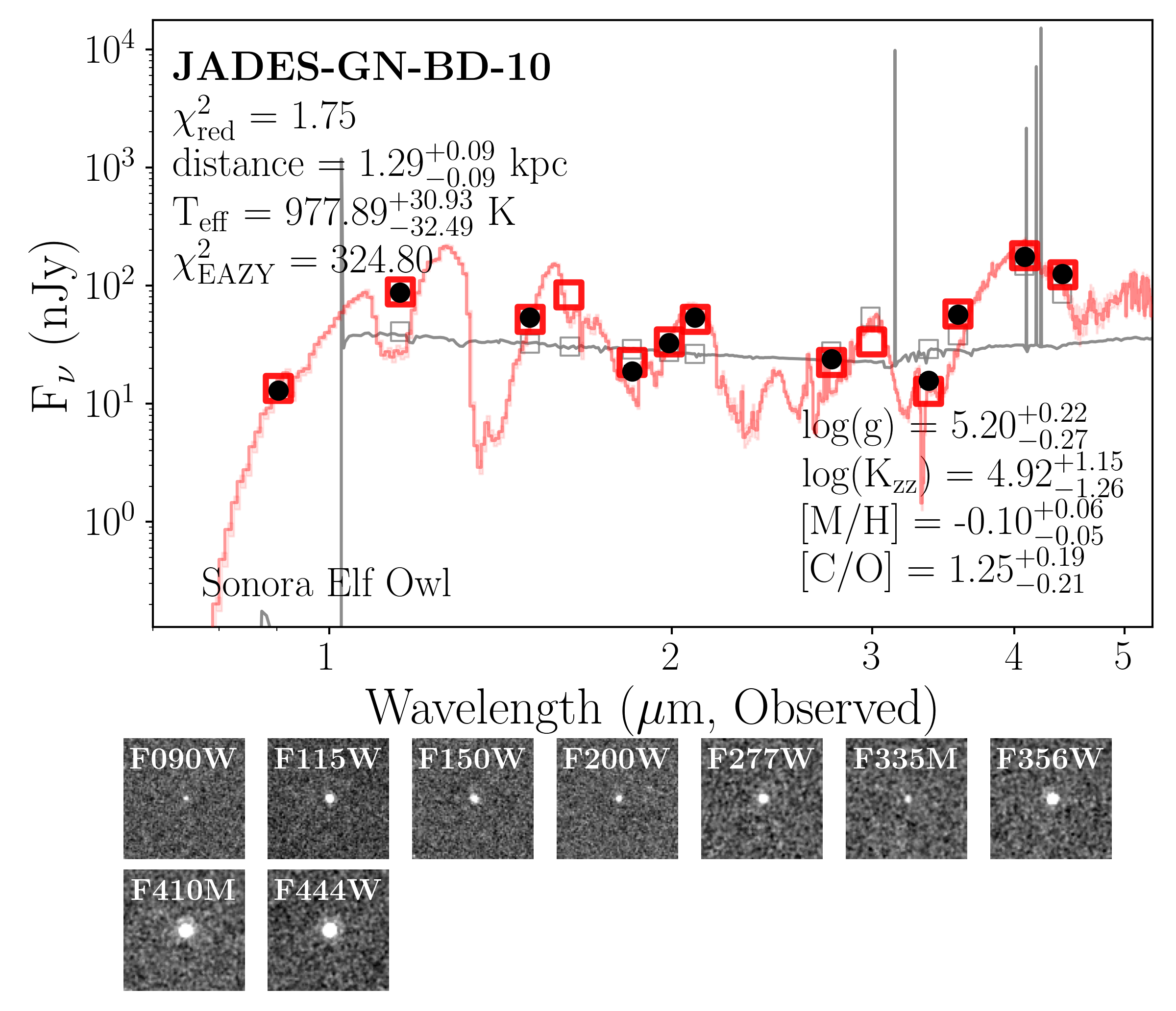}
  \includegraphics[width=0.32\textwidth]{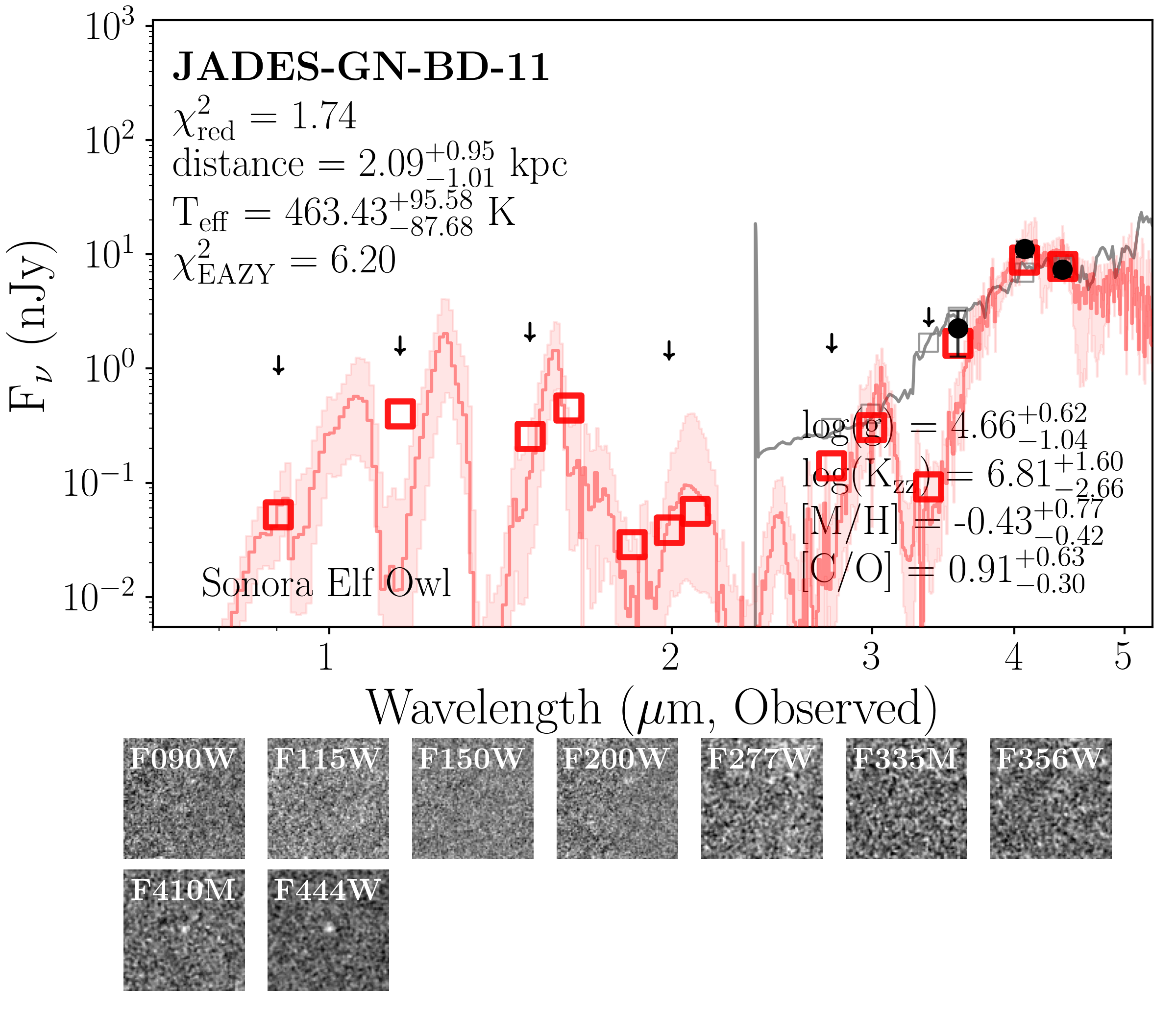}
  \caption{Continued from Figure \ref{fig:SED_plots_pt2}.
  \label{fig:SED_plots_pt3}}
\end{figure*}

\begin{figure*}
  \centering
  \includegraphics[width=0.32\textwidth]{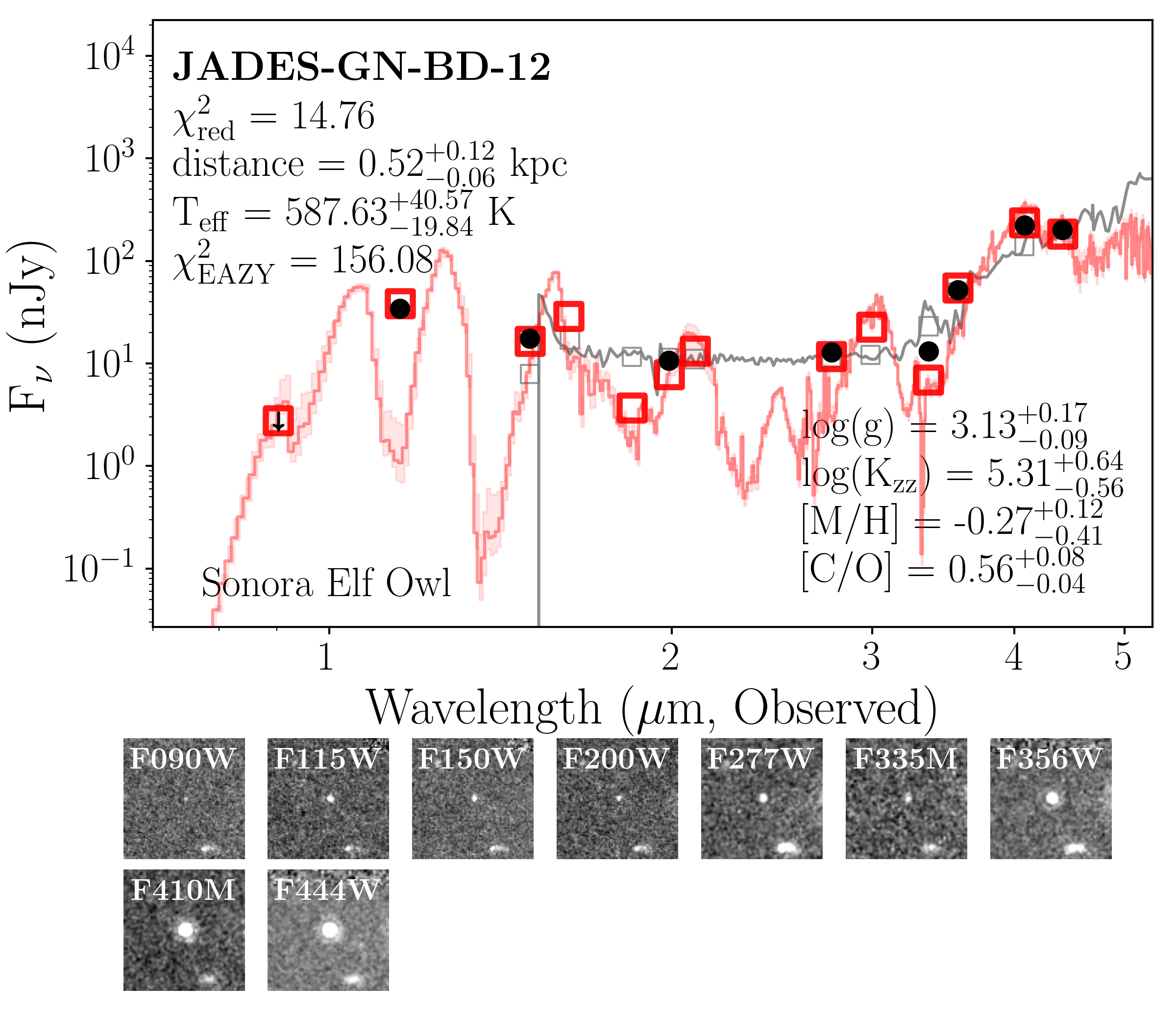}
  \includegraphics[width=0.32\textwidth]{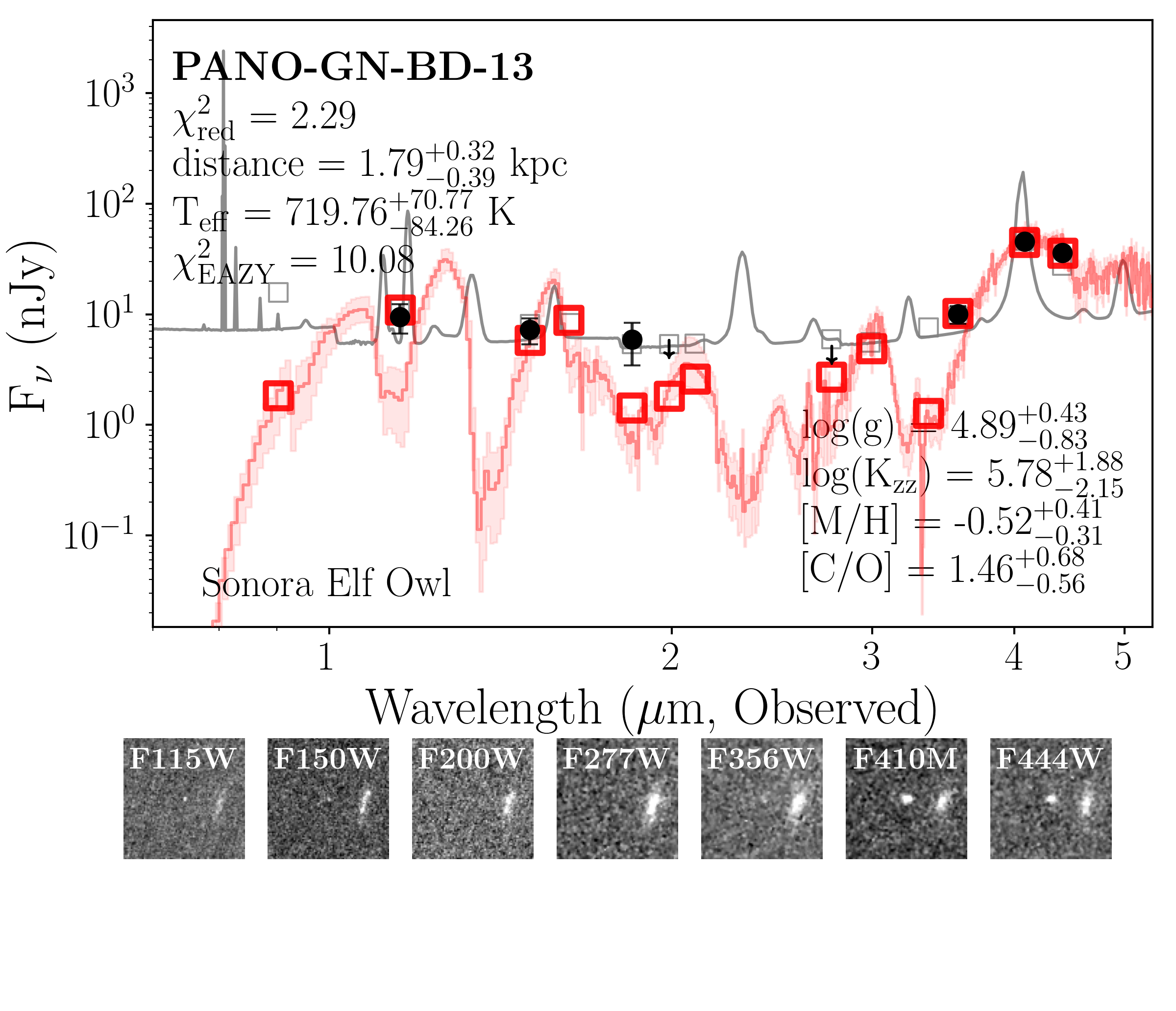}
  \includegraphics[width=0.32\textwidth]{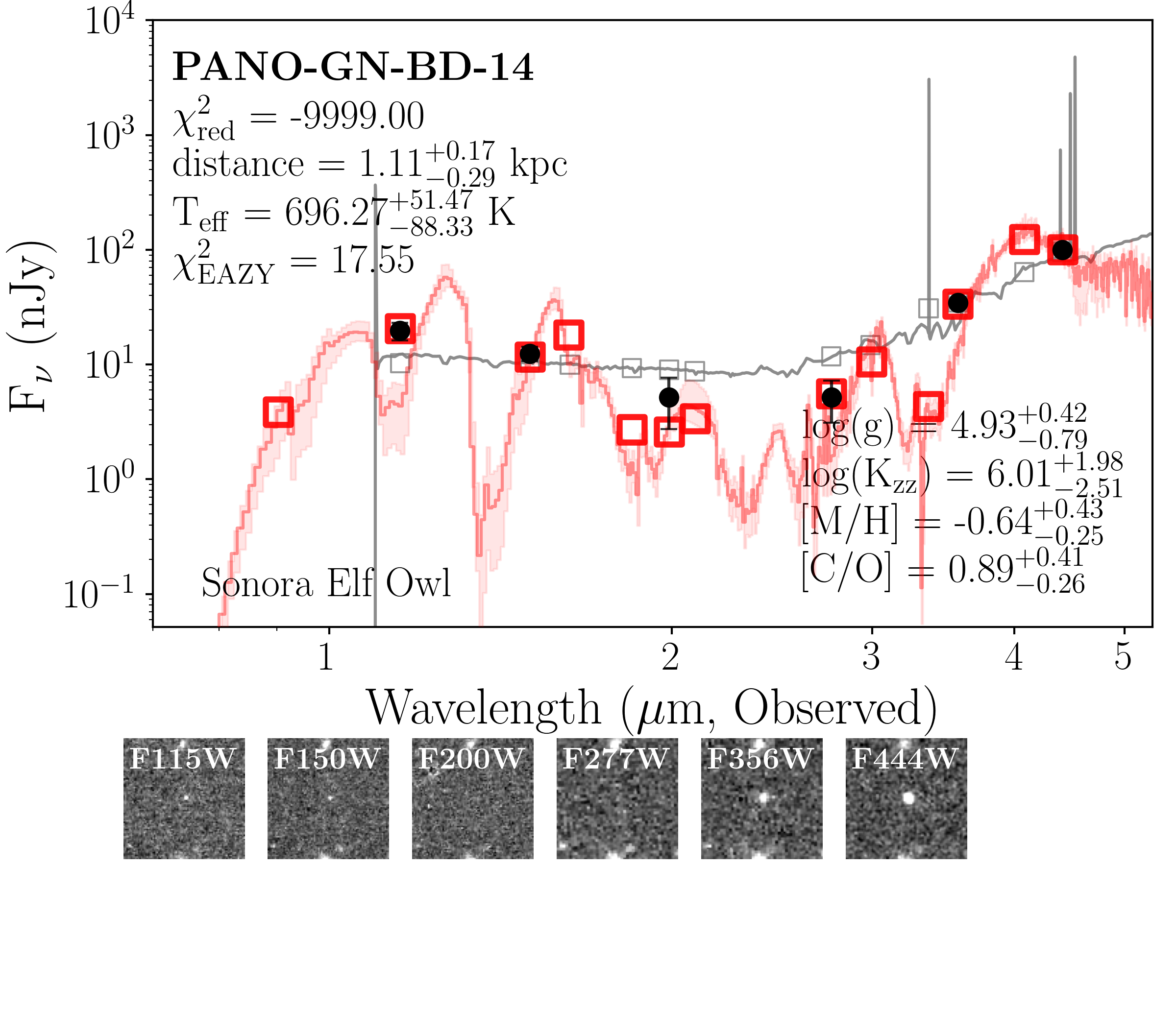}
  \includegraphics[width=0.32\textwidth]{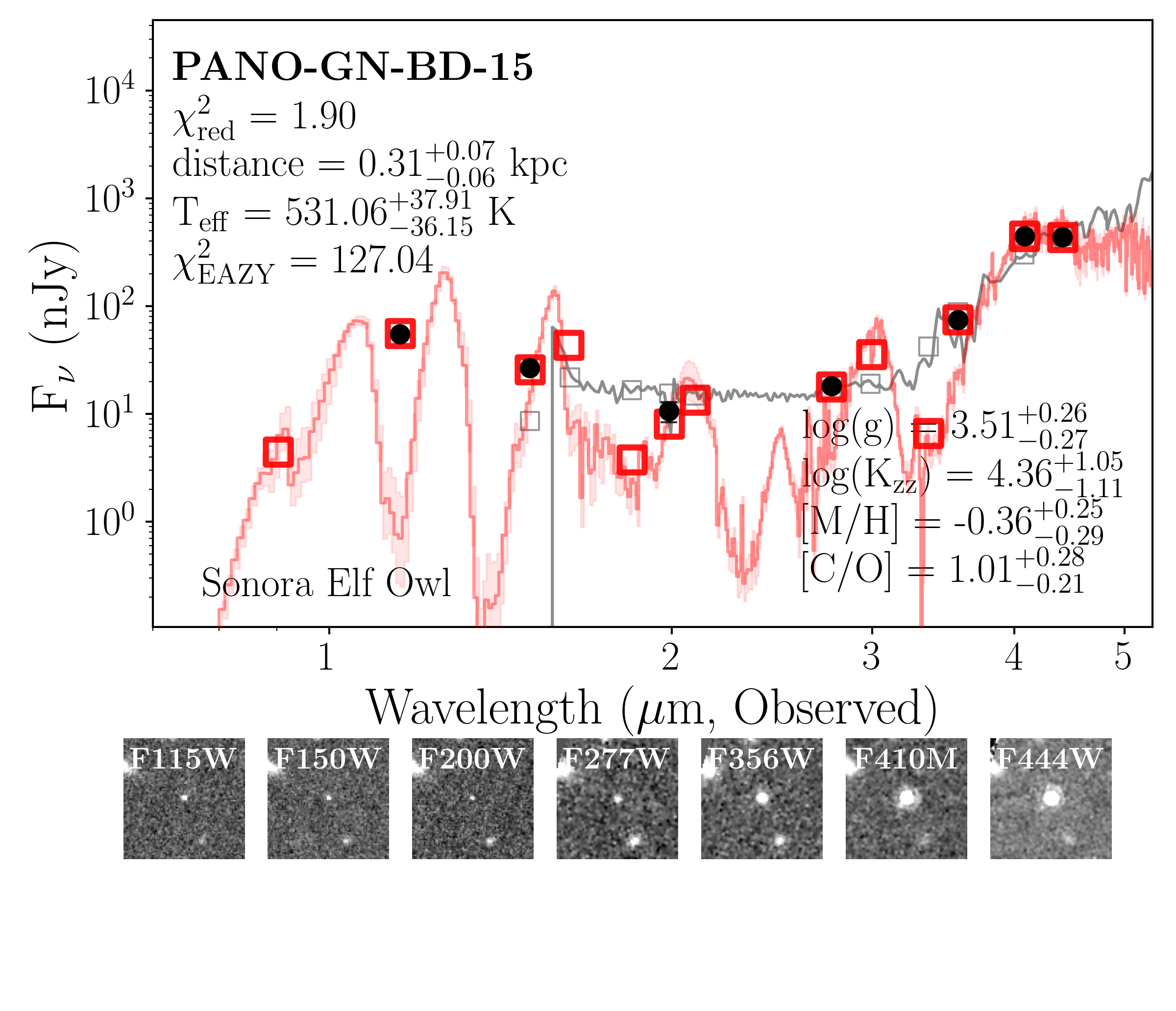}
  \includegraphics[width=0.32\textwidth]{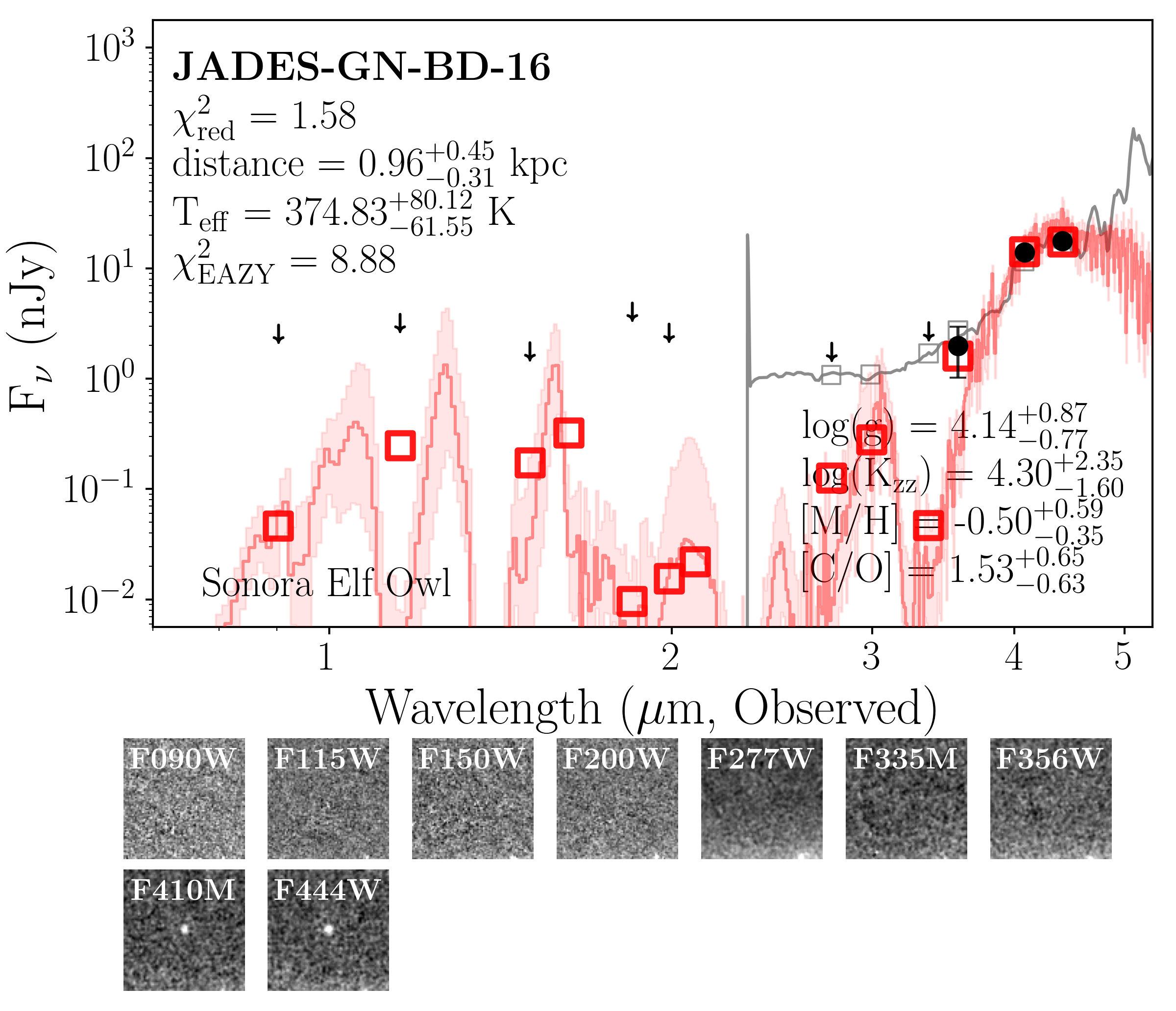}
  \caption{Continued from Figure \ref{fig:SED_plots_pt3}.
  \label{fig:SED_plots_pt4}}
\end{figure*}

\bibliography{ms}{}
\bibliographystyle{aasjournalv7}



\end{document}